%% file: brpap.tex
\documentstyle[twoside,subfigure,12pt,floatfig,psfig]{article}
\newcommand{\eff}{\varepsilon}
\newcommand{\ra}{\rightarrow}
\newcommand{\piz}{\pi^0}

\def\tr{\tau \rightarrow} 
\def\nt{\nu_{\tau}} 
\newcommand{\clse}{e}
\newcommand{\clsm}{\mu}
\newcommand{\h}{h}
\newcommand{\hu}{h\pi^0}
\newcommand{\hd}{h2\pi^0}
\newcommand{\htr}{h3\pi^0}
\newcommand{\hq}{h4\pi^0}
\newcommand{\trh}{3h}
\newcommand{\thu}{3h\pi^0}
\newcommand{\thd}{3h2\pi^0}
\newcommand{\tht}{3h3\pi^0}
\newcommand{\ch}{5h}
\newcommand{\chu}{5h\pi^0}

\newcommand{\clsi}{$e$}
\newcommand{\clsii}{$\mu$}
\newcommand{\clsiii}{$h$}
\newcommand{\clsiv}{$h\pi^0$}
\newcommand{\clsv}{$h2\pi^0$}
\newcommand{\clsvi}{$h3\pi^0$}
\newcommand{\clsvii}{$3h$}
\newcommand{\clsviii}{$3h\pi^0$}
\newcommand{\clsix}{$3h2\pi^0$}
\newcommand{\clsx}{$3h3\pi^0$}
\newcommand{\clsxi}{$5h$}
\newcommand{\clsxii}{$5h\pi^0$}
\newcommand{\clsxiii}{$h4\pi^0$}
\newcommand{\clsxiv}{Class 14 }
\newcommand{\phyi}{$e$}
\newcommand{\phyii}{$\mu$}
\newcommand{\phyiii}{$\pi^-$}
\newcommand{\phyiv}{$\pi^-\pi^0$}
\newcommand{\phyv}{$\pi^-2\pi^0$}
\newcommand{\phyvi}{$\pi^-3\pi^0$}
\newcommand{\phyvii}{$\pi^-\pi^-\pi^+$}
\newcommand{\phyviii}{$\pi^-\pi^-\pi^+\pi^0$}
\newcommand{\phyix}{$\pi^-\pi^-\pi^+2\pi^0$}
\newcommand{\phyx}{$\pi^-\pi^-\pi^+3\pi^0$}
\newcommand{\phyxi}{$3\pi^-2\pi^+$}
\newcommand{\phyxii}{$3\pi^-2\pi^+\pi^0$}
\newcommand{\phyxiii}{$\pi^-4\pi^0$}

\newcommand{\bfg}{\begin{figure}}
\newcommand{\efg}{\end{figure}}
\newcommand{\bitm}{\begin{itemize}}
\newcommand{\eitm}{\end{itemize}}
\newcommand{\bnum}{\begin{enumerate}}
\newcommand{\enum}{\end{enumerate}}
\newcommand{\btbl}{\begin{table}}
\newcommand{\etbl}{\end{table}}
\newcommand{\btbu}{\begin{tabular}}
\newcommand{\etbu}{\end{tabular}}
\newcommand{\beq}{\begin{equation}}
\newcommand{\eeq}{\end{equation}}
\newcommand{\beqn}{\begin{eqnarray}}
\newcommand{\eeqn}{\end{eqnarray}}
\newcommand{\beqns}{\begin{eqnarray*}}
\newcommand{\eeqns}{\end{eqnarray*}}
\newcommand{\vs}{\\[0.3cm]\indent}

\newcommand{\hm}{\hspace{-0.05cm}}

\newcommand{\intl}{\int\limits}
\newcommand{\ointl}{\oint\limits}
\newcommand{\mc}{\multicolumn}
\newcommand{\e}{\epsilon}
\def\NP{{\it Nucl. Phys.}}
\def\PL{{\it Phys. Lett.}}
\def\PR{{\it Phys. Rev.}}

\def\PRL{{\it Phys. Rev. Lett.}}
\def\NIM{{\it Nucl. Inst. Meth.}}
\def\ZP{{\it Z. Phys.}}
\def\EPJ{{\it Eur. Phys. J. }}

\def\CPC{{\it Comp. Phys. Comm.}}
\def\ea{{\it et al.}}
\def\Cl{Coll.}
\def\tauto{$\tau^{-\!\!}\rightarrow\,$}
\def\nut{$\,\nu_\tau$}
\def\pipiz{$ \pi^-\pi^0 $}
\def\pidpiz{$ \pi^-2\pi^0 $}
\def\tpidpiz{$ 2\pi^-\pi^+2\pi^0 $}
\def\pitpiz{$ \pi^-3\pi^0 $}
\def\tpi{$ 2\pi^-\pi^+ $}
\def\tpipiz{$ 2\pi^-\pi^+\pi^0 $}
\def\pitpiz{$ \pi^-3\pi^0 $}

\def\pc{$\%$}

\def\sf{spectral function}
\def\sfs{spectral functions}

\def\as{$\alpha_s$}
\def\asm{$\alpha_s(m_\tau^2)$}

\def\ass{$\alpha_s(s)$}
\def\assz{$\alpha_s(s_0)$}
\def\gluonc{$\langle \frac {\alpha_s}{\pi} GG\rangle$}
\def\ee{$e^+e^-$}

\def\Rt{$R_\tau$}

\def\RtVA{$R_{\tau,V/A}$}

\def\RtVpA{$R_{\tau,V+A}$}

\def\Rts{$R_\tau(s_0)$}

\def\RtVpAs{$R_{\tau,V+A}(s_0)$}

\def\GG{$\langle(\alpha_s/\pi) GG\rangle$}
\def\Osix{$\langle{\cal O}_6\rangle_{V/A}$}
\def\Oeight{$\langle{\cal O}_8\rangle_{V/A}$}
\def\br{branching ratio}
\def\brs{branching ratios}
\def\bfr{branching fraction}

\def\MSbar{$\overline{\rm MS}$}
\def\FOPTCI{$\rm FOPT_{\rm CI}$}

\def\ie{{\it i.e.}} 
 
\def\via{via} 

\def\rs{\raisebox{1.5ex}[-1.5ex]}

\def\rar{\rightarrow}
\begin{document}
\normalsize
\parskip=5pt plus 1pt minus 1pt


\begin{center}
\vspace{2.5cm}
{\Large
\bf
Branching Ratios and Spectral Functions of $\tau$ Decays: 
Final ALEPH Measurements and Physics Implications}\\
\vspace{1cm}
{\large \bf The ALEPH Collaboration}\footnote
{   See next pages for the list of authors.   } \\
\end{center}

\vspace{0.5cm}
	   
\begin{abstract}
The full LEP-1 data set collected with the ALEPH detector at the $Z$ pole 
during 1991-1995 is analysed in order to measure the $\tau$ decay 
branching fractions. The analysis follows the global method
used in the published study based on 1991-1993 data, but  several 
improvements are introduced, especially concerning the treatment of 
photons and $\pi^0$'s. Extensive systematic studies are performed,
in order to match the large statistics of the data sample corresponding 
to over 300\,000 measured and identified $\tau$ decays. Branching 
fractions are obtained for the two leptonic channels and eleven hadronic
channels defined by their respective numbers of charged particles and 
$\pi^0$'s. Using previously published ALEPH results on final states
with charged and neutral kaons, corrections are applied to the hadronic
channels to derive branching ratios for exclusive final states without 
kaons. Thus the analyses of the full LEP-1 ALEPH data are combined to
yield a complete description of $\tau$ decays, encompassing 22 non-strange
and 11 strange hadronic modes. Some physics implications of the results
are given, in particular related to universality in the leptonic charged
weak current, isospin invariance in $a_1$ decays, and the separation of
vector and axial-vector components of the total hadronic rate. Finally,
spectral functions are determined for the dominant hadronic modes and 
updates are given for several analyses. These include: tests of isospin 
invariance between the weak charged and electromagnetic hadronic currents, 
fits of the $\rho$ resonance lineshape, and a QCD analysis of the
nonstrange hadronic decays using spectral moments, yielding the value
$\alpha_s(m^2_\tau) = 0.340 \pm 0.005_{\rm exp} \pm 0.014_{\rm th}$. 
The evolution to the $Z$ mass scale yields 
$\alpha_s(M_Z^2) = 0.1209 \pm 0.0018$. This value agrees well with 
the direct determination from the $Z$ width and provides the most accurate 
test to date of asymptotic freedom in the QCD gauge theory.
\end{abstract}

\vspace{0.cm}
\vfill
\centerline{\it (Submitted to Physics Reports)}
\vspace{1cm}
\thispagestyle{empty}
%
%
\include{authb_plus}

\tableofcontents
\vfill
\pagebreak
\normalsize 

\pagestyle{plain}

\section{Introduction}

Because of its relatively large mass and the simplicity of its decay
mechanism, the $\tau$ lepton offers many interesting, and sometimes unique,
possibilities for testing the Standard Model. These studies involve the
leptonic and hadronic sectors and encompass a wide range of topics, from
the measurement of the leptonic couplings in the weak charged current, 
providing precise universality tests, to a complete investigation of
hadronic production from the QCD vacuum. In the latter case, the $\tau$
decay results have proven to be complementary to those from $e^+e^-$ data,
enabling detailed studies at the fundamental level to be performed, through
the determination of the spectral functions which embody both the rich
hadronic structure seen at low energy and the quark behaviour relevant
in the higher energy regime. The spectral functions play an important 
role in the understanding of hadronic dynamics in the intermediate
energy range and constitute the basic input for QCD studies and for 
evaluating contributions from hadronic vacuum polarization. The latter
are needed for precision tests of the electroweak theory through 
the running of $\alpha$ to the $M_Z$ 
scale and to compute the muon anomalous magnetic moment. 
The topics of interest include: testing the isospin invariance 
between the weak charged and electromagnetic hadronic currents; evaluation 
of chiral sum rules making use of the separate determination of vector and
axial-vector components; and a global QCD analysis including perturbative 
and nonperturbative contributions, offering the possibility of a precise
extraction of the strong coupling at a relatively low energy scale, thus
providing a sensitive test of the running when compared to the value
obtained at the $M_Z$ scale.

A global analysis of the $\tau$ branching ratios was performed 
by the ALEPH Collaboration using the data recorded at LEP through 
the process $e^+e^- \rightarrow Z \rightarrow \tau^+ \tau ^-$ 
in 1990-1991~\cite{aleph01} and in 1991-1993~\cite{aleph13_l,aleph13_h}. 
Specific studies of final states including kaons were published using 
an early data set~\cite{alephk1a,alephk1b}, then with the full 1991-1995 
statistics~\cite{alephk3,alephks,alephkl}, and finally 
compiled~\cite{alephksum}. Nonstrange spectral functions were 
determined earlier~\cite{alephsf1}, then separately for the 
vector~\cite{alephvsf} and the axial-vector~\cite{alephasf} components 
from the 1991-1994 data, leading to a precise determination of
$\alpha_s(m^2_\tau)$ and showing that nonperturbative terms 
are small even at this relatively low scale. The strange
spectral function was likewise extracted, from which it was possible
to determine the strange quark mass~\cite{alephksum}. Finally, a specific 
study of $\eta$ and $\omega$ resonance production in $\tau$ decays 
was carried out~\cite{alepheta}.

In this report a complete and final analysis of $\tau$ decays is presented
using the global method. All data recorded at LEP-I from 1991 to 1995 
with the ALEPH detector 
are used, thus providing an update of the previous results which were based 
on partial data sets. The increase in statistics ---the full sample
corresponds to about 2.5 times the luminosity used in the last published 
global analysis~\cite{aleph13_l,aleph13_h}--- not only allows for a reduction
of the dominant statistical errors but, more importantly, provides a way to
better study possible systematic effects and to eventually correct for them.
Several improvements of the method have been introduced in order to achieve
a better control over the most relevant systematic uncertainties: 
simulation-independent measurement of the $\tau \tau$ 
selection efficiency; improved photon identification
especially at low energy, where the separation between photons from $\tau$
decays and fake photons from fluctuations in hadronic or electromagnetic 
showers is delicate; a new method to correct the Monte Carlo simulation
for the rate of fake photons; and stricter criteria for channels with 
small branching fractions. For consistency and in order to maximally profit 
from the improved analysis all data sets recorded from 1991 to 1995 have
been reprocessed with the latest version of the reconstruction and selection
programs. The results presented in this report thus supersede those
already published in Ref.~\cite{aleph01,aleph13_l,aleph13_h}. 
In the same way the new determination of the nonstrange spectral functions 
replaces earlier results in Ref.~\cite{alephsf1,alephvsf,alephasf}. 
Only the measurements on final states containing kaons, which were 
already based on the full statistics, remain
unchanged~\cite{alephk3,alephks,alephkl,alephksum}.

The new analysis has been carried out independently for the two data sets
taken in 1991-1993 and 1994-1995, respectively. Each data set is associated
to its own Monte Carlo sample describing the corresponding detector 
performance. This method was chosen as it provides an easy way to
compare with the previous analysis of the 1991-1993 data sample and
to check the consistency between the two sets of new results. The latter
comparison is particularly meaningful as the systematic uncertainties are
estimated from the agreement between data and Monte Carlo samples, 
once corrections are applied, and are thus mostly uncorrelated. The final
results are obtained from the combination of both sets.

The paper is organized as follows. After a brief description of the ALEPH
detector, and of the data and simulated samples in Section~\ref{detector}, 
the main analysis tools are presented in Sections~\ref{selection} 
through \ref{brdet}: selection of $\tau \tau$ events and 
determination of non-$\tau$ backgrounds, charged particle identification, 
separation of genuine and fake photons, $\pi^0$ reconstruction, 
decay classification, channel-by-channel adjustment of the number 
of fake photons in the simulation, and branching ratio calculation.
The determination of the different sources of systematic uncertainties 
is the subject of Section~\ref{system}, while Section~\ref{check} provides
some global checks on the results. Corrections are applied to raw results
in order to obtain branching ratios for exclusive channels, as discussed
in Section~\ref{rec_exclu}. The consistency between the results
obtained from the two data sets is examined in Section~\ref{result} 
and the final combined results are given. Section~\ref{phys} 
deals with a few physics topics directly related to the measurement of the
branching ratios, such as universality in the leptonic charged weak current,
the branching ratio into $\rho \nu_\tau$ in the context of vacuum 
polarization calculations, $a_1$ decay fractions into the two 
isospin-related modes $3\pi$ and $\pi 2\pi^0$, and separation of vector and
axial-vector modes. The subject of Section~\ref{sf} is the measurement of 
hadronic spectral functions, the vector part of which is compared to
corresponding results from \ee\ annihilation in Section~\ref{vsf}, providing
tests of the Conserved Vector Current (CVC) hypothesis. Finally, a QCD 
analysis of the vector and axial-vector spectral functions is performed 
in Sections~\ref{qcd} with a determination of the strong coupling at the 
$\tau$ mass scale, $\alpha_s(m^2_\tau)$ and of the small nonperturbative 
contributions.

\section{The ALEPH detector and the data sample}
\label{detector}

A detailed description of the ALEPH detector and its performance can be 
found elsewhere~\cite{alephdet,alephperf}.  Only the features relevant to 
this analysis are briefly mentioned here.

Charged particles are measured by means of three detectors.
The closest detector to the interaction point is a silicon vertex detector
(VDET), which consists of two concentric barrels of double-sided microstrip 
silicon detectors. An inner tracking chamber (ITC), with eight
drift chamber layers, surrounds the vertex detector.
The ITC is followed by a time projection chamber (TPC), a cylindrical
 three-dimensional imaging drift chamber, providing up to 21 space points
for charged particles, and  up to 338 measurements of the ionization
loss, $dE/dx$. Combining the coordinate measurements of these detectors, 
a momentum resolution $\delta p_T/p_T^2 \ = \ 6 \cdot 10^{-4} \
\oplus 5 \cdot 10^{-3}/p_T$ (with $p_T$ in $\mbox{GeV}$)
is achieved in the presence of a 1.5 Tesla magnetic field.

The electromagnetic calorimeter (ECAL), located inside the coil, is
constructed from 45 layers of lead interleaved with proportional wire
chambers. The position and energy of electromagnetic showers are measured
using cathode pads subtending a solid angle of
$0.9^{\circ}\times 0.9^{\circ}$  and connected internally to form projective
towers. Each tower is read out in three segments  with a depth
of 4, 9 and 9 radiation lengths, yielding an energy resolution
$ \delta E/E = 18\%/\sqrt{E} + 0.9\%$ (with E in GeV) for isolated photons
and electrons. The inactive zones of this detector, referred to as cracks, 
represent  2~$\%$ in the barrel and 6~$\%$ in the endcap regions. 
The analysis of the hadronic $\tau$ decays
presented in this paper benefits from the fine granularity and from the
longitudinal segmentation of the calorimeter, which play a crucial role
in the photon and $\pi^0$ reconstruction, and in the rejection of
fake photons.

The hadron calorimeter (HCAL) is composed of the iron of the magnet return
yoke interleaved with 23 layers of streamer tubes giving a digital
hit pattern and has a projective
tower cathode pad readout of hadronic energy with a resolution
of about $85\%/\sqrt{E}$. Outside this calorimeter structure are located
two additional double layers of streamer tubes, providing
three-dimensional coordinates for particles passing through the HCAL.

The trigger efficiency for $\tau$ pair events is measured by comparing
redundant and independent  triggers involving the tracking detectors 
and the calorimeters. The measured trigger efficiency is better than
99.99$\%$  within the selection cuts.

Tau-pair events are simulated by means of a  Monte Carlo program, KORALZ07,
which includes initial state radiation computed
up to order $\alpha^2$ and exponentiated,
and  final state radiative corrections to order $\alpha$~\cite{was}.
The simulation of the subsequent $\tau$ decays also includes single photon
radiation for the decays with up to three hadrons in the final state.
Longitudinal spin correlations are taken into account~\cite{tauola} and 
the $\sin ^2 {\theta _W}$ value is adjusted to be consistent with the 
determination from the measurement of $\tau$ polarization~\cite{alephpol}.
The GEANT-based simulation, with the detector acceptance
and resolution effects, is used to initially evaluate the corresponding
relative efficiencies and backgrounds. It also includes the tracking,
the secondary interactions of hadrons, bremsstrahlung and conversions.
Electromagnetic showers are generated in ECAL according to parameterizations
obtained from test beam data~\cite{alephdet}. For all these effects, detailed
comparisons with relevant data distributions are performed and corrections
to the MC-determined efficiencies are derived, as discussed below for each
specific problem.

The data used in this analysis have been recorded at LEP-1 in 1991-1995. 
The numbers of detected $\tau$ decays are correspondingly 
132\,316 in 1991-1993 and 194\,832 in 1994-1995, for a total of about 
$3.3 \cdot 10^5$. The ratios between Monte Carlo and data statistics 
are 7.3 and 9.7 for the 1991-1993 and 1994-1995 periods, respectively. 
Monte Carlo samples were generated for each year of data taking in order
to follow as closely as possible the status of the detector components.
For convenience and easier reference to previously published results, the
total data and Monte Carlo sets corresponding to the 1991-1993 and 1994-1995 
periods are considered separately. Also, part of the data (approximately
11\% of the $\tau$ sample) was taken off the $Z$ peak in order to measure 
the lineshape. Corresponding Monte Carlo sets were generated to account 
for these conditions.

\section{Selection of $\tau \tau$ events}
\label{selection}
\subsection{The $\tau \tau$ event selector}
\label{select}

The principal characteristics of $\tau \tau$ events in $e^+e^-$ 
annihilation at the $Z$ energy are low multiplicity, 
back-to-back topology and missing
energy. Particles in each event are reconstructed with an energy flow
algorithm~\cite{alephperf} which calculates the visible energy, avoiding
double-counting between the TPC and the calorimeter information. 
The reconstructed event is divided into two hemispheres by a plane 
perpendicular to the thrust axis. The jet
in a given hemisphere is defined by summing the four-momenta of
energy flow objects (charged and neutral). The energies in the two
hemispheres including the energies of photons from final state radiation,
$E_1$ and $E_2$, are useful variables for separating Bhabha, $\mu \mu$
and $\gamma \gamma$-induced events from the $\tau \tau$ sample, while
relatively larger jet masses, wider opening angles, and higher 
multiplicities indicate $Z \rightarrow q \overline{q}$ events. 

All these features are incorporated in a standard selector used extensively
in ALEPH. While more detailed information can be found in earlier
publications~\cite{aleph13_l,aleph13_h,aleph94} the most important cuts
are listed below: 

\begin{enumerate}
  \item The total charged multiplicity should be at least two and no more than eight.
  \item Each hemisphere is required to have at least one charged track.
  \item The scattering angle $\theta^{*}$ in the $\tau \tau$ rest frame, 
calculated using the measured polar angles $\theta _1$  and $\theta _2$ of
the two jets through the relation
$\cos\theta^{*} = \sin(\frac {\theta_1 -\theta_2}{2})/
                  \sin(\frac {\theta_1 +\theta_2}{2})$, should satisfy
$|\cos\theta^*| <0.9$.
  \item The acollinearity angle between the two jets should
exceed $160^{\circ}$.
  \item  The sum of the jet energies should be larger than 
$0.35\times E_{beam} $ and the difference between the transverse momenta 
of the two jets should satisfy
        $ |\Delta p_t| > 0.066\times  E_{beam}$. 
  \item At least one track should extrapolate to the interaction point within 
$\pm 1 $~cm transversally to and $\pm 5 $~cm along the beam axis.
  \item A $\tau$-like hemisphere is defined by its charged multiplicity
equal to one and its invariant mass less than 0.8~GeV. For non 
$\tau$-like hemispheres, the product of the number of energy flow objects 
in each hemisphere should be less than 40 and the sum of the maximal 
opening angle between any two tracks in each hemisphere less than 0.25 rad.
  \item The ratio of the total energy to the beam energy should be smaller 
than 1.6, except for $\mu\mu$-like events (both leading tracks are muons, 
or one is a muon and the energy in the opposite hemisphere is greater than 
$0.9\times E_{beam}$) where this cut is relaxed to 1.8.
For Bhabha-like events (all the charged tracks are electrons), the cut is
tightened to 1.4 if the tangent at the origin to the leading electron 
points to within $\pm 6$~cm of an ECAL crack.
  \item As an additional cut against the residual Bhabha background, 
the total energy measured independently in the wire planes of ECAL has 
to be less than $1.8\times E_{beam} $.
  \item Cuts using tight matching in space and momentum between the tracks 
in opposite hemispheres are used to reject cosmic rays~\cite{aleph13_l}.
\end{enumerate}

In the previous analysis~\cite{aleph13_l,aleph13_h} additional cuts had been 
introduced in order to further reduce the contamination from Bhabha and
$e^+e^- \rar \mu^+\mu^-$ processes. In the present work it was chosen 
to simplify the 
procedure in order to conveniently measure selection efficiencies on the 
data, at the expense of a slightly larger background contamination which 
is anyway also measured in the data sample as explained below. 

\subsection{Measurement of selection efficiencies}
\label{breakmix}

The ``break-mix'' method introduced for the determination of the 
$\tau\tau$ cross section~\cite{aleph94} is used to measure the
selection efficiency of the cuts listed above. A selection of $\tau\tau$ 
events is implemented using very tight cuts on only one hemisphere. Then the
opposite hemisphere corresponds to an essentially unbiased $\tau$ 
decay, which is stored away. Pairs of selected hemispheres are combined 
to construct a $\tau\tau$ event sample built completely from data. 
This sample is used to measure the efficiency of the cuts based only on 
energy and/or topology. 

The same procedure is 
applied to the Monte Carlo sample as well as to the data and the ratio of 
the corresponding efficiencies is taken as a correction to the pure 
Monte Carlo efficiency. In this way the small correlations between 
hemispheres which are neglected in the break-mix data sample are
properly taken into account.

This method is used to determine the efficiencies of cuts 5 to 9 listed
above for both MC and data for large branching ratio channels. Channels
with small branching ratios are corrected
using the same factor as for nearby topologies, with an error properly 
enlarged. Results are given in Section~\ref{selsyst}.
The measured efficiencies are found to be very close to those 
obtained by the simulation, deviations being at most 
at the few per mille level. This situation stems from the facts that the
$\tau$ decay dynamics is ---apart from small branching ratio channels---
very well known, the selection efficiencies are large and the simulation
of the detector is adequate. 

The overall selection efficiency of $\tau\tau$
events is 78.9\%. This value increases to 91.7\% when
the $\tau\tau$ angular distribution is restricted to the detector polar
acceptance (cut 3), giving a better indication for the efficiency of
the cuts designed to exclude non-$\tau\tau$ backgrounds.
In addition, when expressed relatively to each $\tau$ decay, the
selection efficiencies are only weakly dependent on the final state,
with a maximum difference of 10\% amongst the considered decay 
topologies.

\subsection{Estimation of non-$\tau\tau$ backgrounds}
\label{nontau}
 
A method ---developed for the measurement of the 
$\tau$ polarization~\cite{alephpol}--- has been used to measure the 
contributions from the major non-$\tau$ backgrounds: Bhabhas, $\mu^+ \mu^-$ 
pairs, and $\gamma^* \gamma^* \rightarrow e^+ e^-, \mu^+ \mu^-$, and hadrons
events.  The basic idea is to apply cuts to the selected $\tau\tau$ data 
in order to reduce as much as possible the $\tau \tau$ population while 
keeping a high efficiency for the background source under study. 
The procedure does not require an absolute normalization from the Monte 
Carlo simulation of these channels, only a qualitative description of 
the distribution of the discriminating variables.
In the following, `$e$' and `$\mu$' designate charged particles identified 
(see Section~\ref{pid}) as electrons and muons, respectively.

\begin{figure}
   \centerline{\psfig{file=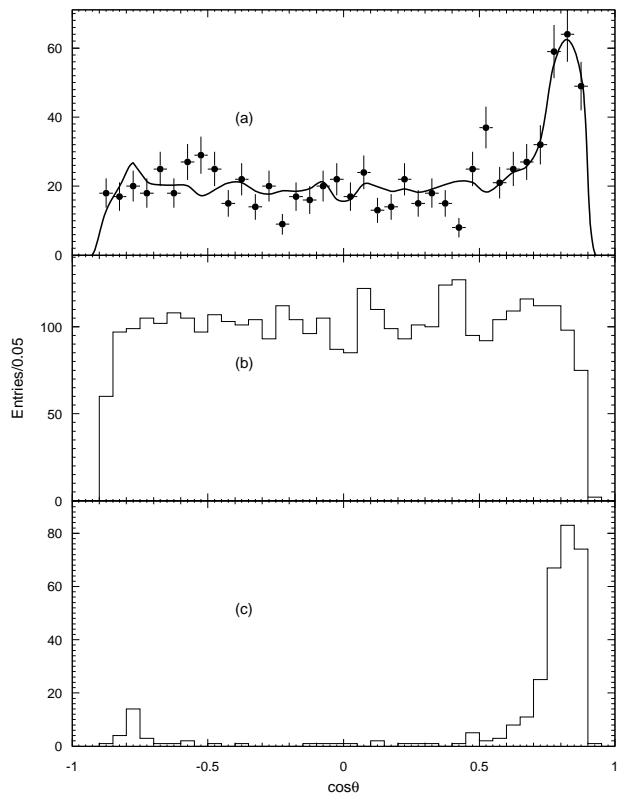,width=80mm}}
   \caption{Determination of the remaining Bhabha contribution in the final
$e$-$e$ sample for 1994-1995 data: $\cos\theta^*$ distributions for (a) data,
(b) $\tau\tau$ simulation, and (c) Bhabha simulation. The solid line
represents a fit of the data with the relative normalization of the 
$\tau\tau$ and Bhabha contributions left free.}
\label{bhabha}
\end{figure}

For Bhabha background, different event topologies are considered in turn,
depending on whether one of the charged particle traverses an inactive area
of the ECAL between detector modules. For $e-e$ and $e-crack$ topologies, 
either the acoplanarity angle is required to be larger than $179^\circ$ 
or the acollinearity angle should be more than $175^\circ$, in which case
the difference of transverse energies is required to be less than 3 GeV. 
The corresponding Bhabha efficiency (for the final sample) is $88\%$ as
determined from the simulation using the UNIBAB generator~\cite{unibab}. 
The angular distribution of the restricted sample is then fitted 
to $\tau \tau$ and Bhabha components (also including a small contribution 
from the other non-$\tau$ backgrounds) from the simulation. Therefore, 
the Bhabha Monte Carlo input is only used to determine the (large) 
selection efficiency and the $cos~\theta^*$ shape inside the final sample, 
not relying on any determination of the absolute Monte Carlo normalization. 
The derived Bhabha contribution has a statistical uncertainty which is
assigned as a systematic error. Several combinations of variables have been 
tried, showing a good stability of the result within its error.
Figure~\ref{bhabha} illustrates the determination for the $e-e$ topology in
the 1994-1995 sample. For the more numerous $e-hadron$ sample, the same 
cuts are used, but they have to be supplemented by an additional 
requirement to suppress
true hadrons as compared to electrons misidentified as hadrons. This is 
achieved by restricting opposite hadrons to have an electron identification
probability larger than 0.01; most of the true hadrons are below this value.

\begin{figure}
   \centerline{\psfig{file=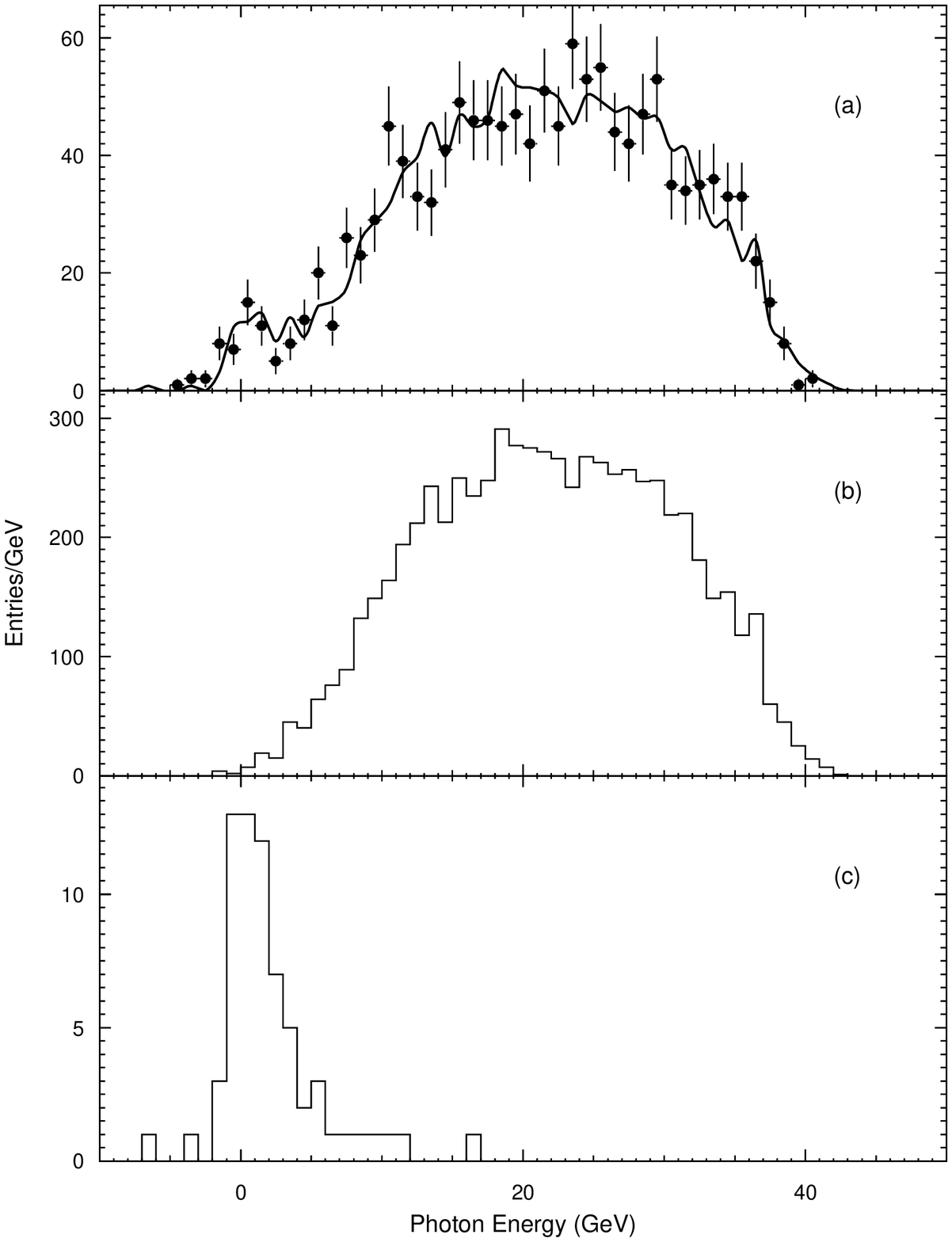,width=80mm}}
   \caption{Determination of the remaining $e^+e^- \rightarrow \mu^+ \mu^-$ 
contribution in the final $\mu$-$\mu$ sample for 1994-1995 data using the
kinematically calculated energy of a hypothetical photon emitted along one 
of the $e^\pm$ momenta: (a) data, (b) $\tau\tau$ simulation, and (c) 
$e^+e^- \rightarrow \mu^+ \mu^-$ simulation. The solid line
represents a fit of the data with the relative normalization of the 
$\tau\tau$ and $\mu$-pair contributions left free.}
\label{mumu}
\end{figure}

A similar technique is used to estimate the $\mu$-pair background. 
Figure~\ref{mumu} shows the corresponding plots for the $\mu-\mu$ topology, 
requiring an acoplanarity angle larger than $178^\circ$, 
for which $90\%$ efficiency for $e e \rar \mu \mu$ events 
in the final sample is achieved. Here, the fitted distribution is
that of the calculated photon energy along the beam for a postulated 
$e e \rar \mu \mu \gamma \gamma_{beam}$ kinematics to take the most general case
compatible with the only information on the two muons. The $e e \rar \mu \mu$ 
background signal is clearly seen for small energies.

\begin{figure}
   \centerline{\psfig{file=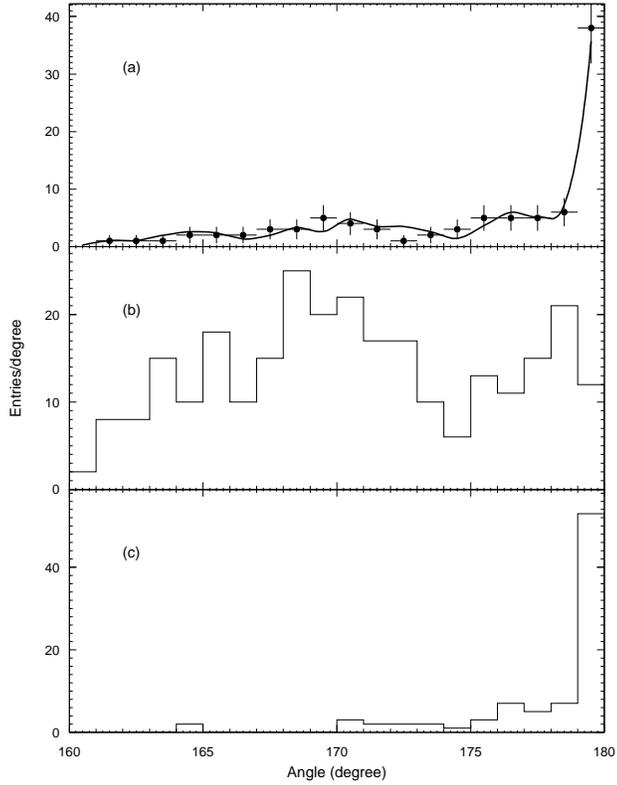,width=80mm}}
   \caption{Determination of the remaining 
$\gamma^* \gamma^* \rightarrow \mu^+ \mu^-$ 
contribution in the final $\mu$-$\mu$ sample for 1994-1995 data: 
acoplanarity angle distributions for (a) data, (b) $\tau\tau$ simulation, 
and (c) $e^+e^- \rightarrow (e^+e^-) \mu^+ \mu^-$ simulation. The solid 
line represents a fit of the data with the relative normalization of the 
$\tau\tau$ and $\gamma\gamma$-induced $\mu$-pair contributions left free. }
\label{ggmm}
\end{figure}

The remaining background from $\gamma^* \gamma^* \rightarrow \mu \mu$ is easily 
found in the $\mu-\mu$ topology after requiring the acollinearity angle to be
smaller than $170^\circ$ leading to an efficiency of $57\%$. The distribution
of the acoplanarity angle shows a distinct signal for the background, 
as seen in Fig.~\ref{ggmm}.

The same procedure is applied for $\gamma^* \gamma^* \rightarrow e e$, but
additional cuts (for example requiring the total event energy to be less
than 35 GeV) must be applied to decrease the Bhabha contribution.
Similar techniques are used to determine the contribution of
$\gamma^* \gamma^* \rightarrow $ hadrons for low multiplicity events.

Cosmic ray background is determined to be negligible~\cite{aleph13_l}. 
All other contributions are estimated with the proper Monte Carlo generators. 
This is the case for four-fermion processes and
hadronic Z decays whose contribution is estimated from the Lund
generator~\cite{lund}. Tests were previously made to ascertain the reliability
of the prediction for the rate of surviving low multiplicity events by
comparing to data~\cite{aleph13_h}, resulting in a systematic uncertainty
included in the analysis.

The non-$\tau$ backgrounds in each channel are listed in Table~\ref{bkg}. 
The contaminations from the different sources are given in 
Table~\ref{bkg_sources} and amount to a total fraction of 
$(1.23 \pm 0.04$)\% in the full data sample.

\begin{table}
\caption{Numbers of non-$\tau$ background events in observed classes 
as defined in Section~\ref{class}.}
\begin{center}
\begin{tabular}{lrr}
\hline\hline
  class   &   1991-93          & 1994-95 \\\hline
 \clsi    &  598 $\pm$   46 &  745 $\pm$   58 \\
 \clsii   &  409 $\pm$   45 &  380 $\pm$   40 \\
 \clsiii  &   93 $\pm$   11 &  100 $\pm$   13 \\
 \clsiv   &  141 $\pm$   22 &  178 $\pm$   26 \\
 \clsv    &   44 $\pm$    9 &   81 $\pm$   16 \\
 \clsvi   &   26 $\pm$    7 &   35 $\pm$    9 \\
 \clsxiii &   12 $\pm$    3 &   19 $\pm$    5 \\
 \clsvii  &   87 $\pm$   20 &  129 $\pm$   30 \\
 \clsviii &   97 $\pm$   23 &  165 $\pm$   39 \\
 \clsix   &   27 $\pm$    7 &   36 $\pm$   10 \\
 \clsx    &   13 $\pm$    4 &   25 $\pm$    7 \\
 \clsxi   &    3 $\pm$    1 &    7 $\pm$    2 \\
 \clsxii  &   16 $\pm$    5 &   21 $\pm$    6 \\
 \clsxiv  &  249 $\pm$   38 &  303 $\pm$   52 \\ \hline
 sum      & 1815 $\pm$   86 & 2224 $\pm$  107 \\
\hline\hline
\end{tabular}
\label{bkg}
\end{center}
\end{table}

\begin{table}
\caption{Estimated contaminations  from the different 
background processes.}
\begin{center}
\begin{tabular}{lc}
\hline\hline
  process   &   contamination ($\times 10^{-3}$)     \\\hline
 Bhabha                  &  4.0\\
 $e e \rar \mu\mu$                &  1.0\\
 two-$\gamma^*$ processes  &  2.2\\
 four-fermion               &  1.1\\
 hadrons                 &  3.9\\
 cosmic rays             &  0.1\\ \hline 
 sum                     &  12.3\\
\hline\hline
\end{tabular}
\label{bkg_sources}
\end{center}
\end{table}

\section{Charged particle identification}

\subsection{Track definition}

A {\it good} track is defined to have a momentum greater than or equal 
to 0.10~GeV (and not smaller than the momentum resolution),
$|\cos(\theta)| \le 0.95$, at least 4 hits in the TPC, and its
minimum distance to the interaction point within
2~cm transversally and 10~cm along the beams.
If a charged track is not a good one, has at least three hits 
in the TPC and  the minimum distance to the interaction point is
within 20~cm transversally to and 40~cm along the beams, then it is 
called a {\it bad} track.

In classifying $\tau$ decays, only good tracks are used, after
removing those identified as electrons which are used to reconstruct
converted photons; the electrons identified as bad tracks
are also included in reconstruction of conversions.

\subsection{The likelihood identification method}
\label{pid}

Charged particle identification plays a crucial role in the measurement
of $\tau$ branching ratios. In this analysis, as in the previous ones,
a likelihood method is used to incorporate the information from the relevant
subdetectors. In this way, each charged particle is assigned a set of 
probabilities from which a particle type is chosen.

Eight discriminating variables are used in the identification procedure:
$dE/dx$ in the TPC, two estimators (transverse and longitudinal) of the 
shower profile in ECAL, the average shower width measured with the HCAL
tubes in the fired planes, the number of fired planes among the last ten,
the energy measured with HCAL pads, the number of hits in the muon
chambers, in a road $\pm 4\sigma$-wide around the track extrapolation, 
where $\sigma$ is the standard deviation expected from multiple scattering,
and finally, the average distance (in units of the multiple-scattering 
standard deviation) of the hits from their expected position 
in the muon chambers.

Probability densities $f^j_i(x_i)$ of discriminating variable $x_i$ 
are determined using the ALEPH simulation for each particle type $j$,
where $j=e,\mu,h$ ($h=$~hadron). No attempt is made in this analysis to
separate kaons from pions in the hadron sample since final states containing
kaons have been previously studied~\cite{alephk3,alephks,alephkl}. Each
charged particle is assigned to the type with the largest global
estimator defined as
\beq
 P^j = \frac {\prod_{i} f^j_i(x_i)} {\sum_j \prod_{i} f^j_i(x_i)}~.
\eeq
The performance of the particle identification has been studied in detail 
using control samples of Bhabha events, $\mu\mu$ pairs, 
$\gamma^*\gamma^*$-induced lepton pairs and hadrons from $\pi^0$-tagged 
$\tau$ decays over the full angular and momentum range. 

\begin{figure}
   \centerline{\psfig{file=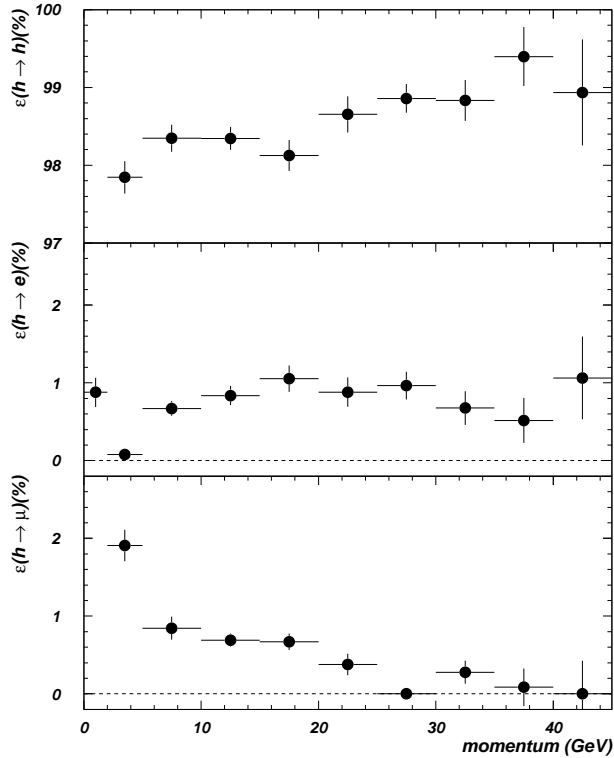,width=80mm}}
   \caption{Hadron identification efficiency and misidentification 
            probability obtained from the $\tau\tau$
            simulation, corrected from data using the control samples, 
            for the 1994-1995 data set. }
\label{ptidh}
\end{figure}

\begin{figure}
   \centerline{\psfig{file=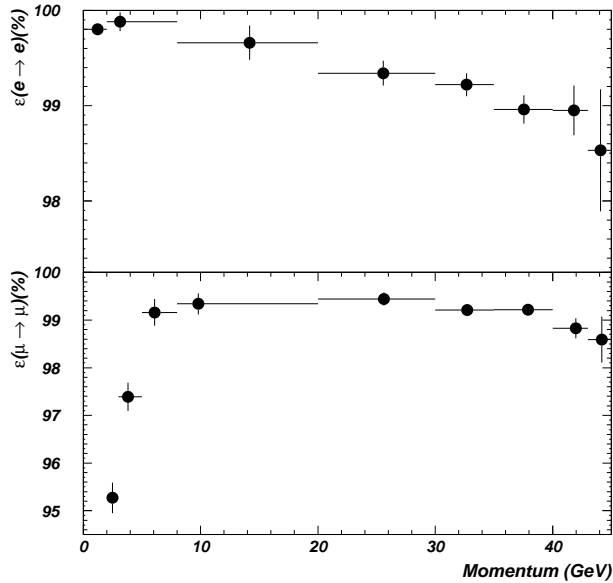,width=80mm}}
   \caption{Lepton identification efficiencies  obtained from the $\tau\tau$
            simulation, corrected from data using the control samples,
            for the 1994-1995 data set. }
\label{ptidl}
\end{figure}

The measurement of the particle identification efficiencies and 
the misidentification probabilities
are described in details in Ref.~\cite{aleph13_l}. Basically the efficiencies
are obtained from the $\tau$ Monte Carlo in order to include the relevant
detector description, but corrected as a function of momentum by the ratio
of data over Monte Carlo efficiencies as measured with the control samples.
This procedure takes care of the slightly different environment 
for a given particle between the various samples. 
Figures~\ref{ptidh} and \ref{ptidl} show the data-corrected 
particle identification efficiencies and misidentification probabilities 
for $e$, $\mu$ and hadron tracks as a function of track momentum.
Significant differences are observed between data and simulation, 
as in previous analyses, emphasizing the necessity of measuring directly 
these quantities on the data.

\section{Photon identification}
\label{photid}

\subsection{The problem of fake photons in ECAL}

The high collimation of $\tau$ decays at LEP energies quite often makes
photon reconstruction difficult, since these photons are close to one
another or close to the showers generated by charged hadrons. Of particular
relevance is the rejection of fake photons which may occur because of
hadronic interactions, fluctuations of electromagnetic showers, or the
overlapping of several showers. These problems reach a tolerable level
thanks to the fine granularity of ECAL, in both transverse and 
longitudinal directions, but they nevertheless require the development 
of proper and reliable methods in order to correctly identify 
photon candidates.  

\begin{figure}
   \centerline{\psfig{file=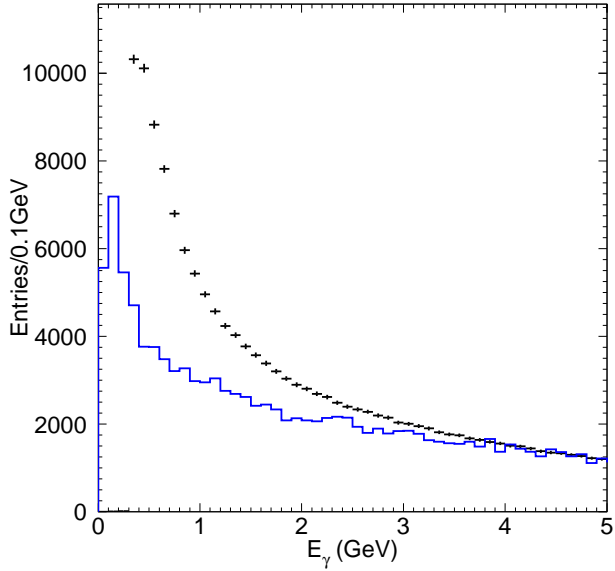,width=80mm}}
   \caption{Energy spectra of reconstructed photons in simulated $\tau\tau$ 
   events (points) and of primary produced photons by KORALZ07 after smearing 
   for energy resolution (histogram). The fake photon contribution at lower 
   energies is clearly indicated by normalizing to the number
   of photons with $E_{\gamma}>5$~GeV). The distribution of primary photons
   is obtained from a statistically reduced sample.}
\label{eg_obs_prd}
\end{figure}

A feeling of the fake photon problem can be obtained from
Fig.~\ref{eg_obs_prd} showing the comparison of simulated 
(at the generator level, after smearing for resolution effects) and 
observed photon energy spectra. 
A clear difference at low energy indicates a large contribution from 
fake photons. Since the fake photons have a complex origin and probably
quite dependent of the details of the detector simulation, the procedure
developed to separate them from the genuine primary photons from $\tau$
decays needs to be calibrated using the data.  

\subsection{Likelihood method for calorimeter photons}

\subsubsection{Photon reconstruction}

The clustering algorithm for the photon reconstruction~\cite{alephperf}
starts with a search for local maxima among the towers in the three ECAL
stacks. The segments of a projective tower which share a face in common
with the local maximum are linked together into a cluster. At the end of the
procedure, every segment of a tower is clustered with its neighbour of
maximal energy. Then a cluster is accepted as a photon candidate if its 
energy exceeds 350~MeV and if its barycentre is at least 2~cm from 
the closest track extrapolation; this distance is slightly modified later 
after corrections for the finite size of the pads are applied.

The energy of the photon is calculated from the energy deposition in the 
four central towers only when the energy distribution is consistent with 
the expectation of a single photon, otherwise the sum of the tower 
energies is taken.

\subsubsection{Method for photon identification}

As presented above, the main problem is the separation between genuine 
photons from $\tau$ decays and the numerous fake photons generated 
either by the hadronic tracks interacting in ECAL, or by the fluctuation 
of the electromagnetic showers. The situation is worse for low energy
photons which are however needed to retain a high $\pi^0$ reconstruction
efficiency. 

A likelihood method is used for discriminating between genuine 
and fake photons. For every cluster, a photon ``probability'' is defined
\beq
 P_{\gamma} = \frac{P^{genuine}}{P^{genuine}+P^{fake}}~,
\eeq
where $P^j$ is the estimator under the photon hypothesis of type $j$.
By definition, genuine photons have a $P_{\gamma}$ value near one while
fake photons have $P_{\gamma}$ near zero. Each estimator $P_j$ is
constructed according to
\beq
 P^j = \prod_{i} {\cal P}^j_i(x_i)
\eeq
where ${\cal P}^j_i$ is the probability density for the photon hypothesis
of type $j$ associated to the discriminating variable $x_i$.

\begin{figure}
   \centerline{\psfig{file=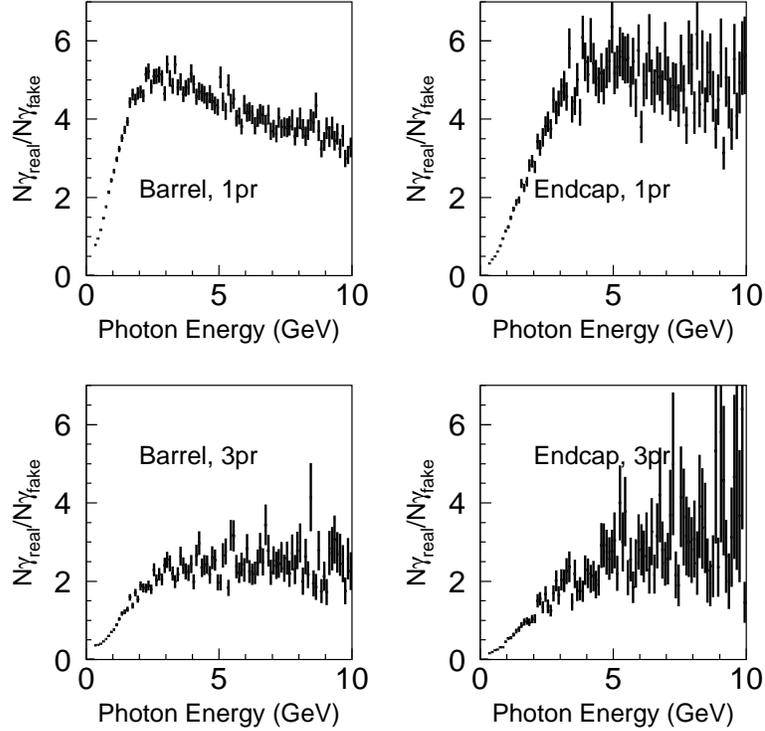,width=100mm}}
   \caption{Genuine-to-fake photon ratio as a function of
   photon energy from simulation, for one- and three-prong $\tau$ decays,
   and in barrel and endcap parts of ECAL.}
\label{rfratio}
\end{figure}

To take into account the ratio of fake to real photons which is strongly
energy-dependent, another quantity is defined as
\beq
 P_{\gamma}^{\prime} = \frac{P^{genuine}}
                         {P^{genuine}+P^{fake}/R_{gf}}~,
\eeq 
where $R_{gf}$ is the ratio of the numbers of genuine and fake photons,
which depends on the photon energy, the number of charged tracks and 
the number of
photons in the hemisphere. This quantity is used in $\pi^0$
reconstruction in order to reduce the fraction of $\pi^0$ candidates
where one of the photons is fake. Figure~\ref{rfratio} shows the 
genuine-to-fake photon ratio as a function of energy as obtained from 
simulation.

\begin{figure}
   \centerline{\psfig{file=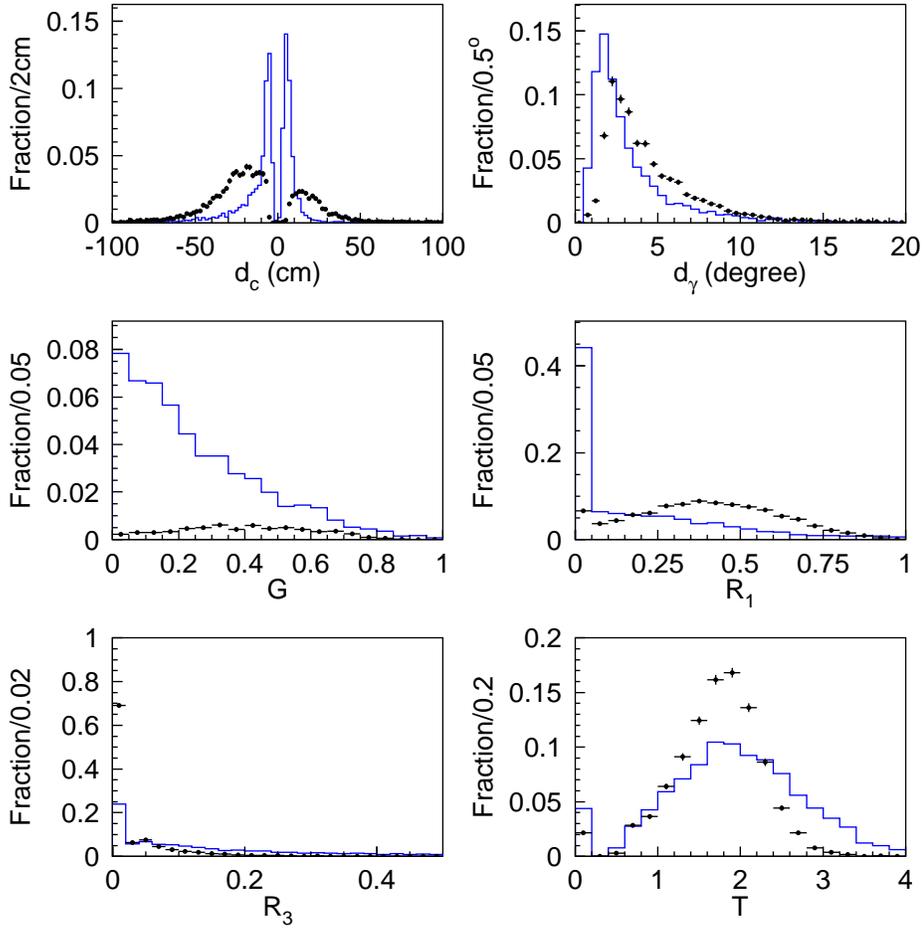,width=120mm}}
   \caption{Reference distributions for fake photons (histogram) and 
   genuine photons (points) with energy between 
   1.0 and 1.5~GeV. The underflows and overflows in each plot are also
   taken into account in the reference distributions.}
\label{refdis}
\end{figure}

Discriminating variables for each photon candidate used are the distance 
to the charged track ($d_c$), the distance 
to the nearest photon ($d_\gamma$), a parameter from the clustering
process resulting from the comparison between the energies in the cluster 
and in the neighbouring cells($G$), the fractions of energy deposition 
in ECAL stacks 1 ($R_1$), 2 ($R_2$) and 3 ($R_3$), and a parameter related 
to the transverse size of the energy distribution ($T$). 
The reference distributions for photons
with energy between 1.0 and 1.5~GeV are shown in Fig.~\ref{refdis}.

\subsubsection{Improvements with respect to the previous analysis}

\begin{figure}
   \centerline{\psfig{file=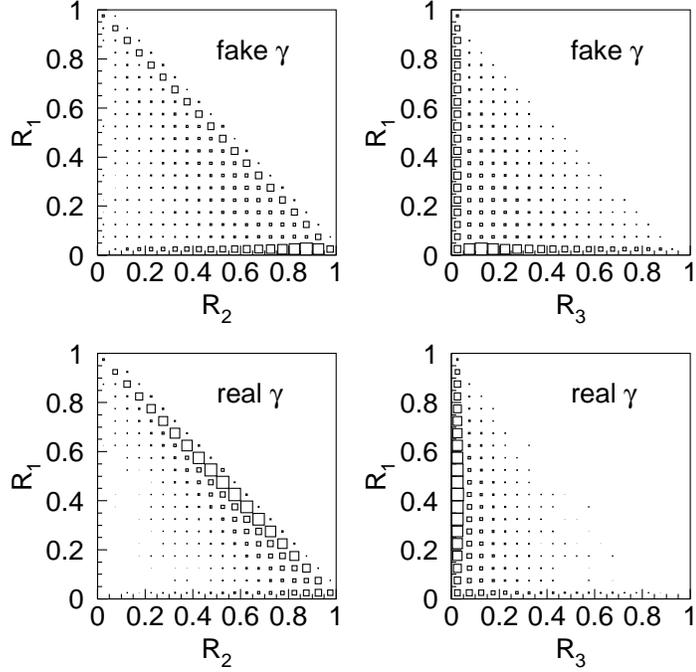,width=90mm}}
   \caption{Correlations between $R_1$ and $R_2$, and between $R_1$ and $R_3$,
   for photons with energy below 3~GeV.}
\label{r1r3}
\end{figure}

Major improvements were introduced at this stage in the analysis compared 
to the previous one~\cite{aleph13_h}:

(1) $R_3$ is used instead of $R_2$. This change was made because 
of the large correlation between $R_2$ and $R_1$ for genuine photons,
whereas the likelihood method ignores correlations between the 
discriminating variables. Also, $R_3$ contributes some
additional discriminating power against fake photons (see Fig.~\ref{r1r3}). 
As a consequence, genuine and fake photons are better separated.

\begin{figure}
   \centerline{\psfig{file=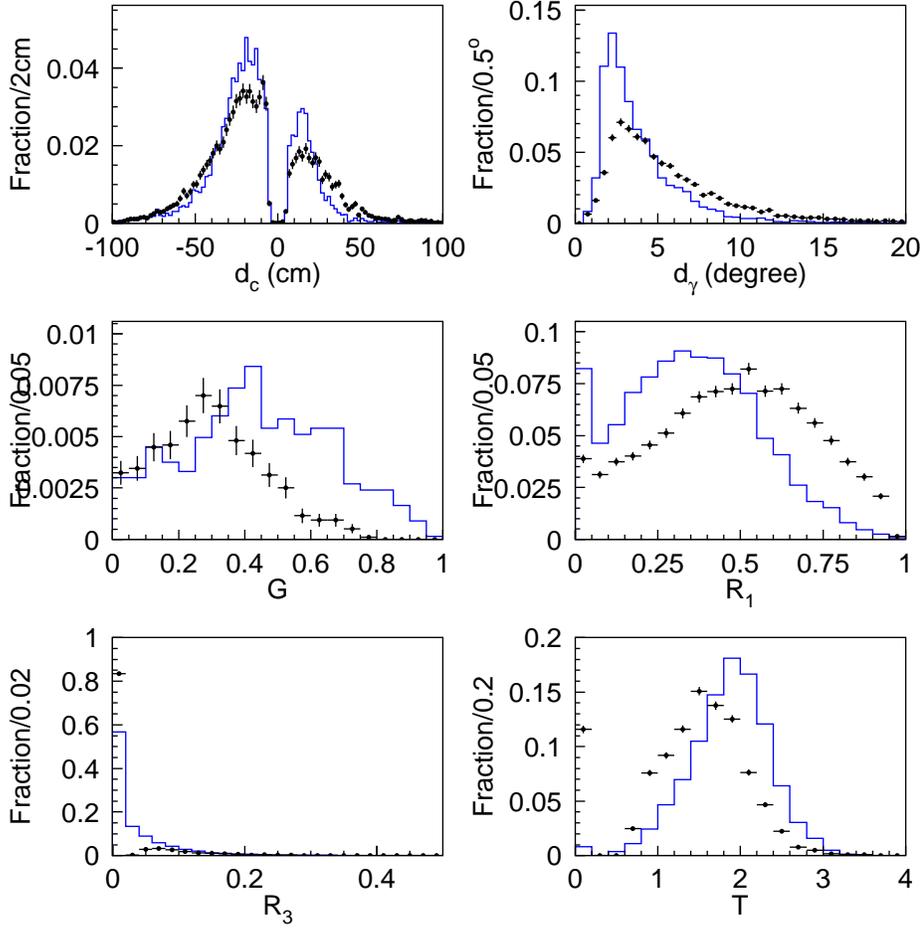,width=120mm}}
   \caption{Comparison of the reference distributions for genuine photons
   in the 0.5-1.0~GeV (points) and 1.5-2.0~GeV (histogram)
   energy ranges.}
\label{bin24}
\end{figure}

(2) Instead of using the same reference distributions for the whole 
energy range as before, energy-dependent reference distributions 
are now used to improve the discrimination between genuine and 
fake photons. This is made necessary because the reference distributions 
show pronounced energy dependence at energies below a few GeV's
(see Fig.~\ref{bin24}).

\begin{figure}
   \centerline{\psfig{file=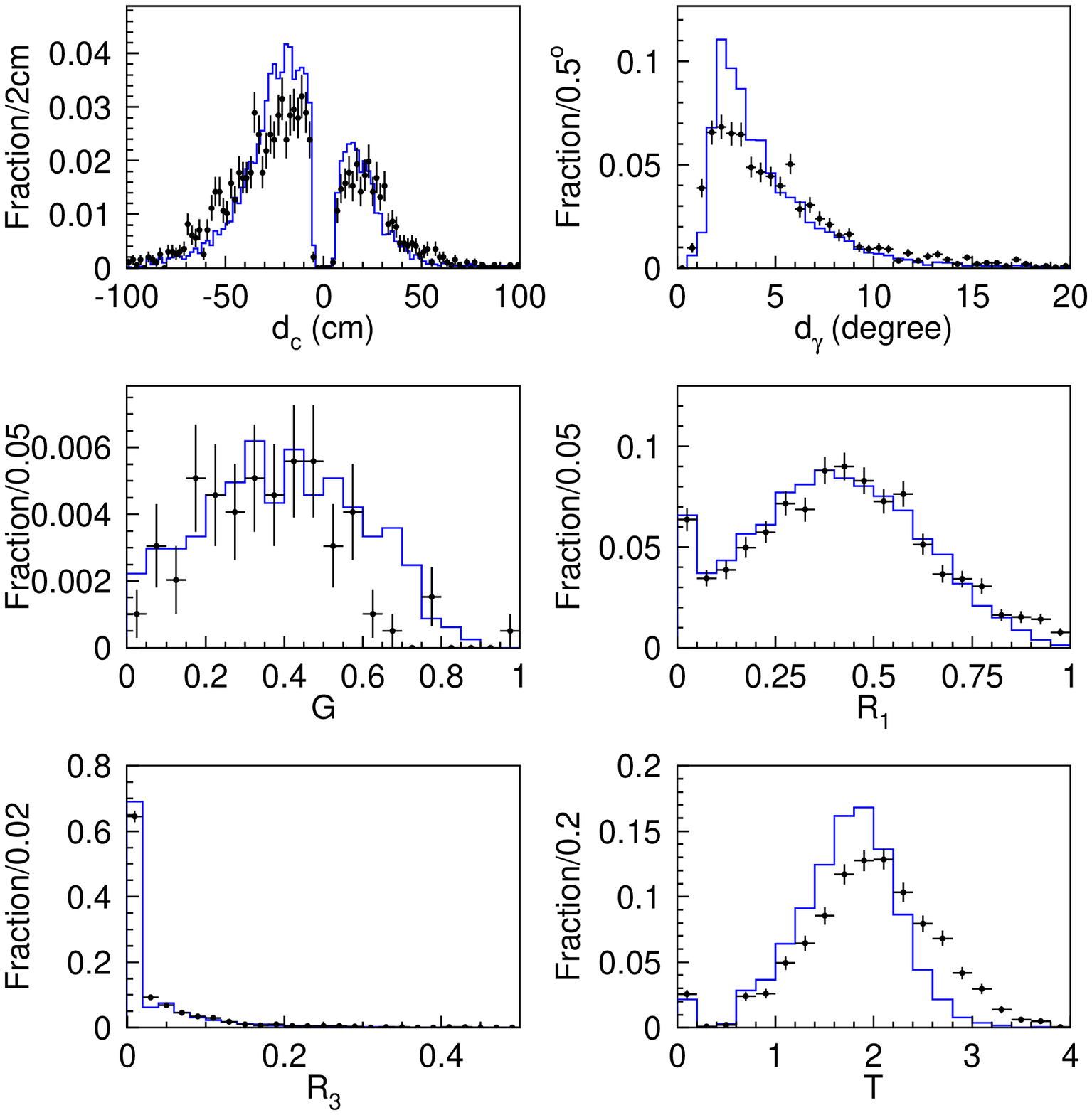,width=120mm}}
   \caption{Comparison of the reference distributions in the 
   barrel (histogram) and endcap (points) parts of ECAL for 
   genuine photons between 1.0 and 1.5~GeV.}
\label{barend}
\end{figure}

\begin{figure}
   \centerline{\psfig{file=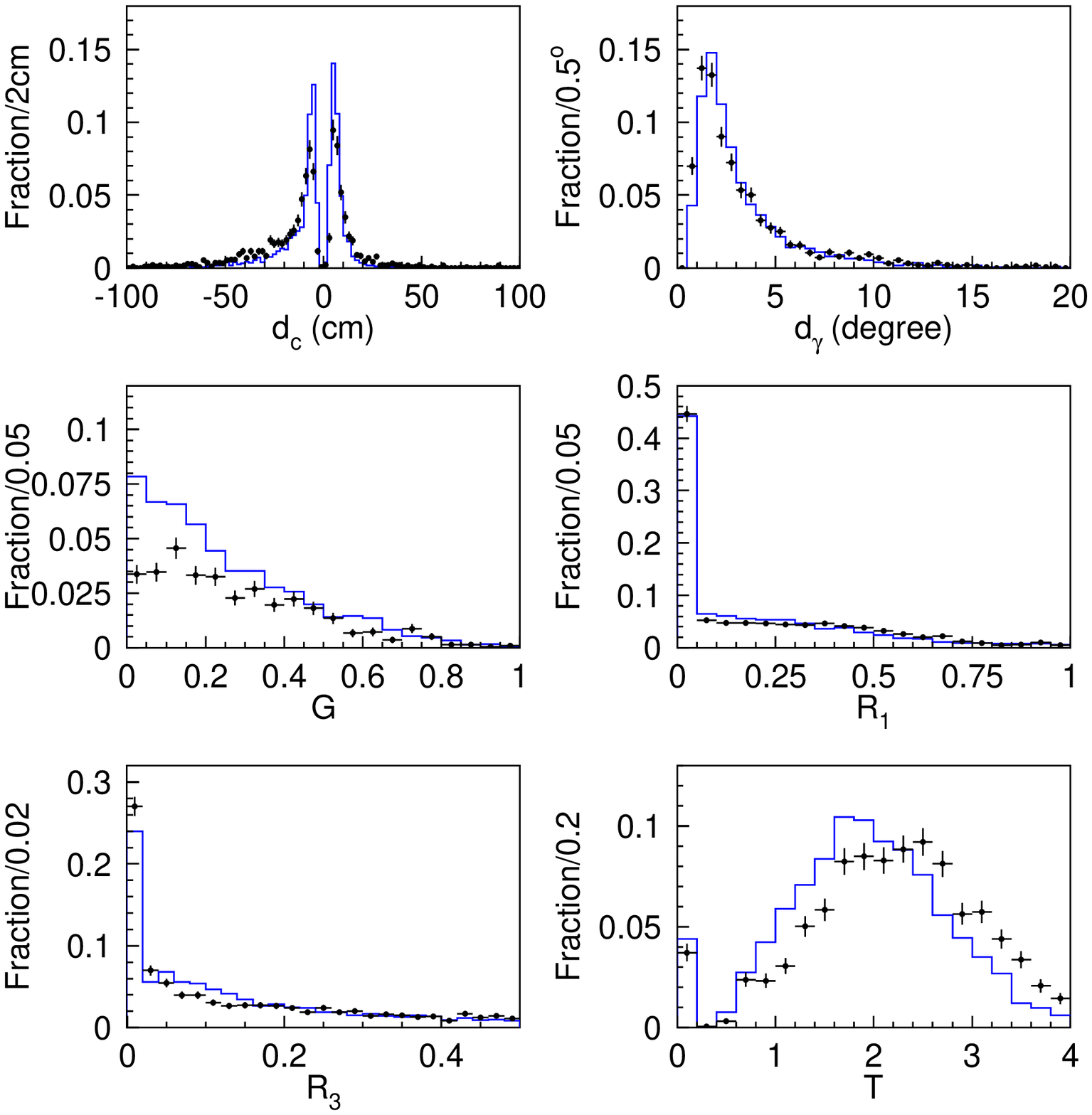,width=120mm}}
   \caption{Comparison of the reference distributions in the 
   barrel (histogram) and endcap (points) parts of ECAL for fake photons 
   between 1.0 and 1.5~GeV.}
\label{barend_fk}
\end{figure}

(3) Large differences are observed between the reference distributions 
of ECAL barrel and endcap photons (Figs.~\ref{barend} and \ref{barend_fk}), 
so different sets of reference distributions are used for photons 
in different regions of the detector.

\begin{figure}
   \centerline{\psfig{file=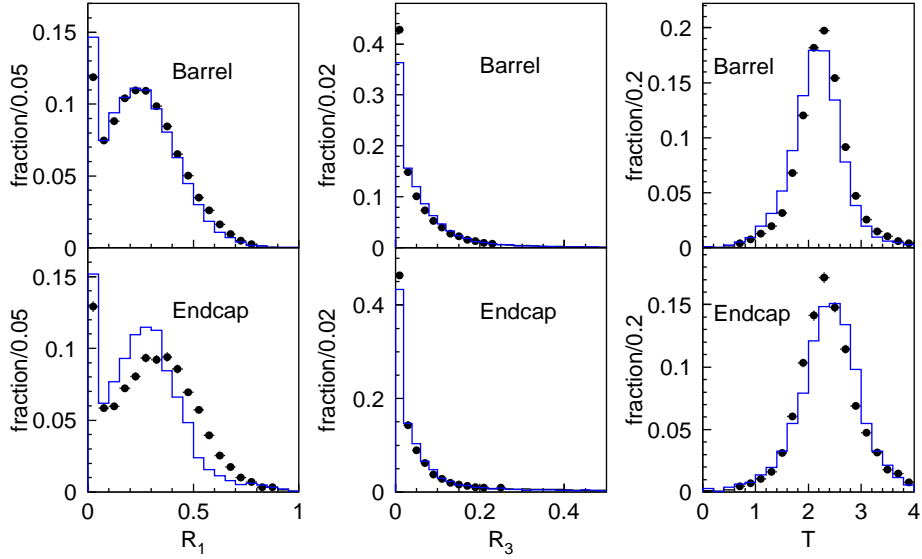,width=120mm}}
   \caption{Comparison of $R_1$, $R_3$ and $T$ distributions 
   in data (points) 
   and simulation (histogram) for photons 
   with energy between 3.0 and 6.0~GeV, in ECAL barrel and
   endcap parts.}
\label{r1r3tra}
\end{figure}

By comparing the distributions of photons at high energy, good
agreement is observed between data and MC photons in $d_c$, $d_\gamma$,
and $G$ distributions, while for $R_1$, $R_3$ and $T$ there are 
discrepancies (Fig.~\ref{r1r3tra}). 
The Monte-Carlo-determined reference distributions are then corrected 
using data. In order to do so, the $R_1$ and $R_3$ data distributions 
are obtained separately for real and fake photons by fitting the 
uncorrelated $d_c$ distributions with fake and genuine photon $d_c$ 
distributions from the simulation. For $T$, the value is shifted and 
smeared according to the difference between data and Monte Carlo.

\subsubsection{Total set of reference distributions}

Fine granularity is introduced when needed in the reference distributions.
Due to the observed strong energy dependence of some distributions, more
energy bins are considered: for genuine photons 7 bins (0.3-0.5, 0.5-1.0,
 1.0-1.5, 1.5-2.0, 2.0-2.5, 2.5-4.0, above 4 GeV) and for fake photons
4 bins (0.3-0.5, 0.5-1.0, 1.0-2.0, above 2 GeV). Different distributions
are set up according to the number of charged tracks (1 and $>1$) and
of photon candidates (1, 2, 3, 4, $>4$ for one prong and 1, 2, $>2$
otherwise). Also the barrel and endcap regions of the calorimeter are
separated so that the different amount of material in front and the 
different geometry is taken into account. 
Thus the photon identification procedure relies on the construction of
768 reference distributions for fake photons and 1344 for genuine photons.

\begin{figure}
   \centerline{\psfig{file=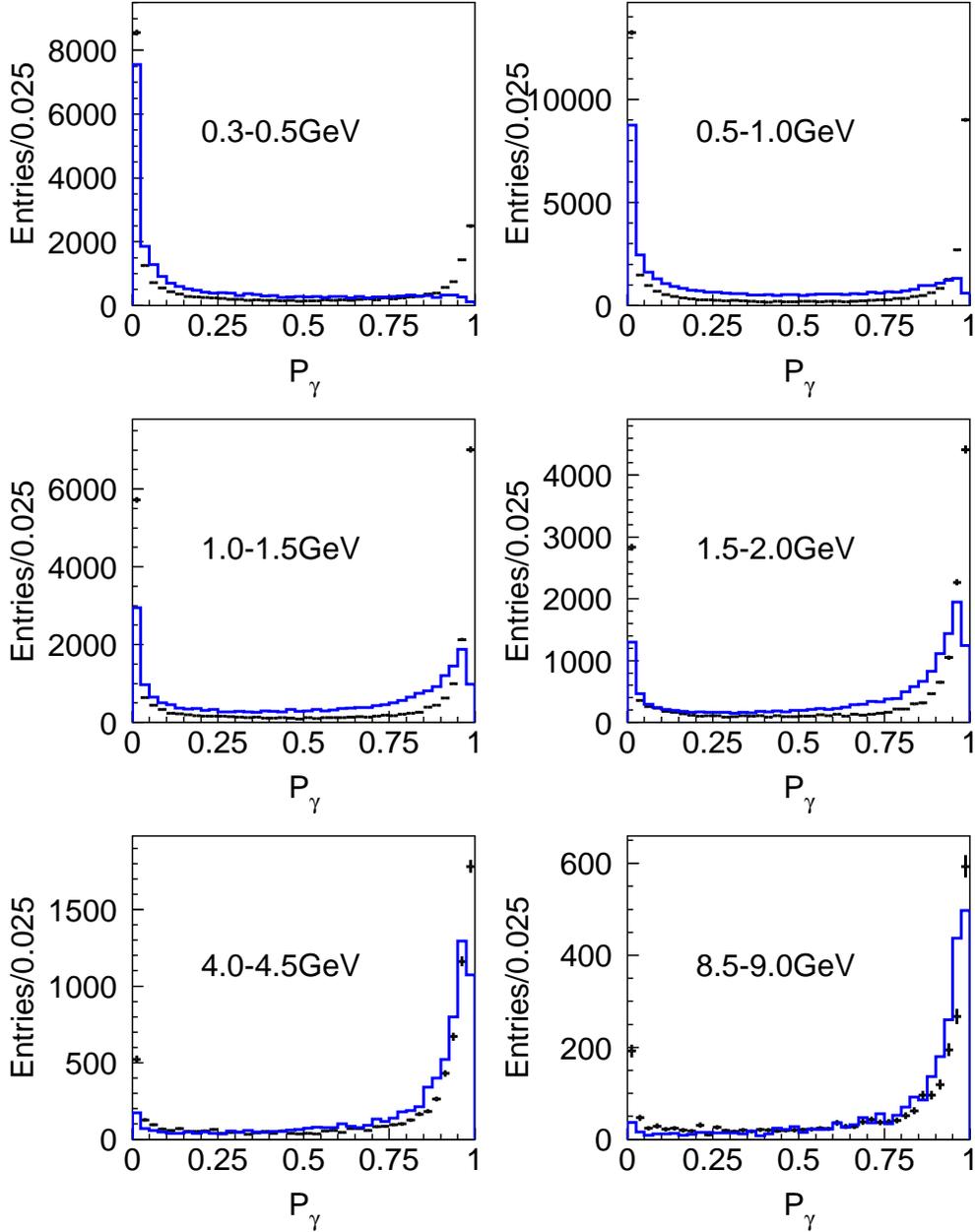,width=130mm}}
   \caption{Comparison of the data probability distributions of a cluster 
to be identified as a genuine photon before (histogram) and after (error 
bars) introducing energy-dependent reference distributions and 
   the $R_3$ variable.}
\label{prbg_old_new}
\end{figure}

Figure~\ref{prbg_old_new} shows the comparison of the photon identification
probability ($P_{\gamma}$) before and after introducing energy-dependent
reference distributions and the $R_3$ variable. A spectacular improvement is
observed in genuine and fake photon discrimination: at low energy, a clear
contribution of genuine photons can be seen, while at high energy, 
a small, but well identified, fake photon component shows up. 
Further checking, along the lines discussed below in Section~\ref{nbfake}, 
shows the validity of the latter signal on the simulated sample.

\subsubsection{Photon energy calibration}
\label{calphot}

\begin{figure}
   \centerline{\psfig{file=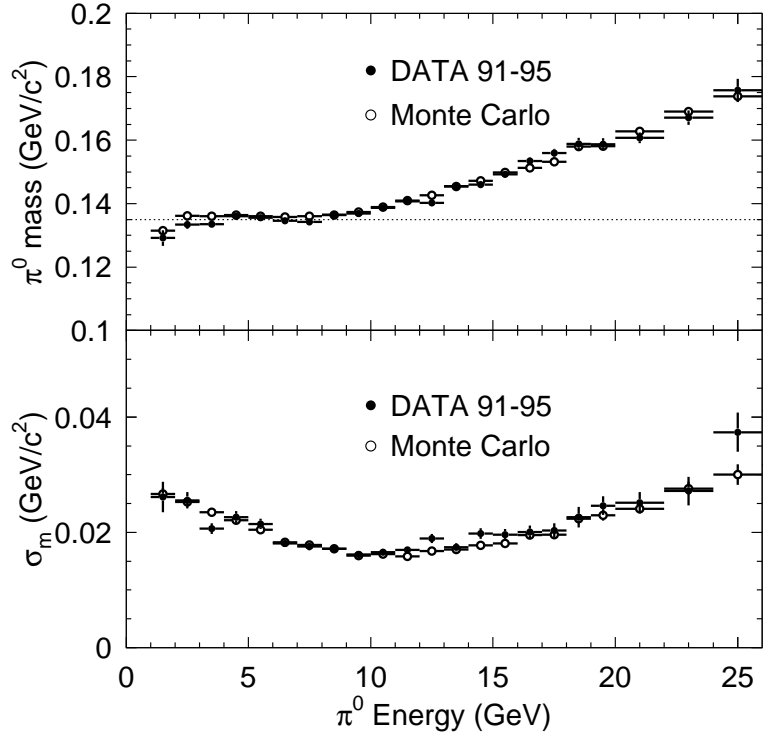,width=100mm}}
   \caption{Reconstructed $\pi^0$ mass (upper plot) and 
   resolution (lower plot) after photon energy and resolution 
   calibration. Good agreement between data and simulation 
   is observed. }
\label{pi0_mass}
\end{figure}

Better photon energy calibration is achieved compared to the previous 
analysis. The procedure, aiming at a relative calibration between data
and simulation, is implemented in several steps. First, a calibration 
is done using electrons from $\tau$ decays,
treating the ECAL information as for a neutral cluster. This method is 
reliable only above 2 GeV because the electron track curvature in the
magnetic field causes the shower to be ill-pointing to the interaction region.
The results are then extrapolated to lower energies. The bias introduced 
by the extrapolation is studied with the simulation comparing 
the true and the extrapolated energies, and the
same correction is applied to data. Then the final calibration is achieved 
by comparing the reconstructed $\pi^0$ mass and resolution as a function 
of energy. The agreement between data and simulation is satisfactory, as
seen in Fig.~\ref{pi0_mass}. 

\subsection{Converted photons}

In order to identify photons which convert inside the tracking volume
all oppositely charged track pairs of a given hemisphere identified
as electrons are considered. These candidates are required to have
an invariant mass smaller than 30~MeV and the minimal distance between
the two helices in the transverse plane must be smaller than 0.5~cm.
Finally, all remaining unpaired charged tracks identified as electrons
are kept as single track photon conversions. These include Compton
scatters or asymmetric conversions where the other track was either 
lost or poorly reconstructed.

\begin{figure}
   \centerline{\psfig{file=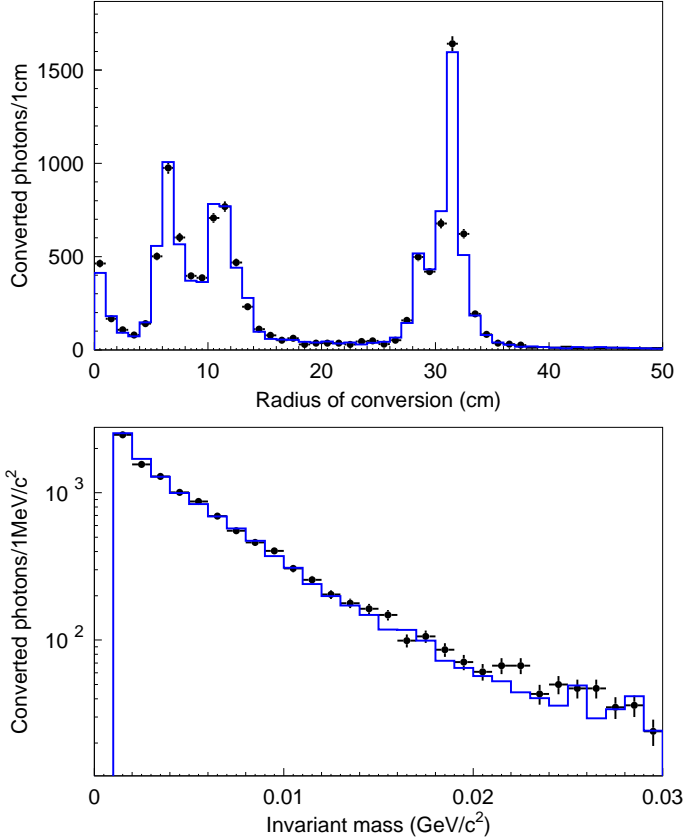,width=90mm}}
   \caption{Comparison of the distributions of the conversion point radius
   and the two-track invariant mass for photon conversions in data 
   (points) and simulation (histogram) for the 1994-1995 data sample. 
   The distribution of material in the detector is well simulated except 
   for a global lower rate in the simulation which is already corrected 
   here.}
\label{conv_pt}
\end{figure}

The $P_{\gamma}$ estimator is naturally set to unity for a converted 
photon. As the conversion rate in simulation is found to be lower 
than that of data by $(7.6 \pm 1.2)$~\% in 1991-1993 and
$(9.3 \pm 1.0)$~\% in 1994-1995, some converted 
photons in data are turned randomly into calorimeter photons and assigned 
a $P_{\gamma}$ value, generated according to the observed distribution for
calorimeter photons at the same energy. Figure~\ref{conv_pt} shows the
comparison of the radial conversion point and the invariant mass of 
two-track conversions between data and simulation. After correcting 
for the conversion rate, good agreement is observed.

\section{$\pi^0$ reconstruction}
\label{pi0rec}

\subsection{Different $\pi^0$ types}

The goal of the $\pi^0$ reconstruction procedure is to achieve a high 
efficiency while keeping the ``fake'' $\pi^0$'s at a reasonably small level.
Three different kinds of $\piz$'s are thus reconstructed: resolved $\piz$
from two-photon pairing, unresolved $\piz$ from merged clusters, and residual
$\piz$ from the remaining single photons after removing radiative,
bremsstrahlung and fake photons. 

\subsection{Resolved $\pi^0$'s}

Since fake and genuine photons are better separated than previously
published~\cite{aleph13_h}, a cut on $P_{\gamma}$ before $\pi^0$ 
reconstruction is introduced to remove a
large fraction of fake photons, while keeping the real photon 
efficiency high. The cut is energy-dependent, 
\( \ln P_{\gamma cut} > 0.15-2.9 \times E_{\gamma} \), with an efficiency
$>$ 90\% at $E_{\gamma}<0.5$~GeV, and $\sim$ 100\% above $2$~GeV.

A $\piz$ estimator $D_{ij}$ is defined~\cite{aleph13_h} to take into 
account the genuine and fake photon ratio at different energies:
\beq
 D_{ij} = P_{\gamma_i}^{\prime}\cdot P_{\gamma_j}^{\prime}
            \cdot P_{\pi^0}~,
\eeq
where $P_{\pi^0}$ is the $\chi^2$ probability of a kinematic 
$\pi^0$-mass constrained fit of the two photons. It should be noted that
$P_{\gamma}^{\prime}$ is used here instead of $P_{\gamma}$, as in the 
earlier analysis, in order to keep the same confidence for $\pi^0$'s 
reconstructed from photons at different energies. Finally, $\pi^0$
candidates are retained if their $D_{ij}$ value  is larger than 
$1\times 10^{-4}$.

In addition, a criterion must be established for choosing among all
the accepted $i-j$ pairs in a multiphoton environment. Configurations
with the maximum number of $\pi^0$'s are retained and among those, the
configuration which maximizes the product of all $D_{ij}$'s is kept.

In the old analysis~\cite{aleph13_h}, a significant difference in 
observed $\pi^0$ mass and resolution in data and simulation was observed. 
After the recalibration of the photon energy and resolution using 
the procedure described above, data and simulation agree reasonably 
well as shown in Fig.~\ref{pi0_mass}.

\subsection{Unresolved $\pi^0$'s}

As the $\pi^0$ energy increases it becomes more difficult to resolve
the two photons and the clustering algorithm may yield a single cluster.
The two-dimensional energy distribution in the plane perpendicular to the
shower direction is examined and energy-weighted moments are computed. 
Assuming only two photons are present, the second moment provides a
measure of the $\gamma\gamma$ invariant mass~\cite{alephperf}. 

\begin{figure}
   \centerline{\psfig{file=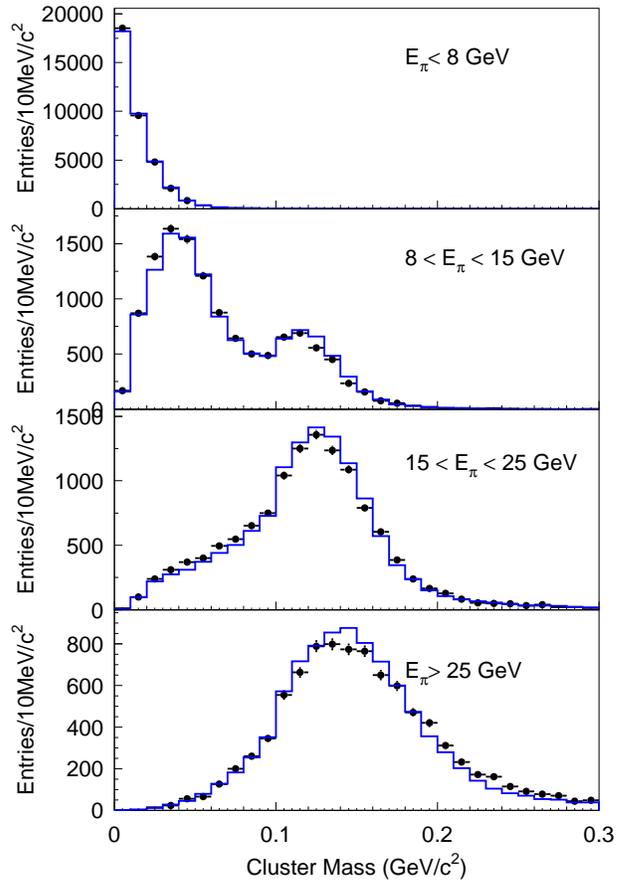,width=80mm}}
   \caption{Comparison of the $\pi^0$ mass distributions of unresolved
   $\pi^0$'s between data (points) and simulation (histogram)
   in different energy ranges. }
\label{mgunres}
\end{figure}

Figure~\ref{mgunres} shows this invariant mass distribution for data
and simulation. Clusters with mass greater than
0.1~GeV are taken as unresolved $\piz$'s. Some discrepancy
is observed between data and Monte Carlo and in order to keep the same 
efficiency a slightly different value for the $\pi^0$ mass cut is applied
in the simulation. It should be noted that this cut 
only affects the definition of unresolved and 
residual $\pi^0$'s, and thus does not change in first order the $\pi^0$
multiplicity, hence the definition of the $\tau$ decay final state. 
Thus it has essentially no influence on the branching ratio
analysis, since both unresolved and residual $\pi^0$s are used
in the $\tau$ decay classification.

\subsection{Residual $\pi^0$'s}

After the pairing and the cluster moment analysis, all the remaining 
photons inside a cone of $30^{\circ}$ around the jet axis are considered.
They come from different sources:
\bitm
\item {\it bremsstrahlung}:  
bremsstrahlung photons radiated along the final charged particles 
in $\tau$ decay, including the contribution from the detector material 
for electrons,
\item {\it radiative}: initial and final state radiation non-collinear
to charged particles in the final state,
\item {\it genuine} photons from $\pi^0$ decays where the partner photon is 
lost because of energy threshold, reconstruction inefficiency, cracks or 
overlap with another electromagnetic or hadronic shower,
\item {\it genuine single} photons, mostly from 
$\omega \rightarrow \pi^0 \gamma$ and $\eta \rightarrow \gamma\gamma$, and
\item {\it fake} photons.
\eitm
With respect to the previous analysis, the cut on $P_{\gamma}$ 
is tightened for residual $\piz$'s since better 
discrimination between real and fake photons is now achieved. The cut used,
\( \ln P_{\gamma cut} > -0.23-0.40 \times E_{\gamma}, \) has a rapidly 
increasing efficiency above 50\% below 0.5~GeV and is about 90\% at 3~GeV.

The radiative and bremsstrahlung photons are selected using the 
same method as described in the previous analysis~\cite{aleph13_h}.
Estimators $P_{brem}$, $P_{rad}$ and $P_{\pi^0 \rightarrow \gamma}$
are calculated to select photons from bremsstrahlung, radiative processes
and from $\pi^0$ decays, respectively. To compute these estimators, the
angle between the photon and the most energetic charged track is used,
in addition to the discriminating variables considered previously for the
photon identification. Radiative and bremsstrahlung photons are not used
in the $\tau$ decay classification discussed below. 

The behaviour of these estimators was studied in Ref.~\cite{aleph13_h}.
The agreement between the number of bremsstrahlung photons in data and
simulation is good, and affects mainly the electron $\tau$ decay channel
where this contribution is important (however it does not affect the
rate). Bremsstrahlung photons in hadronic channels (i.e. radiation along 
the final state charged pions) are at a much lower level and are difficult 
to pick up unambiguously in the data. The estimate of the effect of this 
contribution largely relies on the description of radiation at
the generator level in the Monte Carlo~\cite{photos}. This point 
is addressed in Section~\ref{systrad}.

\section{Decay classification}
\label{class}

Each $\tau$ decay is classified topologically according to the number of 
charged hadrons, the charged particle identification and the number of 
$\piz$'s reconstructed. While for one-prong and five-prong channels 
the exact multiplicity is required, the track number in three-prong channels
is allowed to be 2, 3 or 4, in order to reduce systematic 
effects due to tracking and secondary interactions. Thus 13 classes
are defined as given in Table~\ref{classtable}. In this table, 
the right-most column shows how the different considered $\tau$ decays
contribute to the reconstructed channels as defined. 
Throughout this paper only $\tau^-$ decays are cited and charge-conjugate
modes are implied.

\begin{table}[hbtp]                                                         
\caption{Definition of the reconstructed 
$\tau$ decay classes. All $\tau$ decay modes implemented in the simulation
are specified for each class. The notation $\tau$ stands for 
 $\tau^-$ and the charge conjugate states are implied, while $h$ stands for
any charged hadron ($\pi$ or K).}
\begin{center}
\begin{displaymath}
\begin{tabular}{|c|c|cc|} \hline
Class label & $\begin{array} {c}
 \mbox{Reconstruction} \\ 
 \mbox{criteria} 
\end{array}$ &
\multicolumn{2}{|c|}{Generated $\tau$ decay}   \\ \hline \hline

\rule[0cm]{0cm}{0.5cm}$e$   & 1 $e$ &
 \multicolumn{2}{|c|} {$\tr \ e^- \overline{\nu}_e \ \nt $} \\ \hline

\rule[0cm]{0cm}{0.5cm}$\mu$ & 1 $\mu$ &
 \multicolumn{2}{|c|} {$\tr \ \mu^- \overline{\nu}_\mu \ \nt$} \\ \hline

$h$                & 1 $h$ & $\begin{array} {c}
                   \tr \ \pi^- \ \nt    \\
                   \tr \ K^- \ \nt      \\
                   \tr \ K^{*^-} \ \nt      
                      \end{array}$       &
                      $\begin{array} {c}
                   \tr \ \pi^-  K^0 \overline {K}^0 \ \nt     \\
                   \tr \ K^- K^0 \ \nt                         
                      \end{array}$       \\ \hline

$h \ \pi^0$        & 1 $h$ \ + \ $\pi^0$ & $\begin{array} {c}
                    \tr \ \rho^- \ \nt     \\ 
                    \tr \ \pi^- \pi^0 \overline {K}^0 \ \nt      
                      \end{array}$       &
                     $\begin{array} {c}
                    \tr \ K^- \pi^0 K^0 \ \nt      \\
                    \tr \ K^{*^-} \ \nt      
                      \end{array}$   \\ \hline

$h \ 2\pi^0$       & 1 $h$ \ + \ 2$\pi^0$ & $\begin{array} {c}
                    \tr \ a_1^- \ \nt      \\
                    \tr \ K^{*^-} \ \nt      \\
                    \tr \ K^- \ 2\pi^0  \ \nt      
                       \end{array}$    &
                      $\begin{array} {c}
                    \tr \ \pi^-  \omega  \ \nt  \ ^(\footnotemark{^)}   \\
                    \tr \ \pi^-  K^0 \overline {K}^0 \ \nt     \\
                    \tr \ K^- K^0 \ \nt                         
                       \end{array}$  \\ \hline

$h \ 3\pi^0$       & 1 $h$ \ + \ 3$\pi^0$ & $\begin{array} {c}
                    \tr \ \pi^-  3\pi^0  \ \nt      \\
                    \tr \ \pi^- \pi^0 \overline {K}^0 \ \nt     
                  \end{array}$     &
                  $\begin{array} {c}
                    \tr \ K^- \pi^0 K^0 \ \nt      \\
                    \tr \  \pi^- \pi^0 \eta  \ \nt  \ ^(\footnotemark{^)}
                  \end{array}$     \\ \hline

$h \ 4\pi^0$       & 1 $h$ \ +\ $\geq 4\pi^0$ & $\begin{array} {c}
                    \tr \ \pi^-  4\pi^0 \ \nt  \\
                    \tr \ \pi^-  K^0 \overline {K}^0 \ \nt    
                   \end{array}$      &
                    $\begin{array} {c}
                    \tr \ \pi^- \pi^0 \eta  \ \nt  \ ^(\footnotemark{^)}   
                  \end{array}$ \\ \hline

$3h$               & $2 - 4 h$  & $\begin{array} {c}
                    \tr \ a_1^- \ \nt      \\
                    \tr \ K^{*^-} \ \nt      \\
                    \tr \ K^- \pi^+ \pi^-  \ \nt      
                  \end{array}$         &
                  $\begin{array} {c}
                    \tr \ K^- K^+ \pi^-  \ \nt      \\
                    \tr \ \pi^-  K^0 \overline {K}^0 \ \nt     \\
                    \tr \ K^- K^0 \ \nt                                     
                  \end{array}$  \\ \hline

$3h \ \pi^0$       & $2 - 4 h$  + $\pi^0$ & $\begin{array} {c}
                    \tr \ 2\pi^-\pi^+  \pi^0  \ \nt \ ^(\footnotemark{^)} \\
                    \tr \ \pi^- \pi^0 \overline {K}^0 \ \nt      
                  \end{array}$         &
                  $\begin{array} {c}
                    \tr \ K^- \pi^0 K^0 \ \nt      
                  \end{array}$ \\ \hline

$3h \ 2\pi^0$      & $3 h$  + 2$\pi^0$  & $\begin{array} {c}
                    \tr \ 2\pi^-\pi^+  2\pi^0  \ \nt \   
               ^(\footnotemark{^)}  \\
                    \tr \ \pi^-  K^0 \overline {K}^0 \ \nt     
                  \end{array}$           &
                   $\begin{array} {c}
                    \tr \ \pi^- \pi^0 \eta  \ \nt  \ ^(\footnotemark{^)}
                  \end{array}$
                                \\ \hline

\rule[0cm]{0cm}{0.5cm}$3h \ 3\pi^0$ & $3 h$  + $\geq 3 \pi^0$ &
       \multicolumn{2}{|c|} {$\tr \ 2\pi^-\pi^+  3\pi^0  \ \nt$}  \\ \hline

\rule[0cm]{0cm}{0.5cm}$5h$               & $5h$ &
     $\tr \ 3\pi^-2\pi^+ \ \nt$ & $\tr \ \pi^-  K^0 \overline {K}^0 \ \nt$  
                                \\ \hline

\rule[0cm]{0cm}{0.5cm}$5h \ \pi^0$       & $5h$ + $\pi^0$ &
       \multicolumn{2}{|c|} {$\tr \ 3\pi^-2\pi^+ \pi^0 \ \nt$} 
    \\ \hline \hline

\end{tabular}
\end{displaymath}

\parbox[b]{132mm}{ \footnotesize{$^2$ With
      $\omega \rightarrow \pi^0 \gamma $ }}
\parbox[b]{132mm}{ \footnotesize{$^3$ With $\eta \rightarrow \gamma\gamma$}}
\parbox[b]{132mm}{ \footnotesize{$^4$ With
      $\eta \rightarrow 3\pi^0$ }} 
\parbox[b]{132mm}{ \footnotesize{$^5$ This channel includes  
$\tr \pi \omega \ \nt$ with $\omega \rightarrow \pi^-\pi^+\pi^0$ }} 
\parbox[b]{132mm}{ \footnotesize{$^6$ This channel includes  
$\tr \pi \pi^0 \omega \ \nt$ with $\omega \rightarrow \pi^-\pi^+\pi^0$ }} 
\parbox[b]{132mm}{ \footnotesize{$^7$ With 
$\eta \rightarrow \pi^-\pi^+\gamma$  }} 
 
\label{classtable}
\end{center}

\end{table}

The definition of the leptonic channels requires an identified electron 
or muon with any number of photons. Some cuts on the total final state 
invariant mass are introduced to reduce feedthrough from hadronic 
modes~\cite{aleph13_l}. Also some decays with at least two good electron 
tracks are now included, when one or more of such tracks
are classified as converted photons.

In the previous analysis~\cite{aleph13_h}, the $\thd$,  $\tht$ and $\hq$ 
channels (where $h$ stands for any charged hadron: $\pi$ or K)
suffered from large backgrounds and consequently 
had a low signal-to-noise ratio. 
Most of these backgrounds are due to secondary interactions 
of the hadronic track with material in the inner detector part. 
In order to improve the definition of these channels the following steps 
are taken: (1) require the exact charged multiplicity $n_{ch}=3$ 
(instead of 2, 3, or 4) for $\thd$ and $\tht$, (2) demand a maximum 
impact parameter of charged tracks less than 0.2~cm (instead of 2~cm) 
for $\tht$, and (3) require the number of resolved $\pi^0$s in $\tht$ 
to be 3 or 2, and 4 or 3 in $\hq$.
With these tightened cuts the signal-to-noise ratio improves significantly 
with a small loss of efficiency.

It should be emphasized that all hemispheres from the
selected $\tau\tau$ event sample are classified, except for single tracks 
going into an ECAL crack (but for identified muons) or with a momentum less
than 2 GeV (except for identified electrons and hemispheres with at least
one reconstructed $\piz$). These latter decays, in addition to 
the rejected ones in the $\thd$ or $\tht$ channels are put in a special
class, labelled 14, which then collects all the non-selected 
hemispheres. By definition, the sample in class-14 does not correspond 
to one physical $\tau$ decay mode. In fact, it follows from 
the simulation that this class comprises about 
21\% electron, 27\% muon, 41\% 1-prong hadronic and 11\% 3-prong
$\tau$ decays. However, consideration of class-14 events can test if
the rejected fraction is correctly understood, as discussed later.

The numbers of $\tau$'s classified in each of the 
considered decay channels are listed in Table~\ref{nobs_dt}.

\begin{table}
\caption{Number of reconstructed events in 1991-1993 and 1994-1995 data sets
         in the different considered topologies.}
\begin{center}
\begin{tabular}{lrr}
\hline\hline
Reconstr. class & $n^{obs}_i$ (91-93) & $n^{obs}_i$ (94-95)   \\\hline
 \clsi  &     22405 &     33100 \\ 
 \clsii  &     22235 &     32145 \\ 
 \clsiii  &     15126 &     22429 \\ 
 \clsiv  &     32282 &     49008 \\ 
 \clsv  &     12907 &     18317 \\ 
 \clsvi  &      2681 &      3411 \\ 
 \clsxiii  &       458 &       499 \\ 
 \clsvii  &     11610 &     17315 \\ 
 \clsviii  &      6467 &      9734 \\ 
 \clsix  &      1091 &      1460 \\ 
 \clsx  &       124 &       150 \\ 
 \clsxi  &        60 &       105 \\ 
 \clsxii  &        36 &        59 \\ 
 \clsxiv  &      4834 &      7100 \\ 
\hline
 sum      &   132316 & 194832  \\
\hline\hline
\end{tabular}
\label{nobs_dt}
\end{center}
\end{table}

The KORALZ07 generator~\cite{was} in the Monte Carlo simulation incorporates
all the decay modes considered in Table~\ref{classtable}. 
Since the $h 4\pi^0$ decay channel is not included in the standard 
Monte Carlo, a separate generation was done where one of the produced 
$\tau$ is made to decay into that mode using a phase space model for the 
hadronic final state, while the other $\tau$ is treated
with the standard decay library. The complete behaviour 
between the generated decays and their reconstructed counterparts
using the decay classification is embodied in the efficiency matrix. This
matrix $\eff_{ji}$ gives the probability of a $\tau$ decay generated
in class $j$ to be reconstructed in class $i$. Obtained 
initially using the simulated samples, it is corrected for effects where
data and simulation can possibly differ, such as particle identification,
as discussed previously, and photon identification as affected by the
presence of fake photons. The latter correction is presented next, before
exhibiting the final corrected matrix.

\section{Adjusting the number of fake photons in the simulation}
\label{nbfake}

\begin{figure}
   \centerline{\psfig{file=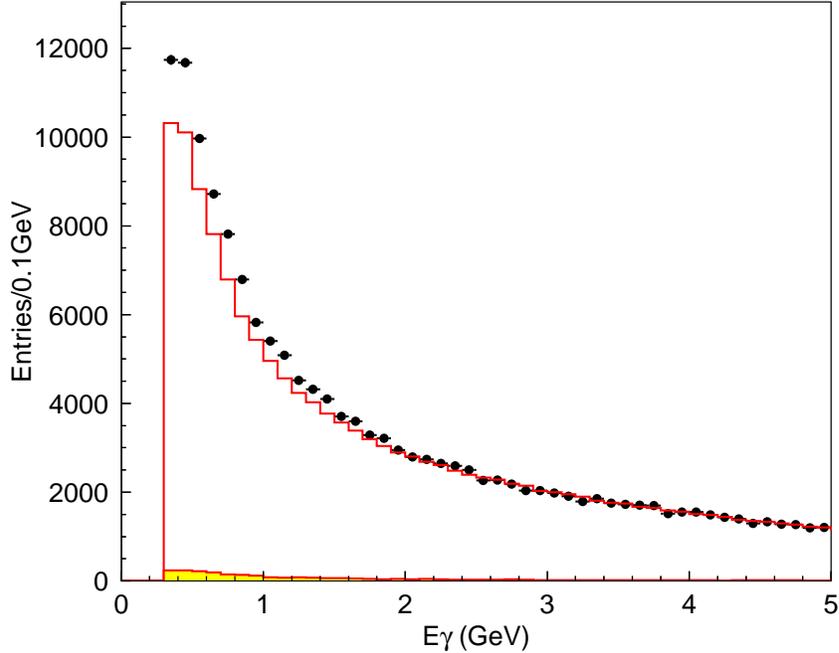,width=110mm}}
   \caption{Energy spectra of observed photons in data (error
   bars) and Monte Carlo (histogram) samples.  When normalized above
   $5$~GeV photon energy the two distributions show a clear discrepancy
   in the low energy region. The shaded histogram shows the contribution from
   non-$\tau$ backgrounds.}
\label{eg_obs_dt_mc}
\end{figure}

The number of fake photons in the simulation does 
not agree well with the rate in data. Such a discrepancy is seen in the 
comparison between the photon energy distributions in data and 
simulation given in Fig.~\ref{eg_obs_dt_mc}: at low energy there is clear 
indication that the rate of simulated fake photons is insufficient.
This effect will affect the classification of the reconstructed 
final states in the Monte Carlo and bias the efficiency matrix constructed 
from this sample. A procedure must be developed to correct for this effect.

\subsection{The method}

Taking explicitly into account the number $k$ of fake photons 
in a $\tau$ decay, the efficiency  matrix can be rewritten as
\beq 
 \eff_{ji} = \sum_{k} \eff_{jik}\cdot \omega_{jk} 
\eeq
where $\eff_{jik}$ is the efficiency for a produced event in class $j$ 
with $k$ fake photons to be reconstructed in class $i$. It can be calculated
from the simulation directly, given the number of fake photon in each event.
Above $\omega_{jk}$ is the fraction of produced class $j$ events with $k$
fake photons. The insufficient fake photons in 
simulation is due to the fact that $\omega_{jk}^{MC}$ is different
from  $\omega_{jk}^{data}$.

Assuming each fake photon is produced randomly, one can get the 
fake photon multiplicity in Monte Carlo by in principle randomly removing 
a fraction of $f$ fake photons in real data sample, {\it i.e.}
\beq
    \omega^{MC} = A \omega^{data} 
\eeq
where
  $\omega^{data}= (\omega_{0}, \omega_{1}, \omega_{2},\ldots , 
  \omega_{n})_{data}$ and
  $\omega^{MC} = (\omega_{0}, \omega_{1}, \omega_{2},\ldots , 
  \omega_{n})_{MC}$ are the fake photon
  multiplicity distributions in data and simulation respectively. The
matrix $A$ is given by
\beq
  A_{kl} = C^{l-k}_{l}f^{l-k}(1-f)^k 
\eeq
 if $k\le l$, otherwise it is 0. Here
  $k,l$ run from 0 up to the maximum number of fake photons.
The above equation holds for all the produced classes, with one parameter 
$f_j$ for each class.

The number of fake photons in data can be expressed as
\beq
 N_i^{f,data} = \sum_j N_j^{produced,data} \sum_k \eff_{jik}\cdot k\cdot
               \sum_l (A^{-1}_j)_{kl}\cdot \omega^{MC}_{jl}
\eeq
$i$ is the index for reconstructed class. In this set of equations, the
numbers $N_i^{f,data}$ are obtained from photon probability fits 
in the reconstructed data samples, 
$N_j^{produced,data}$ from the produced $\tau \tau$ number and
the measured branching ratios (implying an iterative procedure which in fact
converges extremely fast), $\eff_{jik}$ from the simulation with a proper 
procedure to determine $k$ (called matching in the following), 
$\omega^{MC}_{jl}$ from the simulation and matching, 
while $(A^{-1}_j)_{kl}$ depends on the $f_j$ unknowns to be determined by
solving the equations. 

Thus to measure the $f_j$ factors, the number of fake photons in each 
simulated event and the number of fake photons in each reconstructed 
class in data must be determined.

\subsection{Counting fake photons in simulation: the photon matching procedure}
\label{match}

To count the number of fake photons in each simulated event,
the philosophy is to remove all the clusters reconstructed in ECAL 
associated with produced photons at the generator level. 
The remaining clusters are then declared to be fake photons.

To match a produced photon to an ECAL cluster, the cluster with
minimum $\chi^2$ is selected, with
\beq
  \chi^2 = \left(\frac{\delta\theta}{\sigma_{\theta}}\right)^2 +
             \left(\frac{\delta\phi}{\sigma_{\phi}}\right)^2 +
             \left(\frac{\delta E}{\sigma_{E}}\right)^2 
\eeq
where the quantities with $\delta$ correspond to differences between
reconstructed and generated levels, while the $\sigma$ denote the
corresponding expected resolutions.
A matching flag $M$ is assigned to each cluster according 
to the $\chi^2$ value. The photon sample with $M=1$ corresponds to 
small $\chi^2$ and should be almost
exclusively composed of genuine photons, with a small feedthrough from
fake photons. The sample with intermediate $\chi^2$ is still dominated
by genuine photons with a larger fake photon content. At the end of the loop
over the produced (genuine) photons the unmatched clusters are given
a flag $M=-1$, corresponding to an almost pure sample of fake photons with a
small amount of genuine photons.

Many effects affect the matching procedure, leading to an incorrect estimate
of the number of fake photons in the event: energy threshold, ECAL 
non-sensitive regions (cracks, acceptance), merging with other neutral 
clusters, merging with charged tracks, accidental matching.
These effects have been studied very carefully and $\chi^2$ cuts are chosen 
so that the real photon contamination in the fake photon sample is 
the same as the fake photon contamination in 
the real photon sample. In this way the number of fake photons is given
by the number of $M=-1$ photons, facilitating considerably further 
treatment of the fake photon contribution.

\subsection{Fits to data}

\begin{figure}
   \centerline{\psfig{file=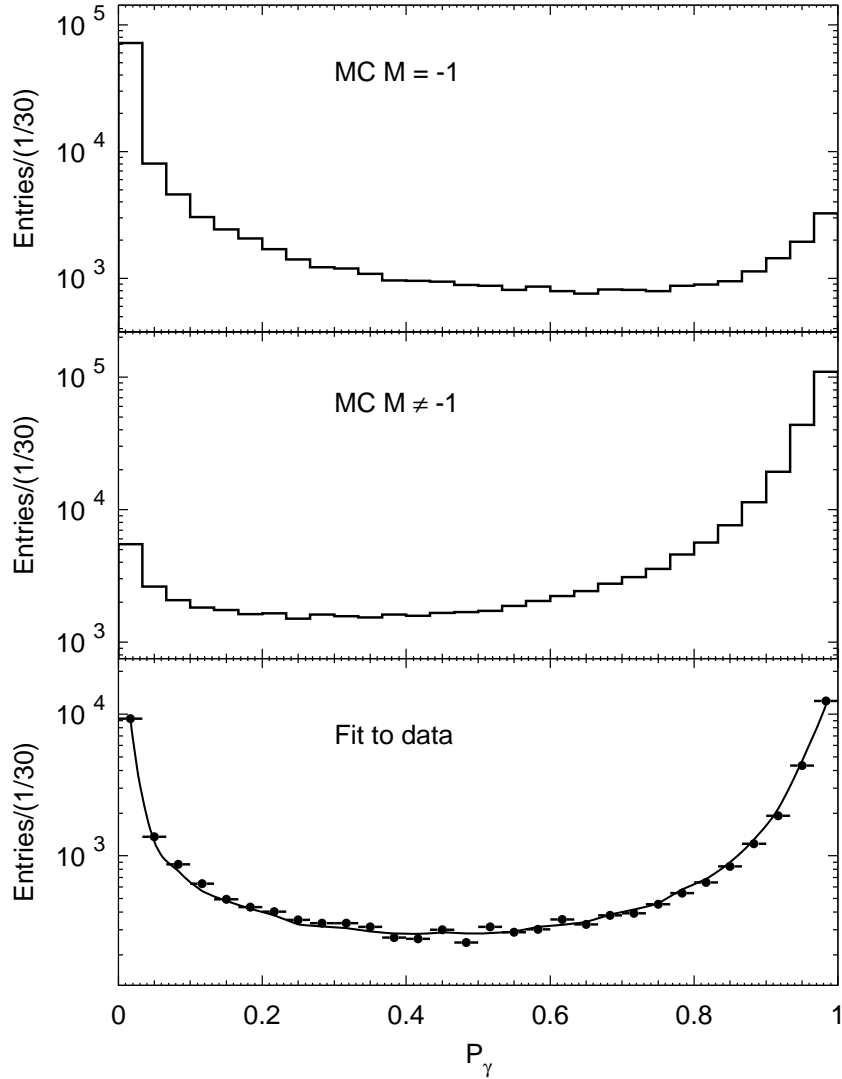,width=110mm}}
   \caption{Fit to the photon probability distribution in observed 
   $\hu$ events for the 1994-1995 data set. From top to bottom, are the 
   probability distributions of fake and real photon samples 
   in the simulation from the matching procedure, and the
   fit to the data distribution using the two Monte Carlo components.}
\label{prbg_fit_dt}
\end{figure}

Using as reference the $P_{\gamma}$ distributions for $M=-1$ and 
$M \ne -1$ of a
reconstructed class in the simulation, the $P_{\gamma}$ distribution
of all photons in a class $i$ of data is fitted with these two components, 
and then corrected by the fraction of contaminated parts in the two 
distributions. Figure~\ref{prbg_fit_dt} shows a typical fit in this 
procedure. The fitted fake photon numbers in each observed class
after correction are listed in Table~\ref{nfake_dt}, with the error 
from the fit and the correction.

\begin{table}
\caption{Number of fake photons in 1991-93 and 1994-1995 data sets.}
\begin{center}
\begin{tabular}{lcc}
\hline\hline
Reconstr. class & 91-93 & 94-95    \\\hline
 \clsi  &   2448 $\pm$   75 &   3552 $\pm$  110 \\ 
 \clsii  &    183 $\pm$   24 &    170 $\pm$   26 \\ 
 \clsiii  &   3718 $\pm$   58 &   5766 $\pm$   98 \\ 
 \clsiv  &   9708 $\pm$  196 &  14844 $\pm$  350 \\ 
 \clsv  &   5941 $\pm$  172 &   8837 $\pm$  237 \\ 
 \clsvi  &   2311 $\pm$   83 &   3085 $\pm$   90 \\ 
 \clsxiii  &    574 $\pm$   35 &    677 $\pm$   35 \\ 
 \clsvii  &   5769 $\pm$   88 &   8639 $\pm$  162 \\ 
 \clsviii  &   4130 $\pm$   98 &   6444 $\pm$  112 \\ 
 \clsix  &   1129 $\pm$   48 &   1432 $\pm$   52 \\ 
 \clsx  &    183 $\pm$   18 &    225 $\pm$   24 \\ 
 \clsxi  &     45 $\pm$    6 &     73 $\pm$   13 \\ 
 \clsxii  &     30 $\pm$    6 &     52 $\pm$    8 \\ 
 \clsxiv  &   1369 $\pm$   43 &   2014 $\pm$   62 \\ 
\hline\hline
\end{tabular}
\label{nfake_dt}
\end{center}
\end{table}

\subsection{An independent check}

The total excess of fake photons in data with
respect to Monte Carlo from the fits to the
$P_{\gamma}$ distributions in each observed class can be checked 
comparing the photon energy spectra of data and Monte Carlo, after 
normalization to the number of $\tau$ events and correcting the simulation
for measured branching ratios. The latter method is 
justified by the fact that fake photons show up as an excess at energies
typically less than 1-2 GeV. The former estimate yields
$7279\pm 334$ and $11419\pm 511$ for 1991-1993 and 1994-1995 data samples
respectively summing up all channels, while the latter yields
$7007\pm 307$ and $10820\pm 376$ for the two data sets. The differences
\beqn 
        \Delta_{91-93} &=& 272 \pm 454 \\
        \Delta_{94-95} &=& 599 \pm 634 
\eeqn
indicate that the number of fake photons estimated from the fit
procedure is reliable.

\subsection{Solving for the fake photon correction factors}

In order to reduce the number of parameters to be determined, 
it is assumed that the correction factors of some channels with 
small branching ratios are the same as those of the nearby
channels with comparable multiplicity. Also the $\mu$ channel is not
considered since it is hardly influenced by fake photons and the 
feedthrough to other channels is very small. Therefore the correction 
factors are only fitted for $\clse$, $\h$, $\hu$, $\hd$, $\trh$ and $\thu$ 
channels. The results are given in Table~\ref{fj_corr}.
They are consistent for the two data/Monte Carlo sets. The resulting
$\chi^2/$~degrees of freedom (DF) is 12.4/5 and 15.2/5 for 1991-1993 and 
1994-1995 data respectively. The final error on the fitted correction factors 
has been properly enlarged to take into account these somewhat large values.

\begin{table}
\caption{Fake photon correction factors for 1991-1993 and 1994-1995 data sets.}
\begin{center}
\begin{tabular}{lcc}
\hline\hline
 produced  class  & 91-93 (\%)  & 94-95 (\%) \\\hline
 $\clse$  &   41.0 $\pm$  1.9 &   36.1 $\pm$  2.2   \\
 $\h$     &    8.3 $\pm$  2.2 &   12.1 $\pm$  2.3   \\
 $\hu$    &   21.7 $\pm$  2.2 &   20.5 $\pm$  2.6   \\
 $\hd$    &   13.0 $\pm$  2.7 &   18.7 $\pm$  2.0   \\
 $\trh$   &   12.9 $\pm$  1.6 &   14.0 $\pm$  1.9   \\
 $\thu$   &   25.1 $\pm$  2.5 &   27.4 $\pm$  1.6   \\
\hline\hline
\end{tabular}
\label{fj_corr}
\end{center}
\end{table}

\subsection{Comparison to the previously used method}

In the published analysis using 1991-1993 data~\cite{aleph13_h} 
a much less sophisticated approach was taken. The deficit of fake photons
in the simulation was determined in a global way to be $(16\pm 8)\%$, 
common to all channels. Since it would have been delicate to generate
extra fake photons, the procedure chosen was to actually do the opposite,
{\it i.e.} randomly kill identified (in the sense of the matching
procedure discussed above) fake photons in the simulation, determine the
new efficiency matrix and use the deviations to correct in the opposite
direction.
   
The current way of dealing with the fake photon problem 
is both more precise and more reliable. The fact that different channels
are treated separately provides a handle on the different origins of the 
fake photons, since, for example, fake photons in the \clsiii ~class
only originate from hadronic interactions, whereas they come from both
hadronic interactions and photon shower fluctuations in the  
\clsiv ~class. Also the previous procedure was dependent on the quality
of the photon matching in simulation and the killing of fake photons
entailed the possibility to mistakenly removing some genuine
photons as well.

\section{Determination of the branching ratios}
\label{brdet}

The branching ratios are determined using
\beqn
 n^{obs}_i - n^{bkg}_i &=& \sum_{j} \eff_{ji} N^{prod}_j\\
 B_j &=& \frac {N^{prod}_j} {\sum_{j} N^{prod}_j}
\eeqn
where
$n^{obs}_i$ is the number of observed $\tau$ candidates in reconstructed 
class $i$, $n^{bkg}_i$ the non-$\tau$ background in reconstructed class 
$i$, $\eff_{ji}$ the efficiency of a produced class $j$ event 
reconstructed as class $i$, and $N^{prod}_j$ the produced events 
number of class $j$. The class numbers $i$, $j$ run from 1 to 14, 
the last one corresponding to the rejected $\tau$ candidates.

The efficiency matrix $\eff_{ji}$ is determined from the Monte Carlo, 
but corrected using data for many effects such as particle identification
efficiency, $\tau\tau$ selection efficiency and fake photon simulation. 
It is given in Table~\ref{eff_matrix2} where the last line shows that the 
selection efficiency remains constant within $\pm$~5\% for all 13 
considered topologies.

\begin{table}
\caption{Efficiency matrix for 1994-95 data (in \%). Generated classes are
given in the first row, and reconstructed classes in the first column.
All the corrections are applied except that for nuclear interactions 
and hadron misidentification in multiprong channels. The last row gives the 
selection efficiency for each produced class.}
\begin{center}
{\tiny
\begin{tabular}{|l|rrrrrrrrrrrrr|}
\hline\hline
     &  $\clse$  &  $\clsm$  &  $\h$     &  $\hu$    &  $\hd$    &  $\htr$   &
$\hq$ & $\trh$   &  $\thu$   &  $\thd$   &  $\tht$   &  $\ch$    &  $\chu$ \\\hline
\clsi &73.26& 0.01& 0.41& 0.45& 0.34& 0.25& 0.74& 0.02& 0.02& 0.05& 0.00& 0.00& 0.00\\
\clsii & 0.01&74.49& 0.63& 0.22& 0.07& 0.21& 0.33& 0.01& 0.01& 0.00& 0.00& 0.00& 0.00\\
\clsiii & 0.25& 0.75&65.03& 3.56& 0.34& 0.06& 0.00& 1.44& 0.10& 0.08& 0.00& 0.80& 0.00\\
\clsiv & 1.02& 0.26& 4.70&68.19&11.31& 2.15& 0.49& 0.48& 1.28& 0.62& 0.05& 0.24& 0.00\\
\clsv & 0.12& 0.01& 0.33& 5.67&57.68&23.13& 7.57& 0.08& 0.39& 1.48& 0.24& 0.04& 0.00\\
\clsvi & 0.01& 0.00& 0.07& 0.41& 6.92&43.06&38.15& 0.01& 0.10& 0.37& 0.71& 0.04& 0.00\\
\clsxiii & 0.00& 0.00& 0.02& 0.05& 0.67& 6.25&25.26& 0.00& 0.02& 0.11& 0.19& 0.00& 0.00\\
\clsvii & 0.01& 0.02& 0.25& 0.07& 0.03& 0.00& 0.00&67.98& 6.77& 0.80& 0.03&22.11& 2.52\\
\clsviii & 0.01& 0.01& 0.22& 0.56& 0.27& 0.06& 0.06& 7.29&58.90&16.53& 4.46& 7.07&16.04\\
\clsix & 0.00& 0.00& 0.04& 0.06& 0.10& 0.08& 0.02& 0.41& 6.02&40.42&25.02& 0.28& 0.65\\
\clsx & 0.00& 0.00& 0.00& 0.00& 0.00& 0.01& 0.05& 0.02& 0.41& 6.19&28.98& 0.00& 0.00\\
\clsxi & 0.00& 0.00& 0.00& 0.00& 0.00& 0.00& 0.00& 0.01& 0.01& 0.00& 0.00&38.70& 4.58\\
\clsxii & 0.00& 0.00& 0.00& 0.00& 0.00& 0.00& 0.00& 0.01& 0.02& 0.03& 0.08& 2.99&38.72\\
\clsxiv & 3.27& 4.17& 6.38& 0.73& 1.08& 1.71& 1.75& 0.80& 3.66& 9.96&13.87& 5.03&9.75\\
\hline\\
sum&77.06&79.72&78.08&79.97&78.81&76.97&74.42&78.56&77.71&76.64&73.64&77.30&72.26\\
\hline\hline
\end{tabular}}
\label{eff_matrix2}
\end{center}
\end{table}

It should be noted that the efficiency matrix $\eff_{ji}$ is independent
of the $\tau$ branching ratios used in the simulation, except for the
subclasses contributing to each defined class as shown in 
Table~\ref{classtable}. The effect depends however on small branching
ratios for final states including kaons and the procedure used for 
this correction relies on the ALEPH measured values~\cite{alephksum}.

The analysis assumes a standard $\tau$ decay description. One could
imagine unknown decay modes not included in the simulation, but since
large detection efficiencies are achieved in the $\tau\tau$ 
selection which is therefore robust, so that these decays would be
difficult to pass unnoticed. An independent measurement of the
branching ratio for undetected decay modes, using a direct search with
a one-sided $\tau$ tag, was done in ALEPH~\cite{aleph_undetect},
limiting this branching ratio to less than 0.11\% at 95\% CL. This
result justifies the assumption that the sum of the branching ratios 
for visible $\tau$ decays is equal to unity.

The equations are conveniently solved by a minimization technique and 
all corrections not yet included in the efficiency matrix are applied.
These small final corrections are discussed in the following section 
on systematic studies. The branching ratios are obtained and listed 
in Table~\ref{br_12345}. The results on the class-14 ``branching ratio''
are discussed in Section~\ref{consist}.

\begin{table}
\caption{Branching ratios (\%) from 1991-1993 and 1994-1995 data sets;
the first error is statistical and the second is systematic.}
\begin{center}
\begin{tabular}{|l|c|c|}
\hline\hline
Topology    &  91-93  & 94-95  \\\hline
 \clsi  &   17.859 $\pm$  0.112 $\pm$  0.058 &   17.799 $\pm$  0.093 $\pm$  0.045 \\ 
 \clsii  &   17.356 $\pm$  0.107 $\pm$  0.055 &   17.273 $\pm$  0.087 $\pm$  0.039 \\ 
 \clsiii  &   12.238 $\pm$  0.105 $\pm$  0.104 &   12.058 $\pm$  0.088 $\pm$  0.083 \\ 
 \clsiv  &   26.132 $\pm$  0.150 $\pm$  0.104 &   26.325 $\pm$  0.123 $\pm$  0.090 \\ 
 \clsv  &    9.680 $\pm$  0.139 $\pm$  0.124 &    9.663 $\pm$  0.107 $\pm$  0.105 \\ 
 \clsvi  &    1.128 $\pm$  0.110 $\pm$  0.086 &    1.229 $\pm$  0.089 $\pm$  0.068 \\ 
 \clsxiii  &    0.227 $\pm$  0.056 $\pm$  0.047 &    0.163 $\pm$  0.050 $\pm$  0.040 \\ 
 \clsvii  &    9.931 $\pm$  0.097 $\pm$  0.072 &    9.769 $\pm$  0.080 $\pm$  0.059 \\ 
 \clsviii  &    4.777 $\pm$  0.093 $\pm$  0.074 &    4.965 $\pm$  0.077 $\pm$  0.066 \\ 
 \clsix  &    0.517 $\pm$  0.063 $\pm$  0.050 &    0.551 $\pm$  0.050 $\pm$  0.038 \\ 
 \clsx  &    0.016 $\pm$  0.029 $\pm$  0.020 &   -0.021 $\pm$  0.023 $\pm$  0.019 \\ 
 \clsxi  &    0.098 $\pm$  0.014 $\pm$  0.006 &    0.098 $\pm$  0.011 $\pm$  0.004 \\ 
 \clsxii  &    0.022 $\pm$  0.010 $\pm$  0.009 &    0.028 $\pm$  0.008 $\pm$  0.007 \\ 
 \clsxiv  &    0.017 $\pm$  0.043 $\pm$  0.042 &    0.099 $\pm$  0.035 $\pm$  0.037 \\ 
\hline\hline
\end{tabular}
\label{br_12345}
\end{center}
\end{table}

\section{Determination of systematic uncertainties}
\label{system}

\subsection{Methodology}

Wherever possible the efficiencies relevant to the analysis have been 
determined using ALEPH data, either directly on the $\tau\tau$ sample
itself or on specifically selected control samples, as for example 
in the case of particle identification. The resulting efficiencies 
are thus measured with known statistical errors. 

In some cases the procedure is less straightforward and involves a model 
for the systematic effect to be evaluated. An important example is given
by the systematics in the simulation of fake photons in ECAL. 
In such cases the evaluation of the systematic error not only takes into
account the statistical aspect, but also some estimate of the systematics
involved in the assumed model. The latter is obtained from studies where
the relevant parameters are varied in a range consistent with the
comparison between data and Monte Carlo distributions.

Quite often a specific cut on a given variable can be directly studied.
The comparison between the corresponding distributions, respectively
in data and Monte Carlo, provides an estimate of a possible discrepancy 
whose effect would be to change the assumed efficiency of the cut. If a
significant deviation is observed, a correction is applied to the
simulation to obtain the nominal branching ratio results, while the error
on the deviation provides the input to the evaluation of the systematic
uncertainty. The analysis is therefore repeated with a full 
re-classification of all the measured $\tau$ decay candidates, changing
the incriminated cut by one standard deviation. Since the new samples
slightly differ from the nominal ones because of feedthrough between the
different channels, the modified results are affected both by the 
systematic change in the variable value and the statistical fluctuation
from the event sample which is uncommon to both samples. In this case
the final systematic uncertainty is obtained by adding linearly the modulus
of the systematic deviation observed and the statistical error from
the monitored independent sample. This procedure is followed in a systematic
way for the studies listed below and its description will of course not be 
repeated each time.

Finally, the systematic deviations for each study are obtained with
their sign in each measured decay channel, thus providing the full
information on the correlations between the results and allowing the 
corresponding covariance matrix to be constructed.

\subsection{Selection of $\tau \tau$ events}
\label{selsyst}

Selection efficiencies of all the dominant channels have been determined
from data using the break-mix method as discussed in Section~\ref{breakmix}.
The efficiencies are very similar for the different data sets. They are 
listed in Table~\ref{sel_45} for the 1994-1995 sample.

\begin{table}
\caption{Selection efficiencies (\%) for $\tau\tau$ events, corresponding to
cuts 5-9 described in Section~\ref{select} , measured by the
break-mix method for the 1994-1995 data.}
\begin{center}
\begin{tabular}{|l|c|c|c|}
\hline\hline
  class   & data (BM)          & Monte Carlo (BM)    & R=data/MC \\\hline
 $\clse$  & 95.70 $\pm$ 0.10 & 95.54 $\pm$ 0.07  & 1.0017 $\pm$ 0.0012 \\
 $\clsm$  & 96.10 $\pm$ 0.10 & 96.20 $\pm$ 0.06  & 0.9990 $\pm$ 0.0012 \\
 $\h$     & 96.03 $\pm$ 0.10 & 95.95 $\pm$ 0.07  & 1.0009 $\pm$ 0.0013 \\
 $\hu$    & 97.07 $\pm$ 0.10 & 97.26 $\pm$ 0.04  & 0.9980 $\pm$ 0.0011 \\
 $\hd$    & 95.36 $\pm$ 0.16 & 95.68 $\pm$ 0.10  & 0.9967 $\pm$ 0.0020 \\
 $\trh$   & 94.77 $\pm$ 0.18 & 95.10 $\pm$ 0.10  & 0.9965 $\pm$ 0.0022 \\
 $\thu$   & 92.53 $\pm$ 0.22 & 92.93 $\pm$ 0.19  & 0.9957 $\pm$ 0.0031 \\
\hline\hline
\end{tabular}
\label{sel_45}
\end{center}
\end{table}

The obtained values are then 
allowed to float around the nominal values with their corresponding errors,
and 1000 toy experiments are generated to estimate the effect 
on the branching ratios. The errors are listed in Table~\ref{sys45} 
in the column `sel'.

\subsection{Non-$\tau$ background}

Most non-$\tau$ background contributions in each $\tau$ decay channel 
have been measured directly in data following the procedure described
in Section~\ref{nontau}. The contamination from hadronic $Z$ decays has 
been determined by Monte Carlo simulation. In that case the relative error 
of non-$\tau$ background is taken as 30\%. Toy experiments are used where 
new values for the non-$\tau$ background are generated according to Gaussian
distributions with the measured error. Branching ratios are reevaluated
in each case and the resulting standard deviations are quoted as errors.
They are given in Table~\ref{sys45} in the column `bkg'.

\subsection{Particle identification}

The same technique is used to estimate the systematic error on particle 
identification and misidentification efficiencies which are obtained
from measurements on control samples (see Section~\ref{pid}) with their 
own statistical and systematic uncertainties. The errors on the branching 
ratios are listed in Table~\ref{sys45} in the column `pid'.

\subsection{Photon detection efficiency}

Photon detection efficiency at low energy and high energy are studied 
by different means.

  \subsubsection{Photon efficiency at low energy}

\begin{figure}
   \centerline{\psfig{file=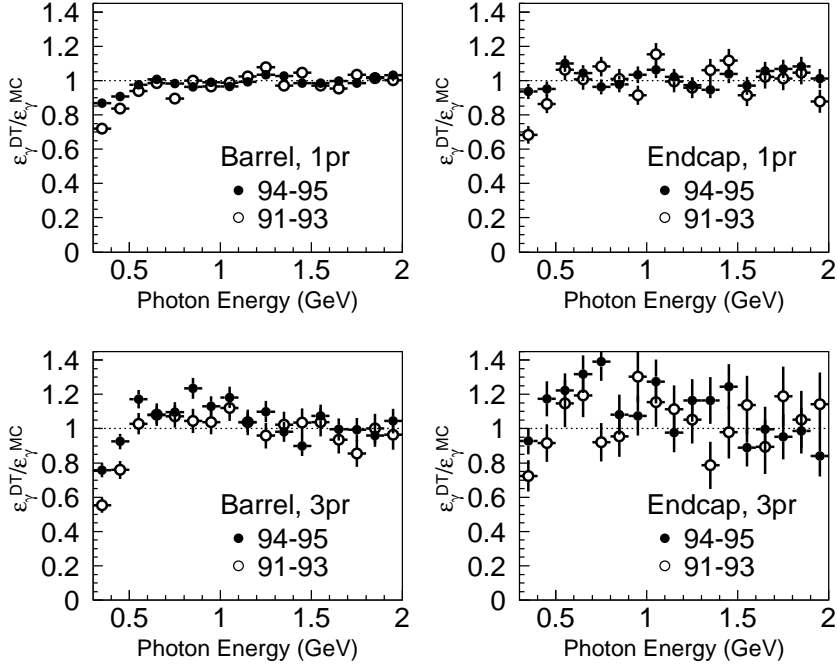,width=110mm}}
   \caption{Relative photon efficiency in data and Monte Carlo for
   1991-1993 (open circles) and 1994-1995 (black dots) data sets.
   The plots are separated for the barrel and endcap sections of the
   calorimeter, and for one- and three-prong $\tau$ decays.}
\label{effg}
\end{figure}

The number of real photons in the data sample is obtained from the fit of 
the $P_{\gamma}$ distribution in each energy bin with the distributions
of genuine and fake photons from Monte Carlo sample. The energy 
spectra of genuine photons for both data and Monte Carlo samples are
then compared in order to obtain the relative photon efficiency 
between data and Monte Carlo (Fig.~\ref{effg}). Some discrepancy 
is observed at low energy, indicating that the efficiency in 
the data sample is different from the simulation. It should be
emphasized that this effect originates almost exclusively from
the detector simulation which was improved for the 1994-1995 sample. 
It is taken into account in the analysis by randomly killing Monte Carlo 
photons in the corresponding energy bins. The error on the efficiency is 
propagated to the final branching ratios and are given in 
Table~\ref{sysg45} in the column `eff'.

\begin{table}
\caption{Systematic error for photon and $\pi^0$ reconstruction 
in 1994-1995 data sample. All numbers are absolute branching ratios
in \%. The labels are defined as follows: photon efficiency at low energy 
(eff), photon efficiency at high energy (dgt), converted photons (cnv),
photon identification efficiency (prb), photon energy calibration (cal),
fake photon correction (fak), $\pi^0$ reconstruction efficiency (dij),
bremsstrahlung and radiative photons (bms). The total $\pi^0$ systematic
uncertainty is given in the rightmost column.}
{\footnotesize
\begin{center}
\begin{tabular}{lccccccccc}
\hline\hline
Topology & eff& dgt& cnv& prb& cal&
                       fak& dij& bms& total $\piz$ \\\hline
 \clsi & 0.006 & 0.002 & 0.005 & 0.000 & 0.004 & 0.003 & 0.005 & 0.003 & 0.011 \\ 
 \clsii & 0.002 & 0.001 & 0.001 & 0.000 & 0.002 & 0.000 & 0.001 & 0.002 & 0.004 \\ 
 \clsiii & 0.022 & 0.029 & 0.006 & 0.056 & 0.009 & 0.011 & 0.002 & 0.019 & 0.071 \\ 
 \clsiv & 0.015 & 0.011 & 0.009 & 0.024 & 0.011 & 0.048 & 0.006 & 0.023 & 0.063 \\ 
 \clsv & 0.033 & 0.013 & 0.013 & 0.041 & 0.018 & 0.048 & 0.025 & 0.038 & 0.089 \\ 
 \clsvi & 0.012 & 0.019 & 0.010 & 0.035 & 0.014 & 0.008 & 0.030 & 0.013 & 0.056 \\ 
 \clsxiii & 0.017 & 0.012 & 0.005 & 0.012 & 0.006 & 0.010 & 0.007 & 0.007 & 0.029 \\ 
 \clsvii & 0.028 & 0.012 & 0.007 & 0.029 & 0.010 & 0.012 & 0.005 & 0.011 & 0.046 \\ 
 \clsviii & 0.004 & 0.003 & 0.010 & 0.014 & 0.006 & 0.020 & 0.010 & 0.017 & 0.033 \\ 
 \clsix & 0.016 & 0.010 & 0.008 & 0.012 & 0.006 & 0.008 & 0.003 & 0.007 & 0.027 \\ 
 \clsx & 0.002 & 0.004 & 0.003 & 0.004 & 0.003 & 0.002 & 0.005 & 0.004 & 0.010 \\ 
 \clsxi & 0.001 & 0.001 & 0.000 & 0.001 & 0.000 & 0.000 & 0.000 & 0.001 & 0.002 \\ 
 \clsxii & 0.001 & 0.000 & 0.001 & 0.001 & 0.001 & 0.000 & 0.001 & 0.001 & 0.002 \\ 
 \clsxiv & 0.008 & 0.004 & 0.004 & 0.002 & 0.003 & 0.002 & 0.004 & 0.006 & 0.013 \\ 
\hline\hline
\end{tabular}
\label{sysg45}
\end{center}}
\end{table}

One could question whether the above effect is indeed 
related to the simulation of photon efficiency or rather 
induced by different physics in data and in simulation.
The fact that the first explanation is the correct one is made clear by the
following observations: (1) the discrepancy occurs only near photon-detection 
threshold with a characteristic shape, (2) the branching ratio values are
adjusted in the Monte Carlo to match those measured in data, 
(3) the effect is observed in all photonic classes, and (4) the effect 
is quantitatively different in the barrel and endcap parts of ECAL.

The efficiency is measured for 1991-1993 and 1994-1995 samples separately,
and the values are found to be different for 1-prong and 
3-prong $\tau$ decays, which is reasonable considering 
the different environment of hadronic interactions in ECAL. 
The efficiencies in the ECAL barrel and endcaps are also measured 
separately, because the photon performance is quite different in the 
endcaps due to more material present in front of the calorimeter.

  \subsubsection{Photon efficiency at high energy}

\begin{figure}
   \centerline{\psfig{file=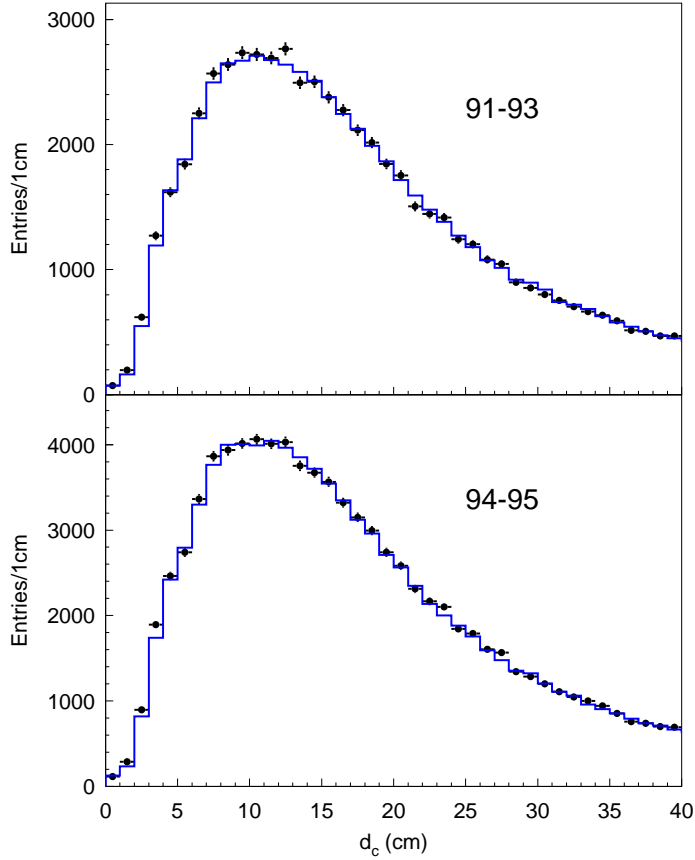,width=90mm}}
   \caption{Comparison of $d_c$ distributions for high energy photons
   ($E_\gamma>3~GeV$) in data (points) and in Monte Carlo (histogram) 
   for 1991-1993 and 1994-1995 data sets. The distributions are normalized 
   to each other for $d_c>8$~cm.}
\label{dc_plt}
\end{figure}

A loss of high energy photons occurs because of the merging 
of their showers into nearby charged track clusters. 
Photon reconstruction requires a cut on $d_c$ at 2~cm in order 
to separate the photon candidate from the charged track cluster. 
The corresponding loss of efficiency depends on the simulation of 
hadronic interactions in ECAL, which
can be checked against data by comparing the actual $d_c$ distributions.
Figure~\ref{dc_plt} shows the comparison for photons with energy 
greater than 3~GeV. After normalization to $ d_c \ge 8$~cm, 
the fraction of photons below $8$~cm is $(0.30\pm 0.15)\%$
more in data for 1991-1993 and is $(0.37\pm 0.13)\%$ more in data 
for 1994-1995.

The observed difference is applied in the analysis and varying it by 
one standard deviation provides an estimate of the systematic error 
induced by the $d_c$ cut.
The errors are listed in Table~\ref{sysg45} in the column `dgt'.

\subsection{Converted photons}

The rate of reconstructed converted photons depends on the 
conversion probability and the misidentification of hadronic tracks 
in a multihadron environment. The misidentification of protons originating
from nuclear interactions should also be considered in this context 
since their energy loss by ionization can be close to that expected 
for electrons in the low momentum range.

\begin{figure}
   \centerline{\psfig{file=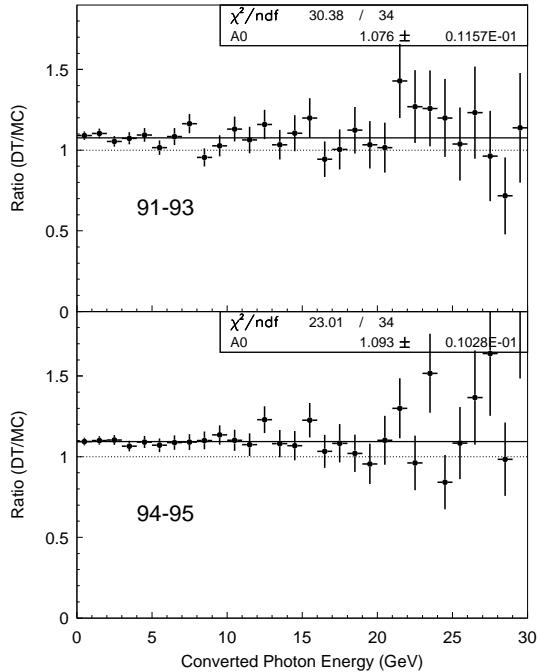,width=70mm}}
   \caption{Ratio of converted photon energy spectra in 
   data and Monte Carlo.}
\label{conv}
\end{figure}

Figure~\ref{conv} shows the ratio of converted photon spectra in data and
Monte Carlo, indicating an inadequate simulation of the rate of 
converted photons. Since in the analysis converted photons are given 
an identification probability of 1, the net effect is an efficiency 
increase for photons in data. To correct for this, some converted 
photons in the data sample are randomly declared to be calorimeter 
photons according to the observed excess and given an identification
probability value according to the corresponding ECAL photon
distribution. Their fate in the analysis thus follows that of 
the ordinary calorimetric photons which can be removed when probability 
cuts are applied. 

The error on the conversion rate ratio is propagated to branching 
fractions through a full reanalysis. The corresponding systematic errors
are listed in Table~\ref{sysg45} in the column `cnv'.

\begin{figure}
   \centerline{\psfig{file=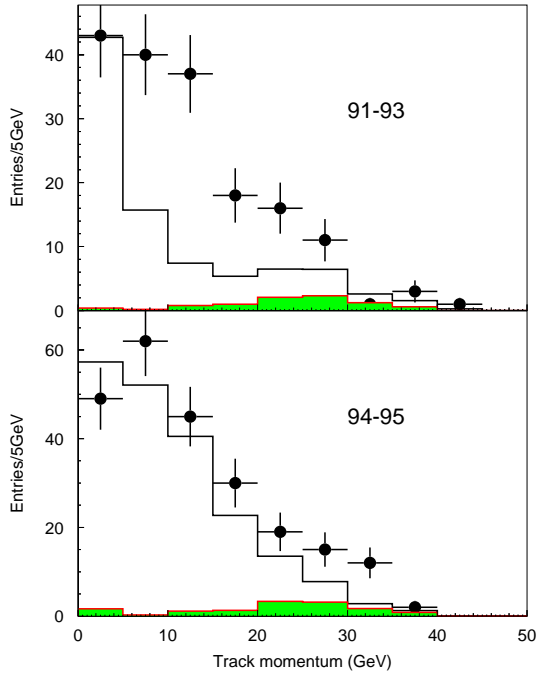,width=70mm}}
   \caption{Comparison of hadron misidentification in multihadronic final
   states (see text for details). Data are given by points and 
   Monte Carlo by histogram, while the shaded
   histogram shows the contribution from non-$\tau$ background. A clear
   improvement is observed in the 1994-1995 simulation.}
\label{conv2}
\end{figure}

The misidentification of hadrons to electrons in a multihadron 
environment is measured by comparing events reconstructed as 
$\thu$ with two tracks identified as hadrons and one reconstructed 
converted photon: this sample is dominated by produced $\trh$ 
events with one hadronic track misidentified as electron. 
Comparing the event numbers in data and Monte Carlo after normalization
to the total $\tau$ event number, the rate of hadron misidentification
is found to be $(0.06\pm 0.03)\%$ lower in simulation
for the 1994-1995 data sample, corresponding to an absolute change in the
$\trh$ fraction of $(+0.010\pm 0.005)\%$.
A bigger difference is observed in the 1991-1993 data sample where
the misidentification probability is found to be $(0.25\pm 0.04)\%$
lower in the Monte Carlo. This problem only affects feedthrough between 
nearby 3-prong channels, the largest correction being to the $\trh$ channel,
$(+0.034\pm 0.007)\%$, while the effect on
$\thu$ and $\thd$ modes is $(-0.019\pm 0.004)\%$ and 
$(-0.015\pm 0.003)\%$, respectively. The found difference is illustrated
in Fig.~\ref{conv2} where a comparison of the data and Monte Carlo 
distributions of the momenta of the converted photons is shown.
The error of this correction is included under `cnv' in
Table~\ref{sysg45}.

\begin{figure}
   \centerline{\psfig{file=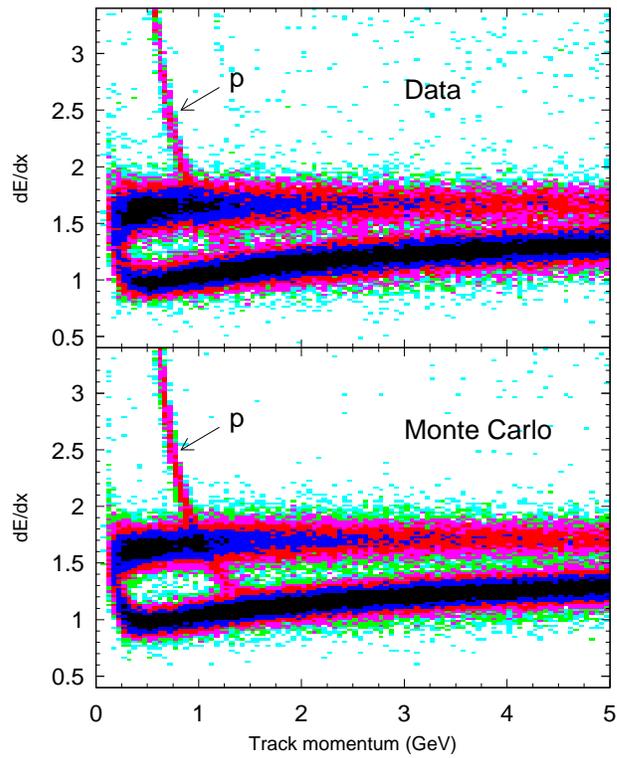,width=80mm}}
   \caption{Measured $dE/dx$ as a function of momentum for all 
   (good and bad) tracks in the selected $\tau$ sample: 
   the proton band from nuclear interactions is clearly observed, 
   and predicted by the simulation.}
\label{ddx}
\end{figure}

Proton tracks are observed when hadrons interact with the material of
the inner detectors. Since at low energy, only $dE/dx$ information is used
for particle identification, protons can be misidentified as electrons for
momenta between 0.9 and 1.1~GeV, as seen in Fig.~\ref{ddx}. 
By selecting tracks with large $dE/dx$ at even lower momenta so that 
protons dominate the sample, the proton rates can be compared
in data and in simulation directly. Good agreement is observed and the
systematic error on $\tau$ branching ratios from proton tracks in the
photon conversions is negligible.

\subsection{Photon identification efficiency}

\begin{figure}
   \centerline{\psfig{file=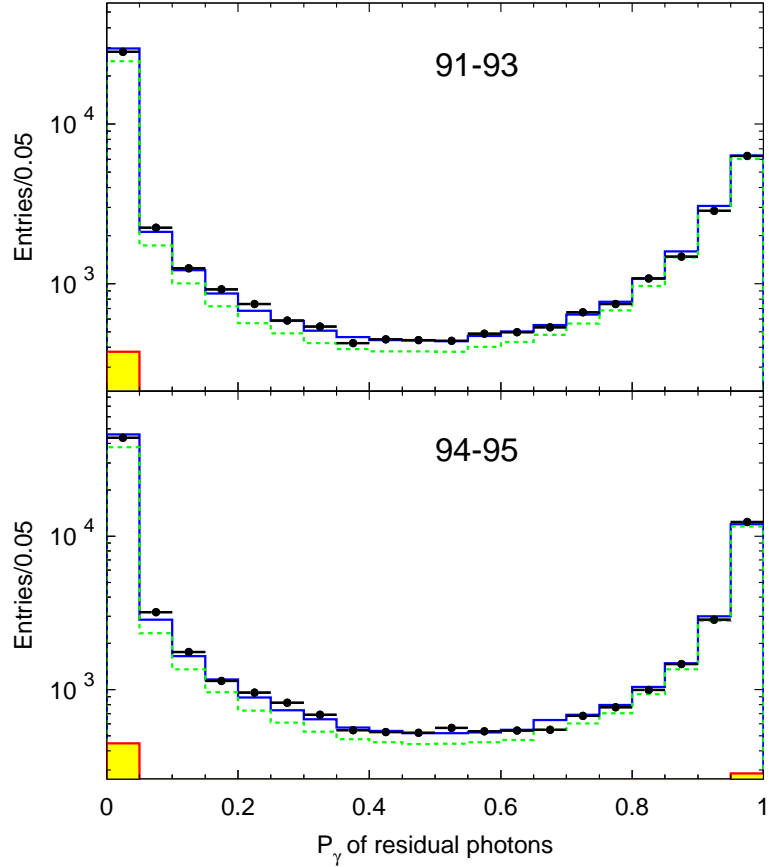,width=100mm}}
   \caption{Comparison of photon identification probability distributions in 
   data (points) and Monte Carlo (solid histogram, after fake photon 
   correction) for residual photons in the 0.3-3.0~GeV energy range. 
   The dashed histogram shows the simulated distribution before fake
   photon correction, and the shaded histogram indicates the contribution of
   non-$\tau$ background.}
\label{prbg_dt_mc}
\end{figure}

Since the reference distributions for photon identification 
are not exactly the same for data and Monte Carlo (even after corrections), 
the same probability cut will have different efficiencies for the two cases.
Figure~\ref{prbg_dt_mc} shows the comparison of $P_{\gamma}$ distributions 
between data and simulation for residual photons in the 0.3-3.0~GeV range: 
good agreement is observed overall, however some small shape 
differences can be seen. A comparison of the same distributions 
is performed for photons in small energy slices, after correcting 
the branching ratios in the Monte Carlo and for the non-$\tau$ background. 
The numbers of photons surviving the cut are compared, and the efficiency 
is found to be a little larger in the simulation. This effect 
is then corrected in the analysis by using a slightly tighter cut for 
the simulation, corresponding to the observed difference in the numbers 
of accepted photons. The systematic errors on branching ratios appear 
in Table~\ref{sysg45} in the column `prb'.

\subsection{Photon energy calibration}

The photon energy scale relative to the simulation is calibrated using 
high energy electrons from $\tau$ decays as discussed in 
Section~\ref{calphot}. The parameterization of the $E/p$ ratio is
extrapolated to low energy for both data and Monte Carlo. 
The error on the calibration factors is propagated to
branching ratios and are listed in Table~\ref{sysg45}(`cal').

\subsection{Fake photon correction}

The error on the fake photon correction comes from the uncertainty
in measuring the total number of fake photons in each observed class 
for data, and in determining the fake photon multiplicity in each 
generated Monte Carlo event, using the matching procedure described
in Section~\ref{match}. The latter has a negligible effect
on branching ratios since the dependence of the efficiency matrix 
on the fake photon multiplicity is weak. The fake photon numbers are 
determined from fits to the $P_{\gamma}$ distributions, and the errors 
are obtained in the global fit of the fake photon correction factors. 
These errors are then propagated to branching ratios. 
To reduce the total number of parameters
the correction factors on low branching ratio channels are assumed
to be equal to those of the nearby high branching ratio channels, thus
somewhat increasing the $\chi^2$ of the fits. The fitted errors
are then enlarged by the $\sqrt{\chi^2/DF}$ values in each channel.
The errors are reported in Table~\ref{sysg45} in the column `fak'.

\subsection{ $\piz$ reconstruction efficiency}

\begin{figure}
   \centerline{\psfig{file=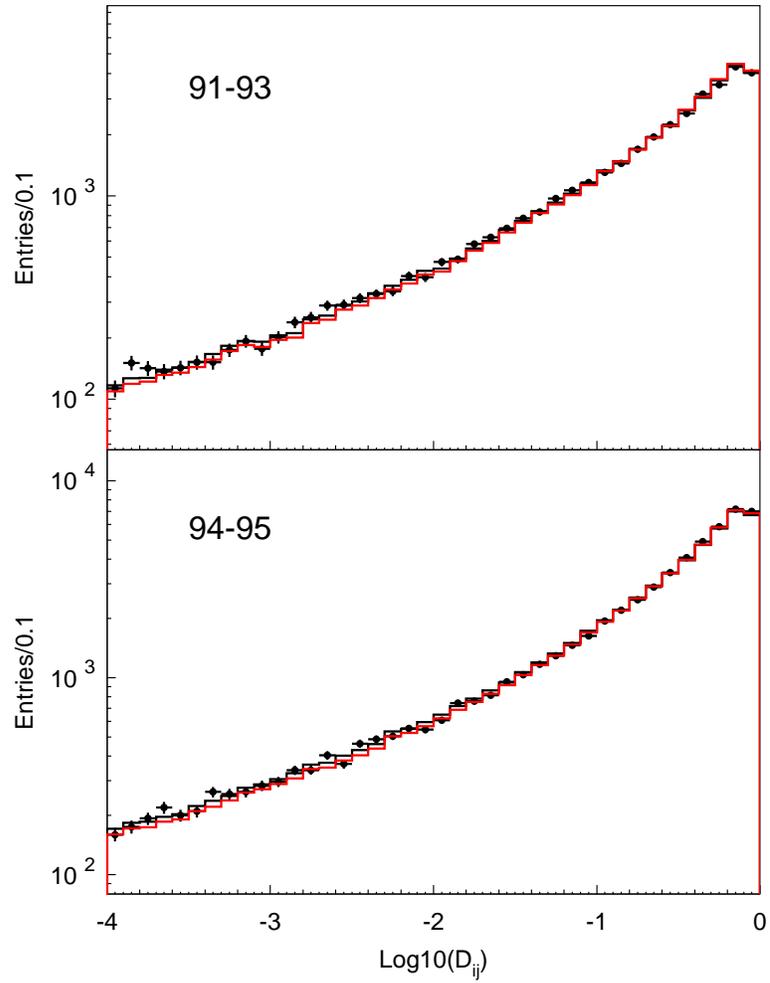,width=100mm}}
   \caption{Comparison of $D_{ij}$ distributions in data (points)
   and simulation (dark histogram) for 1991-1993 and 1994-1995 data sets. 
   The light histograms show the distribution before fake photon correction.}
\label{dij}
\end{figure}

To estimate the uncertainty on the reconstruction of the resolved
$\pi^0$'s, a comparison of the $D_{ij}$ (Section~\ref{pi0rec}) 
distributions is shown in Fig.~\ref{dij} 
after branching ratio, fake photon, and non-$\tau$ background
corrections are applied in the simulation. Good agreement is observed.
The agreement on the reconstruction efficiencies of
resolved $\pi^0$'s  in data and Monte Carlo is tested with a precision 
of 0.17\% and 0.14\% for the 1991-1993 and 1994-1995 samples respectively. 
They can be reproduced by changing the value of the $D_{ij}$ cut by
$0.14\times 10^{-4}$ and $0.12\times 10^{-4}$ for 1991-1993 and 
1994-1995 respectively, the nominal cut being set at $1\times 10^{-4}$. 
The effect of changing the cut by this amount in the simulation was taken 
as the systematic error, listed in Table~\ref{sysg45} in the column `dij'.

\subsection{ Simulation of bremsstrahlung and radiative photons}
\label{systrad}

\begin{figure}
   \centerline{\psfig{file=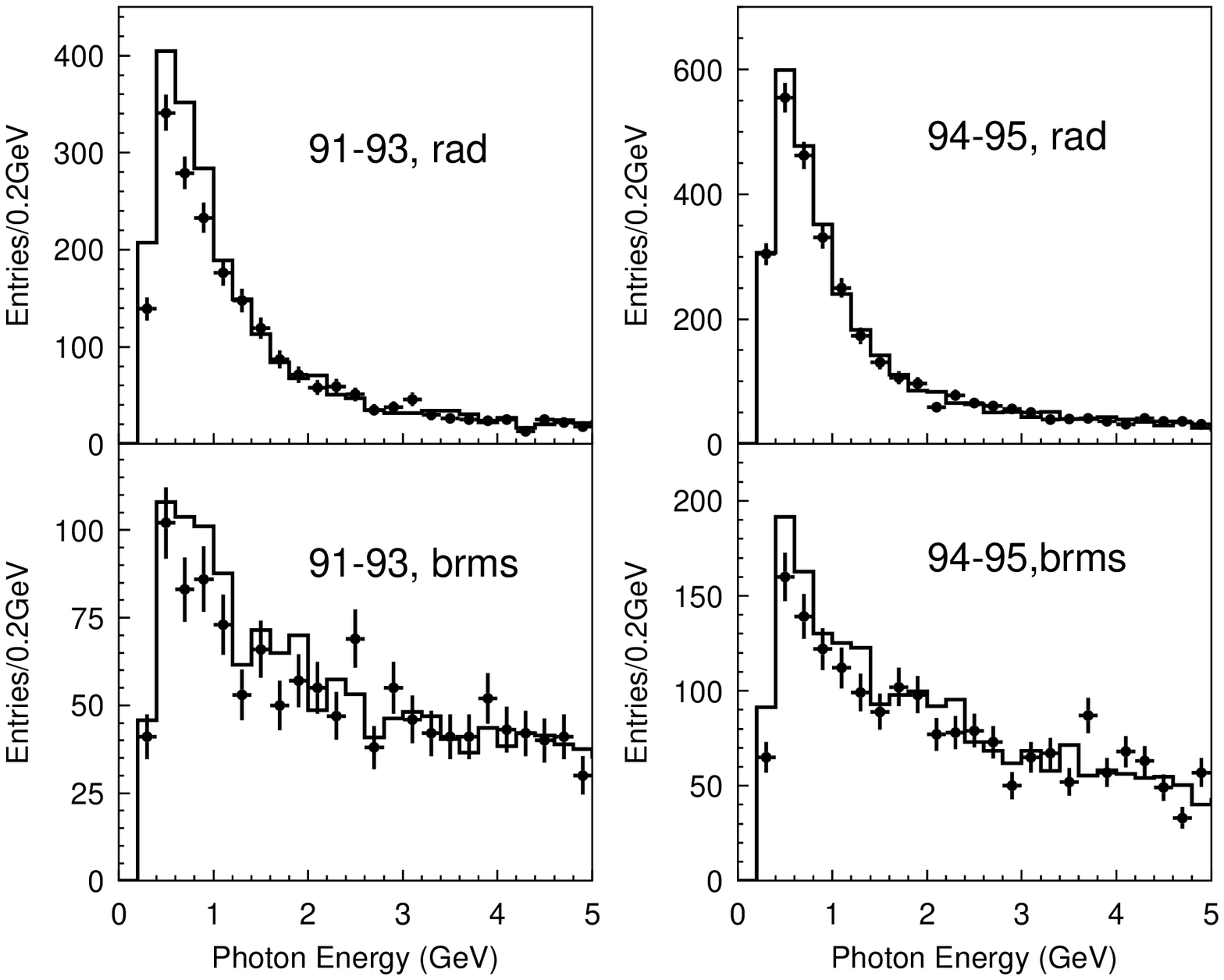,width=100mm}}
   \caption{Comparison of measured energy spectra of bremsstrahlung
   and radiative photons in data (points)
   and Monte Carlo (histogram) for 1991-1993 and 1994-1995 data sets. }
\label{brmsrd}
\end{figure}

\begin{figure}
   \centerline{\psfig{file=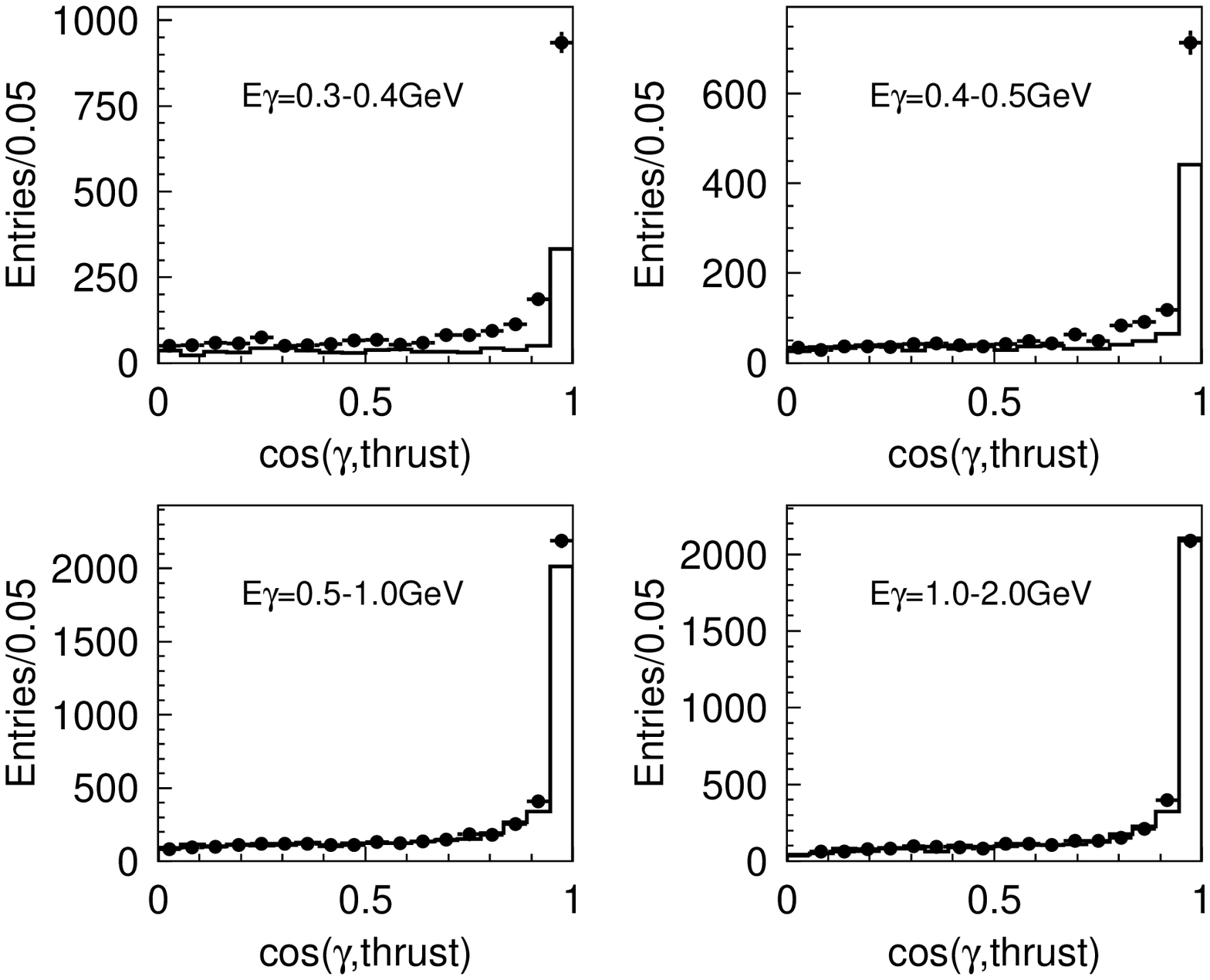,width=100mm}}
   \caption{Comparison of the angle between produced radiative photons 
   and jet axis in 1991-1993 (points) and 1994-1995 (histogram) 
   simulated samples for different energy slices.}
\label{rdmc}
\end{figure}

Figure~\ref{brmsrd} shows the comparison of the energy spectra of
selected radiative and bremsstrahlung photons in data and simulation
after normalization to the total number of $\tau$ events. Reasonable 
agreement is found except for the radiative photon spectrum in the 1991-1993 
sample, where more photons are found in the simulation. 
A comparison of 1991-1993 and 1994-1995 simulated samples clearly 
shows that too many radiative photons were produced in 1991-1993 
at the generator level (Fig.~\ref{rdmc}). To correct for this, 
photons matched as radiative in the 1991-1993 Monte Carlo 
sample are deleted randomly according to the 1994-1995 
distribution, which agrees with data. The angle between the radiative 
photon and the jet axis is also taken into account in this procedure. 

The statistical error in the measurement of the numbers of radiative 
and bremsstrahlung photons is propagated to the $\tau$ branching ratios and 
listed in Table~\ref{sysg45} in the column `bms'.

It turns out that most of the bremsstrahlung photons originate from radiation
in the detector material by electrons in the decay 
$\tau^- \rightarrow \nu_\tau e^- \overline{\nu}_e$. The comparison in
Fig.~\ref{brmsrd} shows that the description of radiation is adequate in the
simulation. However, any difference at this level would only affect the
electron momentum spectrum, but not the decay classification.

The situation is more delicate for radiative photons in hadronic decay 
channels. From simulation, it is found that only 23\% of these photons
are recognized as such, while the purity in the identified radiative
photon sample is about 55\% (others photons are either fake or originating
from $\piz$ decays). These values are weakly dependent on the decay channel.
Since the simulation is relied upon for the unobserved radiative photons, it
is important to validate the radiation procedure in the simulation.

In KORALZ07~\cite{was} a model for radiation based on PHOTOS~\cite{photos} 
is implemented: charged particles independently emit photons according
to a bremsstrahlung probability. A test of this model can be performed in the 
$\pi\piz$ channel where a complete calculation of radiative corrections 
exists~\cite{ecker}.
If a photon is radiated by either the $\rho$ or the charged $\pi$, 
then the final state can be identified as $\pi\piz$ if the radiative 
photon is classified as a bremsstrahlung or radiative photon, or undetected. 
Otherwise such a decay would be (wrongly) classified in the $\pi2\piz$ 
class and the corresponding bias should be corrected for through the 
simulation. It is thus important to verify the radiation model against 
the full calculation. A good agreement is found for the radiative rate for 
photon energies $E_\gamma^*$ above 12~MeV in the $\tau$ centre-of-mass frame,
comparable to the detection threshold of 350~MeV in the $e^+e^-$ frame: 
the simulation using PHOTOS yields a radiative branching ratio of 
$(2.91 \pm 0.04)\times 10^{-3}$ to be compared to a value of 
$2.9\times 10^{-3}$ from the exact calculation. 
In both cases, the radiative yield includes photons emitted 
by the $\tau$ lepton and by the charged $\rho$ and $\pi$ in the decay. 
The simplified bremsstrahlung probability is however no longer adequate 
for hard radiation: the corresponding values for  $E_\gamma^* > 300$~MeV are
$(0.90 \pm 0.09)\times 10^{-4}$ for the simulation 
and $1.8\times 10^{-4}$ for the calculation. 
Although significant, such a difference has no practical influence
in this analysis, since it only affects a negligible part of the radiative
photon spectrum. The comparison of the total radiative yield above detection 
threshold leads to an absolute systematic error of 0.005\% on the branching ratio
in the $\pi\piz$ channel, a factor of 3 to 4 smaller than the estimated uncertainties
quoted in Table~\ref{sysg45} from the direct data/simulation 
comparison of the radiative photon spectra. The latter values are therefore 
retained as final uncertainties for this source.

\subsection{ Simulation of $\piz$ Dalitz decays}

The simulation of the $\piz$ Dalitz decays is checked at the generator
level. Good agreement is found with the world average branching ratio.
The quality of the simulation of $\piz$ Dalitz decays can also be
checked in the distribution of the reconstructed photon conversion 
point near interaction point which agree well with data  
(Fig.~\ref{conv_pt}). The possible systematic effect in the simulation 
of $\piz$ Dalitz decays on the $\tau$ branching ratios is found 
to be negligible.

\begin{table}
\caption{Total systematic errors for branching ratios measured from the 
     1994-1995 data sample. All numbers are absolute in per cent.
     The labels are defined as follows: photon and $\piz$ reconstruction
     ($\piz$), event selection efficiency (sel), non-$\tau$ background 
     (bkg),  charged particle identification (pid), 
     secondary interactions (int), tracking (trk), Monte Carlo dynamics
     (dyn), Monte Carlo statistics (mcs), total systematic
     uncertainty (total).}
{\footnotesize
\begin{center}
\begin{tabular}{lccccccccc}
\hline\hline
Topology    &  $\piz$ & sel & bkg & pid & int & 
              trk & dyn & mcs & total \\\hline
 \clsi & 0.011 & 0.021 & 0.029 & 0.019 & 0.009 & 0.000 & 0.000 & 0.015 & 0.045 \\ 
 \clsii & 0.004 & 0.020 & 0.020 & 0.021 & 0.008 & 0.000 & 0.000 & 0.015 & 0.039 \\ 
 \clsiii & 0.071 & 0.016 & 0.010 & 0.022 & 0.022 & 0.014 & 0.000 & 0.019 & 0.083 \\ 
 \clsiv & 0.063 & 0.027 & 0.019 & 0.011 & 0.045 & 0.009 & 0.000 & 0.027 & 0.090 \\ 
 \clsv & 0.089 & 0.021 & 0.014 & 0.004 & 0.007 & 0.003 & 0.040 & 0.028 & 0.105 \\ 
 \clsvi & 0.056 & 0.012 & 0.015 & 0.000 & 0.008 & 0.001 & 0.008 & 0.030 & 0.068 \\ 
 \clsxiii & 0.029 & 0.005 & 0.011 & 0.000 & 0.015 & 0.000 & 0.000 & 0.019 & 0.040 \\ 
 \clsvii & 0.047 & 0.021 & 0.018 & 0.004 & 0.012 & 0.014 & 0.006 & 0.015 & 0.059 \\ 
 \clsviii & 0.033 & 0.017 & 0.029 & 0.002 & 0.041 & 0.009 & 0.007 & 0.018 & 0.066 \\ 
 \clsix & 0.027 & 0.008 & 0.015 & 0.000 & 0.009 & 0.003 & 0.012 & 0.014 & 0.038 \\ 
 \clsx & 0.010 & 0.012 & 0.002 & 0.000 & 0.002 & 0.001 & 0.010 & 0.006 & 0.019 \\ 
 \clsxi & 0.002 & 0.000 & 0.002 & 0.000 & 0.000 & 0.001 & 0.000 & 0.003 & 0.004 \\ 
 \clsxii & 0.002 & 0.000 & 0.006 & 0.000 & 0.000 & 0.000 & 0.000 & 0.002 & 0.007 \\ 
 \clsxiv & 0.013 & 0.003 & 0.022 & 0.002 & 0.024 & 0.000 & 0.000 & 0.011 & 0.037 \\ 
\hline\hline
\end{tabular}
\label{sys45}
\end{center}}
\end{table}

\subsection{ Nuclear interactions}

Secondary interactions of charged hadrons produced from $\tau$ decays 
with material in the inner part of the detector are studied 
with the help of an enriched sample.
Interaction events are characterized by the presence of many bad tracks 
and proton tracks. By requiring more than three bad
tracks in the hemisphere or at least one proton track identified
with energy loss information, interaction events are selected
with an efficiency of about 50\% and a purity of 70\%, as estimated
from the simulation.

\begin{figure}
   \centerline{\psfig{file=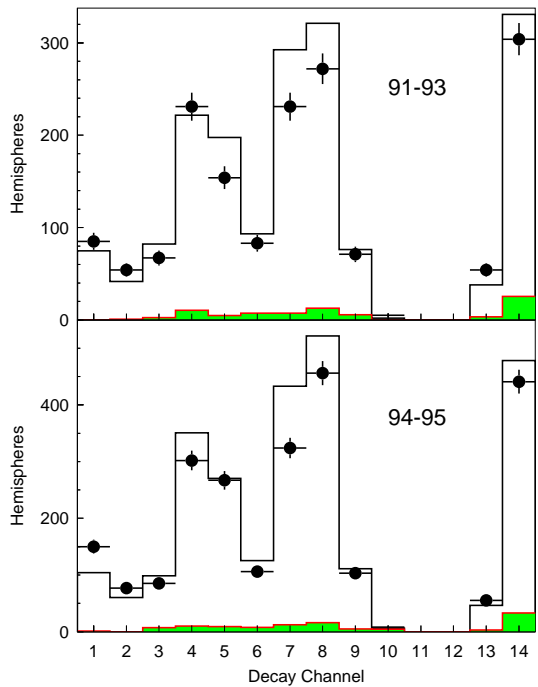,width=70mm}}
   \caption{Comparison of the numbers of selected $\tau$ decays with 
   secondary interactions for each channel (see Table~\ref{classtable}
   in data (points) and Monte Carlo (histogram)). 
   The shaded part gives the contribution from non-$\tau$ background,
   mainly from $Z\rightarrow q\bar{q}$ events. The secondary tracks found
   in leptonic channels (classes 1 and 2) originate from the opposite
   (hadronic) decay.}
\label{inter}
\end{figure}

Figure~\ref{inter} shows the comparison of data to Monte Carlo of the 
observed $\tau$ decays in each reconstructed class for this 
interaction-rich sample, after normalizing to the total number of $\tau$
events. Globally it is found that the MC overestimates the
rate for secondary interactions by about $10\%$, with some 
channel-to-channel variation.

By comparing the total number of interaction events in data and simulation,
the difference of interaction rates is measured and corrected, modifying
the efficiency matrix slightly. The branching ratios are then
reevaluated with this new efficiency matrix. The differences
between the new branching ratios and those without correction
are taken as the conservative estimate of the systematic error
from the simulation of nuclear interactions. This procedure is justified
by the fact that, although the rate differences can be well measured, it 
does not ensure that the interaction dynamics is properly simulated. 
As shown in Fig.~\ref{inter}, even after the applied global rate 
correction the agreement between data and Monte Carlo, although
much improved, remains not very satisfactory.

\begin{figure}
   \centerline{\psfig{file=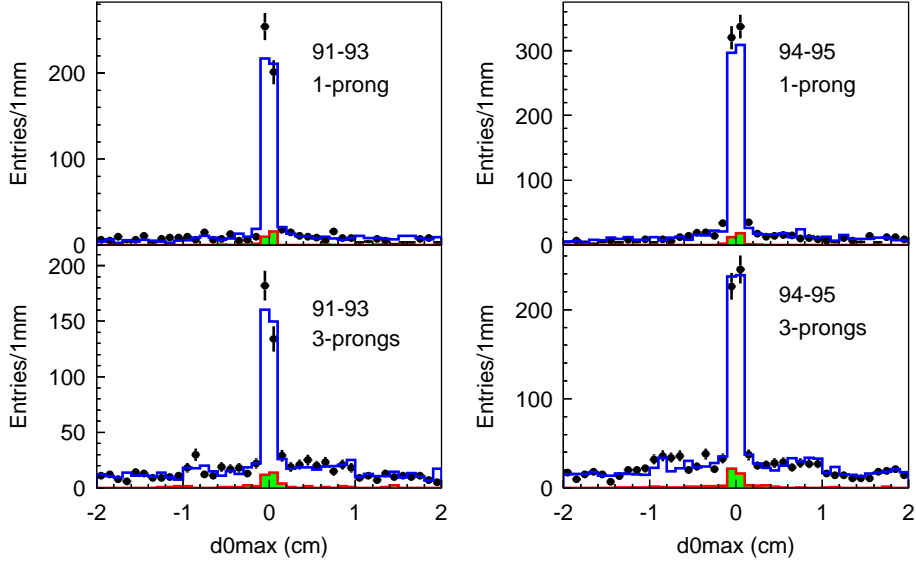,width=120mm}}
   \caption{Comparison of the distributions of the maximum $d_0$ value for
   selected $\tau$ decays with secondary interactions in data (points) 
   and Monte Carlo (histogram), the shaded part shows the contribution from 
   non-$\tau$ background, mainly from $Z\rightarrow q\bar{q}$ events.
   The global interaction rate difference has been corrected for.}
\label{d0max}
\end{figure}

Another approach to study the influence of secondary interactions
relies on the distribution of the distance in the transverse plane 
between the extrapolated track and the interaction point. 
Figure~\ref{d0max} shows the corresponding comparison of data to simulation
(in three-prong channels, the largest of the three distances is plotted). 
After the global interaction rate correction, good agreement
is observed for both one-prong and three-prong channels for 1991-1993
and 1994-1995 data sets, thus corroborating the findings of the other
study.

The errors are listed in Table~\ref{sys45} in the column `int'.

\subsection{ Tracking}

The only situation where tracking efficiency is relevant 
occurs  in a three-prong decay when two of the tracks are lost. 
This effect can be studied by looking at events
with same sign hemispheres. In order to reduce the effect of secondary 
interactions and asymmetric photon conversions, the requirements of 
no proton track in both hemispheres, less than four bad tracks and no track
classified as converted photon, are made. In this way $121\pm 9$ events 
are obtained in the Monte Carlo sample after normalizing to the total 
$\tau$ events in the 1994-1995 data sample.
The efficiency of measuring same sign events resulting from tracking is 
found to be 55\% in simulation with a purity of 60\%,
the remainder being dominated by
one-prong $\tau$ decays with nuclear interactions. The corresponding
number from the data sample is $180\pm 13$, significantly larger
than the rate seen in the simulation. Taking into account the 
corrected simulation rate of nuclear interactions as determined 
in the preceding section, the data sample produces an estimated 
excess of $(81\pm 24)$\% same sign hemispheres
compared to the simulation. This value corresponds to a tracking
efficiency difference of $(0.3\pm 0.1)$\% between data and simulation.
This correction is applied to the efficiency matrix element corresponding
to produced three-prong $\tau$ decays reconstructed as 1-prong,
which increases by about 40\%,
only about half of the three-prong $\tau$ decays reconstructed as
one-prong are due to tracking efficiency, the another half originating from
$K^0_s\rightarrow \pi^+\pi^-$ decays with their secondary vertex far from 
the interaction point. The efficiency of three-prong to three-prong decays
is decreased accordingly. The error on the tracking efficiency is
propagated to the $\tau$ branching ratio measurement. 
The largest correction in the branching ratios due to tracking 
is in the $\h$ channel, which decreases by $(0.047\pm 0.014)$\% 
absolutely, while the $\trh$ channel correspondingly 
increases by the same amount.

The same analysis gives ($104\pm 7$) same sign hemisphere events in 
1991-1993 Monte Carlo sample, while ($109\pm 10$) events are observed in
data, thus these are in good agreement. Following the same procedure, the excess of
misreconstructed events in the data sample is $(9\pm 21)$\%. No correction is 
taken into account this time and the systematic uncertainty is evaluated 
in the same way as for 1994-1995 sample.

The errors are shown in Table~\ref{sys45} in the column `trk'.

\subsection{ Dynamics}

Uncertainties in the dynamics of the hadronic $\tau$ decays can lead to
systematic biases when computing the efficiency matrix from the
simulation. The fact that selection efficiencies are large and weakly
dependent on the hadronic invariant mass is an important factor limiting
those biases. Nevertheless systematic checks have been performed,
comparing the measured mass distributions in all reconstructed channels
to their simulated counterparts. 

Some small differences are observed in the $3h$ and $h 2\piz$ channels.
However no new evaluation of this systematic uncertainty was done 
for this analysis as it is not dominant. The result is taken from the 
published paper on 1991-1993 data~\cite{aleph13_h} which was 
based on detailed tests of several models describing the 
$a_1$ resonance. Similarly, systematic uncertainties
were derived for the higher multiplicity modes by comparing decay models
with different resonance contributions in the final states. Finally the
effect of $\tau$ polarization on the efficiency matrix is estimated to
be negligible.

The corresponding errors are listed in Table~\ref{sys45} in the column `dyn'.

\subsection{ Monte Carlo statistics}

Multinomial fluctuations for each generated class in the Monte Carlo sample 
using the measured fractions of events in each reconstructed class are
included in the efficiency matrix uncertainties. Generating 1000 sets of
efficiency matrices, the RMS of the branching ratio values are obtained 
and taken as the uncertainty from the finite Monte Carlo statistics. It
is shown in Table~\ref{sys45} in the column `mcs'.

\subsection{ Total systematic errors}

Assuming that all the above systematics from specific sources are 
uncorrelated, the total systematic error is computed for each channel. 
The final values are listed in the last columns of Table~\ref{sys45}. 
Correlations exist between different channels for a given
source of systematics: the correlation matrix for the total systematic
uncertainties is given in Section~\ref{combires}.

\section{Global systematic checks}
\label{check}

After taking into account all the corrections to the simulation 
(measured $\tau$ branching ratio values, fake photon simulation, \ldots) 
and including non-$\tau$ backgrounds, some distributions
are compared in data and Monte Carlo in order to provide global checks
of the consistency between data and simulation. Of course such tests 
cannot be taken literally as they could reveal some difference in the
$\tau$ decay dynamics in some channel, however such effects are expected
to be small as most $\tau$ decay channels are well understood from the
basic principles of the Standard Model. This comparison is done for
the distributions of charged particle momenta, of photon and $\pi^0$ 
energies and of the hadronic invariant mass spectra for each
$\tau$ decay mode.

\subsection{Charged particle momentum distributions}

\begin{figure}
   \centerline{\psfig{file=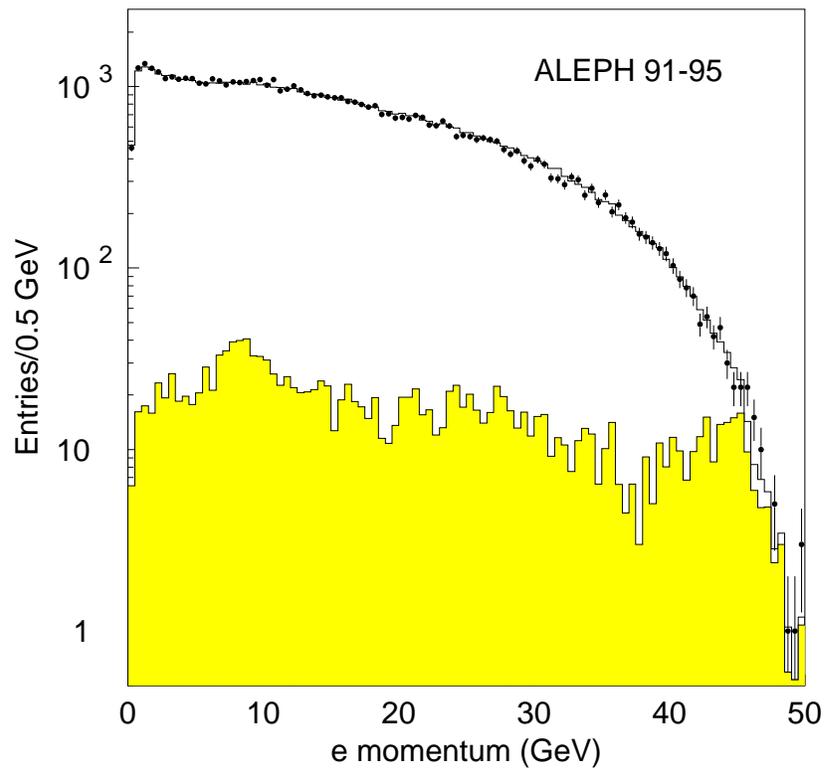,width=120mm}}
   \caption{Comparison of electron momentum spectra in data (points)
   and simulation (histogram) for the full 1991-1995 data sample. 
   The shaded histogram is the contribution of non-$\tau$ background.}
   \label{elmom}
\end{figure}

\begin{figure}
   \centerline{\psfig{file=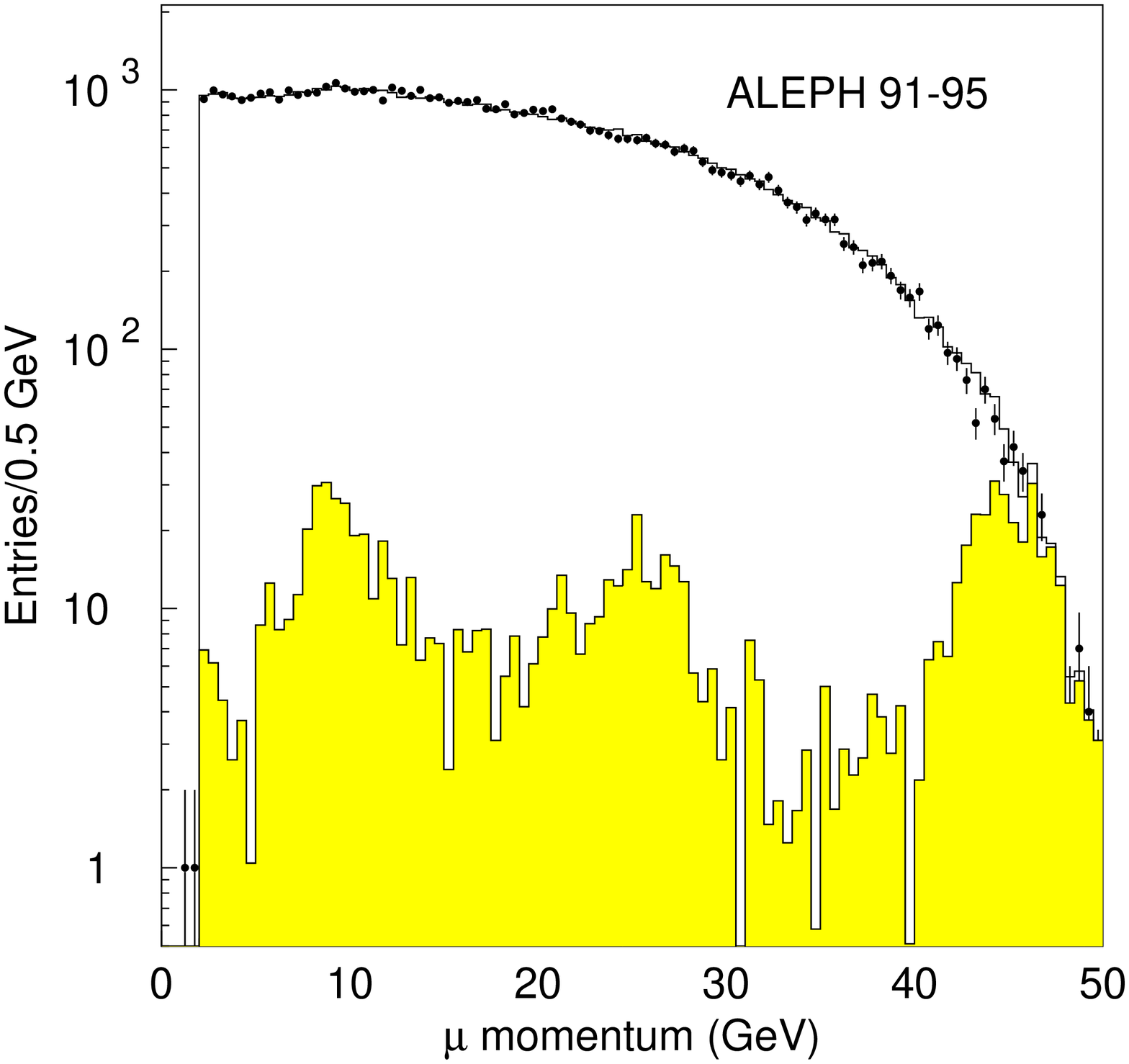,width=120mm}}
   \caption{Comparison of muon momentum spectra in data (points)
   and simulation (histogram) for the full 1991-1995 data sample. 
   The shaded histogram is the contribution of non-$\tau$ background.}
   \label{mumom}
\end{figure}

The only accessible observable in the leptonic channels is the lepton
momentum. Figures~\ref{elmom} and \ref{mumom} show the respective track 
momentum distributions for electrons and muons. They are in good 
agreement with the simulation including backgrounds and the Standard
Model spectrum. 

\begin{figure}
   \centerline{\psfig{file=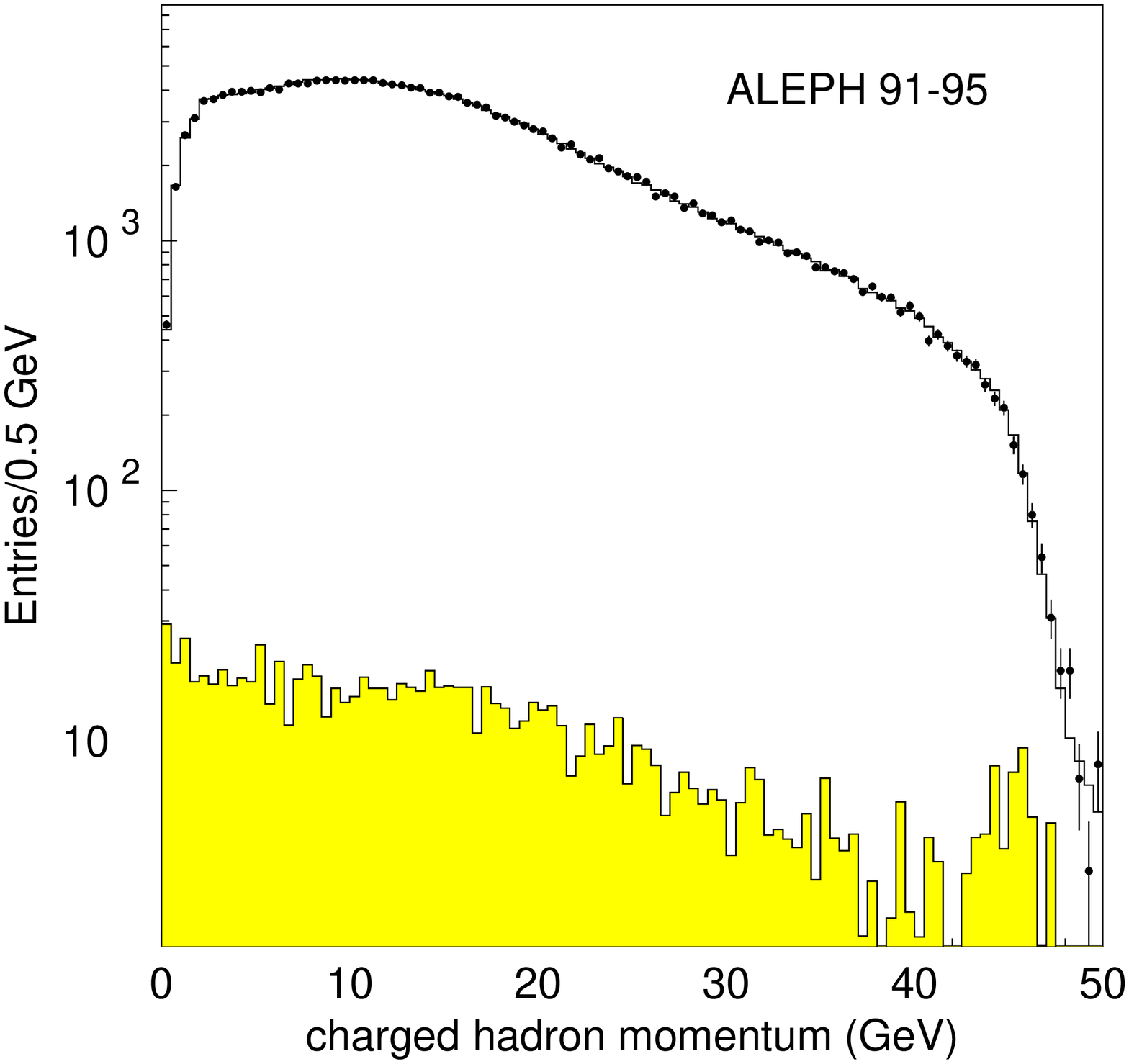,width=120mm}}
   \caption{Comparison of charged hadron momentum spectra in data (points)
   and simulation (histogram) for the full 1991-1995 data sample. 
   The shaded histogram is the contribution of non-$\tau$ background.}
   \label{hmom}
\end{figure}

The charged hadron momentum distribution is shown in Fig.~\ref{hmom}.
Good agreement is observed between data and the simulation.

\subsection{Photon energy distribution}

\begin{figure}
   \centerline{\psfig{file=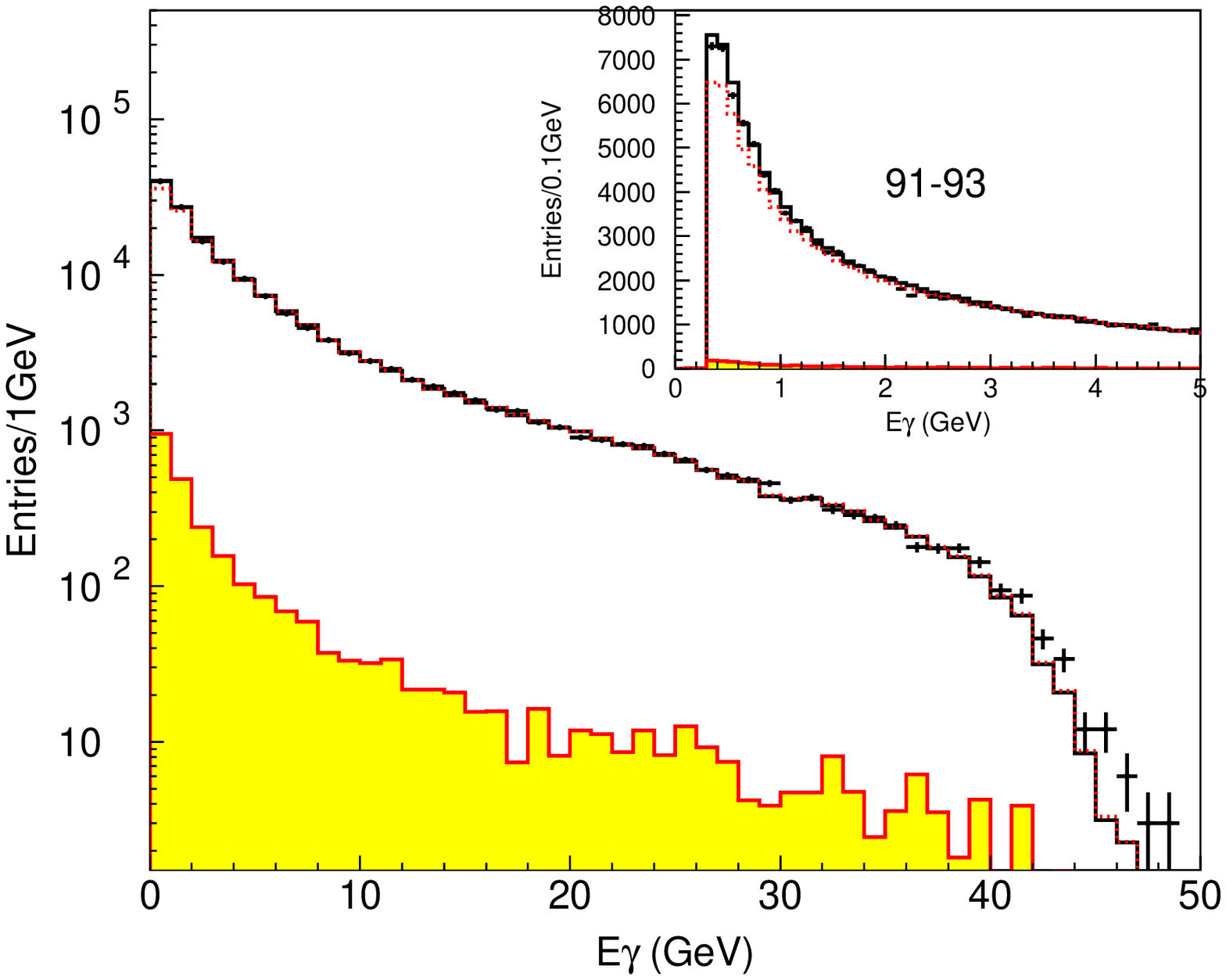,width=120mm}}
   \centerline{\psfig{file=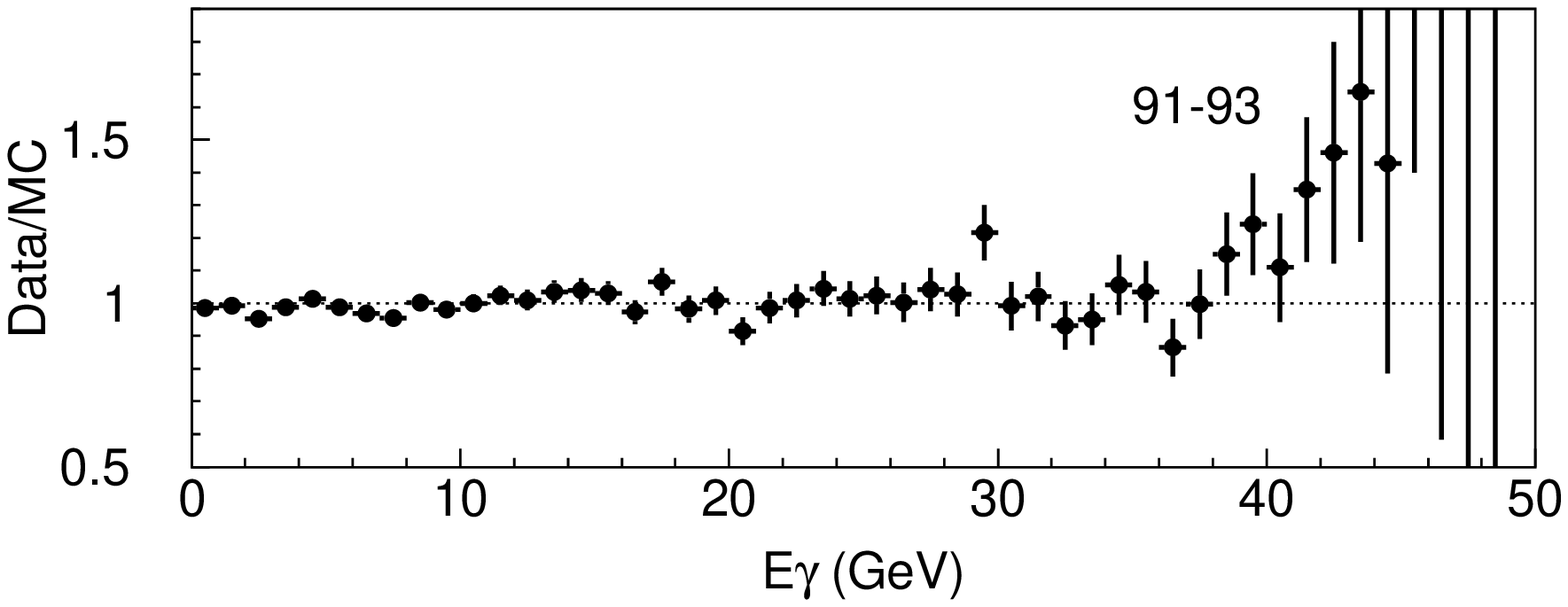,width=120mm}}
   \caption{Comparison of photon energy spectra for all $\tau$
   events in data (points) and simulation (histogram) after
   all the corrections for 1991-1993 data set. The dotted line shows the
   simulated distribution before correction and the shaded histogram is
   the contribution of non-$\tau$ background.}
\label{eg_corr1}
\end{figure}

\begin{figure}
   \centerline{\psfig{file=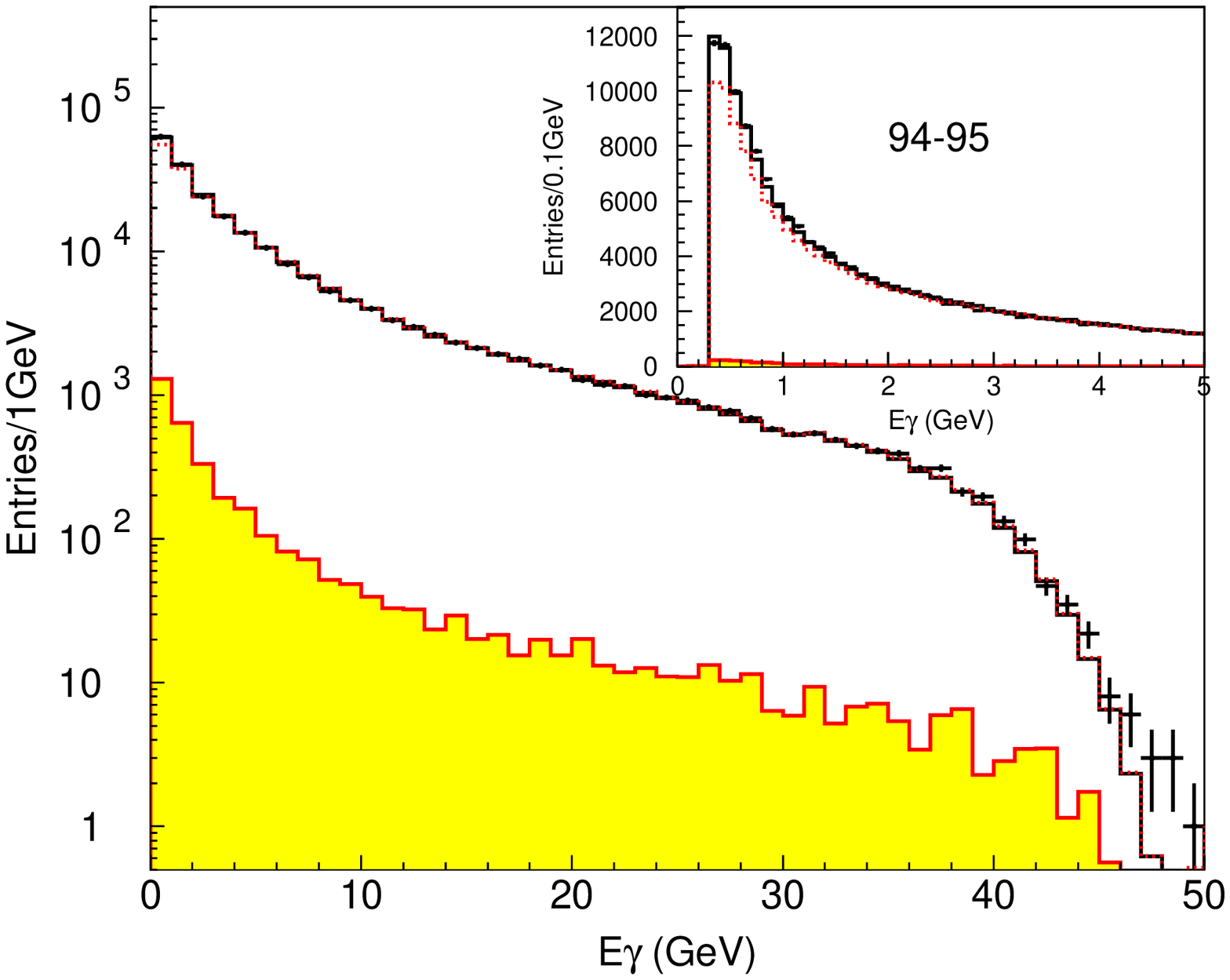,width=120mm}}
   \centerline{\psfig{file=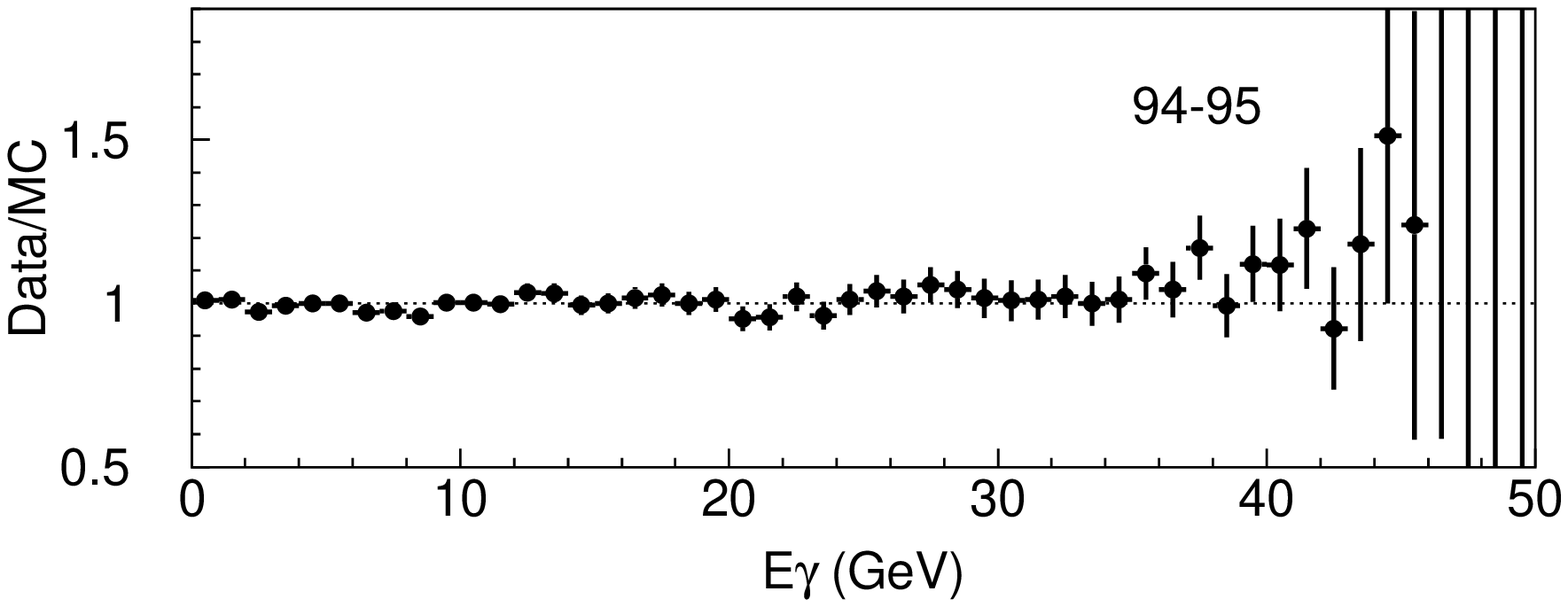,width=120mm}}
   \caption{Comparison of photon energy spectra for all $\tau$
   events in data (points) and simulation (histogram) after
   all the corrections for 1994-1995 data set. The dotted line shows the
   simulated distribution before correction and the shaded histogram is
   the contribution of non-$\tau$ background.}
\label{eg_corr4}
\end{figure}

The comparison of photon energy spectra in data and Monte Carlo is shown
in Figs.~\ref{eg_corr1} and \ref{eg_corr4} for 1991-1993 and 1994-1995
data sets respectively. Good agreement is observed except for the largest
energies. This effect is understood as coming from the correction of
saturation and leakage which is applied to single clusters in data, while
no such correction is done in the simulation. This
procedure is incorrect for unresolved $\piz$'s not recognized as such,
since the correction is applied to the resulting shower rather than to the 
two merged individual photonic showers, leading to an overestimate of the
$\piz$ energy. This shortcoming has no effect on the branching ratio
measurement. The agreement between data and Monte Carlo at low energy
after fake photon correction shows that the procedure is reliable,
since no energy information was used in the fake photon correction 
procedure.

\subsection{$\pi^0$ energy distribution}

\begin{figure}
   \centerline{\psfig{file=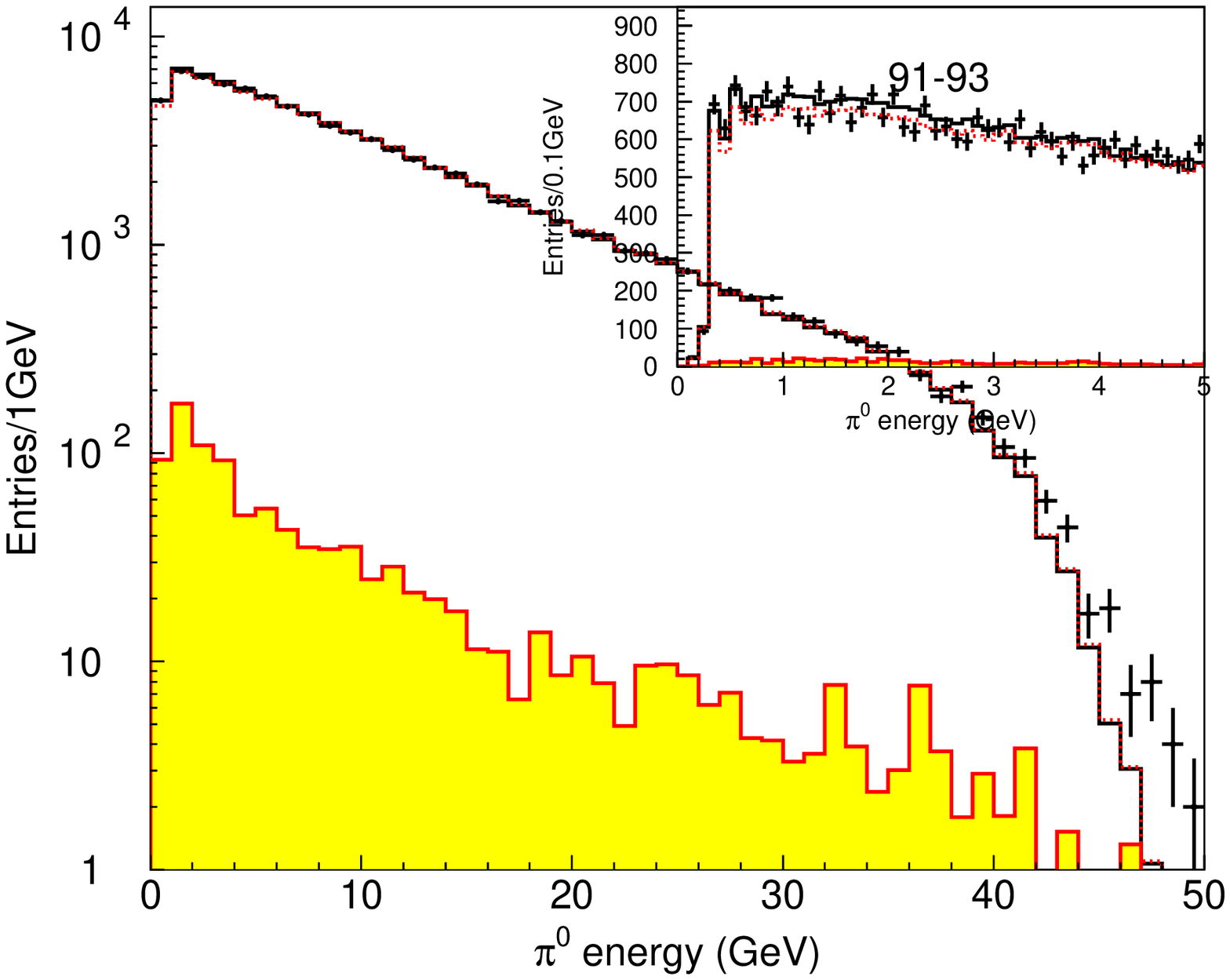,width=120mm}}
   \centerline{\psfig{file=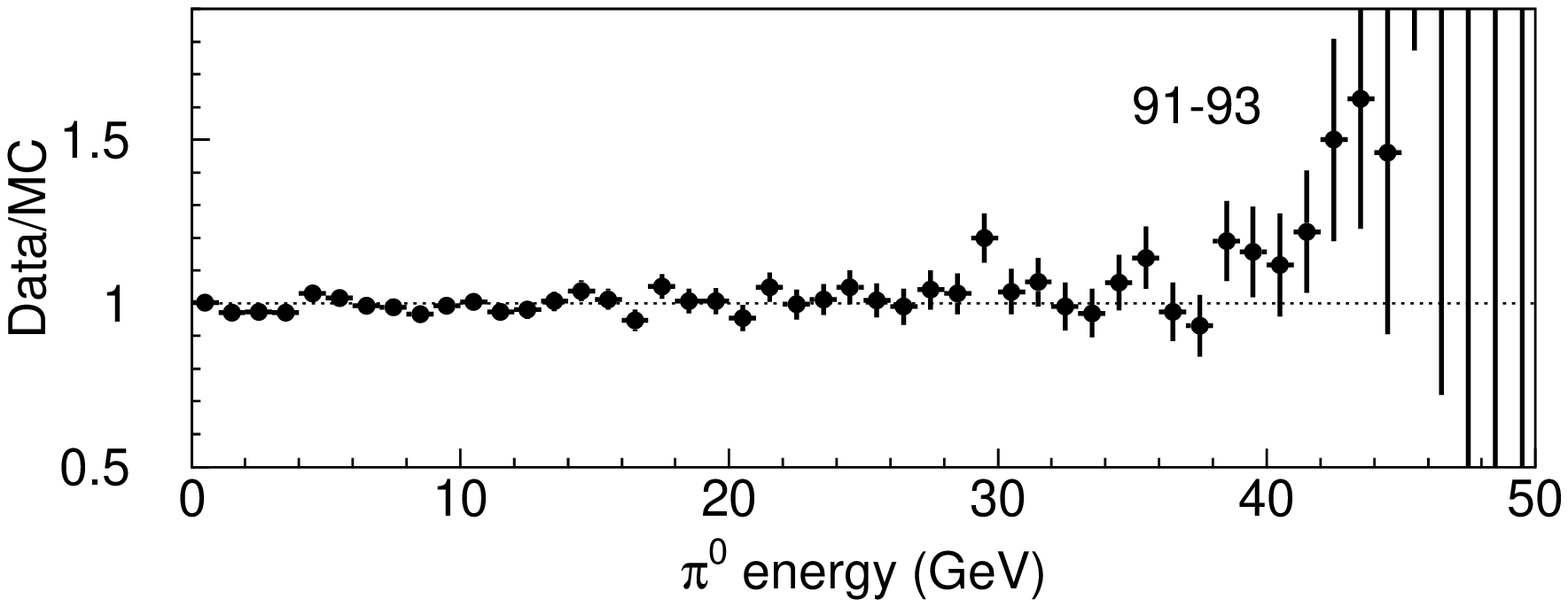,width=120mm}}
   \caption{Comparison of $\piz$ energy spectra for all $\tau$
   events in data (points) and simulation (histogram) after
   all the corrections for 1991-1993 data set. The dotted line shows the
   simulated distribution before correction and the shaded histogram is
   the contribution of non-$\tau$ background.}
\label{epi_corr1}
\end{figure}

\begin{figure}
   \centerline{\psfig{file=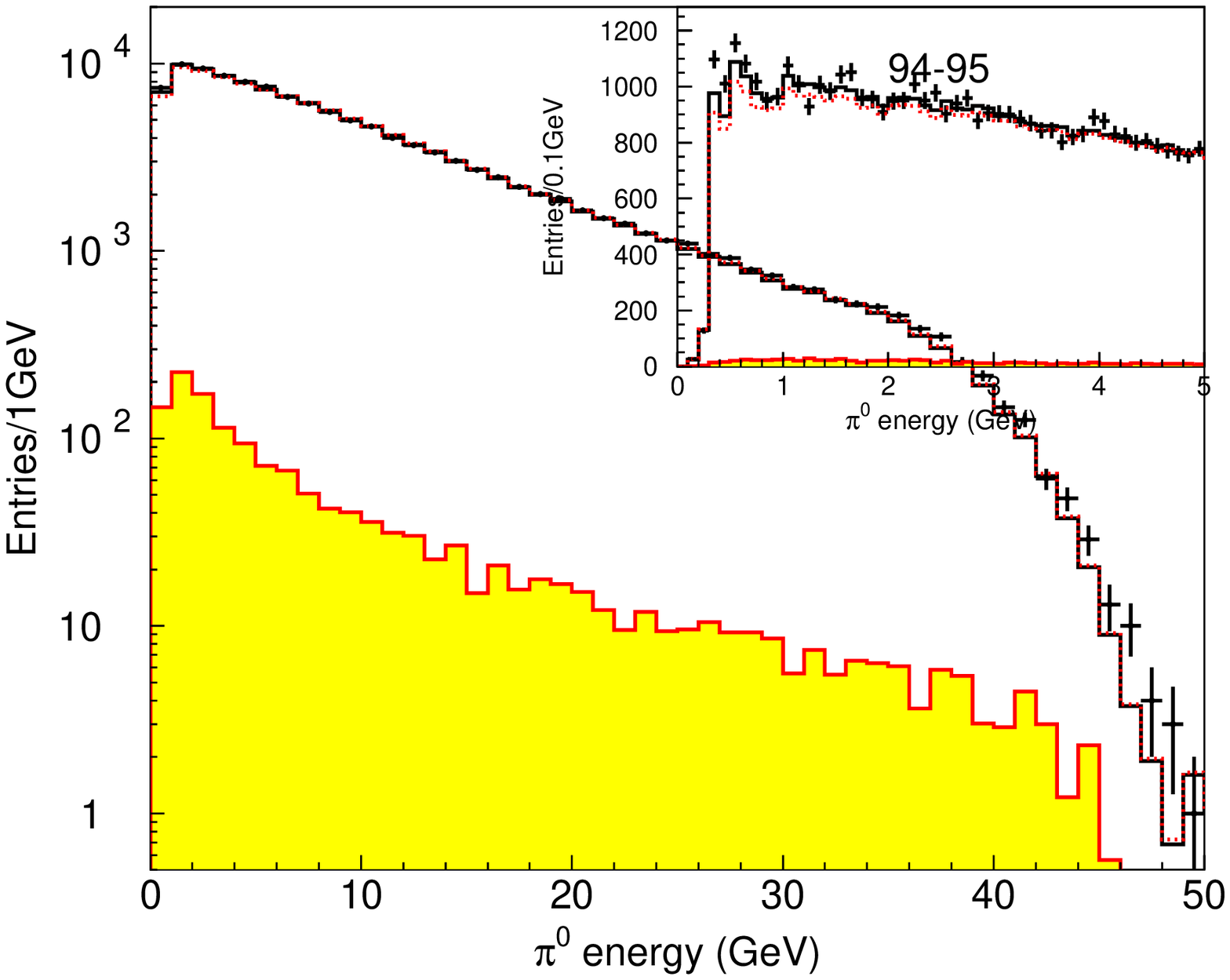,width=120mm}}
   \centerline{\psfig{file=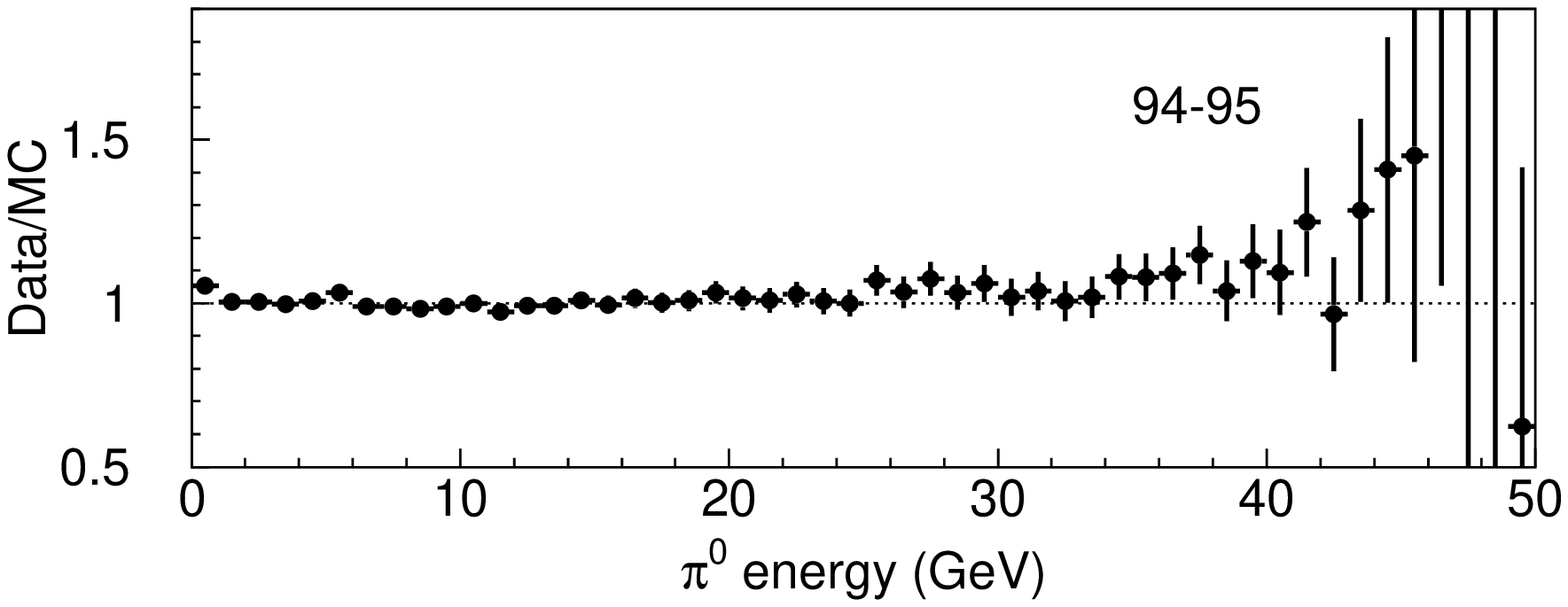,width=120mm}}
   \caption{Comparison of $\piz$ energy spectra for all $\tau$
   events in data (points) and simulation (histogram) after
   all the corrections for 1994-1995 data set. The dotted line shows the
   simulated distribution before correction and the shaded histogram is
   the contribution of non-$\tau$ background.}
\label{epi_corr4}
\end{figure}

The comparison of $\pi^0$ energy spectra in data and Monte Carlo is shown
in Figs.~\ref{epi_corr1} and \ref{epi_corr4} for 1991-1993 and 1994-1995
data sets respectively. Good agreement is observed except 
at energies below 1~GeV for the 1994-1995 data and for the
highest energies. The former discrepancy is within the quoted systematic
uncertainty for $\gamma - \pi^0$ reconstruction. As for high energy, the 
explanation is the same as for single photons since,
at high energy, unresolved $\piz$'s dominate the sample.

\subsection{Invariant mass spectra in multihadron channels}

\begin{figure}
   \centerline{\psfig{file=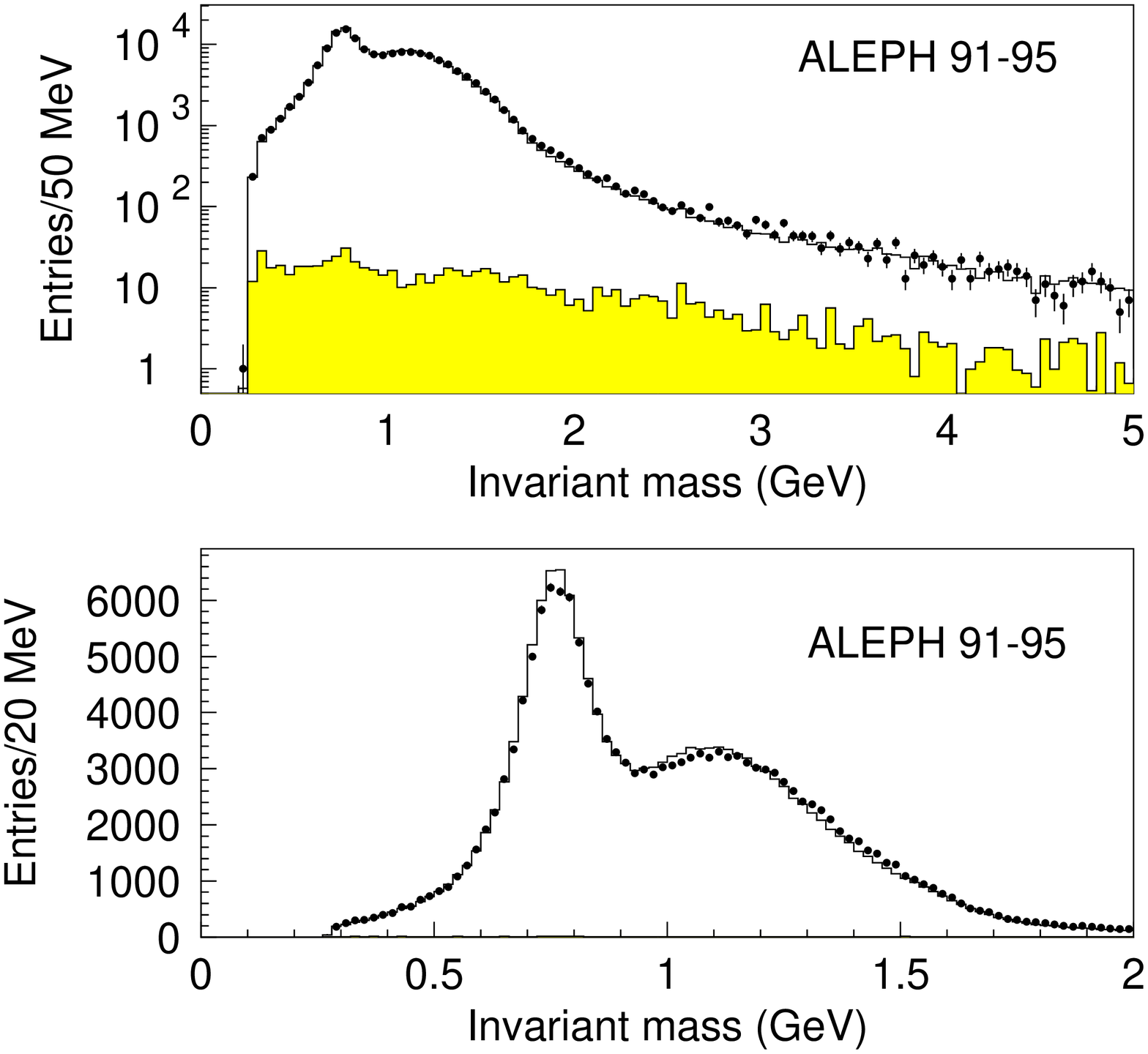,width=160mm}}
   \caption{Comparison of mass spectra (assuming all charged particles
   are pions) in data (points)
   and simulation (histogram) after fake photon correction for all the
   hadronic $\tau$ decays (except single hadron) in 1991-1995 data sample.
   The shaded histogram is the contribution of non-$\tau$ background.
   The same plots are displayed in logarithmic
   (top) and linear (bottom) vertical scales.}
\label{spec_corr1}
\end{figure}

The comparison of mass spectra for all $\tau$ decays in data
and Monte Carlo is shown in Fig.~\ref{spec_corr1}
for the full 1991-1995 data set. As noticed earlier
perfect agreement cannot be expected since the Monte Carlo simulation 
of the dynamics in some decay modes may not be accurate. Even in prominent
channels like $\pi \piz$, $3 \pi$ and $\pi 2\piz$, the precise shape of
the mass distribution is not predicted by theory and the phenomenology
of the corresponding spectral functions does, on the contrary,
rely on precise measurements (see Section~\ref{sf} on the
measurement of spectral functions from this analysis). 
The long tail above the $\tau$ mass in the distribution is dominated
by secondary interaction of hadronic tracks with inner detector 
material. Globally, the agreement is reasonable and confirms that the 
simulation of secondary interactions in the detector is adequate.

\begin{figure}
   \centerline{\psfig{file=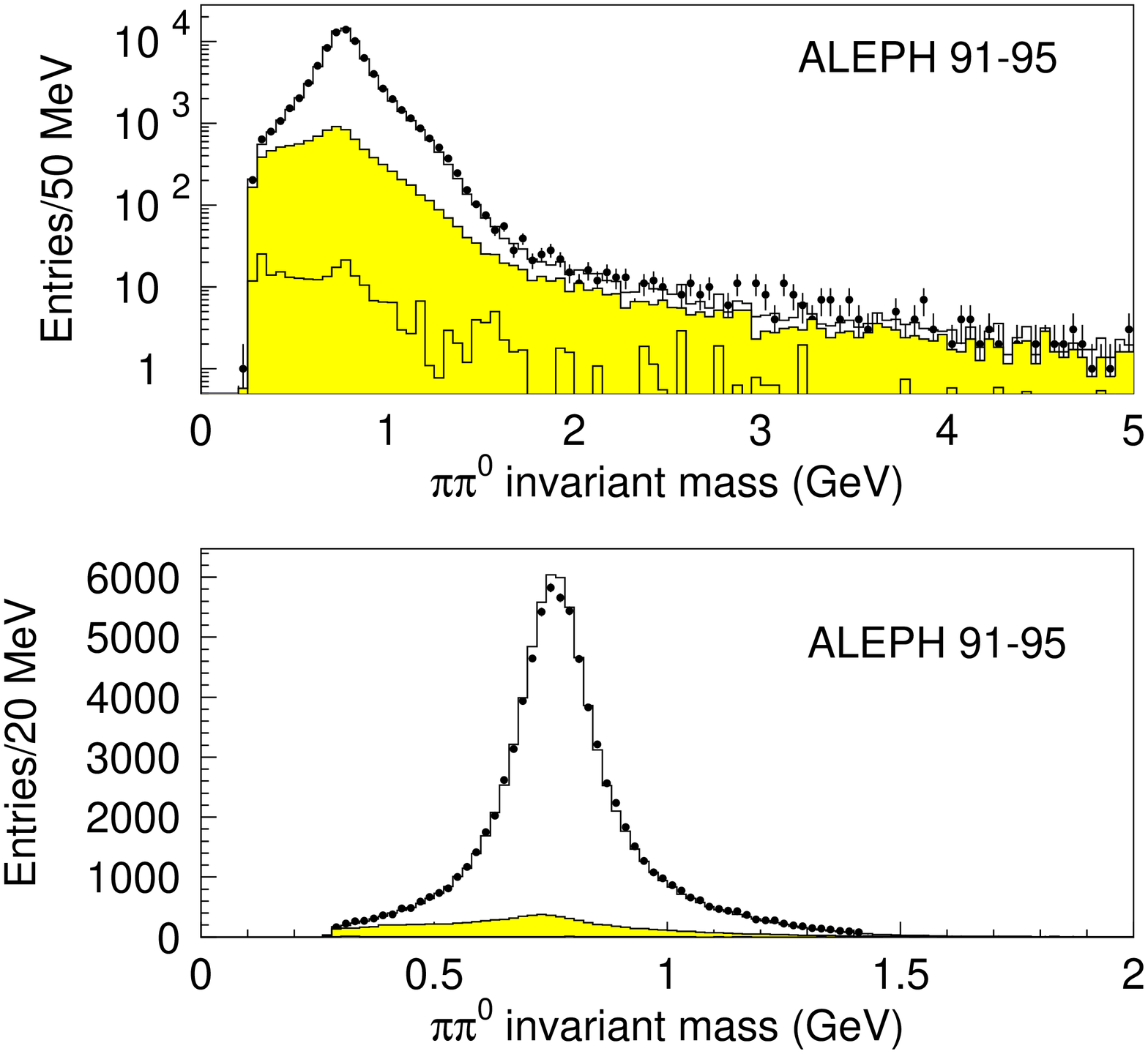,width=160mm}}
   \caption{Comparison of hadronic mass spectra in data (points)
   and simulation (histogram) after fake photon correction for the
   \clsiv sample in 1991-1995 data. The lower shaded histogram is
   the contribution of non-$\tau$ background and the upper shaded
   is from $\tau$ feedthrough. The plots are displayed in logarithmic
   (top) and linear (bottom) vertical scales.}
\label{spec_4}
\end{figure}

The comparison of mass spectra for reconstructed \clsiv ~events in
data and Monte Carlo is shown in Fig.~\ref{spec_4}
for the full 1991-1995 data set, with the Monte Carlo
predicted feedthrough from other channels and non-$\tau$ backgrounds.

\section{From reconstructed classes to exclusive modes}
\label{rec_exclu}

So far branching fractions have been determined in 13 classes
corresponding to major $\tau$ decay modes. However, as shown in
Table~\ref{classtable}, these classes still contain the contributions
from final states involving kaons. The latter are coming from 
Cabibbo-suppressed $\tau$ decays or modes with a $K\overline{K}$ pair,
both characterized by small branching ratios compared to the nonstrange
modes without kaons.

Complete analyses of $\tau$ decays involving 
neutral or charged kaons have been performed by ALEPH on the full 
LEP 1 data~\cite{alephk3,alephks,alephkl}. They are summarized in
Ref.~\cite{alephksum} where measurements with 
$K^0_S$ or $K^0_L$ are combined.
The ALEPH analyses have provided measurements of branching ratios 
of modes with kaons containing up to 4 hadrons in the final states. 
Thus they are fully adequate to cover the needs of the present analysis 
of the nonstrange modes.

The $\tau$ decays involving $\eta$ or $\omega$ mesons also require 
special attention in this analysis because of their electromagnetic
decay modes. Indeed the final state classification relies in part 
on the $\piz$ multiplicity, thereby assuming that all photons ---except
those specifically identified as bremsstrahlung or radiative---
originate from $\piz$ decays. Therefore the non-$\piz$ photons from
$\eta$ and $\omega$ decays are treated as $\piz$ candidates in the
analysis and the systematic bias introduced by this effect must be
evaluated. The corrections are based on specific measurements by ALEPH 
of $\tau$ decay modes containing those mesons~\cite{alepheta}. Thus
the final results correspond to exclusive branching ratios obtained
from the values measured in the topological classification, 
corrected by the removed contributions from $K$, $\eta$ and $\omega$  
modes measured separately, taking into account through the Monte Carlo
their specific selection and reconstruction efficiencies to enter the
classification. This delicate bookkeeping takes into account all the
major decay modes of the considered mesons~\cite{pdg2004}, including
the isospin-violating $\omega \rightarrow \pi^+\pi^-$ decay mode.
The main decay modes considered are $\pi \omega$, $\pi \pi^0 \omega$
and $\pi \pi^0 \eta$ with branching fractions of 
$(2.26 \pm 0.18)\times 10^{-2}$, $(4.3 \pm 0.5)\times 10^{-3}$, 
and $(1.80 \pm 0.45)\times 10^{-3}$~\cite{alepheta}, 
respectively. The first two values are derived from 
the branching ratios for the $3\pi \pi^0$ and $3\pi 2\pi^0$ 
modes obtained in this analysis and the measured $\omega$ 
fractions of $0.431 \pm 0.033$ from ALEPH~\cite{alepheta} 
and the average value, $0.78 \pm 0.06$, from ALEPH~\cite{alepheta} 
and CLEO~\cite{cleoomega}, respectively.

Some much smaller contributions with $\eta$ and $\omega$ have been 
identified and measured by CLEO~\cite{cleoeta3pi} with the decay modes 
$\tau \rightarrow \nu_\tau \eta\pi^-\pi^+\pi^-$ ($(2.4 \pm 0.5)\times 10^{-4}$), 
$\tau \rightarrow \nu_\tau \eta\pi^-2\piz$ ($(1.5 \pm 0.5)\times 10^{-4}$),
$\tau \rightarrow \nu_\tau \omega\pi^-\pi^+\pi^-$ ($(1.2\pm 0.2)\times 10^{-4}$),
and $\tau \rightarrow \nu_\tau \omega\pi^-2\piz$ ($(1.5 \pm 0.5)\times 10^{-4}$).
Even though the corrections from these channels are very small they have
been included for the sake of completeness. Finally, another very small
correction has been applied to take into account the $a_1$ 
radiative decay into $\pi \gamma$ with a branching fraction of 
$(2.1 \pm 0.8)\times 10^{-3}$ obtained from Ref.~\cite{zielinski}.

The corrections used to obtain exclusive branching ratios for the listed 
nonstrange modes are given in Table~\ref{subbr}. 

\begin{table}
\caption{Corrections for the exclusive nonstrange branching ratios.
`QE' and `E' denote the quasi-exclusive (with kaons, 
$\omega$ and $\eta$ included) and exclusive modes, respectively. 
Treating the $\omega$ and $\eta$
contributions separately is made necessary because of their large
radiative ($i.e.$ with photons not originating from $\pi^0$'s) modes.}
\begin{center}
\begin{tabular}{llc}
\hline\hline
  QE class & E class &  correction to BR (\%)    \\\hline
 \clsi &  \phyi  &       -0.000 $\pm$  0.000 \\
 \clsii &  \phyii  &     -0.000 $\pm$  0.000 \\
 \clsiii &  \phyiii  &   -1.341 $\pm$  0.040 \\
 \clsiv &  \phyiv  &     -0.756 $\pm$  0.038 \\
 \clsv &  \phyv  &       -0.408 $\pm$  0.030 \\
 \clsvi &  \phyvi  &     -0.236 $\pm$  0.032 \\
 \clsxiii &  \phyxiii  & -0.085 $\pm$  0.016 \\
 \clsvii &  \phyvii  &   -0.770 $\pm$  0.057 \\
 \clsviii &  \phyviii  & -1.994 $\pm$  0.100 \\
 \clsix &  \phyix  &     -0.480 $\pm$  0.071 \\
 \clsx &  \phyx  &       -0.032 $\pm$  0.006 \\
 \clsxi &  \phyxi  &     -0.026 $\pm$  0.004 \\
 \clsxii &  \phyxii  &   -0.012 $\pm$  0.002 \\
\hline\hline
\end{tabular}
\label{subbr}
\end{center}
\end{table}

\section{Results}
\label{result}

\subsection{Overall consistency test}
\label{consist}

Rejected $\tau$ hemispheres because of charged particle identification
cuts are placed in class 14; these cuts include the 2 GeV minimum momentum 
and the ECAL-crack veto for some one-prong modes, and
the strict definition of higher multiplicity channels. 
As already emphasized, this sample does not correspond to a nominal
$\tau$ decay mode and should be explained by all other measured fractions
in the other classes and the efficiency matrix. Thus the determination of
a hypothetical signal in this class is a measure of the level of 
consistency achieved in the analysis.

For this determination the efficiency of the possible signal in class 14
is taken to be 100\%. The results, already shown in Table~\ref{br_12345}
separately for the 1991-1993 and 1994-1995 data sets, are consistent 
and are combined to give $B_{14}= (0.066 \pm (0.027)_{stat} 
\pm (0.021)_{syst,c} \pm (0.025)_{syst,unc})\%$, where the last two errors
refer to the common and uncommon uncertainties from the two data sets.
With a combined error of $0.042\%$ this value is consistent with zero
and provides a nontrivial check of the overall procedure
at the 0.1\% level for branching ratios. It is interesting to
note that this value coincides, approximately and accidentally, with
the limit achieved of 0.11\% at 95\% CL in a direct search for 
``invisible'' decays not selected in the 13-channel classification.

In the following it is assumed that all $\tau$ decay modes 
have been properly considered at the 0.1\% precision level and
no physics contribution beyond standard $\tau$ decays is further allowed.
Thus the quantity $B_{14}$ is now constrained to be zero.

It can be further noticed that this analysis provides a branching ratio 
in the $3\pi 3\piz$ class which is consistent with zero for both
1991-1993 and 1994-1995 data sets (see Table~\ref{br_12345}).
The result is therefore given as an upper limit at 95\% CL
\beq
 B_{3\pi 3\piz} < 4.9\times 10^{-4}
\eeq
consistent with the measurement made by CLEO~\cite{cleo6pi} yielding
$B_{3\pi 3\piz} = (2.2 \pm 0.5)\times 10^{-4}$. The final state is dominated by
$\eta$ and $\omega$ resonances~\cite{cleo6pi} and using other channels allows 
a lower limit to be obtained for this branching ratio, 
$(2.6 \pm 0.4)\times 10^{-4}$. 
In the following a value of $(3 \pm 1)\times 10^{-4}$ is used as input for this 
channel and the global analysis is performed in terms of the remaining 
12 defined channels which are refitted. As for other channels proper
subtractions are made for the contributions of modes with $\eta$ which
are listed separately.

\subsection{Comparison of 1991-1993 and 1994-1995 results}

Since the same procedure is applied for the analyses of 1991-1993 
and 1994-1995 data the results must be consistent within the statistical 
errors of data and Monte Carlo.
Table~\ref{9193_9495} shows the list of the differences of branching ratios
with their expected fluctuations. Good agreement is observed with a $\chi^2$
of 8.6 for 11 DF. 

\begin{table}
\caption{Differences of branching ratios between 1991-1993 and 
1994-1995 data samples; only the statistical errors in data and 
Monte Carlo are considered.}
\begin{center}
\begin{tabular}{lr}
\hline\hline
 class & $\Delta$ BR (\%)    \\\hline
 \clsi &      0.040 $\pm$  0.148 \\
 \clsii &     0.061 $\pm$  0.140 \\
 \clsiii &    0.145 $\pm$  0.139 \\
 \clsiv &    -0.186 $\pm$  0.197 \\
 \clsv &      0.018 $\pm$  0.178 \\
 \clsvi &    -0.105 $\pm$  0.143 \\
 \clsxiii &   0.065 $\pm$  0.076 \\
 \clsvii &    0.163 $\pm$  0.128 \\
 \clsviii &  -0.201 $\pm$  0.120 \\
 \clsix &     0.046 $\pm$  0.078 \\
 \clsxi &     0.006 $\pm$  0.018 \\
 \clsxii &   -0.006 $\pm$  0.014 \\
\hline\hline
\end{tabular}
\label{9193_9495}
\end{center}
\end{table}

In conclusion the two independent data and Monte Carlo samples 
give consistent results. The 1994-1995 results confirm the trend of
larger $h \pi^0$ and $3h \pi^0$, and smaller $h$ and $3h$ branching ratios 
compared to the previous analyses, as observed in 1991-1993 data sample.

\subsection{Final combined results}
\label{combires}

Finally the two sets of results are combined. Using only 
statistical or total weights ---in the latter taking into account correlated
errors from dynamics and secondary interactions--- gives almost identical 
results. The final results obtained with the total weights are shown in
Table~\ref{finalBR}. 

\begin{table}
\caption{Combined results for the exclusive branching ratios (B) 
         for modes without kaons. The contributions from channels with 
         $\eta$ and $\omega$ are given separately, the latter only for the 
         electromagnetic $\omega$ decays. All results are from this 
         analysis, unless explicitly stated. The ``estimates'' 
         are discussed in the text.}
\begin{center}
\begin{tabular}{lrc}
\hline\hline
 mode & B $\pm\sigma_{\hbox{stat}} \pm \sigma_{\hbox{syst}}$
  [\%] \\\hline
 \phyi &    17.837 $\pm$  0.072 $\pm$  0.036 &\\
 \phyii &    17.319 $\pm$  0.070 $\pm$  0.032 &\\
 \phyiii &    10.828 $\pm$  0.070 $\pm$  0.078 &\\
 \phyiv &    25.471 $\pm$  0.097 $\pm$  0.085 &\\
 \phyv &     9.239 $\pm$  0.086 $\pm$  0.090 &\\
 \phyvi &      0.977 $\pm$  0.069 $\pm$  0.058 &\\
 \phyxiii &      0.112 $\pm$  0.037 $\pm$  0.035 &\\
 \phyvii &     9.041 $\pm$  0.060 $\pm$  0.076 &\\
 \phyviii &     4.590 $\pm$  0.057 $\pm$  0.064 &\\
 \phyix &      0.392 $\pm$  0.030 $\pm$  0.035 &\\
 \phyx &      0.013 $\pm$  0.000 $\pm$  0.010 & estimate\\
 \phyxi &      0.072 $\pm$  0.009 $\pm$  0.012 &\\
 \phyxii &      0.014 $\pm$  0.007 $\pm$  0.006 &\\
$\pi^- \pi^0 \eta$ & 0.180 $\pm$ 0.040 $\pm$ 0.020 & ALEPH~\cite{alepheta}\\
$\pi^- 2\pi^0 \eta$ & 0.015 $\pm$ 0.004 $\pm$ 0.003 & CLEO~\cite{cleoeta3pi}\\
$\pi^- \pi^- \pi^+ \eta$ & 0.024 $\pm$ 0.003 $\pm$ 0.004 & CLEO~\cite{cleoeta3pi}\\
$a_1^- (\rightarrow \pi^- \gamma)$ & 0.040 $\pm$ 0.000 $\pm$ 0.020 & estimate \\
$\pi^- \omega (\rightarrow \pi^0 \gamma, \pi^+ \pi^-)$ & 0.253 $\pm$ 0.005 $\pm$ 0.017 & ALEPH~\cite{alepheta}\\
$\pi^- \pi^0 \omega (\rightarrow \pi^0 \gamma, \pi^+ \pi^-)$ & 0.048 $\pm$ 0.006 $\pm$ 0.007 & ALEPH~\cite{alepheta} + CLEO~\cite{cleoomega}\\
$\pi^- 2\pi^0 \omega (\rightarrow \pi^0 \gamma, \pi^+ \pi^-)$ & 0.002 $\pm$ 0.001 $\pm$ 0.001 & CLEO~\cite{cleoeta3pi}\\
$\pi^- \pi^- \pi^+\omega (\rightarrow \pi^0 \gamma, \pi^+ \pi^-)$ & 0.001 $\pm$ 0.001 $\pm$ 0.001 & CLEO~\cite{cleoeta3pi}\\
\hline\hline
\end{tabular}
\label{finalBR}
\end{center}
\end{table}

The branching ratios obtained for the different channels are
correlated with each other. On one hand the statistical fluctuations
in the data and the Monte Carlo sample are driven by the multinomial
distribution of the corresponding events, producing well-understood
correlations. On the other hand the systematic effects also induce
significant correlations between the different channels.
All the systematic studies were done keeping track of the correlated
variations in the final branching ratio results, thus allowing a proper
propagation of errors. The full covariance matrices from statistical 
(from data) and systematic origins are given 
in Tables~\ref{covmat_stat} and \ref{covmat_syst}.

\begin{table}
\caption{Correlation matrix of the statistical errors on the 
branching fractions.}
\begin{center}
\small
\begin{tabular}{l|rrrrrrrrrrrr}
\hline\hline
  &\clsii &\clsiii &\clsiv &\clsv &\clsvi &\clsxiii &\clsvii &\clsviii &\clsix &\clsx &\clsxi &\clsxii
  \\\hline

\clsi & -0.21&-0.15&-0.25&-0.09&-0.01& 0.00&-0.15&-0.10& 0.03&-0.06& 0.00& 0.01\\
\clsii &  1.00&-0.13&-0.21&-0.07&-0.06& 0.00&-0.09&-0.07& 0.00&-0.02& 0.00&-0.04\\
 \clsiii &     & 1.00&-0.31&-0.02& 0.01&-0.06&-0.12&-0.06&-0.02& 0.01&-0.01& 0.02\\
 \clsiv &     &     & 1.00&-0.40& 0.05& 0.00&-0.11&-0.06&-0.02& 0.00&-0.04&-0.04\\
\clsv &      &     &     & 1.00&-0.51& 0.26&-0.09& 0.01&-0.07& 0.06&-0.01& 0.03\\
 \clsvi &     &     &     &     & 1.00&-0.75& 0.01&-0.03& 0.05&-0.02&-0.01& 0.01\\
 \clsxiii &     &     &     &     &     & 1.00&-0.02&-0.02&-0.03& 0.01& 0.02&-0.03\\
 \clsvii &     &     &     &     &     &     & 1.00&-0.33& 0.08&-0.05&-0.04& 0.00\\
  \clsviii &    &     &     &     &     &     &     & 1.00&-0.45& 0.19&-0.02&-0.02\\
  \clsix &    &     &     &     &     &     &     &     & 1.00&-0.65& 0.03& 0.02\\
  \clsx &    &     &     &     &     &     &     &     &     & 1.00&-0.01&-0.04\\
  \clsxi &    &     &     &     &     &     &     &     &     &     & 1.00&-0.24\\
  \clsxii &    &     &     &     &     &     &     &     &     &     &     & 1.00\\
\hline\hline
\end{tabular}
\label{covmat_stat}
\end{center}
\end{table}
\normalsize

\begin{table}
\caption{Correlation matrix of the systematic errors on the 
branching fractions.}
\begin{center}
\small
\begin{tabular}{l|rrrrrrrrrrrr}
\hline\hline
  &\clsii &\clsiii &\clsiv &\clsv &\clsvi &\clsxiii &\clsvii &\clsviii &\clsix &\clsx &\clsxi &\clsxii
  \\\hline

\clsi & -0.17&-0.01& 0.02& 0.01& 0.03&-0.08&-0.17&-0.22&-0.05& 0.02& 0.00& 0.00\\
\clsii &  1.00& 0.05& 0.09&-0.03& 0.02&-0.13&-0.11&-0.24&-0.06& 0.01& 0.03&-0.04\\
 \clsiii &     & 1.00& 0.36&-0.29&-0.32&-0.42& 0.34&-0.40&-0.40&-0.07& 0.16&-0.09\\
 \clsiv &     &     & 1.00&-0.35&-0.02&-0.33& 0.01&-0.54&-0.26& 0.02& 0.11&-0.06\\
\clsv &      &     &     & 1.00&-0.01& 0.13&-0.24& 0.07& 0.13& 0.06&-0.13& 0.03\\
 \clsvi &     &     &     &     & 1.00&-0.13&-0.29&-0.02& 0.15& 0.09&-0.06& 0.04\\
 \clsxiii &     &     &     &     &     & 1.00&-0.14& 0.34& 0.27& 0.00&-0.12&-0.05\\
 \clsvii &     &     &     &     &     &     & 1.00&-0.03&-0.16&-0.11& 0.17&-0.06\\
  \clsviii &    &     &     &     &     &     &     & 1.00& 0.04&-0.03&-0.07&-0.01\\
  \clsix &    &     &     &     &     &     &     &     & 1.00&-0.14&-0.09& 0.07\\
  \clsx &    &     &     &     &     &     &     &     &     & 1.00&-0.02& 0.02\\
  \clsxi &    &     &     &     &     &     &     &     &     &     & 1.00&-0.26\\
  \clsxii &    &     &     &     &     &     &     &     &     &     &     & 1.00\\
\hline\hline
\end{tabular}
\label{covmat_syst}
\end{center}
\end{table}
\normalsize

\subsection{An independent analysis of the leptonic branching ratios}
\label{ep}

A dedicated analysis of the leptonic branching ratios has been performed
as a by-product of one of the two methods used by ALEPH to measure the
$\tau$ polarization and polarization asymmetry~\cite{alephpol}. It identifies
and selects directly single $\tau$ hemispheres with leptons without a full
selection of other $tau$ decay modes. In order to normalize the selected 
samples, the number of produced $\tau$ pairs is derived using the 
$e^+e^- \rightarrow \tau^+ \tau^-$ cross section measured by 
ALEPH~\cite{aleph94} and the precise determination of luminosity using
small-angle Bhabha scattering. 

Particle identification is performed using a method quite similar to the one 
used in the global analysis and described in Section~\ref{pid} but using 
a totally independent code. The performances of the two procedures are very
close. Similarly, the treatment of photons and the separation between genuine
and fake photons follows similar philosophies, but it is implemented 
separately.

Hemispheres with $\tau$ decay leptonic candidates are selected in two steps.
In the preselection step, events with acollinear ``jets''
($\cos \theta_{\rm acol} < -0.9$) are retained with a loose lepton 
identification at least on one side. Then, strict cuts are applied to
select electron and muon hemispheres, by rejecting hadronic $\tau$ decays
(electron or muon identification, veto if a $\piz$ is reconstructed in the 
same hemisphere) and non-$\tau$ background (using methods similar to those 
described in Section~\ref{nontau}). The selection efficiencies are about
69\% and 75\% for electronic and muonic decays, respectively. 
The corresponding values for the $\tau$ decay feedthrough contaminations
are 1.0\% and 0.8\%, and 1.5\% and 0.4\% for the non-$\tau$ background, 
respectively. The muon identification in this method allows the detection of
lower muon momenta down to 1.3~GeV, compared to 2~GeV in the global method.
A total of 48882 electron and 50782 muon hemispheres are thus selected.
In this method the largest systematic 
uncertainty originates from the $\tau$-pair normalization and is completely
correlated for the electron and muon channels.
The results of this specific analysis are:

\beqn
B(\tau \rightarrow \nu_\tau e \overline{\nu}_e) &=& (17.778 \pm 0.080 \pm 0.049)\%~,\\
B(\tau \rightarrow \nu_\tau \mu \overline{\nu}_\mu) &=& (17.299 \pm 0.077 \pm 0.045)\%~,
\eeqn
where the first errors are statistical and the second systematic.

Although both analyses share the same data and have both large selection 
efficiencies they have been carried out in a completely different and 
independent way, from the selection of candidates to the identification
method of the final state. The comparison thus provides a valuable check.

The samples of events have been compared and the correlations found to be
0.861 for electrons and 0.896 for muons. Part of the Monte Carlo statistics
is also non common. An important effect could come from the normalization
of the independent method using the luminosity-derived number of produced
$\tau\tau$ events, which is common to both electron and muon analyses.
 
The results of the independent analysis are consistent with the present
values of the leptonic branching ratios. The differences found,
$\Delta B_e  = (0.059 \pm 0.051 \pm 0.029)) \%$ and
$\Delta B_\mu= (0.020 \pm 0.042 \pm 0.028) \%$,
are well within the expectation of the respective noncommon errors. 
The second quoted errors are from the normalization of $\tau\tau$ events
and are completely correlated.

Since the results are in agreement they could be combined. However, the 
resulting improvement is not visible for the statistical error and small
for the (non-dominant) systematic uncertainty. Furthermore, it is very 
cumbersome to average the leptonic branching fractions alone in the framework
of the global analysis where all branching ratios are derived in a consistent 
way. For these reasons the final results are taken from the global analysis,
while the independent analysis provides a meaningful check for the leptonic
channels.

\section{Discussion of the results}
\label{phys}

\subsection{Comparison with other experiments}

\begin{figure}
   \centerline{\psfig{file=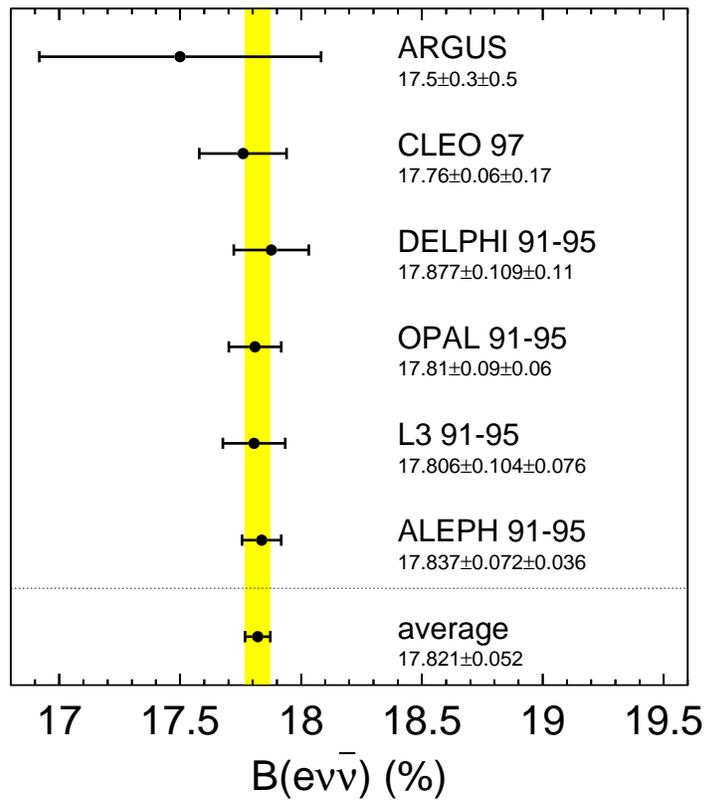,width=120mm}}
   \caption{Comparison of ALEPH measurement with published precise results
   from other experiments for $\tau \ra e \nu \bar{\nu}$. 
   References for other experiments are
   ARGUS~\cite{argus_be}, CLEO~\cite{cleo_be}, DELPHI~\cite{delphi_be},
   OPAL~\cite{opal_be}, L3~\cite{l3_be}.}
   \label{bre}
\end{figure}

\begin{figure}
   \centerline{\psfig{file=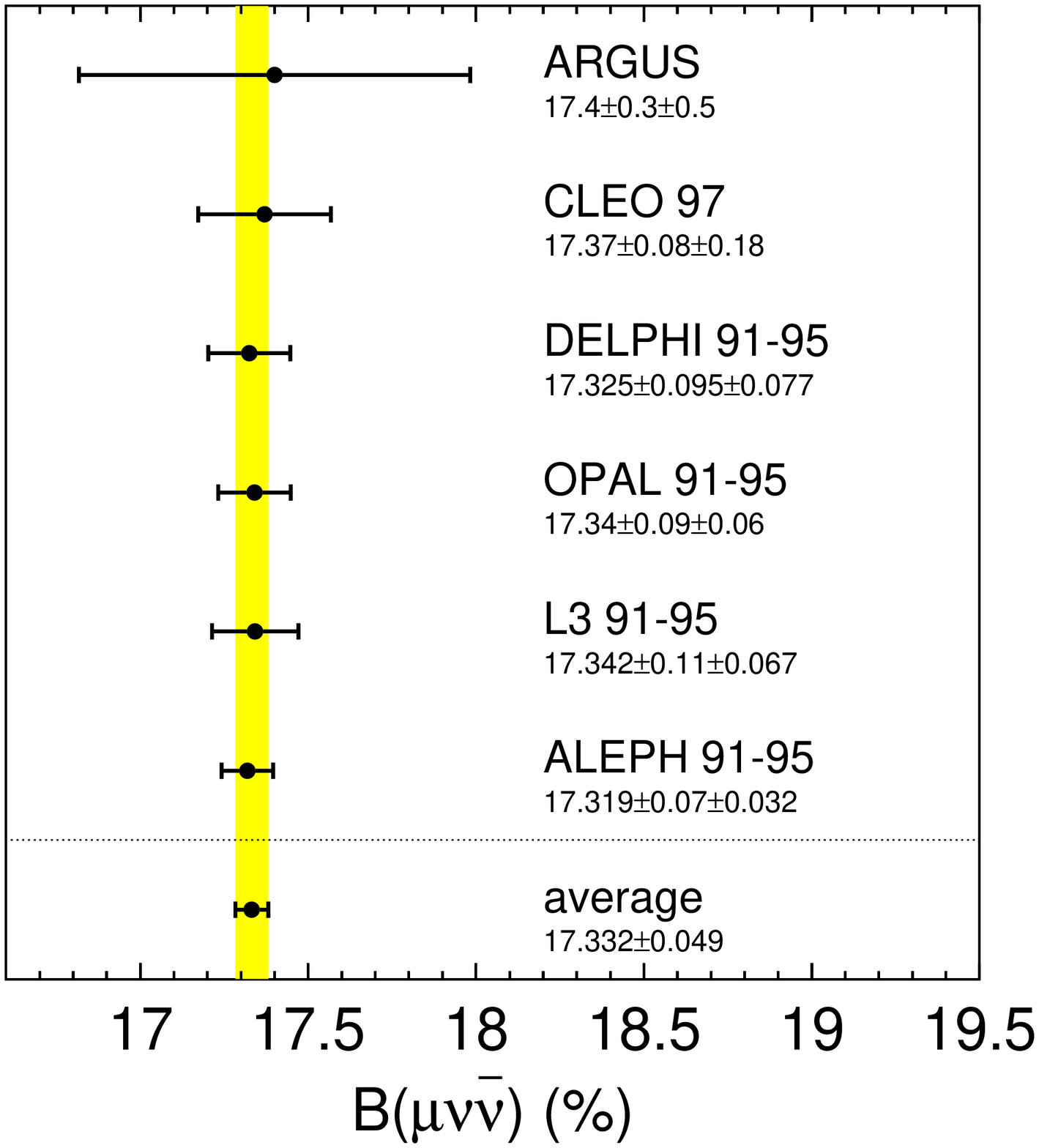,width=120mm}}
   \caption{Comparison of ALEPH measurement with published precise results
   from other experiments for $\tau \ra \mu \nu \bar{\nu}$. 
   References for other experiments are
   ARGUS~\cite{argus_be}, CLEO~\cite{cleo_be}, DELPHI~\cite{delphi_be},
   OPAL~\cite{opal_bmu}, L3~\cite{l3_be}.}
   \label{brmu}
\end{figure}

\begin{figure}
   \centerline{\psfig{file=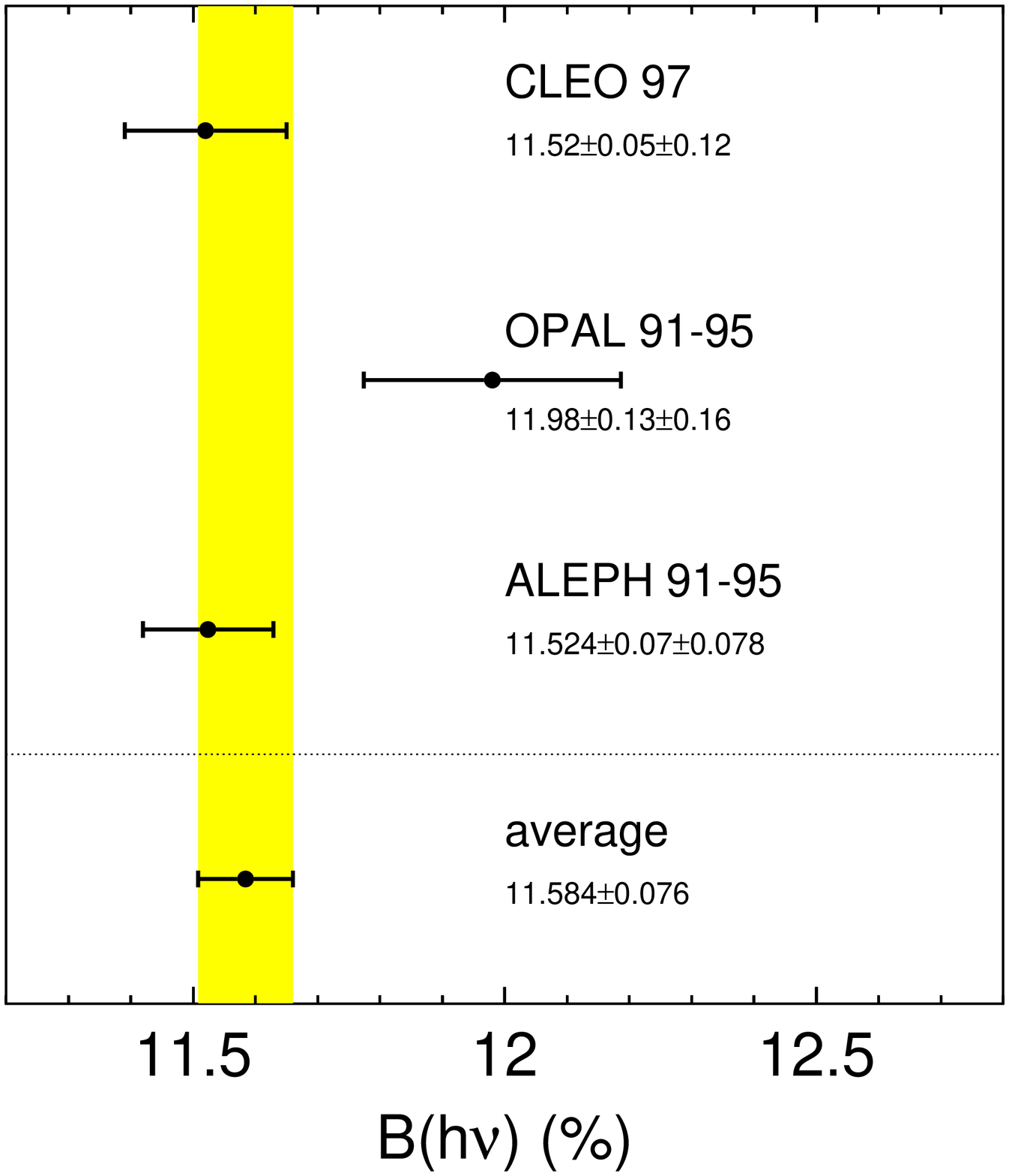,width=120mm}}
   \caption{Comparison of ALEPH measurement with published precise results
   from other experiments for $\tau \ra h \nu$ (sum of $\pi \nu$ and
   $K \nu$). References for other experiments are
   CLEO~\cite{cleo_be}, OPAL~\cite{opal_bh}.}
   \label{brh}
\end{figure}

\begin{figure}
   \centerline{\psfig{file=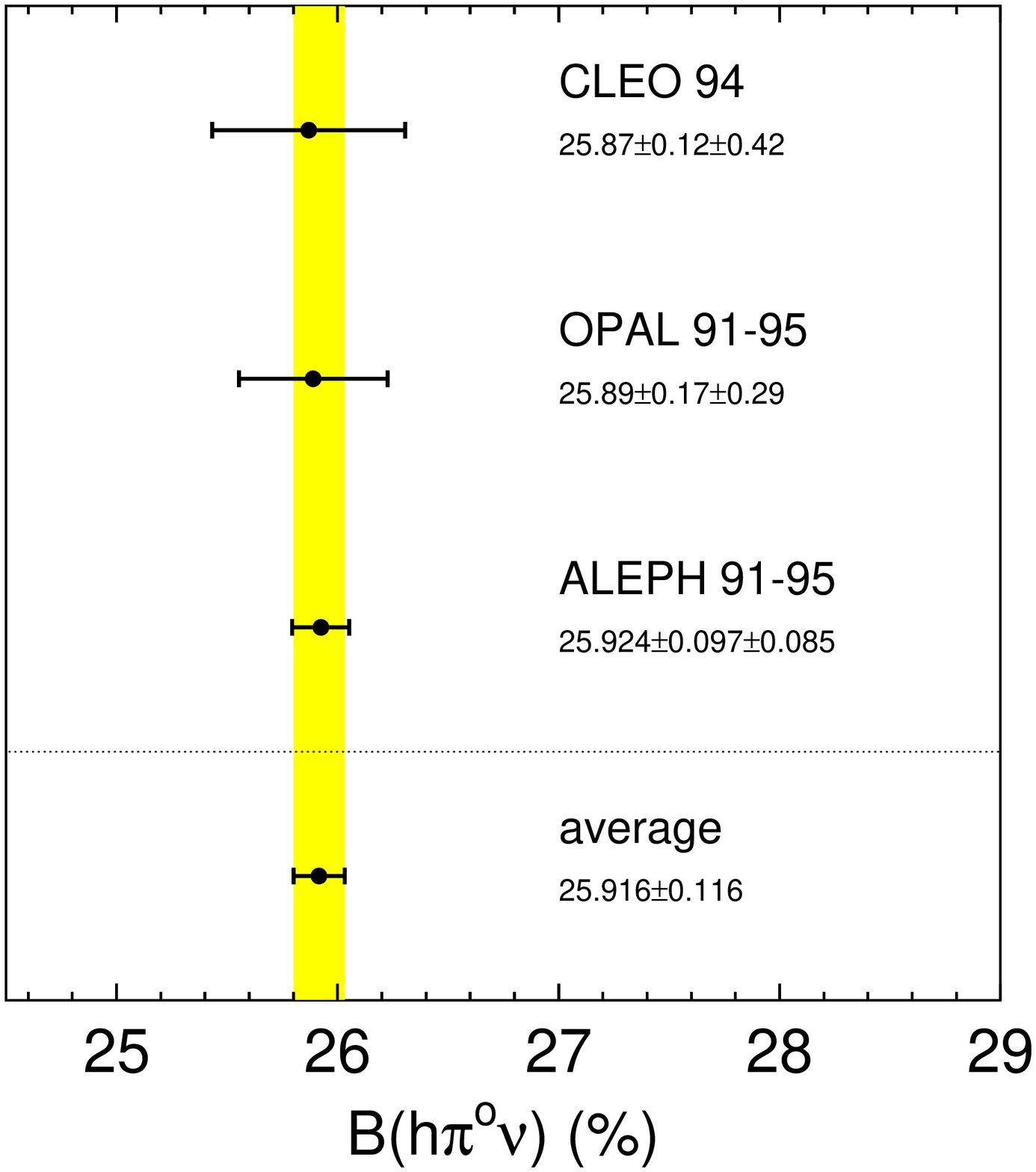,width=120mm}}
   \caption{Comparison of ALEPH measurement with published precise results
   from other experiments for $\tau \ra h\piz \nu $ (sum of $\pi\piz \nu$ and
   $K\piz \nu$). References for other experiments are
   CLEO~\cite{cleo_bhpi0}, OPAL~\cite{opal_bh}.}
   \label{brhpi0}
\end{figure}

\begin{figure}
   \centerline{\psfig{file=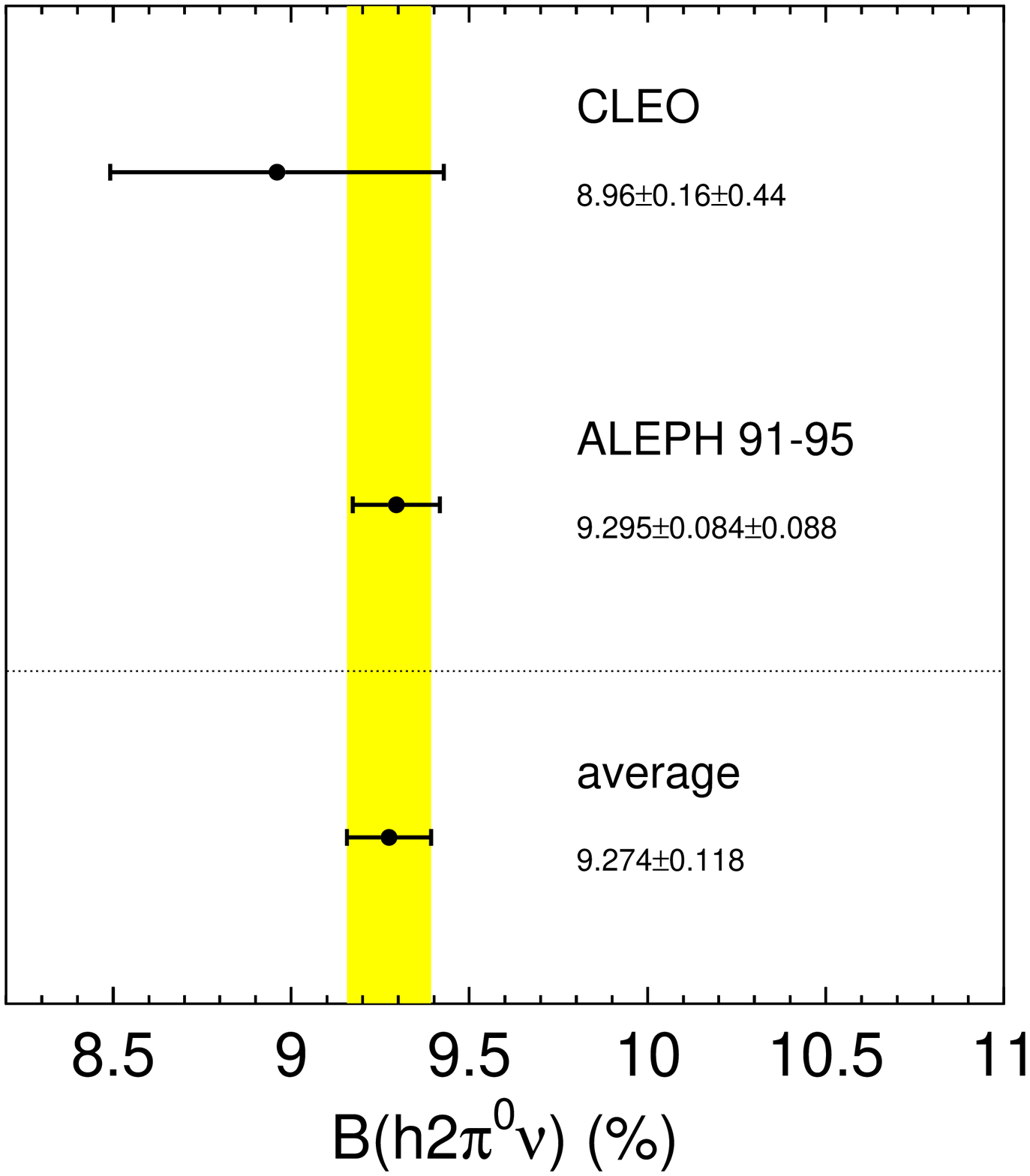,width=120mm}}
   \caption{Comparison of ALEPH measurement with published precise results
   from other experiments for $\tau \ra h2\piz \nu $ (sum of $\pi 2\piz \nu$ 
   and $K 2\piz \nu$). References for other experiments are
   CLEO~\cite{cleo_bh2pi0}.}
   \label{brh2pi0}
\end{figure}

\begin{figure}
   \centerline{\psfig{file=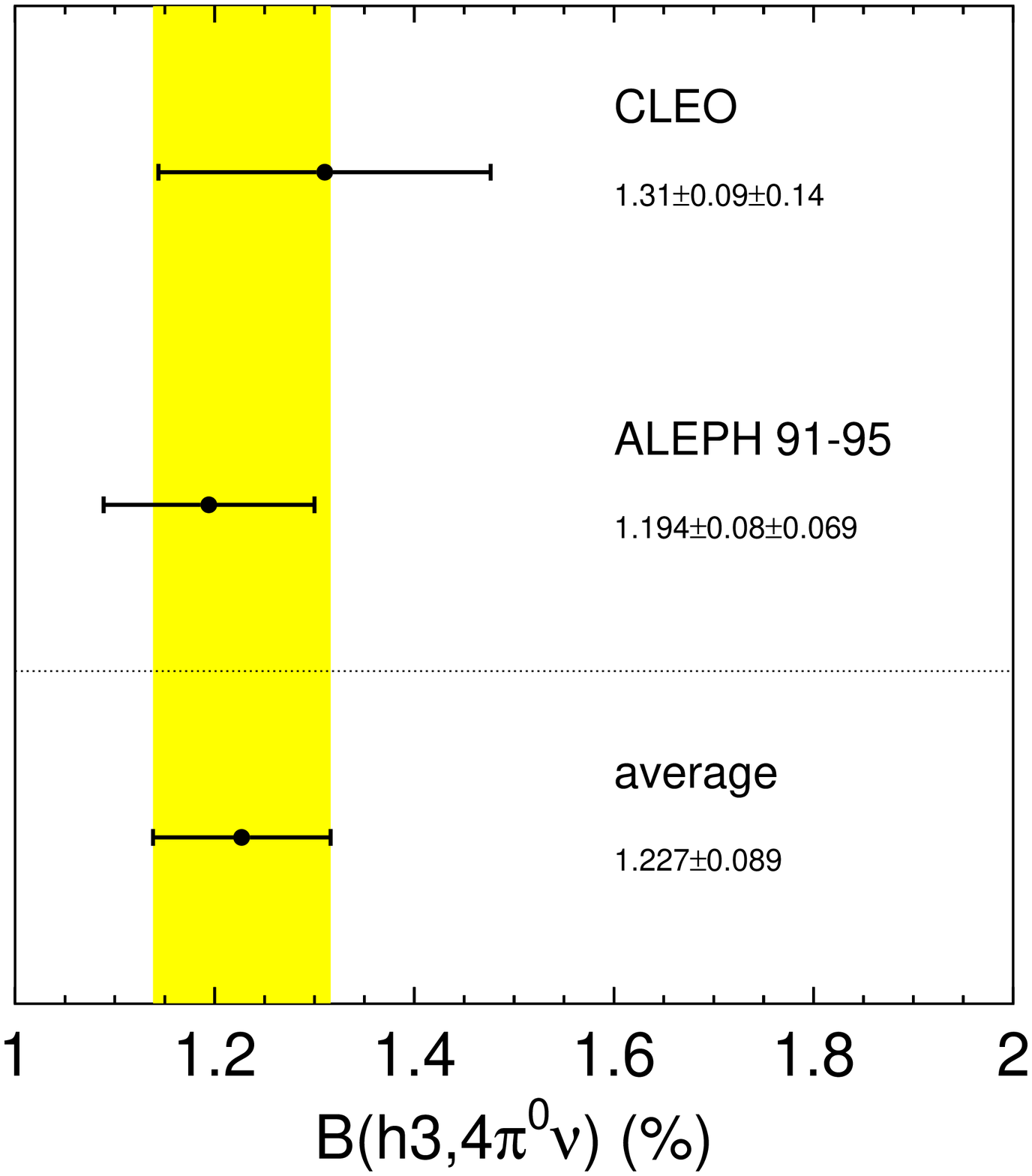,width=120mm}}
   \caption{Comparison of ALEPH measurement with published precise results
   from other experiments for $\tau \ra h (3,4)\piz \nu $ (sum of 
   $\pi (3,4)\piz \nu$ and $K (3,4)\piz \nu$). References for other experiments
   are CLEO~\cite{cleo_bh2pi0}.}
   \label{brh3-4pi0}
\end{figure}

\begin{figure}
   \centerline{\psfig{file=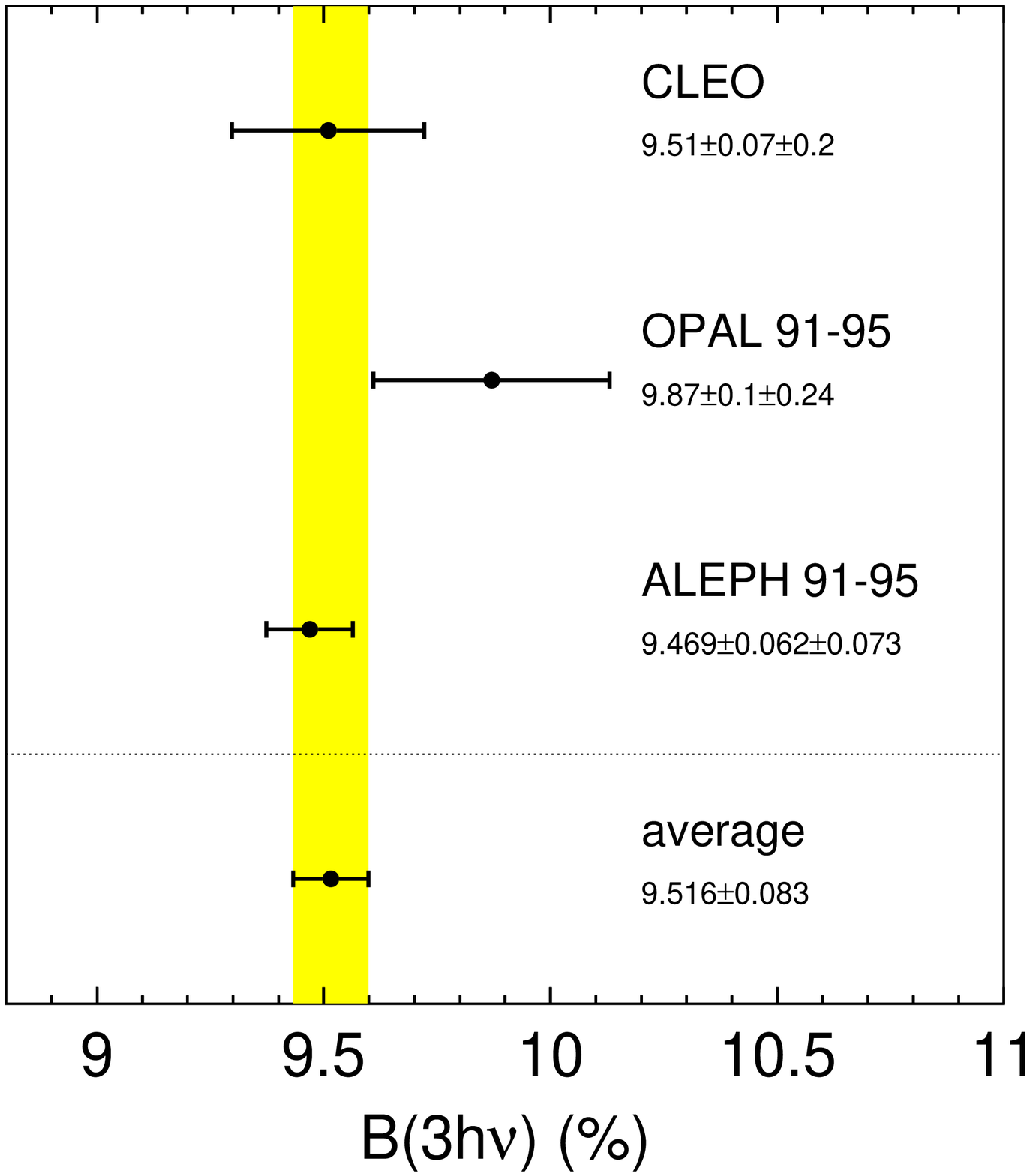,width=120mm}}
   \caption{Comparison of ALEPH measurement with published precise results
   from other experiments for $\tau \ra 3h \nu $ (sum of $3\pi \nu$,
   $K 2\pi \nu$ and $2K \pi \nu$). References for other experiments are
   CLEO~\cite{cleo_b3h}, OPAL~\cite{opal_b3h}.}
   \label{br3h}
\end{figure}

\begin{figure}
   \centerline{\psfig{file=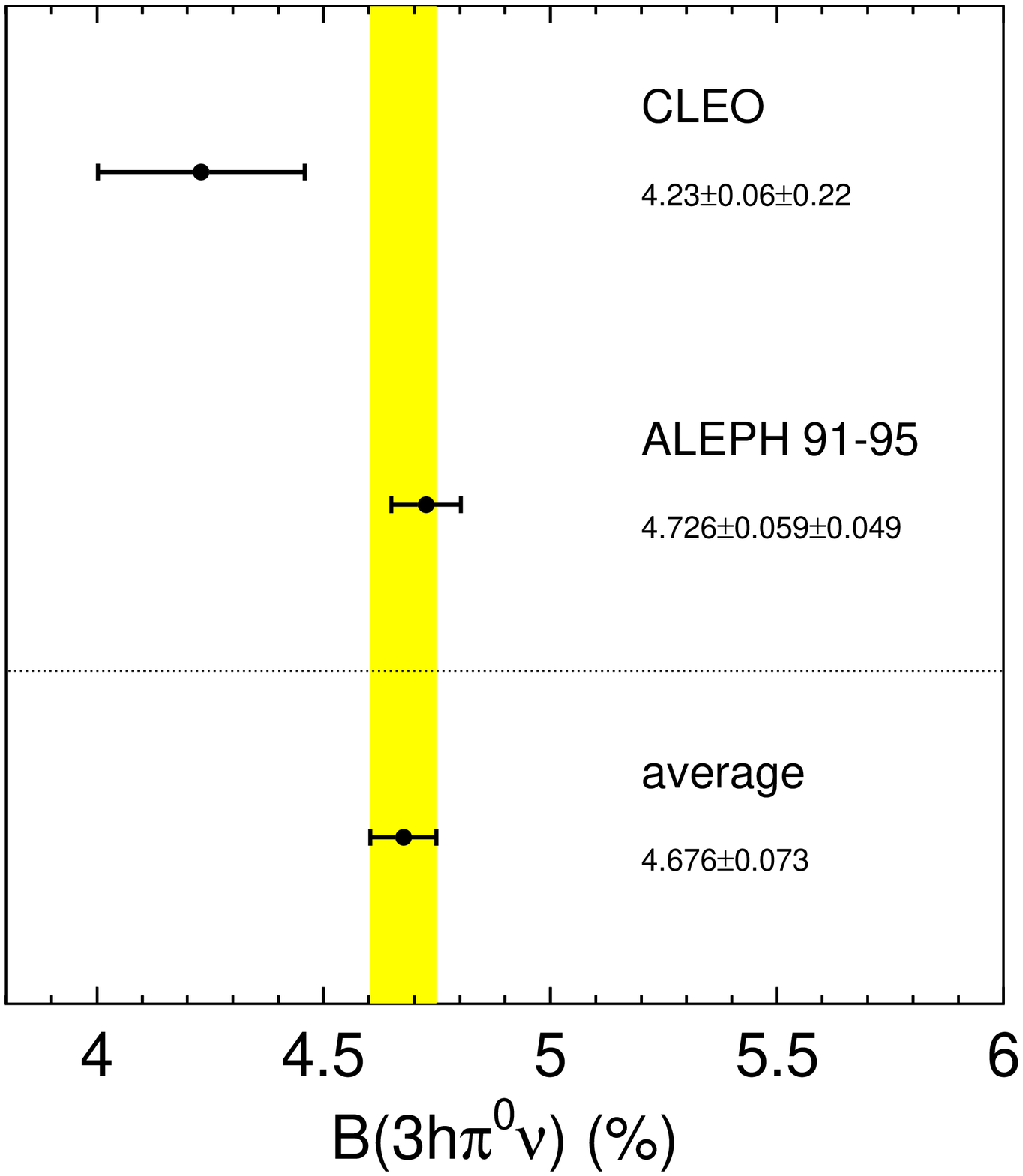,width=120mm}}
   \caption{Comparison of ALEPH measurement with published precise results
   from other experiments for $\tau \ra 3h\piz \nu $ (sum of $3\pi \piz \nu$, 
   $K 2\pi \nu$ and $2K \pi \nu$). References for other experiments are
   CLEO~\cite{cleo_b3h}.}
   \label{br3hpi0}
\end{figure}

\begin{figure}
   \centerline{\psfig{file=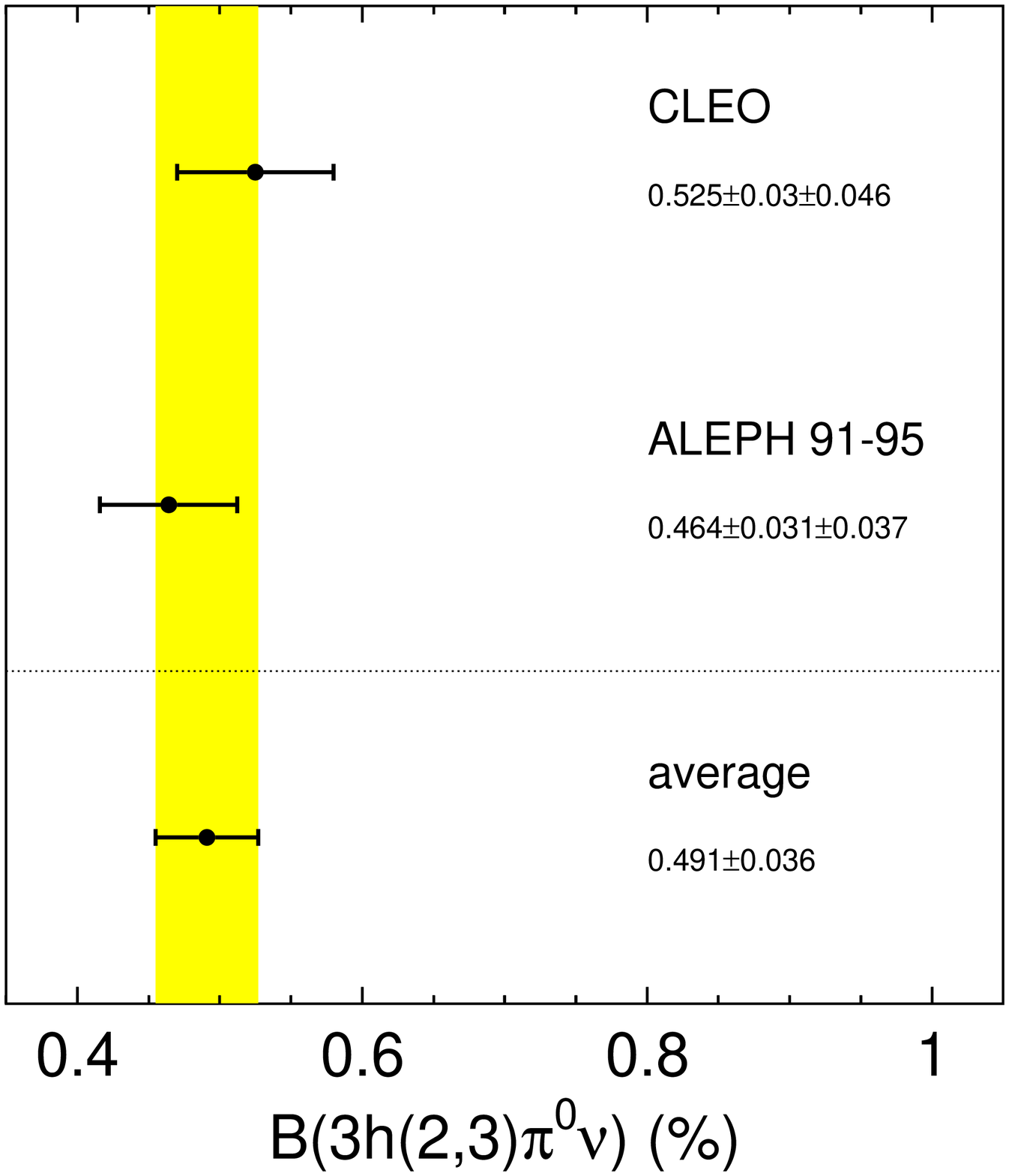,width=120mm}}
   \caption{Comparison of ALEPH measurement with published precise results
   from other experiments for $\tau \ra 3h (2,3)\piz \nu $ (sum of 
   $3\pi (2,3)\piz \nu$, $K 2\pi (2,3)\piz \nu$ and $2K \pi (2,3)\piz \nu$).
   References for other experiments are
   CLEO~\cite{cleo_b3h2-3pi0}.}
   \label{br3h2-3pi0}
\end{figure}

\begin{figure}
   \centerline{\psfig{file=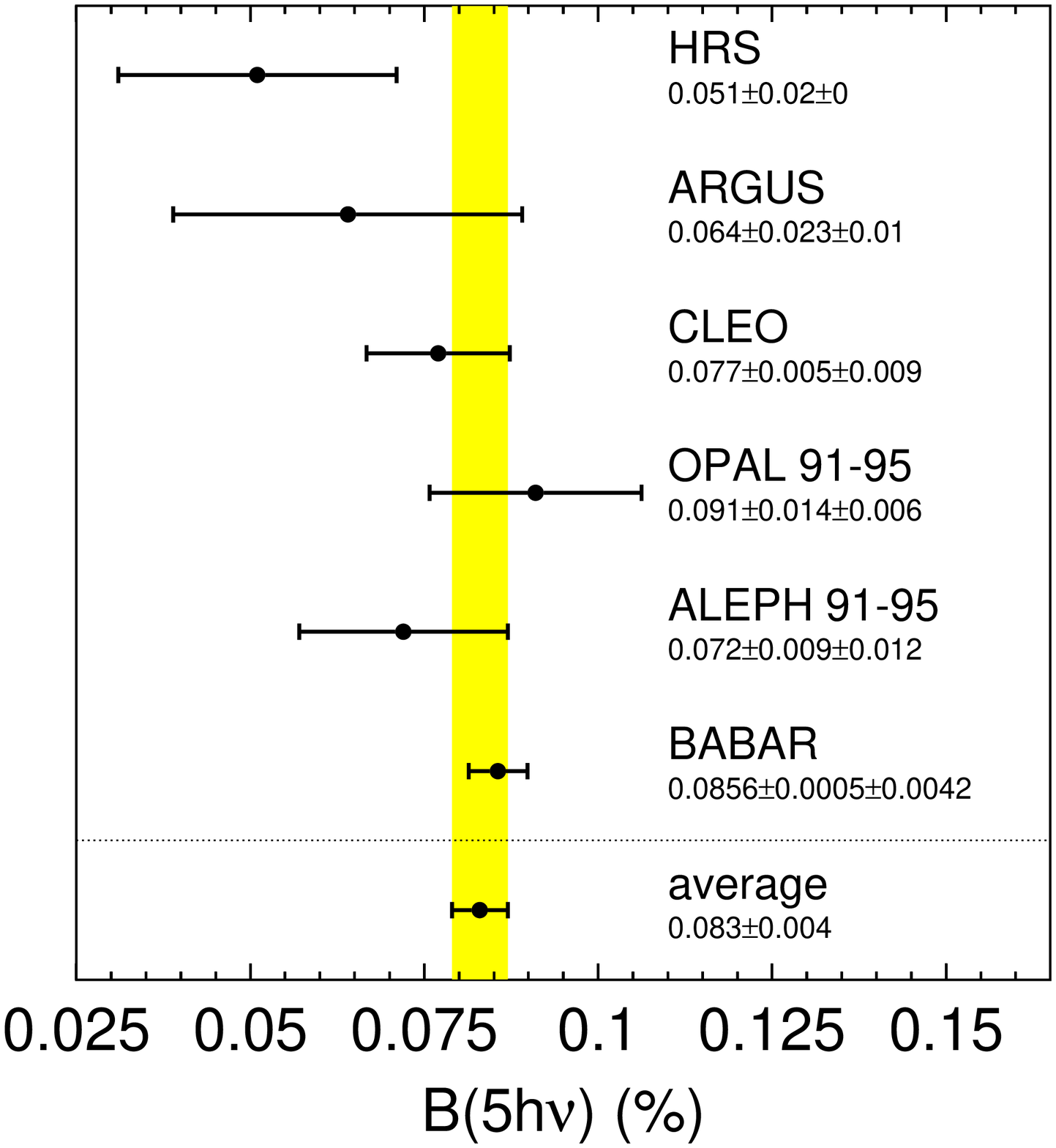,width=120mm}}
   \caption{Comparison of ALEPH measurement with published precise results
   from other experiments for $\tau \ra 5h \nu$. References for other 
   experiments are HRS~\cite{hrs_b5h}, ARGUS~\cite{argus_b5h},
   CLEO~\cite{cleo_b5h}, OPAL~\cite{opal_b5h}, BABAR~\cite{babar_b5h}.}
   \label{br5h}
\end{figure}

\begin{figure}
   \centerline{\psfig{file=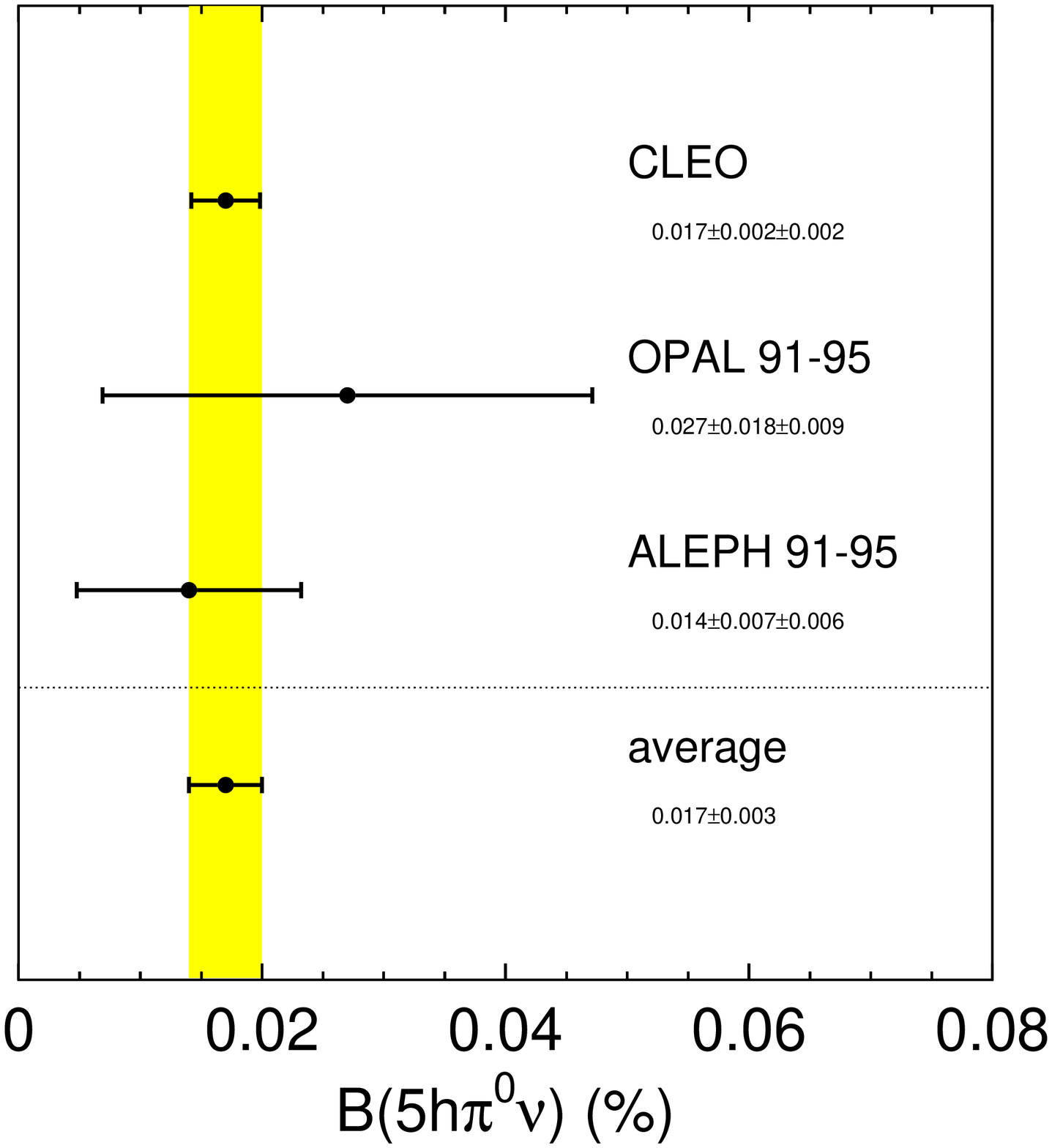,width=120mm}}
   \caption{Comparison of ALEPH measurement with published precise results
   from other experiments for $\tau \ra 5h\piz \nu$.
   References for other experiments are
   CLEO~\cite{cleo_b5hpi0}, OPAL~\cite{opal_b5h}.}
   \label{br5hpi0}
\end{figure}

As the results from other experiments are most often given without $K-\pi$
separation for charged hadrons, comparison is made summing branching ratios
with charged kaons and pions. Also contributions from modes with $\eta$ are
included in the relevant hadronic final states.
Figures~\ref{bre}, \ref{brmu}, \ref{brh}, \ref{brhpi0}, \ref{brh2pi0}, 
\ref{brh3-4pi0}, \ref{br3h},  \ref{br3hpi0}, \ref{br3h2-3pi0},
\ref{br5h} and \ref{br5hpi0} show that the results of this analysis are 
in good agreement with those from other experiments. 
In all these cases, ALEPH achieves the best precision, except for
branching ratios under 0.1\%, such as $\tau \ra 5h \nu$ and
$\tau \ra 5h \piz \nu$, where CLEO results are more precise because
of their higher statistics.

A meaningful comparison can be performed between the exclusive fractions 
and the topological branching ratios $B_i$, where $i$ refers to the charged 
particle multiplicity in the decay. Even though the latter have 
essentially no physics interest, their determination can constitute 
a valuable cross check as they depend only on selection efficiency, 
tracking, handling of secondary interactions and electron identification 
for photon conversions, and not on photon identification.
The results from this analysis can be compared in this way with a dedicated 
analysis recently performed by DELPHI~\cite{delphitopol}. 
Summing up appropriately (both analyses assume a negligible
contribution from hadronic multiplicities larger than five, in agreement
with the 90\% CL limit by CLEO~\cite{cleo7pr}, $B_7 < 2.4 \times 10^{-6}$), 
one gets
\beqn
 B_3 &=& (14.652 \pm 0.067 \pm 0.086) \%~, \\
 B_5 &=& (0.093 \pm 0.009 \pm 0.012) \%~, 
\eeqn
in good agreement with the DELPHI values,
$B_3 = (14.569 \pm 0.093 \pm 0.048) \%$ and 
$B_5 = (0.115 \pm 0.013 \pm 0.006) \%$.
The rather small systematic uncertainty in the DELPHI results reflects 
the fact that a sharper study of hadronic interactions can be
performed when only charged particles are considered in the analysis.
In addition the modes with $K^0_s \rightarrow \pi^+ \pi^-$ decays are
subtracted statistically here, rather than trying to identify them on an 
event-by-event basis.

\subsection{Universality in the leptonic charged current}
\label{uni}

\subsubsection{$\mu-e$ universality from the leptonic branching ratios}

In the standard V-A theory with leptonic coupling $g_l$ at the
$W l \overline{\nu}_l$ vertex, the $\tau$ leptonic partial width can be 
computed, including radiative corrections~\cite{marciano-sirlin} and
safely neglecting neutrino masses:
\beq
\Gamma(\tau \rightarrow \nu_\tau l \overline{\nu}_l (\gamma)) =
 \frac {G_\tau G_l m^5_\tau}{192 \pi^3} f\left(\frac {m^2_l}{m^2_\tau}\right)
 \delta^\tau_W \delta^\tau_\gamma~,
\eeq
where
\beqn
 G_l &=& \frac {g^2_l}{4 \sqrt{2} M^2_W}~,  \nonumber \\
 \delta^\tau_W &=& 1 + \frac {3}{5} \frac {m^2_\tau}{M^2_W}~, \nonumber \\
 \delta^\tau_\gamma &=& 1+\frac {\alpha(m_\tau)}{2\pi}
 \left(\frac {25}{4}-\pi^2\right)~, \nonumber \\
 f(x) &=& 1 -8x +8x^3 -x^4 -12x^2 {\rm ln}x~. 
\eeqn
Numerically, the W propagator correction and the radiative corrections
are small:
$\delta^\tau_W =1+2.9\times 10^{-4}$ and 
$\delta^\tau_\gamma=1-43.2\times 10^{-4}$.

Taking the ratio of the two leptonic branching fractions, a direct test
of $\mu-e$ universality is obtained. The measured ratio
\beq
  \frac {B_\mu} {B_e} = 0.9709 \pm 0.0060 \pm 0.0029
\eeq
agrees with the expectation 
of 0.97257 when universality holds. Alternatively
the measurements yield the ratio of couplings
\beq
  \frac {g_\mu} {g_e} = 0.9991 \pm 0.0033 
\eeq
which is consistent with unity.

This result is in agreement with the best test of $\mu-e$ universality
of the W couplings obtained in the comparison of the rates for
$\pi \rightarrow \mu \overline{\nu}_\mu$ and 
$\pi \rightarrow e \overline{\nu}_e$ decays, where the results 
of the two most accurate experiments~\cite{triumf,psi} can be averaged
to yield $\frac {g_\mu} {g_e} = 1.0012 \pm 0.0016 $. The results have
comparable precision, but it should be pointed out that they are in fact
complementary. The $\tau$ result given here probes the coupling to
a transverse W (helicity $\pm$1) while the $\pi$ decays measure the
coupling to a longitudinal W (helicity 0). It is indeed conceivable
that either approach could be sensitive to different nonstandard
corrections to universality.

Since $B_e$ and $B_\mu$ are consistent with $\mu-e$ universality their
values can be combined, taking common errors into account, into 
a consistent leptonic branching ratio for a massless lepton 
(the electron, since $f(\frac {m^2_e}{m^2_\tau}) = 1$ within $10^{-6}$)
\beq
B^{(m_l=0)}_l = (17.822 \pm 0.044 \pm 0.022 )\%~,
\eeq
where the first error is statistical and the second systematic.

\subsubsection{Tests of $\tau-\mu$ and  $\tau-e$ universality}

Comparing the rates for 
$\Gamma(\tau \rightarrow \nu_\tau e \overline{\nu}_e (\gamma))$,
$\Gamma(\tau \rightarrow \nu_\tau \mu \overline{\nu}_\mu (\gamma))$ 
and $\Gamma(\mu \rightarrow \nu_\mu e \overline{\nu}_e (\gamma))$
provides direct tests of the universality of $\tau-\mu-e$ couplings.
Taking the relevant ratios with calculated radiative corrections, 
one obtains
\beqn
 \left(\frac {g_\tau}{g_\mu}\right)^2 &=& \frac {\tau_\mu}{\tau_\tau} 
 \left(\frac{m_\mu}{m_\tau}\right)^5 B_e
 \frac {f(\frac {m^2_e}{m^2_\mu})}{f(\frac {m^2_e}{m^2_\tau})} 
 \Delta_W \Delta_\gamma~, \\
 \left(\frac {g_\tau}{g_e}\right)^2 &=& \frac {\tau_\mu}{\tau_\tau} 
 \left(\frac{m_\mu}{m_\tau}\right)^5 B_\mu 
 \frac {f(\frac {m^2_e}{m^2_\mu})}{f(\frac {m^2_\mu}{m^2_\tau})} 
 \Delta_W \Delta_\gamma~,
\eeqn 
where $f(\frac {m^2_e}{m^2_\mu}) = 0.9998$,
$\Delta_W = \frac {\delta^\mu_W}{\delta^\tau_W} = 1 - 2.9\times 10^{-4}$,
$\Delta_\gamma = \frac {\delta^\mu_\gamma}{\delta^\tau_\gamma} 
= 1 + 8.5\times 10^{-5}$,
and $\tau_l$ is the lepton $l$ lifetime.

From the present measurements of $B_e$, $B_\mu$, the $\tau$ 
mass~\cite{pdg2004}, $m_\tau = (1777.03^{+0.30} _{-0.26})$~MeV (dominated
by the BES result~\cite{besmtau}), the $\tau$ lifetime~\cite{pdg2004},
$\tau_\tau = (290.6 \pm 1.1)$ fs and the other 
quantities from Ref.~\cite{pdg2004}, universality can be tested:
\beqn
 \frac {g_\tau}{g_\mu} &=& 
 1.0009 \pm 0.0023(B_e) \pm 0.0019(\tau_\tau) \pm 0.0004(m_\tau)~, \\
 \frac {g_\tau}{g_e} &=& 
 1.0001 \pm 0.0022(B_\mu) \pm 0.0019(\tau_\tau) \pm 0.0004(m_\tau)~.
\eeqn

\subsubsection{$\tau-\mu$ universality from the pionic branching ratio}
\label{pi_uni}

The measurement of $B_\pi$ also permits an independent test of 
$\tau-\mu$ universality through the relation
\beq
 \left(\frac{g_\tau}{g_\mu}\right)^2 = 
     \frac{B_\pi}{B_{\pi \rightarrow \mu \overline{\nu}_\mu}}
     \frac{\tau_\pi}{\tau_\tau} \frac{2 m_\pi m_\mu^2}{m_\tau^3}
     \left(\frac{1-\frac{m_\mu^2}{m_\pi^2}}
                {1-\frac{m_\pi^2}{m_\tau^2}}\right)^2
      \frac {1}{\delta_{\tau / \pi}}~,
\label{eq_pi_uni}
\eeq
where the radiative correction~\cite{decker-fink} amounts to
$\delta_{\tau / \pi} = 1.0016 \pm 0.0014$.
Using the world-averaged values for the
$\tau$ and $\pi$ ($\tau_\pi$) lifetimes, and the branching ratio for the decay
$\pi \rightarrow \mu \nu$~\cite{pdg2004}, the present result
for $B_\pi$, one obtains
\beq
  \frac {g_\tau} {g_\mu} = 
0.9962 \pm 0.0048(B_\pi) \pm 0.0019(\tau_\tau) \pm 0.0002(m_\tau) \pm 0.0007(rad)~,
\eeq
comparing the measured value, $B_\pi = (10.828 \pm 0.105$)\%,
to the expected one assuming universality, ($10.910 \pm 0.042$)\%.

The two determinations of $\frac {g_\tau}{g_\mu}$ obtained from $B_e$
and $B_\pi$ are consistent with each other and can be combined to yield
\beq
  \frac {g_\tau} {g_\mu} = 
1.0000 \pm 0.0021(B_e,B_\pi) \pm 0.0019(\tau_\tau) \pm 0.0004(m_\tau)~.
\eeq
Universality of the $\tau$ and $\mu$ charged-current couplings holds
at the 0.29\% level with about equal contributions from the present 
determination of $B_e$ and $B_\pi$, and the world-averaged value 
for the $\tau$ lifetime.

\begin{figure}
   \centerline{\psfig{file=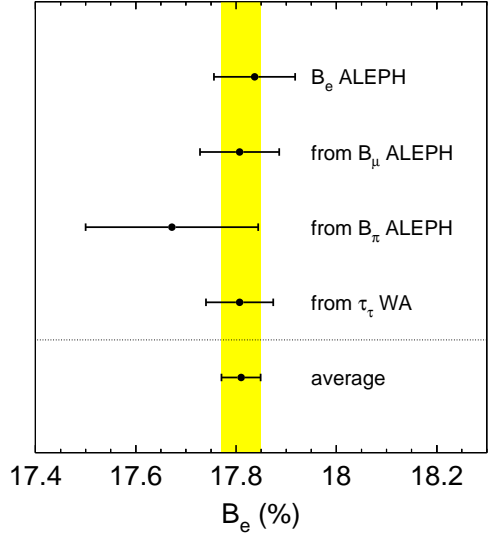,width=80mm}}
   \caption{The measured value for $B_e$ compared to predictions from
  other measurements assuming leptonic universality. The vertical band
  gives the average of all determinations.}
   \label{allbe}
\end{figure}

The consistency of the present branching ratio measurements with leptonic
universality is displayed in Fig.~\ref{allbe} where the result for $B_e$
is compared to computed values of $B_e$ using as input $B_\mu$ (assuming
$e - \mu$ universality), $\tau_\tau$ and $\tau_\mu$ ($\mu - \tau$ 
universality), and $B_\pi$ and $\tau_\pi$ ($\mu - \tau$ universality). 
All values are consistent and yield the average
\beq
 B_e^{\rm universality} = (17.810 \pm 0.039)\%~.
\label{uni_be}
\eeq

\subsection{The $\pi\piz$ branching ratio in the context 
of $a_\mu^{had}$}

The $\pi\piz$ final state is dominated by the $\rho$ resonance as 
demonstrated in Fig.~\ref{spec_4}. Its mass distribution ---or better, 
the corresponding spectral function, see Section~\ref{sf}--- 
is a basic ingredient of vacuum polarization calculations,
such as used for computing the hadronic contribution to the
anomalous magnetic moment of the muon $a_\mu^{had}$. In this case the
$\rho$ contribution is dominant (71\%) and therefore controls the final 
precision of the result. It was proposed in Ref.~\cite{adh} to use the
spectral functions obtained from the measurement of hadronic $\tau$ decays
in order to improve the precision of the prediction for $a_\mu^{had}$.
The calculation was later improved with the help of QCD constraints for
energies above the $\tau$ mass~\cite{dh97a} and even below~\cite{dh97b}.

The normalization of the spectral function is provided by the
branching fraction $B_{\pi\piz}$. The present world average is
completely dominated by the published ALEPH result~\cite{aleph13_h}. 
The new result given here is larger by 0.68\%, thus one can expect
a slightly larger contribution to $a_\mu^{had}$. 

\begin{figure}[t]
   \centerline{\psfig{file=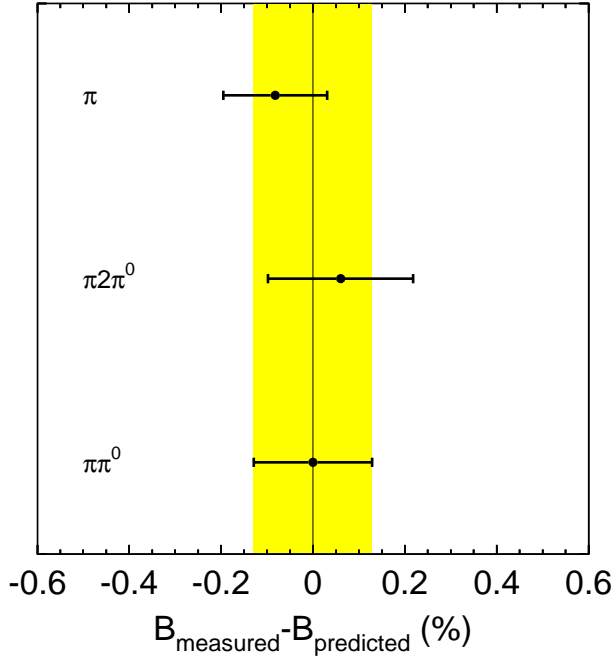,width=80mm}}
   \caption{The differences between the measured values for $B_{\pi}$ and
  $B_{\pi 2\piz}$ and their respective predictions from leptonic universality
  and isospin symmetry with $\pi$-mass breaking, compared to the precision
  on $B_{\pi \piz}$.}
   \label{brho_check}
\end{figure}

A new evaluation~\cite{dehz} of $a_\mu^{had}$ was available, using the 
preliminary spectral functions from the present analysis, the published CLEO 
results~\cite{cleo_2pi} and new results from $e^+e^-$ annihilation 
from CMD-2~\cite{cmd2_old}. Revision of the CMD-2 results~\cite{cmd2_new}
prompted a re-evaluation~\cite{dehz03}, which revealed a disagreement 
between the $\tau$ and $e^+e^-$ spectral functions (see Section~\ref{sf}).
Whereas the $\tau$ estimate leads to a prediction of the muon magnetic
moment in agreement with the latest most precise measurement from the BNL 
experiment E-821~\cite{bnlg-2}, the predicted value using only \ee\ data 
lies 2.4 standard deviations below the measurement.  
In view of this situation, it is important to check all the ingredients, 
in particular the determination of the branching ratio $B_{\pi\piz}$. 
As most of the systematic uncertainty in $B_{\pi\piz}$ 
comes from $\gamma / \piz$ reconstruction, it is helpful to cross
check the results in the ``adjacent'' hadronic modes, {\it i.e.} the
$\pi$ and $\pi 2\piz$ channels. This is possible if universality in the 
weak charged current is assumed, leading to an absolute prediction of 
$B_{\pi}$ using as input the $\tau$ and $\pi$ lifetimes 
(Section~\ref{pi_uni}, Eq.~(\ref{eq_pi_uni})), and by 
computing $B_{\pi 2\piz}$ assuming isospin symmetry from the measurement of 
$B_{3\pi}$ which is essentially uncorrelated with $B_{\pi\piz}$ 
(Section~\ref{a1}).
The comparison, shown in Fig.~\ref{brho_check}, does not point to a
systematic bias in the determination of $B_{\pi\piz}$ within the quoted 
uncertainty.

\subsection{$a_1$ decays to $3 \pi$ and $\pi 2 \pi^0$}
\label{a1}

With the level of precision reached it is interesting to compare the rates
in the $3 \pi$ and $\pi 2 \pi^0$ channels which are dominated
by the $a_1$ resonance. The dominant $\rho \pi$ intermediate state leads
to equal rates, but a small isospin-breaking effect is expected from
different charged and neutral $\pi$ masses, slightly favouring the  
$\pi 2 \pi^0$ channel, as discussed below.

A recent CLEO partial-wave analysis of the $\pi 2 \pi^0$ final 
state~\cite{cleoa1} has shown that the situation is in fact much more 
complicated with many intermediate states, in particular involving 
isoscalars, amounting to about 20\% of the total rate and producing 
strong interference effects. A good description of the $a_1$ decays 
was achieved in the CLEO study, which can be applied to the $3 \pi$ final 
state, predicting~\cite{cleoa1} a ratio of the rates $3 \pi$/$\pi 2 \pi^0$
equal to 0.985. This value, which includes known isospin-breaking from
the pion masses, turns out to be in good agreement with the measured value
from this analysis which shows the expected trend
\beq
 \frac{B_{3 \pi}}{B_{\pi 2 \pi^0}} = 0.979 \pm 0.018~.
\eeq

\subsection{Separation of vector and axial-vector contributions}
\label{vasep}

\subsubsection{Strange and nonstrange contributions}
\label{nonstrange}

From the complete analysis of the $\tau$ branching ratios presented in
this paper, it is possible to determine the nonstrange vector ($V$) 
and axial-vector ($A$) contributions to the total $\tau$ hadronic width, 
conveniently expressed in terms of their ratios to the leptonic width, 
called $R_{\tau,V}$ and $R_{\tau,A}$, respectively. The determination of 
the strange counterpart $R_{\tau,S}$ is already published~\cite{alephksum}.

The ratio \Rt\ for the total hadronic width is calculated from the leptonic 
\brs\ alone, and eventually from the electronic \br\ alone,
\beqn
  R_\tau&=&\frac{1-B_e-B_\mu}{B_e}
           =\frac{1}{B_e}-1.9726
          \nonumber \\
        &=& 3.642 \pm 0.012~. 
\eeqn
taking for $B(\tau^-\rightarrow e^-\,\bar{\nu}_e\nu_\tau)$ the value obtained
in Section~\ref{uni} assuming universality in the leptonic weak current.
All $R$ values given below are rescaled so that the
sum of the hadronic and the leptonic branching ratios, the latter computed 
using the ``universal'' value given in Eq.~\ref{uni_be}), add up to unity.
The deviations introduced in this way are extremely small, less than one tenth
of the experimental error, but this procedure guarantees the consistency 
between all values.
Using the ALEPH measurement of the strange branching ratios~\cite{alephksum},
supplemented by the small contribution from the $K^{*-} \eta$ channel
measured by CLEO~\cite{cleoksteta}, 
\beqn
   B_S       &=& (2.85 \pm 0.11)~\%~,\nonumber \\
   R_{\tau,S}&=& 0.1603\pm0.0064~,
\eeqn
the following result is obtained for the nonstrange component
\beqn
\label{eq_rtauvpa}
    B_{V+A}        &=& (62.01 \pm 0.14)~\%~,\nonumber \\
    R_{\tau,V+A}   &=& 3.482 \,\pm\, 0.014~.
\eeqn

Separation of $V$ and $A$ components in hadronic final states with only pions
is straightforward. Isospin invariance relates the spin-parity of such 
systems to their number of pions: $G$-parity =1 (even number) corresponds to
vector states, while $G=-1$ (odd number) tags axial-vector states. This
property places a strong requirement on the efficiency of $\piz$
reconstruction, a constraint that was emphasized in this analysis.

\subsubsection{V and A separation in modes with a $K \overline{K}$ pair}
\label{kkbar}

Modes with a $K \overline{K}$ pair are not in general eigenstates 
of $G$-parity and contribute to both $V$ and $A$ channels. While the decay 
to $K^- K^0$ is pure vector, the situation is a priori not clear in the
$K \overline{K} \pi$ channel, observed in three charged modes:  
$K^- K^+ \pi^-$, $K^- K^0 \pi^0$ and $K^0 \overline{K}^0 \pi^-$.
Three sources of information exist on the possible $V$/$A$ content in this
decay mode:\\
\indent {\it (i)} in the ALEPH analysis of $\tau$ decay modes with 
kaons~\cite{alephksum}, an estimate of the vector contribution was obtained
using the available $e^+e^-$ annihilation data in the  $K \overline{K} \pi$
channel, extracted in the $I=1$ state. In fact, this contribution was found 
to be very small, yielding a branching ratio consistent with zero,
$B_{\rm CVC} (\tau \rightarrow \nu_\tau [K \overline{K} \pi]_V) = 
(0.26 \pm 0.39) \times 10^{-3}$, corresponding to an axial fraction,
$f_A ({\rm CVC})= 0.94^{+0.06}_{-0.08}$.\\
\indent {\it (ii)} The CVC/ALEPH result is corroborated by a partial-wave 
and lineshape analysis of the $a_1$ resonance from $\tau$ decays in the 
$\nu_\tau \pi^- 2\pi^0$ mode by CLEO~\cite{cleoa1}. 
The observation through unitarity of the 
opening of the $K^* K$ decay mode of the $a_1$ is claimed and a branching 
ratio $B(a_1 \rightarrow K^* K) = (3.3 \pm 0.5)\%$ obtained. Knowing the
decay rate of $\tau \rightarrow \nu_\tau a_1$, it is easy to see that such 
a result completely saturates, and even exceeds, the total rate for the
$K \overline{K} \pi$ channel. The corresponding axial fraction turns out to be
$f_A ({\rm CLEO} ~3\pi) = 1.30 \pm 0.24$.\\
\indent {\it (iii)} A new piece of information, also contributed by CLEO, 
but conflicting with the two previous results, was recently 
published~\cite{cleo_kkpi}. It is based on a partial-wave analysis in the
$K^- K^+ \pi^-$ channel using two-body resonance production and including
many possibly contributing channels. A much smaller axial contribution is
obtained, $f_A ({\rm CLEO} ~KK\pi) = 0.56 \pm 0.10$.

\begin{figure}[t]
   \centerline{\psfig{file=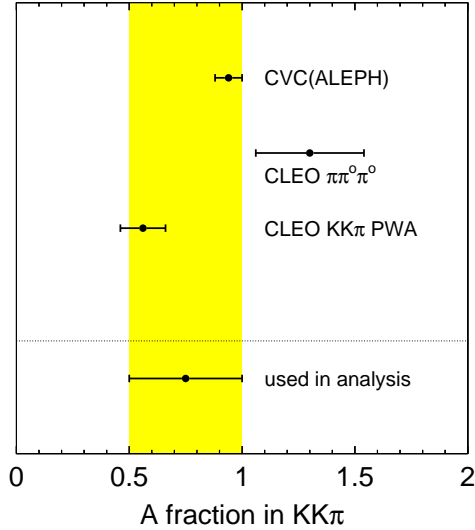,width=80mm}}
   \caption{The axial-vector fraction $f_A$ of the hadrons in the 
            $\tau \rightarrow \nu_\tau K \overline{K} \pi$ decay mode from
            three independent determinations: ALEPH~\cite{alephksum} using CVC
            and $e^+e^-$ data, CLEO~\cite{cleoa1} with a partial-wave analysis
            of $a_1 \rightarrow \pi^- 2\pi^0$ decays, and 
            CLEO~\cite{cleo_kkpi} with a direct partial-wave analysis of the
            $\tau \rightarrow \nu_\tau K^+ K^- \pi^-$ decays. The shaded band
            corresponds to the conservative range used in this analysis.}
   \label{va_kkpi}
\end{figure}

The three determinations are displayed in Fig.~\ref{va_kkpi}. Since they are
not consistent, a conservative value of $f_A = 0.75 \pm 0.25$ is assumed.
It encompasses the range allowed by the previous results and still represents
some progress over our former analyses~\cite{alephvsf,alephasf} where a value 
of $0.5 \pm 0.5$ was used. For the decays into $K \overline{K}\pi\pi$ no 
information is available in this respect and the same conservative fraction of 
$0.5 \pm 0.5$ is assumed.

All the present measurements and previous ones for modes with a 
$K \overline{K}$ pair are summarized in a complete list of determined
branching ratios for nonstrange hadronic final states up to six hadrons,
and four with kaons. The results are given in Table~\ref{completeBR}.

\subsubsection{Total V and A contributions in the non-strange sector}
\label{total_va}

The total nonstrange vector and axial-vector contributions 
obtained in this analysis are:
\beqn
\label{eq_rtauv}
   B_V      &=& (31.82 \pm 0.18 \pm 0.12)~\%~,\\
  R_{\tau,V}&=& 1.787 \pm 0.011 \pm 0.007~,\\
\label{eq_rtaua}
   B_A      &=& (30.19 \pm 0.18 \pm 0.12)~\%~,\\
  R_{\tau,A}&=& 1.695 \pm 0.011 \pm 0.007~,
\eeqn
where the second errors reflect the uncertainties in the $V/A$ separation
in the channels with $K \overline{K}$ pairs.
Accounting for the correlations between the respective uncertainties, one
obtains the difference between the vector and axial-vector components,
which is physically related to the amount of nonperturbative 
contributions in the nonstrange hadronic $\tau$ decay width 
(see Section~\ref{qcd}):
\beqn
\label{eq_rtauvma}
    B_{V-A}       &=& (1.63 \pm 0.34 \pm 0.24)~\%~,\\
    R_{\tau,V-A}  &=& 0.092 \pm 0.018 \pm 0.014~,
\eeqn
where again the second error has the same meaning 
as in Eqs.~(\ref{eq_rtauv}) and (\ref{eq_rtaua}). The ratio
\beq
\label{eq_rvmavpa}
    \frac {R_{\tau,V-A}} {R_{\tau,V+A}}     = 0.026 \pm 0.007
\eeq
is a measure of the relative importance of nonperturbative contributions.

These values yield the normalization of the corresponding spectral functions,
which will be determined later for the QCD analysis of hadronic $\tau$
decays and vacuum polarization calculations. Compared to the previous 
analysis, the vector/axial-vector separation is significantly improved 
and will result in better determined nonperturbative contributions.

\subsection{Summary of all measured branching ratios}

A summary list of all $\tau$ branching ratios measured by ALEPH using the 
full LEP-1 statistics is given in Table~\ref{completeBR}.

\begin{table} 
\caption{A summary list of ALEPH branching ratios (\%). Statistical and
systematical uncertainties have been combined in quadrature. The labels $V$,
 $A$ and $S$ refer to the nonstrange vector and axial-vector, and strange
components, respectively. CLEO results on channels with small branching
fractions have been used as indicated.}
\begin{center}
\setlength{\tabcolsep}{1.5pc}
\begin{tabular}{lrrc}
\hline\hline
 mode & B [\%] & \\\hline
   \phyi &    17.837 $\pm$  0.080 &&\\
   \phyii &    17.319 $\pm$  0.077 &&\\
\hline
   \phyiii &    10.828 $\pm$  0.105 &A&\\
   \phyiv &    25.471 $\pm$  0.129  &V&\\
   \phyv &     9.239 $\pm$  0.124  &A&\\
   \phyvi &      0.977 $\pm$  0.090  &V&\\
   \phyxiii &      0.112 $\pm$  0.051  &A&\\
   \phyvii &     9.041 $\pm$  0.097  &A&\\
   \phyviii &     4.590 $\pm$  0.086  &V&\\
   \phyix &      0.392 $\pm$  0.046  &A&\\
   \phyx &      0.013 $\pm$  0.010  &V &estimate\\
   \phyxi &      0.072 $\pm$  0.015  &A&\\
   \phyxii &      0.014 $\pm$  0.009  &V&\\
   $\pi^- \pi^0 \eta$ & 0.180 $\pm$ 0.045  &V& \\
   $(3\pi)^-  \eta$ & 0.039 $\pm$ 0.007  &A& CLEO~\cite{cleoeta3pi}\\
   $a_1^- (\ra \pi^- \gamma)$ & 0.040 $\pm$ 0.020  &A& estimate \\
   $\pi^- \omega (\ra \pi^0 \gamma, \pi^+ \pi^-)$ & 0.253 $\pm$ 0.018     &V& \\
   $\pi^- \pi^0 \omega (\ra \pi^0 \gamma, \pi^+ \pi^-)$ & 0.048 $\pm$
   0.009  &A& + CLEO~\cite{cleoomega}\\
   $(3\pi)^- \omega (\ra \pi^0 \gamma, \pi^+ \pi^-)$ & 0.003 $\pm$
   0.003  &V& CLEO~\cite{cleoeta3pi}\\
\hline
   $K^- K^0 $ &  0.163 $\pm$ 0.027 &V&\\
   $K^- \pi^0 K^0 $    &  0.145 $\pm$ 0.027  & $(75\pm25)\%$ A&\\
   $\pi^- K^0 \overline{K^0}$
                &  0.153 $\pm$ 0.035  &$(75\pm25)\%$ A&\\
   $K^- K^+ \pi^- $    &  0.163 $\pm$ 0.027  &$(75\pm25)\%$ A&\\
   $(K\overline{K}\pi\pi)^-$ &  0.050 $\pm$ 0.020   &$(50 \pm 50)\%$ A&\\
\hline
   $K^- $ &  0.696 $\pm$ 0.029 &S&\\
   $K^- \pi^0 $    &  0.444 $\pm$ 0.035  &S&\\
   $\overline{K^0} \pi^-$ &  0.917 $\pm$ 0.052  &S&\\
   $K^- 2\pi^0 $    &  0.056 $\pm$ 0.025  &S&\\ 
   $K^- \pi^+ \pi^- $    &  0.214 $\pm$ 0.047  &S&\\
   $\overline{K^0} \pi^- \piz$ &  0.327 $\pm$ 0.051  &S&\\
   $(K 3\pi)^- $ &  0.076 $\pm$ 0.044 &S&\\
   $K^- \eta$  &  0.029 $\pm$ 0.014 &S&\\
   $K^- \omega$  &  0.067 $\pm$ 0.021 &S&\\
   $K^{*-} \eta$  &  0.029 $\pm$ 0.009 &S& CLEO~\cite{cleoksteta}\\
\hline\hline
\end{tabular}
\label{completeBR}
\end{center}
\end{table}

\section{Determination of hadronic spectral functions}
\label{sf}

\subsection{Spectral functions}

The \sf\ $v_1$ ($a_1$, $a_0$), where the subscript 
refers to the spin $J$ of the hadronic system, is here def\/ined
for a nonstrange vector (axial-vector) hadronic $\tau$ decay channel 
${V^-}$\nut\ (${A^-}$\nut). The \sf\ is obtained by dividing the normalized 
invariant mass-squared distribution $(1/N_{V/A})(d N_{V/A}/d s)$ 
for a given hadronic mass $\sqrt{s}$ by the appropriate kinematic 
factor
\beqn
\label{eq_sf}
   v_1(s) 
   &\equiv&
           \frac{m_\tau^2}{6\,|V_{ud}|^2\,S_{\mathrm{EW}}}\,
              \frac{B(\tau^-\rightarrow {V^-}\,\nu_\tau)}
                   {B(\tau^-\rightarrow e^-\,\bar{\nu}_e\nu_\tau)} \nonumber \\
   & & \hspace{2cm}        
              \times\frac{d N_{V}}{N_{V}\,ds}\,
              \left[ \left(1-\frac{s}{m_\tau^2}\right)^{\!\!2}\,
                     \left(1+\frac{2s}{m_\tau^2}\right)
              \right]^{-1}\hspace{-0.3cm}, \\[0.2cm]
   a_1(s) 
   &\equiv&
           \frac{m_\tau^2}{6\,|V_{ud}|^2\,S_{\mathrm{EW}}}\,
              \frac{B(\tau^-\rightarrow {A^-}\,\nu_\tau)}
                   {B(\tau^-\rightarrow e^-\,\bar{\nu}_e\nu_\tau)} \nonumber \\
   & & \hspace{2cm}        
              \times\frac{d N_{A}}{N_{A}\,ds}\,
              \left[ \left(1-\frac{s}{m_\tau^2}\right)^{\!\!2}\,
                     \left(1+\frac{2s}{m_\tau^2}\right)
              \right]^{-1}\hspace{-0.3cm}, \\[0.2cm]
   a_0(s) 
   &\equiv& 
           \frac{m_\tau^2}{6\,|V_{ud}|^2\,S_{\mathrm{EW}}}\,
              \frac{B(\tau^-\rightarrow {\pi^-}\,\nu_\tau)}
                   {B(\tau^-\rightarrow e^-\,\bar{\nu}_e\nu_\tau)}
              \frac{d N_{A}}{N_{A}\,ds}\,
              \left(1-\frac{s}{m_\tau^2}\right)^{\!\!-2}\,
              \hspace{-0.3cm},
\label{eq_spect_fun}
\eeqn
where $|V_{ud}|=0.9746\pm0.0006$~\cite{dehz03} denotes the CKM 
weak mixing matrix element and $S_{\mathrm{EW}}=1.0198\pm0.0006$ 
accounts for electroweak radiative corrections~\cite{marciano-sirlin}, 
as discussed in Section~\ref{vsf_cvc}. Due to the conserved vector current, 
there is no $J=0$ contribution to the vector \sf, 
while the only contribution to $a_0$ is assumed to be from 
the pion pole. It is connected \via\ PCAC to the pion decay constant, 
$f_\pi$ through $a_{0,\,{\pi}}(s)=4\pi^2 f_\pi^2\,\delta(s-m_\pi^2)$.
The \sfs\ are normalized by the ratio of the vector/axial-vector 
\bfr\ $B(\tau^-\rightarrow {V^-/A^-}\,\nu_\tau)$ to the \bfr\ of 
the massless leptonic, \ie, electron, channel. For the latter, the
value obtained in Section~\ref{uni} using universality is used, {\it i.e.}
$B(\tau^-\rightarrow e^-\,\bar{\nu}_e\nu_\tau)=(17.810\pm0.039)\%$.
\vs
Using unitarity and analyticity, the \sfs\ 
are connected to the imaginary part of the two-point correlation (or 
hadronic vacuum polarization) functions~\cite{bnp,pichtau94}
$\Pi_{ij,U}^{\mu\nu}(q) \equiv i\int d^4x\,e^{iqx}
\langle 0|T(U_{ij}^\mu(x)U_{ij}^\nu(0)^\dag)|0\rangle=(-g^{\mu\nu}q^2
+q^\mu q^\nu)\,\Pi^{(1)}_{ij,U}(q^2)$ $+q^\mu q^\nu\,\Pi^{(0)}_{ij,U}(q^2)$
of vector $(U_{ij}^\mu\equiv V_{ij}^\mu=\bar{q}_j\gamma^\mu q_i)$
or axial-vector 
($U_{ij}^\mu\equiv A_{ij}^\mu=\bar{q}_j\gamma^\mu\gamma_5 q_i$) 
colour-singlet quark currents in corresponding quantum states and
for time-like momenta-squared $q^2>0$. Lorentz decomposition 
is used to separate the correlation function into its $J=1$ and $J=0$ 
parts. Thus, using the def\/inition~(\ref{eq_sf}), one identif\/ies 
for non-strange quark currents 
\beqn
\label{eq_imv}
   {\rm Im}\,\Pi^{(1)}_{\bar{u}d,V}(s)
   &=& \frac{1}{2\pi}v_1(s)~,\nonumber \\
   {\rm Im}\,\Pi^{(1)}_{\bar{u}d,A}(s)
   &=& \frac{1}{2\pi}a_1(s)~,\nonumber \\
   {\rm Im}\,\Pi^{(0)}_{\bar{u}d,A}(s)
   &=& \frac{1}{2\pi}a_0(s)~,
\eeqn
which provide the basis for comparing short-distance theory with data.

\subsection{The unfolding procedure}

The measurement of the $\tau$ \sfs\ def\/ined in Eq.~(\ref{eq_spect_fun})
requires the determination of the invariant mass-squared 
distributions, obtained from the experimental distributions after 
unfolding from the ef\/fects of measurement distortion. The 
unfolding procedure used in this analysis follows a method published 
in Ref.~\cite{unfold}, which has been extensively used in previous
publications~\cite{alephvsf,alephasf}. It is based on the regularized inversion 
of the detector response matrix, obtained from the Monte Carlo simulation, 
using the Singular Value Decomposition (SVD) technique. The regularization 
function applied minimizes the average curvature of the distribution. 
The optimal choice of the regularization strength is found by means of the 
Monte Carlo simulation where the true distribution is known. 

In order to measure exclusive \sfs, individual unfolding procedures 
with specif\/ic detector response matrices and regularization parameters 
are applied for each $\tau$ decay channel $X$ considered. An iterative
procedure is applied to correct the Monte Carlo spectral functions
used to subtract the feed across between the modes. Each \sf\ is
determined in 140 mass-squared bins of equal width (0.025 GeV$^2$). 

\subsection{Specific systematic studies}

All systematic uncertainties concerning the decay classif\/ication 
are contained in the covariance matrix of 
the branching ratios obtained in Section~\ref{system}. Therefore
only the systematic ef\/fects af\/fecting the shape of the mass-squared 
distributions, and not its normalization, need to be examined here.

The general procedure is the same as for the branching ratio analysis.
All effects affecting the decay classification and the calculation of the
hadronic invariant mass are considered in turn. Comparisons of data and 
Monte Carlo distributions are made and the corresponding biases are 
corrected for, as discussed in Section~\ref{system}, while the uncertainty 
in the correction is taken as input for the calculation of the systematic 
uncertainty. In the case of the spectral functions, the whole analysis 
including the unfolding procedure is repeated, for each systematic effect.
This generates new mass distributions under the systematic change which
are compared bin-by-bin to the nominal ones, hence providing the full
$140~\times~140$ covariance matrix of the \sf\ for the studied effect.
In this process it was found necessary to smooth the mass distributions,
before and after applying the systematic effect, in order to construct
stable covariance matrices.

Following the systematic studies for the determination of the branching 
ratios, extensive studies are performed to determine the uncertainties
at the level of the spectral functions. They include the effects from 
the photon and $\pi^0$ energy calibration and resolution, the photon 
detection efficiency (especially in the threshold region above 350~MeV), 
the shapes of the identification probability distribution, the estimate 
of the number of fake photons, the proximity in the calorimeter of other 
photon showers and of energy deposition by charged particles, and the
separation between radiative and $\pi^0$ decay photons for residual single
photons. The photon and $\piz$ reconstruction constitutes the main part 
of the uncertainty for the determination of \sfs.

Similarly, the ef\/fects of momentum calibration and resolution 
uncertainties in the reconstruction of charged tracks are checked, 
accompanied by tests of the reconstruction ef\/f\/iciency of highly 
collimated multi-prong events, and the simulation of secondary nuclear 
interactions.

In addition, systematic errors introduced by the unfolding procedure are
tested by comparing known, true distributions to their corresponding 
unfolded ones.

F\/inally, systematic errors due to the limited Monte Carlo
statistics and to uncertainties in the \brs\ are added.

\subsection{Spectral functions for nonstrange exclusive modes}

\subsubsection{The results}

Before unfolding the mass distributions, the $\tau$ and non-$\tau$ 
backgrounds are subtracted. In the case of $\tau$ feedthrough the
Monte Carlo distributions normalized to the measured branching 
fractions are used. 
Since the \sfs\ are measured for nonstrange exclusive final states, the
contributions from strange modes classified in the same topology are
subtracted using their Monte Carlo \sfs\ normalized by the measured
branching fractions~\cite{alephksum}. The measurement of the strange \sfs\ 
has been published elsewhere~\cite{alephksum}. 

\begin{figure}[t]
   \centerline{\psfig{file=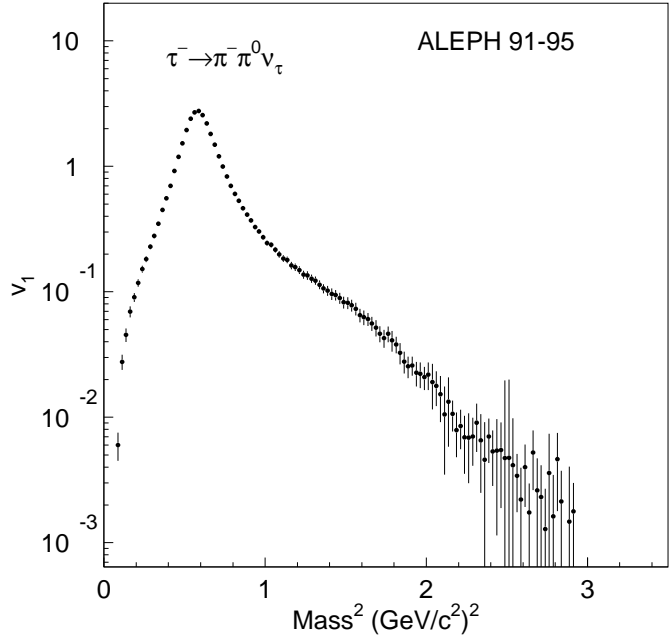,width=100mm}}
\caption[.]{\label{sf_2pi} 
    The spectral function for the $\pi \pi^0$ hadronic mode.}
\end{figure}

\begin{figure}
   \centerline{\psfig{file=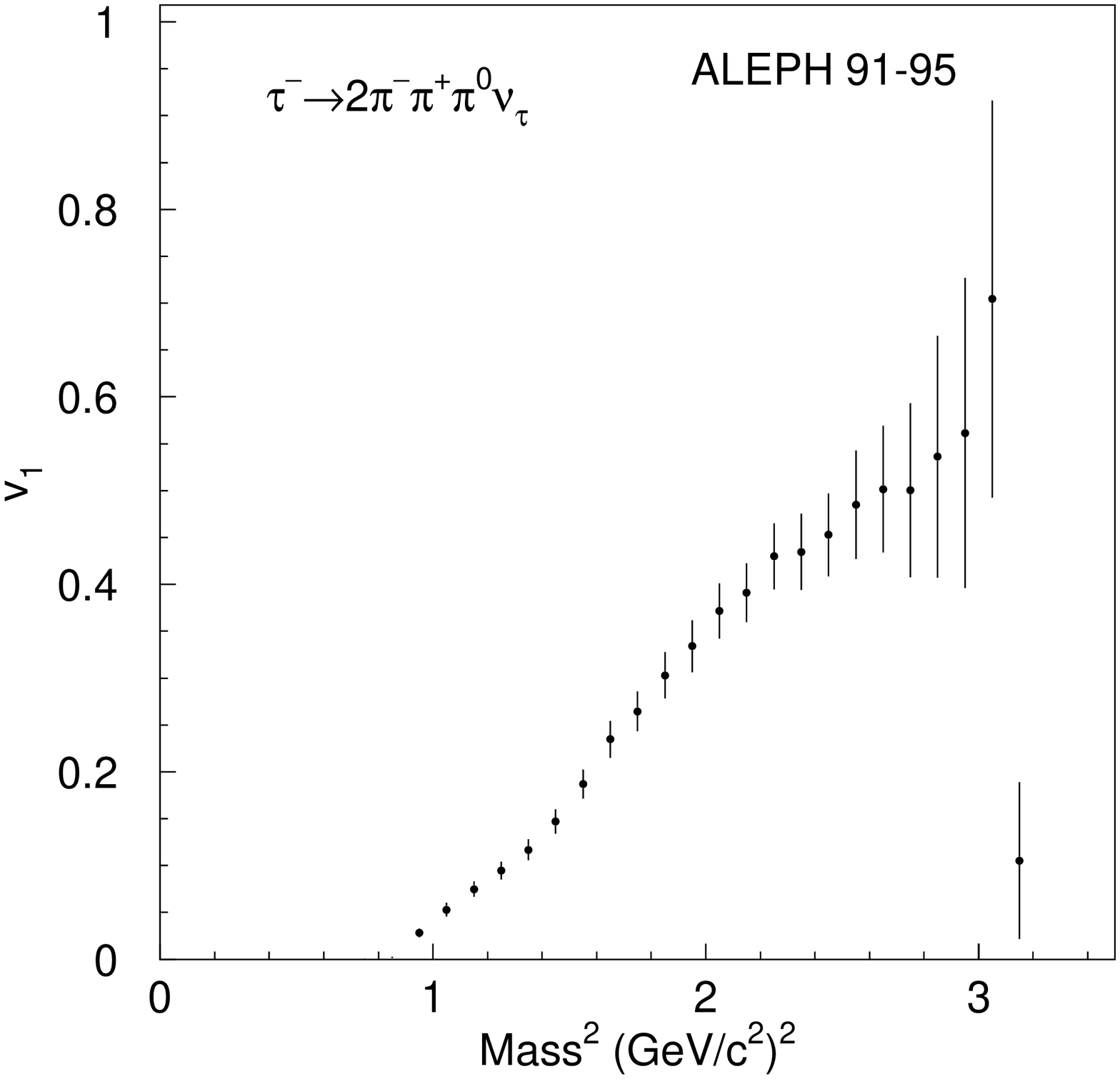,width=90mm}}
\caption[.]{\label{sf_3pipi0} 
    The spectral function for the $3\pi \pi^0$ hadronic mode. While the 
original spectral function is determined in 0.025 GeV$^2$ bins, it is 
rebinned here to 0.1 GeV$^2$ for easier reading.}
\end{figure}

\begin{figure}
   \centerline{\psfig{file=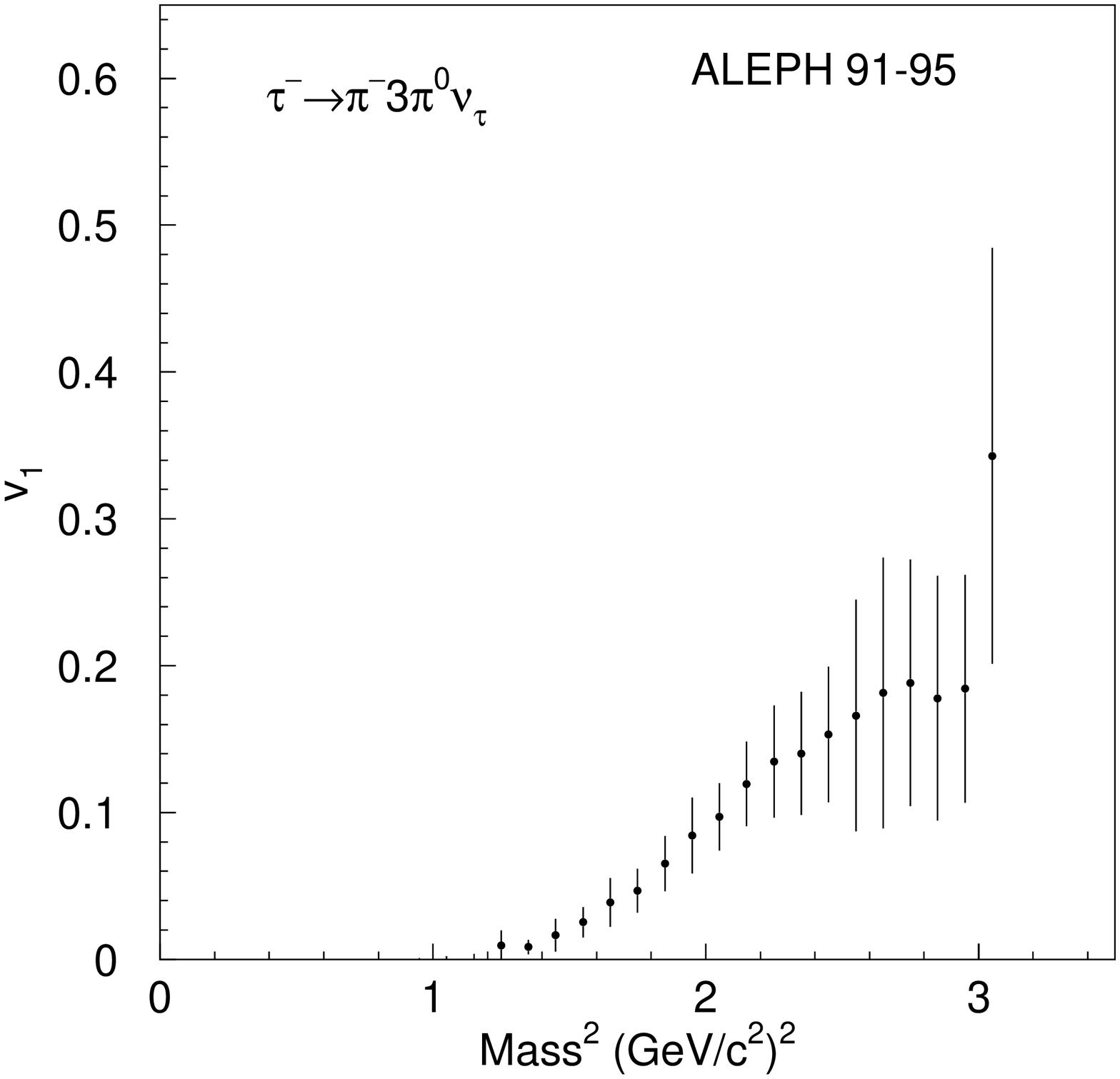,width=90mm}}
\caption[.]{\label{sf_pi3pi0} 
    The spectral function for the $\pi 3\pi^0$ hadronic mode. While the 
original spectral function is determined in 0.025 GeV$^2$ bins, it is 
rebinned here to 0.1 GeV$^2$ for easier reading.}
\end{figure}

The vector \sfs\ of the dominant two- and four-pion modes are given 
in Figs.~\ref{sf_2pi}, \ref{sf_3pipi0} and \ref{sf_pi3pi0}. 
The error bars shown correspond to the diagonal elements of the covariance matrix. 
They include both statistical and systematic uncertainties. 
It should be pointed out that the unfolding generates strong 
correlations between neighbouring bins. This information is contained 
in the covariance matrix that is determined for each \sf~\footnote{The 
corrected invariant mass-squared distributions from this analysis and 
their covariance matrices are publicly accessible 
at the website http://aleph.web.lal.in2p3.fr/tau/specfun 
for the inclusive as well as the main exclusive channels.}.

\begin{figure}
   \centerline{\psfig{file=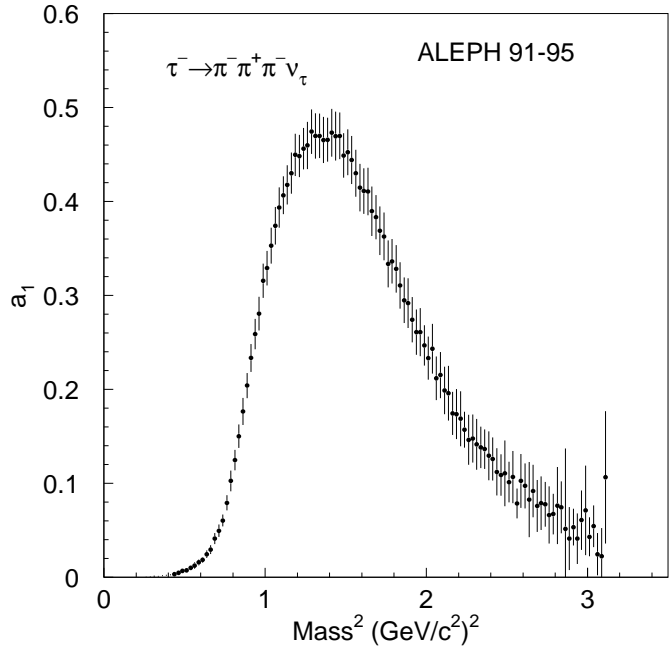,width=100mm}}
\caption[.]{\label{sf_3pi} 
    The spectral function for the $3\pi$ hadronic mode.}
\end{figure}

\begin{figure}
   \centerline{\psfig{file=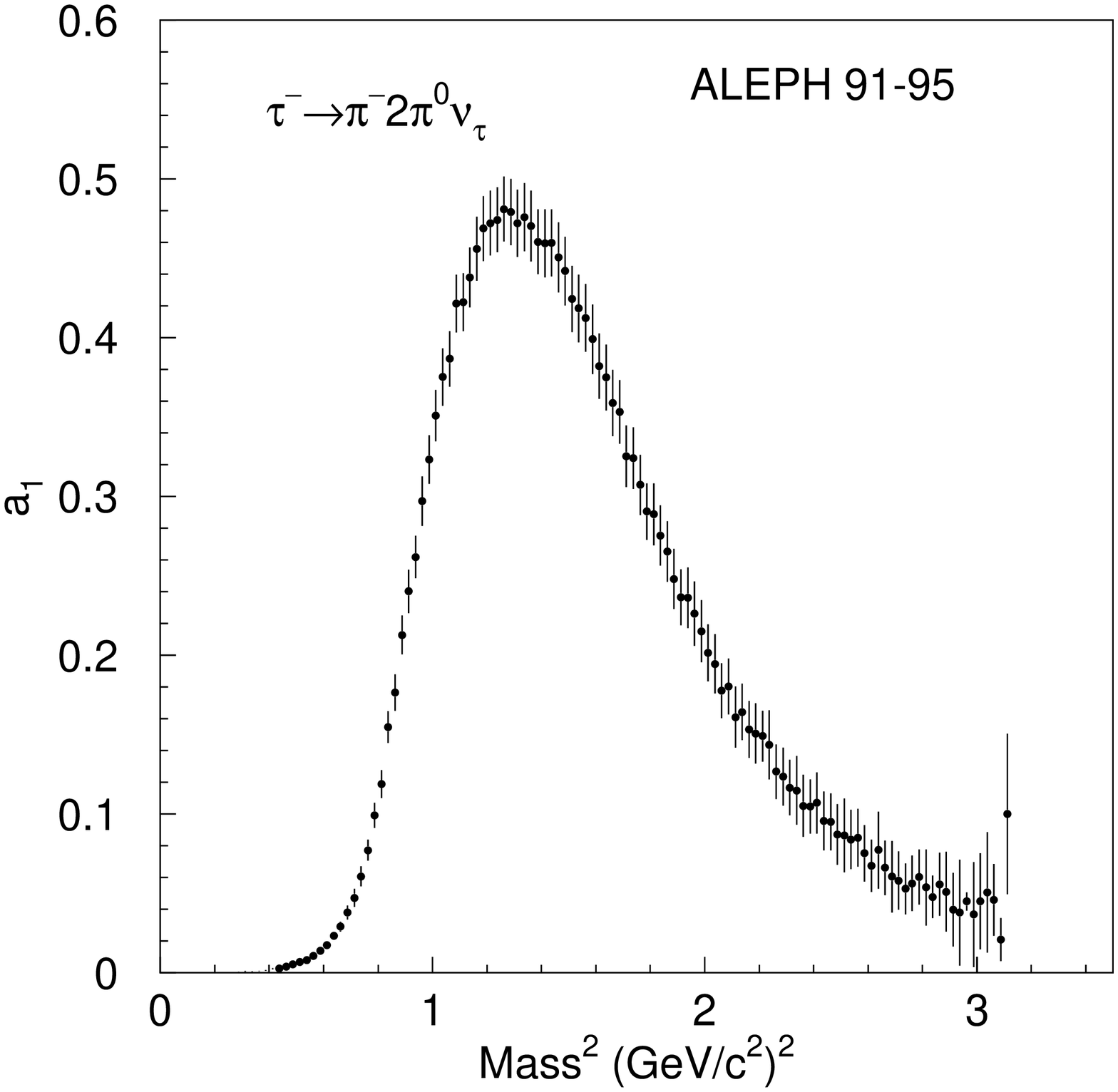,width=100mm}}
\caption[.]{\label{sf_pi2pi0} 
    The spectral function for the $\pi 2\pi^0$ hadronic mode.}
\end{figure}

\begin{figure}
   \centerline{\psfig{file=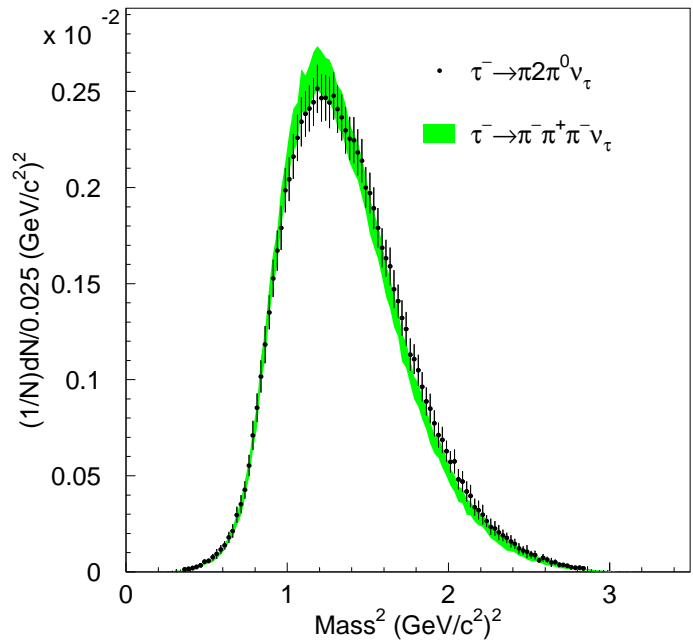,width=100mm}}
\caption[.]{\label{sf_3pi_pi2pi0} 
    Comparison of the corrected mass spectra for the $3\pi$ and $\pi 2\pi^0$ 
    hadronic modes.}
\end{figure}

Figures~\ref{sf_3pi} and \ref{sf_pi2pi0} show the unfolded \tpi\nut\ 
and \pidpiz\nut\ \sfs\ which are the dominant contributions 
to the axial-vector \sf. Both spectra are dominated by the $a_1$ resonance 
and indeed the two distributions have consistent shape, while their 
normalization was already observed to agree, as expected from
isospin invariance with a small correction from the slightly
different charged and neutral pion masses. The comparison is seen
in Fig.~\ref{sf_3pi_pi2pi0}.

\begin{figure}
   \centerline{\psfig{file=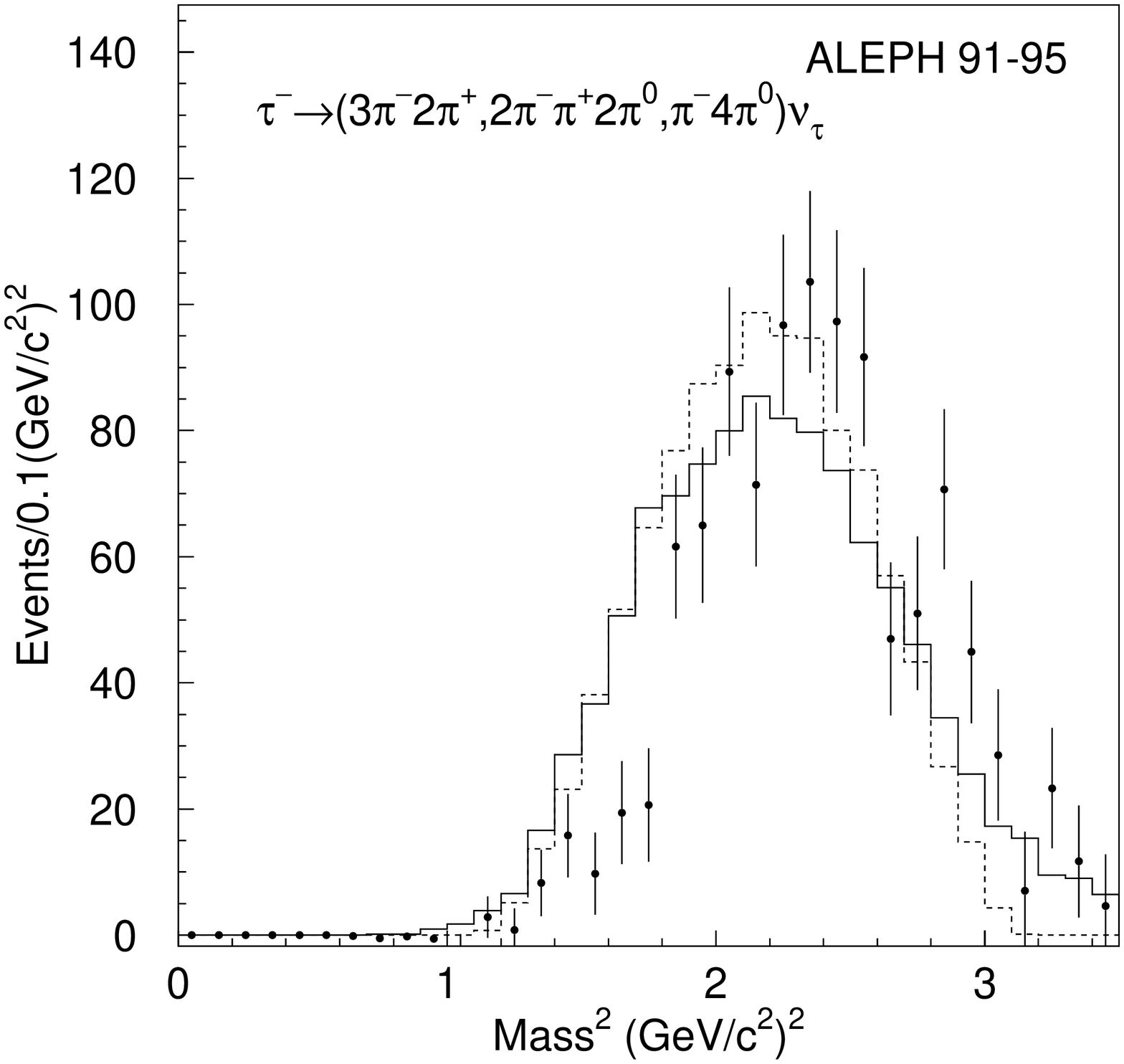,width=100mm}}
\caption[.]{\label{mass2_5pi} 
    Comparison of the background-subtracted mass-squared spectrum in the 
5-pion modes ($\pi^- 4\pi^0 \nu_\tau$, $2\pi^- \pi^+ 2\pi^0 \nu_\tau$
and $3\pi^- 2\pi^+ \nu_\tau$) with the phase-space model used in the
TAUOLA decay generator~\cite{tauola}. The dashed histogram is the generated 
distribution, while the solid histogram is the reconstructed spectrum
with the detector response which can be directly compared to the data
(points). A clear disagreement is observed. }
\end{figure}

The 5-pion channels, $\pi^- 4\pi^0 \nu_\tau$, $2\pi^- \pi^+ 2\pi^0 \nu_\tau$
and $3\pi^- 2\pi^+ \nu_\tau$, have small branching ratios and relatively
large feedthrough background (except for the 5-charged-pion mode). Their
combined mass-squared distribution, shown in Fig.~\ref{mass2_5pi}, does
not agree with the TAUOLA decay simulation in the KORALZ07 
generator~\cite{tauola}. The experimental spectrum is harder than the 
phase-space model used in TAUOLA, in the absence of prior experimental 
information on the dynamics in these decay modes.

\subsubsection{Comparison with other experiments}

Spectral functions for some specific final states have been determined
by CLEO and OPAL. Since the data are not presented with the same binning
and furthermore plotted sometimes against the mass rather than the 
mass-squared, the comparison must involve some treatment of the original 
data.

\begin{figure}
   \centerline{\psfig{file=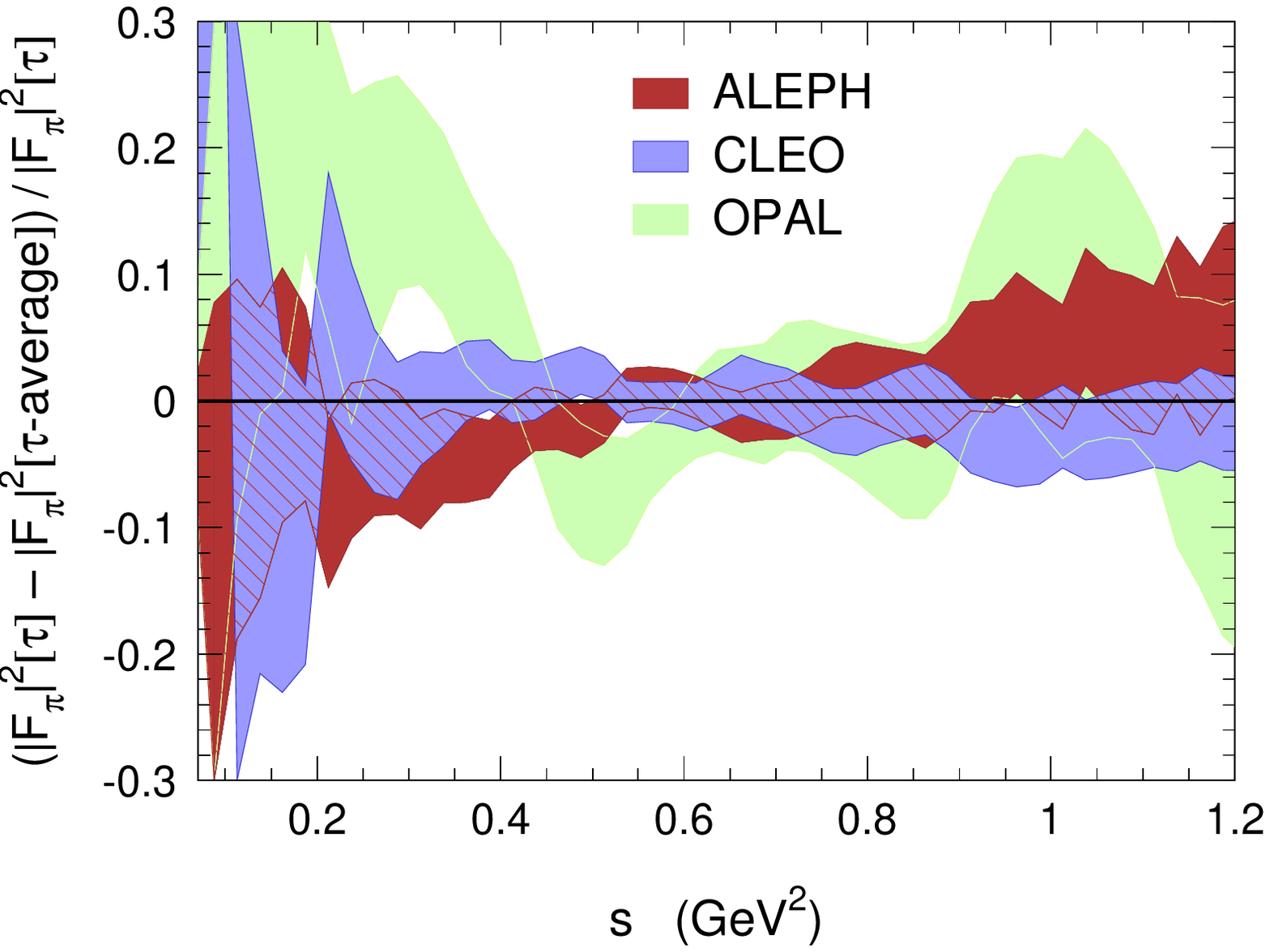,width=120mm}}
\vspace{0.6cm}
   \centerline{\psfig{file=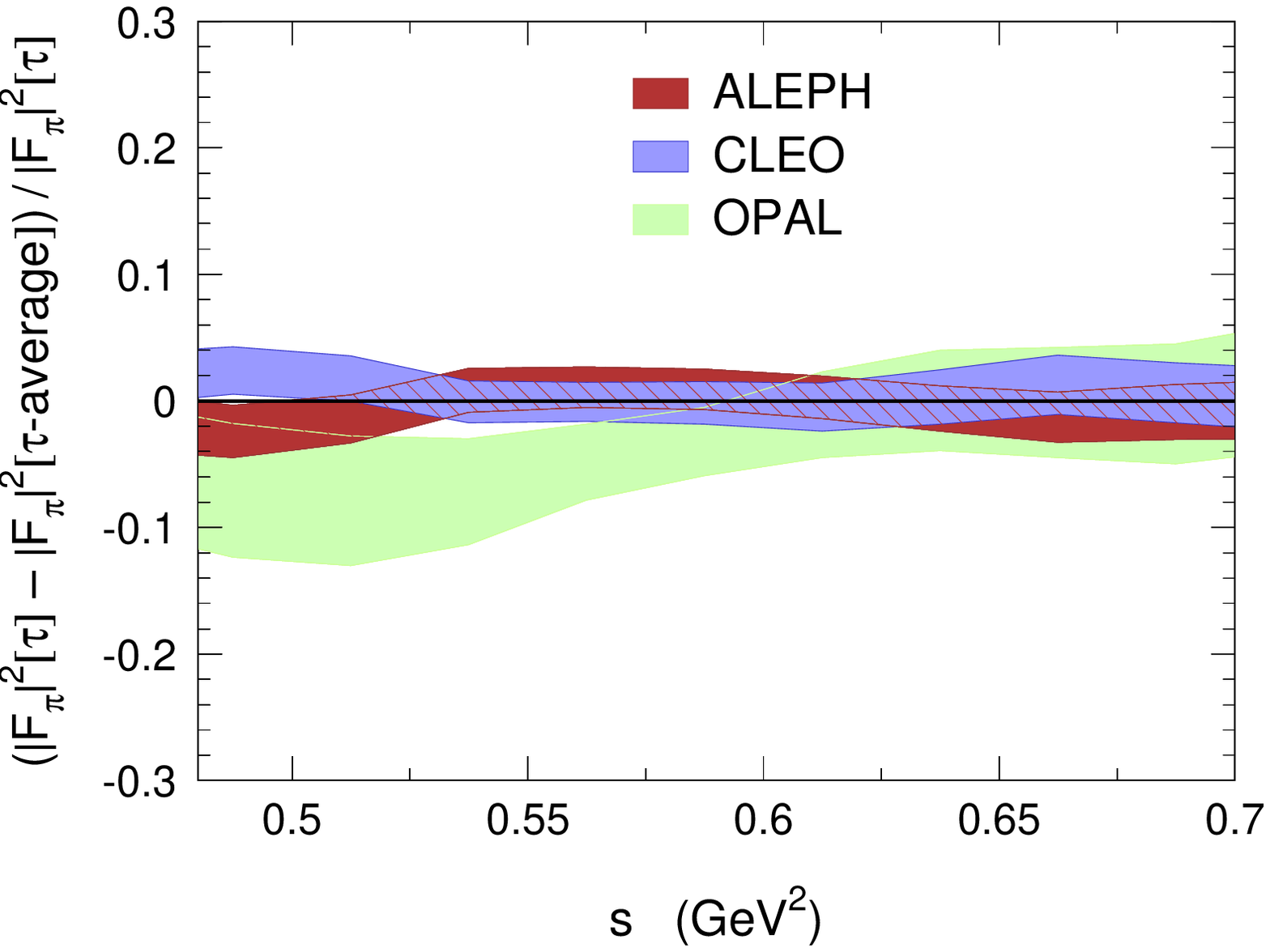,width=120mm}}
\caption[.]{\label{comp_tau_2pi} Relative comparison of the $\pi^+\pi^-$ 
        \sfs\ extracted from $\tau$ data from different experiments, 
	expressed as a ratio to the average $\tau$ \sf. The lower figure 
        emphasizes the $\rho$ region. For CLEO only statistical errors 
        are shown.} 
\end{figure}

The results for the $\pi\pi^0$ \sf\ from ALEPH, CLEO~\cite{cleo_2pi}, 
and OPAL~\cite{opal_2pi} are compared in Fig.~\ref{comp_tau_2pi} in the $\rho$ 
resonance region. For this comparison, each data set is normalized to the weighted 
mean of the three measurements. The different data are in agreement within their 
quoted uncertainties, when taking into account the correlation between different 
mass bins, particularly the strong statistical correlation between neighbouring 
points due to the unfolding procedure. It should be pointed out that each \sf\ is 
normalized to the world-average branching ratio for $\tau \ra \nu_\tau \pi \pi^0$ 
that is dominated by the ALEPH result. Hence the present exercise should be 
understood as a test of the shape of the \sf. However, if the branching ratio 
determined by each experiment (see Section~\ref{sec_brcvc}) is used instead, 
the \sfs\ are still in agreement, albeit with larger uncertainties in the case 
of CLEO and OPAL. The two most precise results from ALEPH and CLEO do agree well. 
The statistics is comparable in the two cases, however due to a flat
acceptance in ALEPH and a strongly increasing one in CLEO, ALEPH data are more 
precise below the $\rho$ peak, while CLEO is more precise above.

\subsection{Results on inclusive $V$, $A$ and $V \pm A$ spectral functions}

\subsubsection{The vector spectral function}

\begin{figure}
   \centerline{\psfig{file=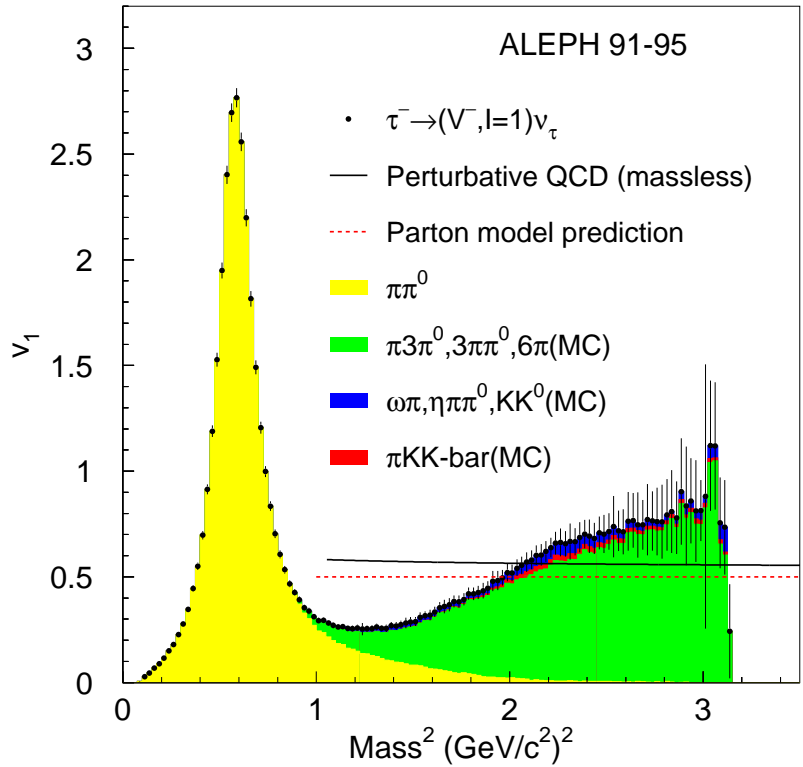,width=120mm}}
\caption[.]{\label{fig_vsf} 
            The total vector \sf. The shaded areas indicate the 
            contributions from the exclusive $\tau$ vector channels,
            where the shapes of the contributions labelled {\small `MC'} 
            are taken from the Monte Carlo simulation. The lines show 
            the predictions from the naive parton model and from massless 
            perturbative QCD using $\alpha_s(M_Z^2)=0.120$.}
\end{figure}

The inclusive $\tau$ vector \sf\ with its most important contributions is 
shown in F\/ig.~\ref{fig_vsf}. The dashed line depicts the naive parton 
model prediction while the massless perturbative QCD prediction~\cite{3loop} 
using $\alpha_s(M_Z^2)=0.120$ (solid line) lies roughly 25\pc\ lower
than the data at $m_\tau^2$. Although the statistical power of the data
is weak near the kinematic limit, the trend of the spectral function
clearly indicates that the asymptotic region is not reached.
\vs
The two- and four-pion f\/inal states are measured exclusively, while the 
six-pion state is only partly measured. The total six-pion \br\
has been determined in~\cite{alephvsf} using isospin symmetry. 
However, one has to account for the fact that the six-pion channel 
is contaminated by isospin-violating \tauto$\eta$\,\tpi\nut, 
$\eta$\,\pidpiz\nut\ decays, as reported by the CLEO 
Collaboration~\cite{cleoeta3pi}.

The small fraction of the $\omega\,\pi^-$\nut\ decay channel that is not
reconstructed in the four-pion f\/inal state is added using the simulation.
Similarly, one corrects for $\eta \pi^-\piz$\nut\ decay modes where
$\eta$ decays into pions. For the $\eta\rightarrow2\gamma$ mode, the $\tau$ 
decay is classif\/ied in the $h 3\pi^0 \nu_\tau$ f\/inal state, since the two-photon 
mass is inconsistent with the $\piz$ mass and consequently each photon is 
reconstructed as a $\piz$.

The $K^-\,K^0$\nut\ mass distribution is taken entirely from the 
simulation. The vector contributions in the $K\overline{K}\pi$ and
$K\overline{K}\pi\pi$ modes are taken from Section~\ref{vasep}. 
The corresponding \sfs\ are obtained from the Monte Carlo simulation. 
\vs
The invariant mass spectra of the small contributions 
labelled {\small `MC'} in F\/igs.~\ref{fig_vsf} and \ref{fig_asf} are 
taken from the Monte Carlo simulation accompanied by a channel-dependent 
systematic error of up to $50\%$ of the bin entry.

\subsubsection{The axial-vector spectral function}

In complete analogy to the vector \sf, the inclusive axial-vector
\sf\ is obtained by summing up the exclusive axial-vector \sfs\ 
with the addition of small unmeasured modes taken from the 
Monte Carlo simulation.

The small fraction of the $\omega \pi^-\piz$\nut\ decay channel 
that is not accounted for in the \tpidpiz\nut\ f\/inal state is added
from the simulation. Also considered are the axial-vector 
$\eta\,(3\pi)^-$\nut\ f\/inal states~\cite{cleoeta3pi}. 
CLEO observed that the dominant part of this mode issues from the 
\tauto${\rm f}_1(1285)\pi^-$ intermediate state, with
$B($\tauto\/f$_1\pi^-$\nut$)=(0.068\pm0.030)$\pc, measured in the 
f$_1\rightarrow\eta\,\pi^+\pi^-$ and f$_1\rightarrow\eta\,\pi^0\pi^0$ 
decay modes~\cite{cleoeta3pi}. Since the f$_1$ meson is isoscalar, 
the \brs\ relate as $B(\tau^-\rightarrow\eta\,$\tpi\nut$)=2\times 
B(\tau^-\rightarrow\eta\,$\pidpiz\nut$)$. The distributions are taken 
from the ordinary six-pion phase space simulation accompanied 
by large systematic errors. As discussed for the vector \sf, 
the $K\overline{K}\pi$ 
f\/inal states contribute dominantly to the inclusive axial-vector 
\sf, with full anticorrelation to the inclusive vector \sf. 
Invariant mass distributions for these channels are taken from the simulation.

\begin{figure}
   \centerline{\psfig{file=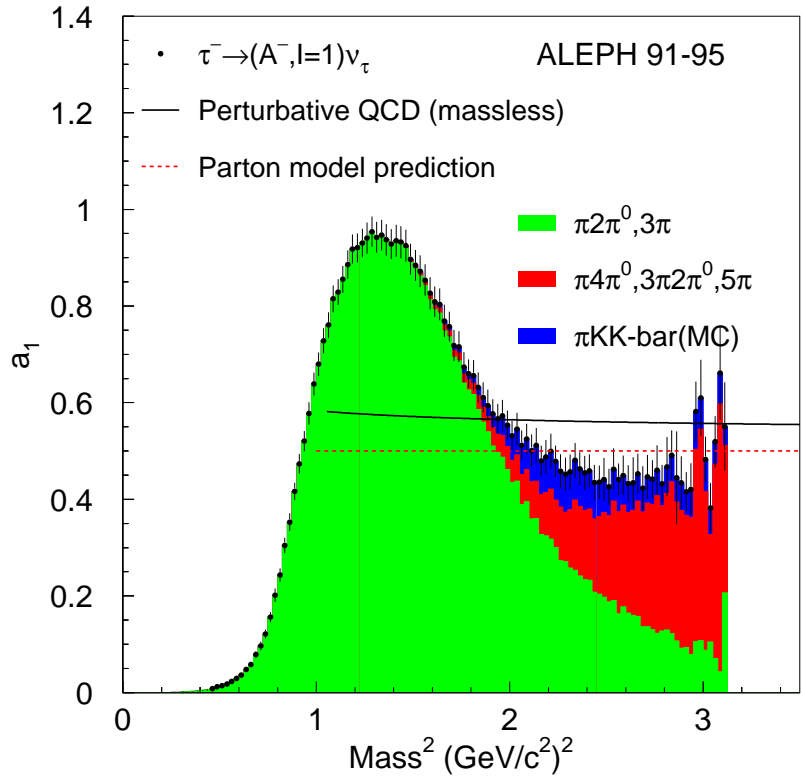,width=120mm}}
\caption[.]{\label{fig_asf} 
            The total axial-vector \sf. The shaded areas indicate the 
            contributions from the exclusive $\tau$ vector channels,
            where the shapes of the contributions labelled {\small `MC'} 
            are taken from the Monte Carlo simulation. The lines show 
            the predictions from the naive parton model and from massless 
            perturbative QCD using $\alpha_s(M_Z^2)=0.120$.}
\end{figure}

The total inclusive axial-vector \sf\ is plotted in 
F\/ig.~\ref{fig_asf} together with the naive parton model and the 
massless, perturbative QCD prediction. One observes that the asymptotic 
region is apparently not reached at the $\tau$ mass scale. 

\subsubsection{The $V \pm A$ spectral functions}

For the total $(v_1+a_1)$ hadronic \sf\ one does not have to distinguish 
the current properties of the non-strange hadronic $\tau$ decay channels. 
Hence the mixture of all contributing non-strange f\/inal states is 
measured inclusively using the following procedure. 

\begin{figure}
   \centerline{\psfig{file=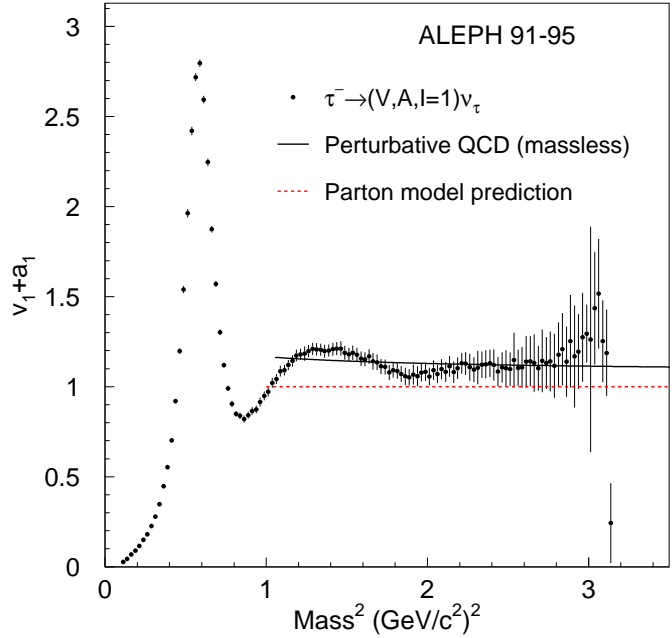,width=100mm}}
  \caption[.]{\label{sf_vpa}
           The inclusive vector plus axial-vector 
           ($v_1+a_1$) \sf\ and predictions from the 
           parton model and from massless perturbative QCD using
           $\alpha_s(M_Z^2)=0.120$.}
\end{figure}

The two- and three-pion f\/inal states dominate and their exclusive 
measurements are added with proper accounting for the correlations. The 
remaining contributing topologies are treated inclusively, \ie, without 
separation of the vector and axial-vector decay modes. This reduces the
statistical uncertainty. The ef\/fect of the feedthrough between $\tau$
f\/inal states on the invariant mass spectrum is described by the Monte 
Carlo simulation and thus corrected in the data unfolding. In this 
procedure the simulated mass distributions are iteratively corrected
using the exclusive vector/axial-vector unfolded mass spectra. 
Another advantage of the inclusive $(v_1+a_1)$ 
measurement is that one does not have to separate the 
vector/axial-vector currents of the K${\overline{\mathrm K}}\pi$ and 
K$\overline{\mathrm K}\pi\pi$ modes. The ($v_1+a_1$) \sf\ is depicted in 
F\/ig.~\ref{sf_vpa}. The improvement in precision when comparing 
to a sum of the two parts (F\/ig.~\ref{fig_vsf} and F\/ig.~\ref{fig_asf}) 
is obvious at higher mass-squared.

\begin{figure}
   \centerline{\psfig{file=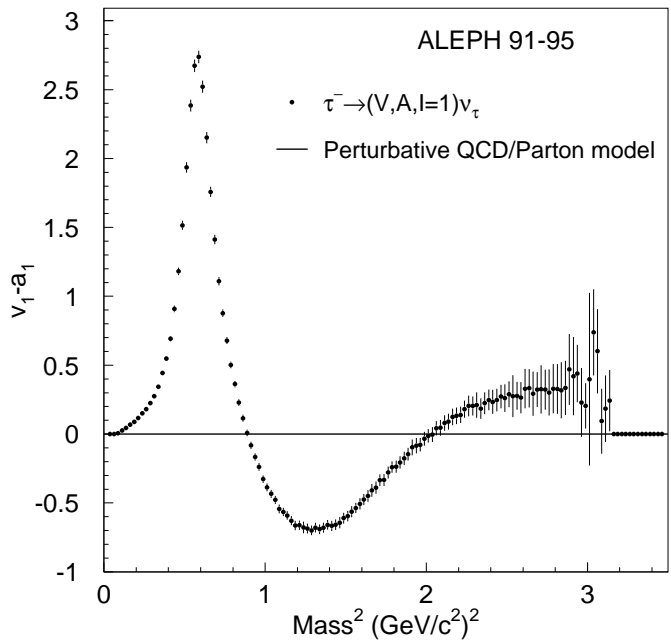,width=100mm}}
  \caption[.]{\label{sf_vma}
           The vector minus axial-vector ($v_1-a_1$)
           \sf. In the parton model as well as in perturbative QCD vector 
           and axial-vector contributions are degenerate.}
\end{figure}

One clearly sees the oscillating behaviour of the \sf\ but, unlike 
the vector/axial-vector \sfs, it does approximately reach the asymptotic 
limit predicted by perturbative QCD at $s\rightarrow m_\tau^2$. Also,
the $V+A$ \sf, including the $\delta$-function $\pi$ contribution,
exhibits the features expected from global quark-hadron duality: 
despite the huge oscillatory effects from $\pi$, and $\rho$, $a_1$ and 
$\rho'$ hadron resonances, the \sf\ qualitatively averages out to the quark
contribution from perturbative~\cite{svz}. This observation is the 
physical basis for the quantitative QCD analysis performed 
in Section~\ref{qcd}. 
\vs 
In the case of the $(v_1-a_1)$ \sf, uncertainties on the $V/A$ 
separation are reinforced due to their anticorrelation.
In addition, anticorrelations in the branching ratios
between $\tau$ f\/inal states with adjacent numbers of pions 
increase the errors. The ($v_1-a_1$) \sf\ is shown in 
F\/ig.~\ref{sf_vma}. The oscillating behaviour of the 
respective $v_1$ and $a_1$ \sfs\ is emphasized and the
asymptotic behaviour is clearly not reached at $m_\tau^2$.
However again here, the oscillation generated by the hadron resonances
qualitatively averages out to zero, as predicted by perturbative 
QCD.

\section{The vector spectral functions and tests of CVC}
\label{vsf}

\subsection{General remarks on spectral functions in $\tau$ decays and
$e^+e^-$ annihilation}
\label{vsf_cvc}

Using the optical theorem and in the limit of isospin invariance (implicit
in the hypothesis of CVC---Conserved Vector Current) 
the \sf\ of a $\tau$ decay mode $X^-$\nut, where the hadronic system 
$X$ is in a vector state, is related to the \ee\ annihilation cross 
section of the corresponding isovector f\/inal state $X^0$:
\beq
\label{eq_cvc}
 \sigma_{e^+e^-\rightarrow X^0}^{I=1} \:=\:
 \frac{4\pi\alpha^2}{s}\,v_{1,\,X^-}~,
\eeq
with the electromagnetic f\/ine structure constant $\alpha$.

Since the breaking of isospin symmetry is expected at some level, 
in particular from electromagnetic effects,
it is useful to carefully write down all the factors involved in the 
comparison of \ee\ and $\tau$ spectral functions in order to make
explicit the possible sources of CVC breakdown. For the dominant
$\pi\pi$ spectral functions, one has on the \ee\ side
\begin{eqnarray}
\sigma (e^+e^-\longrightarrow \pi^+\pi^-)&=&\frac{4\pi\alpha^2}{s} v_0(s)~,\\
    v_0(s)&=&\frac {\beta_0^3(s)} {12 \pi} |F^0_\pi(s)|^2 ,\nonumber
\label{ee_ff}
\end{eqnarray}
where $\beta_0^3(s)$ is the threshold kinematic factor and $F^0_\pi(s)$
the pion form factor. On the $\tau$ side, the physics is contained in the
hadronic mass distribution through 
\begin{eqnarray}
\frac {1}{\Gamma} \frac {d \Gamma}{ds}
 (\tau \longrightarrow \pi^-\pi^0 \nu_\tau) &=& \nonumber \\
         & & \hspace{-2.7cm}
                    \frac {6 \pi |V_{ud}|^2 S_{EW}}{m_\tau^2}
                    \frac {B_e}{B_{\pi \pi^0}} C(s) v_-(s)~, \\
      v_-(s) &=& \frac {\beta_-^3(s)} {12 \pi} |F^-_\pi(s)|^2 ~,\nonumber \\
      C(s) &=& \left(1- \frac {s}{m_\tau^2}\right)^2 \nonumber
       \left(1 + \frac {2 s}{m_\tau^2}\right)~.
\label{tau_ff}
\end{eqnarray}
SU(2) symmetry implies $v_-(s) =  v_0(s)$. The threshold functions $\beta_{0,-}$ 
are defined by
\beq
 \beta_{0,-}=\beta(s,m_{\pi^-},m_{\pi^{0,-}})~,
\eeq
where
\beq
 \beta(s,m_1,m_2)=\left[\left(1-\frac{(m_1+m_2)^2}{s}\right)
                     \left(1-\frac{(m_1-m_2)^2}{s}\right)\right]^{1/2}~.
\eeq

In this analysis of $\tau$ decays the rate is measured inclusively
with respect to radiative photons, {\it i.e.}, for 
$\tau \rightarrow \nu_\tau \pi \piz (\gamma)$. The measured spectral function
is thus $v_-^*(s) =  v_-(s)~G(s)$, where $G(s)$ is a radiative correction.

Several levels of SU(2) breaking can be identified.
\begin{itemize}
\item {\it Electroweak radiative corrections to $\tau$ decays} are contained 
in the $S_{\mathrm{EW}}$ factor~\cite{marciano-sirlin,braaten} which is 
dominated by short-distance effects. As such it is expected to be weakly 
dependent on the specific hadronic final state, as verified in the 
$\tau \rightarrow (\pi, K) \nu_\tau$ channels~\cite{decker-fink}. 
Recently, detailed calculations have been performed for the $\pi \pi^0$
channel~\cite{ecker}, which also confirm the relative smallness of the
long-distance contributions. The total correction is
\beq
 S_{EW} = \frac {S_{EW}^{\rm had} S_{EM}^{\rm had} }{S_{EM}^{\rm lep}}
\eeq
where $S_{EW}^{\rm had}$ is the leading-log short-distance electroweak
factor (which vanishes for leptons) and $S_{EM}^{\rm had,lep}$ are the
nonleading electromagnetic corrections. The latter corrections are
calculated in Ref.~\cite{braaten} at the quark level and in
Ref.~\cite{ecker} at the hadron level for the $\pi \pi^0$ decay mode,
and in Refs.~\cite{marciano-sirlin,braaten} for leptons. The total 
correction amounts~\cite{dehz} to $S_{EW}^{\rm inclu} = 1.0198 \pm 0.0006$ 
for the inclusive hadron decay rate and 
$S_{EW}^{\pi \pi^0} = (1.0232 \pm 0.0006)~G_{EM}^{\pi \pi^0}(s)$ 
for the $\pi \pi^0$ decay mode, where $G_{EM}^{\pi \pi^0}(s)$ is an
$s$-dependent radiative correction~\cite{ecker}.
\item {\it The pion mass splitting} breaks isospin symmetry in 
the spectral functions~\cite{adh,czyz} since $\beta_-(s) \neq \beta_0(s)$.
\item Isospin symmetry is also broken in {\it the pion form factor}
~\cite{adh,ecker} from the $\pi$ mass splitting.
\item A similar effect is expected from {\it the $\rho$ mass splitting}. 
The theoretical expectation~\cite{bijnens} gives a limit ($<0.7$~MeV), 
but this is only a rough estimate. Hence the question must be investigated
experimentally, the best approach being the explicit comparison
of $\tau$ and $e^+e^-$ $2\pi$ spectral functions, after correction for 
the other isospin-breaking effects. No correction for $\rho$ mass splitting
is applied initially.
\item Explicit {\it electromagnetic decays} such as 
$\pi \gamma$, $\eta \gamma$, $l^+l^-$ and $\pi \pi \gamma$ introduce
small differences between the widths of the charged and neutral $\rho$'s.
\item Isospin violation in the strong amplitude through the {\it mass 
difference between u and d quarks} is expected to be negligible.
\item An obvious, but large correction must be applied to the $\tau$ \sf\
to introduce the effect of $\rho-\omega$ mixing, only present in the
neutral channel. This correction is computed using the parameters determined
in the \ee\ experiments in their fits of the $\pi^+ \pi^-$ lineshape taking
into account the $\rho-\omega$ interference~\cite{cmd2_new}. 
\end{itemize}

\subsection{Comparison with $e^+e^-$ results}

\subsubsection{The $2\pi$ spectral function}

\begin{figure}[p]
   \centerline{\psfig{file=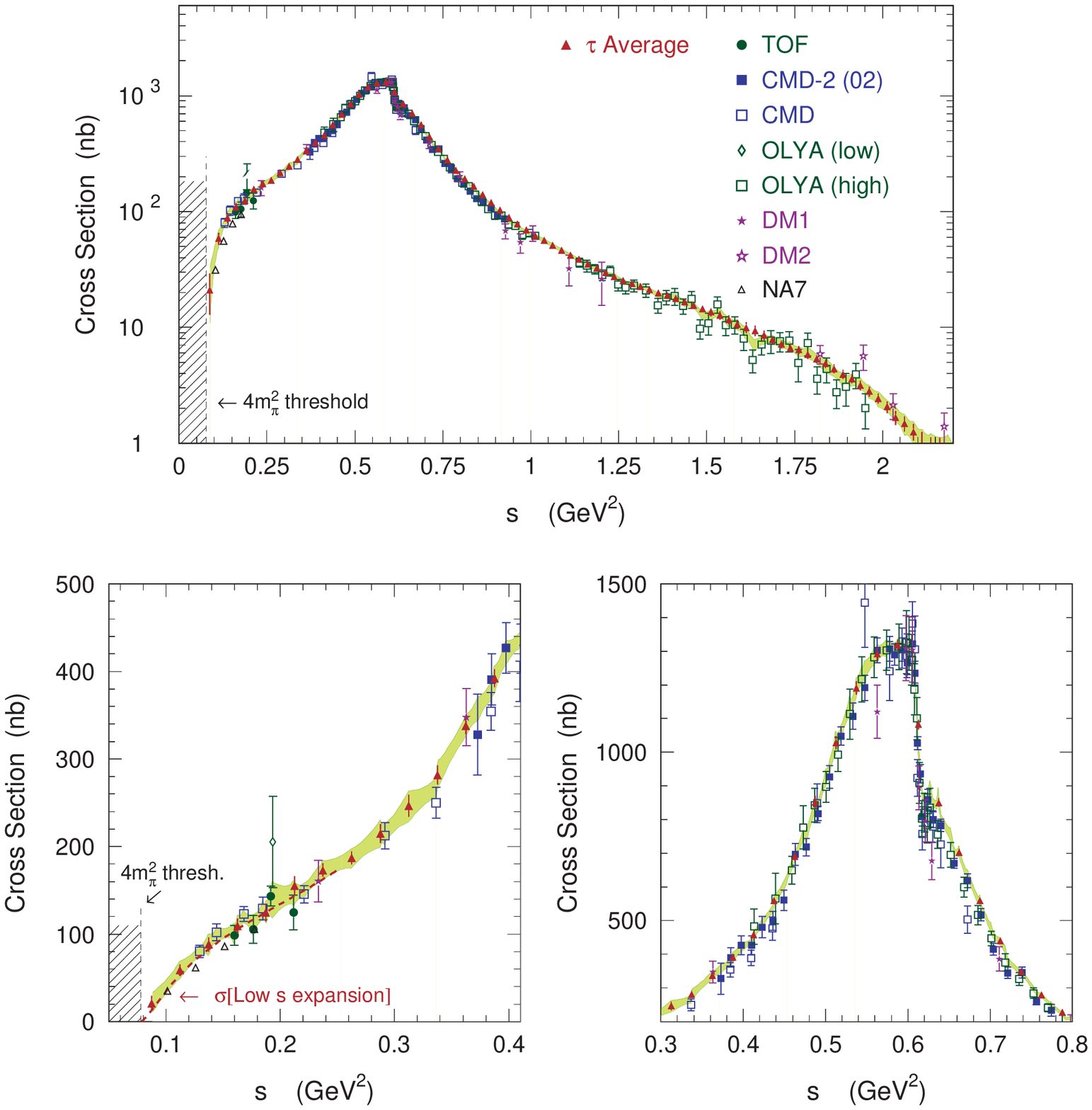,width=160mm}}
\caption[.]{  Comparison of the $\pi^+\pi^-$ \sfs\
    	from \ee\  and isospin-breaking corrected $\tau$ data from ALEPH,
        CLEO, and OPAL 
	expressed as \ee\ cross sections. The band indicates the 
	combined \ee\ and $\tau$ result within $1\sigma$ errors.
	It is given for illustration purpose only. The complete references for
        the \ee\ data are given in Refs.~\cite{dehz,dehz03} and \cite{na7}.}
\label{tau_ee_2pi}
\end{figure}

\begin{figure}[p]
   \centerline{\psfig{file=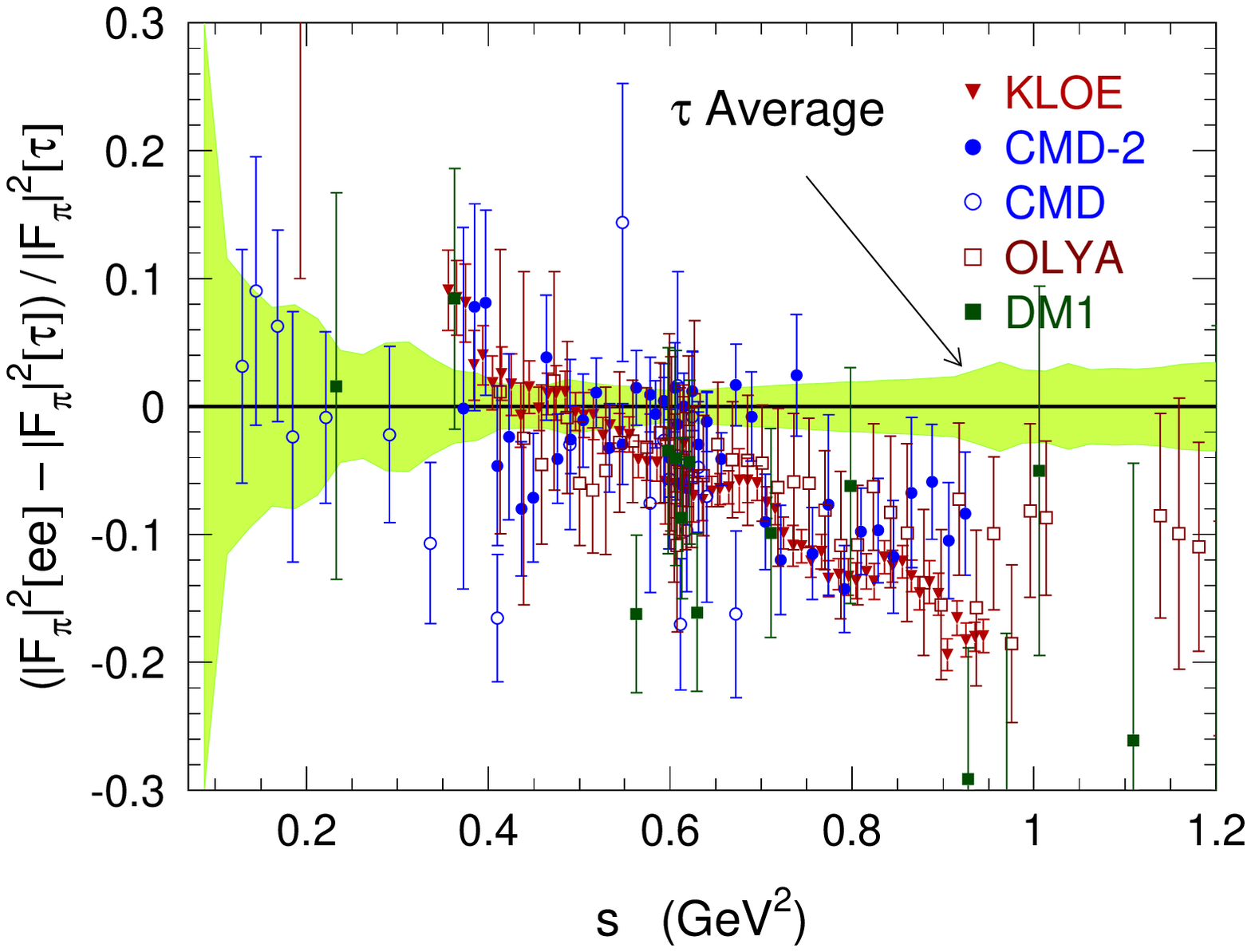,width=120mm}}

   \centerline{\psfig{file=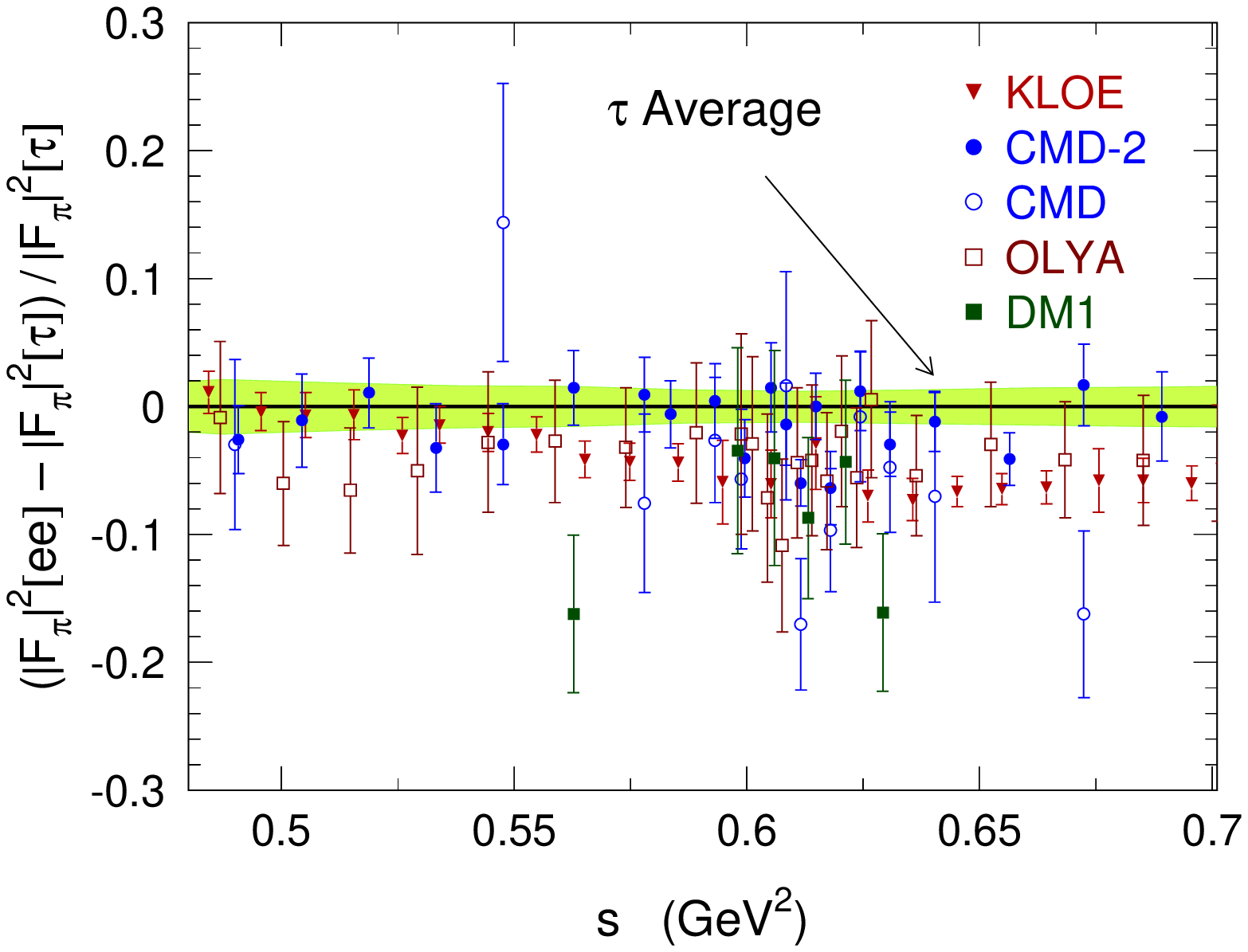,width=120mm}}
\caption[.]{  Relative comparison of the $\pi^+\pi^-$ \sfs\
    	from \ee\  and isospin-breaking corrected $\tau$ data from ALEPH, 
	CLEO, and OPAL 
        expressed as a ratio to the $\tau$ \sf. The band shows the 
        uncertainty on the latter function. The complete references for
        the \ee\ data are given in Refs.~\cite{dehz,dehz03} and \cite{kloe}.}
\label{comp_eetau_2pi}
\end{figure}

Figure~\ref{tau_ee_2pi} shows the comparison for the $2\pi$ \sfs\ obtained 
by averaging the results of the present analysis and the published data
from CLEO~\cite{cleo_2pi} and OPAL~\cite{opal_2pi}. The \ee\ data
are taken from TOF~\cite{tof_2pi}, OLYA~\cite{olya_2pi}, CMD~\cite{cmd_2pi},
CMD-2~\cite{cmd2_new}, DM1~\cite{dm1_2pi}, DM2~\cite{dm2_2pi}, and
NA7~\cite{na7}. The most 
precise results come from CMD-2 and have undergone recently a major revision 
due to previously incorrect radiative corrections~\cite{cmd2_old}.
Visually, the agreement seems satisfactory, however the large dynamical range 
involved does not permit an accurate test. To do so, the \ee\ data are plotted
as a point-by-point ratio to the $\tau$ \sf\ in Fig.~\ref{comp_eetau_2pi}, 
also showing the recent accurate data from KLOE~\cite{kloe} 
using the radiative return technique.
The data are found to be consistent below and around the $\rho$ peak, 
while a discrepancy is observed for masses larger than 0.85~GeV.

\subsubsection{The $4\pi$ spectral functions}

The \sf\ measurements of the $\tau$ 
vector current f\/inal states \pitpiz\ and \tpipiz\ are 
compared to the cross sections of the corresponding \ee\ annihilation 
into the isovector states $2\pi^- 2\pi^+$ and $\pi^- \pi^+ 2\pi^0$. Using 
Eq.~(\ref{eq_spect_fun}) and isospin invariance 
the following relations hold:
\begin{eqnarray}
\label{cvc_4pi}
 \sigma_{e^+e^-\rightarrow\pi^+\pi^-\pi^+\pi^-}^{I=1} 
        & \:=\: &
             2\cdot\frac{4\pi\alpha^2}{s}\,
             v_{1,\,\pi^-\,3\pi^0\,\nu_\tau}~, \\[0.3cm]
\label{cvc_2pi2pi0}
 \sigma_{e^+e^-\rightarrow\pi^+\pi^-\pi^0\pi^0}^{I=1} 
        & \:=\: &
             \frac{4\pi\alpha^2}{s}\,
             \left[v_{1,\,2\pi^-\pi^+\pi^0\,\nu_\tau} 
                  \:-\:
                     v_{1,\,\pi^-\,3\pi^0\,\nu_\tau}
             \right]~.
\end{eqnarray}

\begin{figure}
   \centerline{\psfig{file=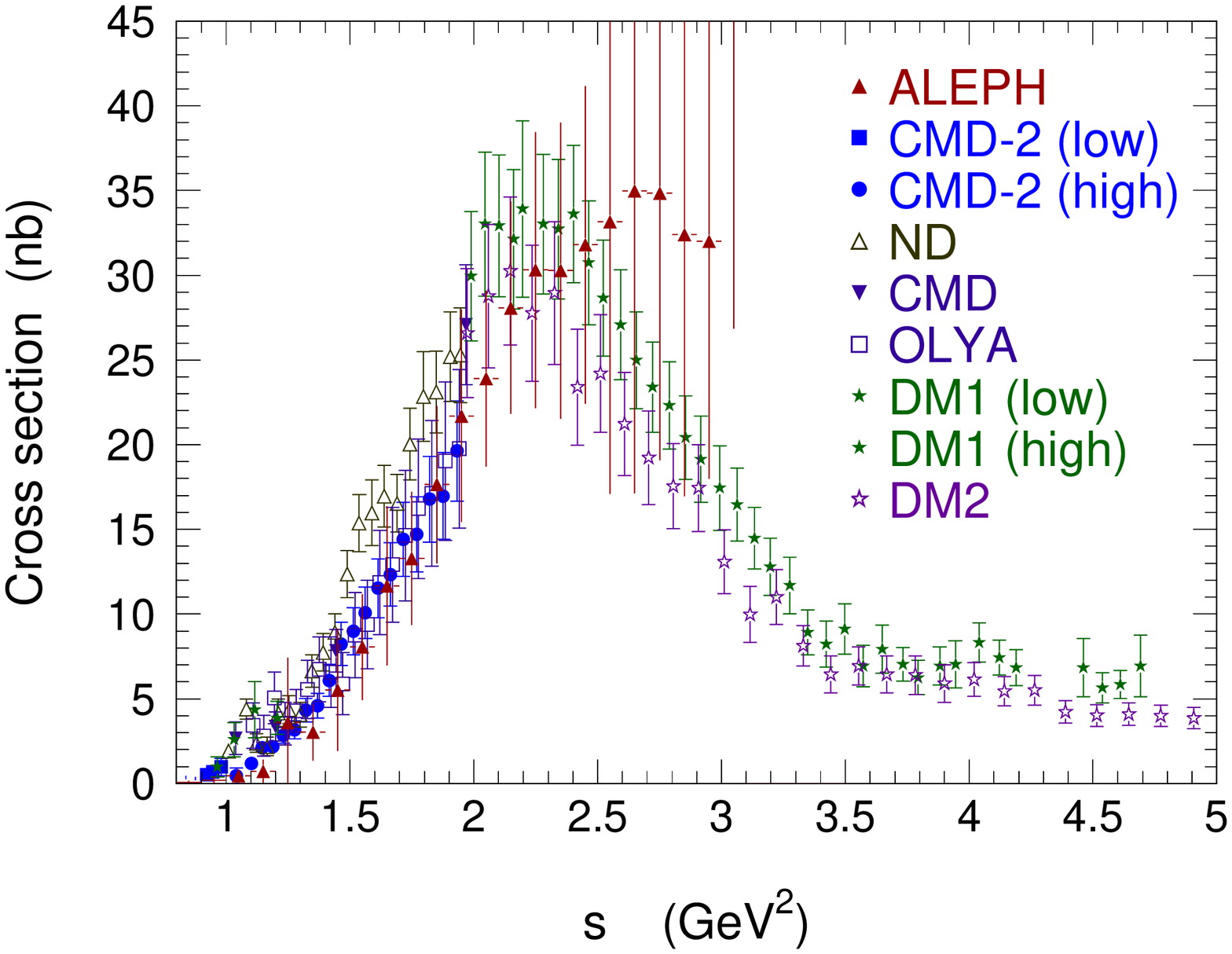,width=120mm}}
\caption{ Comparison of the $2\pi^+2\pi^-$ \sfs\
    	from \ee\ and isospin-breaking corrected $\tau$ data,
	expressed as \ee\ cross sections. The complete references for
        the \ee\ data are given in Refs.~\cite{dehz,dehz03}.}
\label{comp_4pi_eetau}
\end{figure}

\begin{figure}
   \centerline{\psfig{file=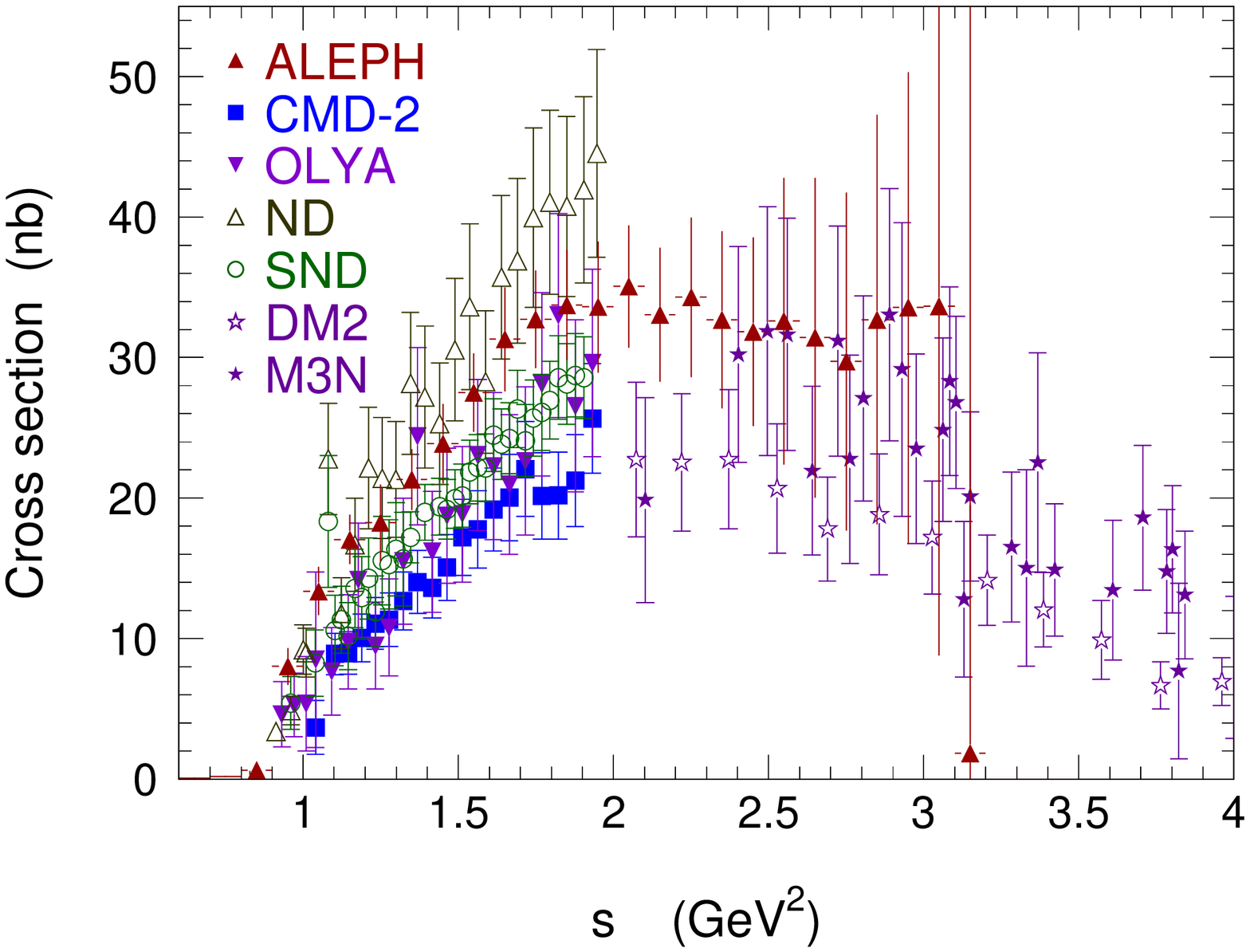,width=120mm}}
\caption{ Comparison of the $\pi^+\pi^- 2\pi^0$ \sfs\
    	from \ee\ and isospin-breaking corrected $\tau$ data,
	expressed as \ee\ cross sections. The complete references for
        the \ee\ data are given in Refs.~\cite{dehz,dehz03}.}
\label{comp_2pi2pi0_eetau}
\end{figure}

The comparison of the $4\pi$ cross sections is given 
in Fig.~\ref{comp_4pi_eetau}
for the $2\pi^+2\pi^-$ channel and in Fig.~\ref{comp_2pi2pi0_eetau} for 
$\pi^+\pi^-2\pi^0$, where the latter suffers from
large differences between the results from the various \ee\ 
experiments. The $\tau$ data, combining two measured \sfs\ according
to Eq.~(\ref{cvc_2pi2pi0}) and corrected for isospin breaking originating 
from the charged-vs-neutral pion mass difference\cite{czyz}, 
lie somewhat in between with large
uncertainties above 2~GeV$^2$ because of the lack of statistics and a large 
feedthrough background in the $\tau\rightarrow\nu_\tau\,\pi^-3\pi^0$
mode. In spite of these difficulties the  $\pi^-3\pi^0$ \sf\ is in agreement
with \ee\ data as can be seen in Fig.~\ref{comp_4pi_eetau}. It is clear
that intrinsic discrepancies exist among the \ee\ experiments and that a
quantitative test of CVC in the $\pi^+\pi^-2\pi^0$ channel is
premature.

\subsubsection{Branching Ratios in $\tau$ Decays and CVC}
\label{sec_brcvc}

It is possible to compare the \ee\ and $\tau$ spectral functions 
in a more quantitative way by calculating weighted integrals
over the mass range of interest up to the $\tau$ mass. 
One convenient choice is provided by the $\tau$ branching fractions 
which involve as a weight the kinematic factor 
$(1-\frac{s}{m_\tau^2})^2(1+\frac{2s}{m_\tau^2})$
coming from the $V-A$ charged current in $\tau$ decay. It is then 
possible to directly compare the measured $\tau$ branching ratios 
to their prediction through isospin invariance (CVC) using as input 
the \ee\ isovector \sfs.
 
Using the ``universality-improved'' branching fraction given in
Eq.~(\ref{uni_be}), the results for the main channels are given
in Table~\ref{tab_brcvc}. The errors quoted for the CVC values are split
into uncertainties from ({\it i}) the experimental input (the \ee\ 
annihilation cross sections) and the numerical integration procedure,
({\it ii}) additional radiative corrections applied to some of the \ee\ 
data (see Ref.~\cite{dehz} for details), and ({\it iii}) 
the isospin-breaking corrections when relating $\tau$ and \ee\ \sfs. 
\vs
\begin{table}[t]
\caption{\label{tab_brcvc}
	Branching fractions of $\tau$ vector decays into 	
	2 and 4 pions in the final state. Second column: present ALEPH 
	results. Third column: inferred from \ee\ spectral
	functions using the isospin 
	relations~(\ref{eq_cvc},\ref{ee_ff},\ref{tau_ff},\ref{cvc_4pi},
        \ref{cvc_2pi2pi0}) and
        correcting for isospin breaking following Refs.~\cite{dehz,dehz03}. 
	Experimental errors, including uncertainties on the integration 
        procedure, and theoretical (missing radiative corrections for \ee,
	and isospin-breaking corrections and $V_{ud}$ for $\tau$)
	are shown separately.
	Right column: differences between the
	direct measurements in $\tau$ decays and the CVC evaluations,
        where the separate errors have been added in quadrature.}
\begin{center}
\setlength{\tabcolsep}{0.75pc}
{\small
\begin{tabular}{lrrr} \hline 
&&& \\[-0.3cm]
		& \mc{3}{c}{Branching fractions  (in \%)} \\
\rs{~~~~~Mode} 	& \mc{1}{c}{$\tau$ ALEPH} 	
		& \mc{1}{c}{$e^+e^-$ via CVC} & $\Delta(\tau-e^+e^-)$ 
\\[0.15cm]
\hline
&&& \\[-0.3cm]
\mc{1}{l}{$~\tau^-\ra\nu_\tau\pi^-\pi^0$}
		& $25.47 \pm 0.13$ 
		& $24.52 \pm \underbrace{0.26_{\rm exp}	
			\pm 0.11_{\rm rad}\pm 0.12_{\rm SU(2)}}_{0.31}$ 
		& $+0.95 \pm 0.33$ 
	\\[0.7cm]
\mc{1}{l}{$~\tau^-\ra\nu_\tau\pi^-3\pi^0$}
		& $ 0.98 \pm 0.09$ 
		& $ 1.09 \pm \underbrace{0.06_{\rm exp}
		        \pm 0.02_{\rm rad}\pm 0.05_{\rm SU(2)}}_{0.08}$ 
		& $-0.11 \pm 0.12$ 
	\\[0.7cm]
\mc{1}{l}{$~\tau^-\ra\nu_\tau2\pi^-\pi^+\pi^0$}
		& $ 4.59 \pm 0.09$ 
		& $ 3.63 \pm \underbrace{0.19_{\rm exp}
			\pm 0.04_{\rm rad}\pm 0.09_{\rm SU(2)}}_{0.21}$ 
		& $+0.96 \pm 0.23$ 
	\\[0.7cm]
 \hline
\end{tabular}
}
\end{center}
\end{table} 
\begin{figure}[t]
\centerline{\psfig{file=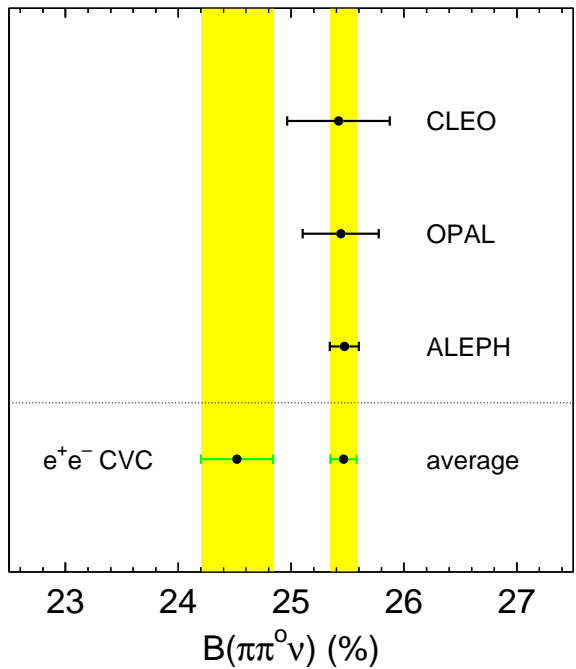,width=100mm}}
\caption[.]{ The measured branching ratios for 
       $\tau\rightarrow\nu_\tau\pi^-\pi^0$ compared to the prediction
       from the $e^+e^-\ra\pi^+\pi^-$ \sf\ applying the isospin-breaking
       correction factors discussed in Section~\ref{vsf_cvc}.
       The measured branching ratios are from ALEPH,
       CLEO~\cite{cleo_bhpi0} and OPAL~\cite{opal_bh}.
       The OPAL result was obtained from their $h \pi^0$ branching 
       ratio, reduced by the small $K \pi^0$ contribution 
       measured by ALEPH~\cite{alephksum} and CLEO~\cite{cleo_kpi0}.}
\label{fig_cvc_2pi}
\end{figure}
As expected from the preceding discussion, a large discrepancy is 
observed for the $\tau\rightarrow \nu_\tau\,\pi^-\pi^0$ 
branching ratio, with a difference of 
$(0.95\pm0.13_\tau\pm0.26_{\rm ee}\pm0.11_{\rm rad}
\pm0.12_{\rm SU(2)})\%$, 
where the uncertainties are from the $\tau$ branching ratio, 
\ee\ cross sections, \ee\ missing radiative corrections and isospin-breaking 
corrections (including the uncertainty on $V_{ud}$), respectively. 
Adding all errors in quadrature, the effect represents a 2.9~$\sigma$ 
discrepancy. Since the disagreement between \ee\ and $\tau$ \sfs\ is 
more pronounced at energies above 0.85~GeV, a smaller discrepancy is expected
in the calculation of the hadronic contribution to the muon anomalous 
magnetic moment because, in this case, the spectral function is weighted
by a steeply falling kernel $K(s)$. 

Besides the result from the present analysis, a similar comparison can be 
performed using results from other experiments. The values for the $\tau$ 
branching ratios involve measurements~\cite{cleo_bhpi0,opal_bh}
given without charged hadron identification, {\it i.e.} for the
$h\pi^0\nu_\tau$, $h3\pi^0\nu_\tau$ and $3h\pi^0\nu_\tau$ final states. 
The corresponding channels with charged kaons have been 
measured~\cite{alephksum,cleo_kpi0} and their contributions are subtracted 
out in order to obtain the pure pionic modes. The comparison is
displayed in Fig.~\ref{fig_cvc_2pi} where it is clear that ALEPH, CLEO, 
and OPAL all separately disagree with the \ee-based CVC result.
\vs
The situation in the $4\pi$ channels is different. Agreement is observed 
for the $\pi^-3\pi^0$ mode within a relative accuracy of $12\%$, however
the comparison is not satisfactory for the $2\pi^-\pi^+\pi^0$ mode. 
In the latter case, the relative difference is very large, 
$(23\pm6)$\%, compared to a reasonable level of isospin symmetry 
breaking. As such, it rather points to experimental problems that have
to be investigated, which are emphasized by the scatter observed among 
the different \ee\ results. In this respect, it is to be noted that new 
preliminary results from the SND and CMD-2 experiments have been recently 
presented~\cite{eidelman_tau04}; while the new data contradict earlier 
results from the same experiments, they are in good agreement with the 
ALEPH spectral functions.

\subsection{Fits to the $\pi\piz$ spectral function}

\subsubsection{Procedure}

The $\pi\piz$ spectral function is dominated by the wide $\rho$ resonance,
parametrized in the forthcoming fits following Gounaris-Sakurai~\cite{gounarissak} 
(GS). The GS parametrization takes into accounts analyticity and unitarity properties.  
The fits make use of the covariance matrix taking into account the 
correlations between different mass bins. 

If one assumes vector dominance, the pion form factor is given by interfering
amplitudes from the known isovector meson resonances $\rho(770)$, 
$\rho(1450)$ and $\rho(1700)$ with relative strengths 1, $\beta$, and $\gamma$. 
Although one could expect from the quark model that $\beta$ and $\gamma$ are real
and respectively negative and positive, the phase of $\beta$, $\phi_\beta$ is left 
free in the fits, while the much smaller $\gamma$ is assumed to be real for lack 
of precise experimental information at large masses. Taking into account 
$\rho - \omega$ interference, one writes
\beq\label{fpi_vdm}
F^{I=1,0}_\pi(s) 
                 \:=\; 
                    \frac{{\mathrm BW}_{\rho(770)}(s)\,
                    \frac{1+\delta\,{\mathrm BW}_{\omega(783)}(s)}
                         {1+\delta} \:+\:                            
                             \beta\,{\mathrm BW}_{\rho(1450)}(s) \:+\: 
                             \gamma\,{\mathrm BW}_{\rho(1700)}(s)}
                         {1\:+\:\beta\:+\:\gamma}~,
\eeq
with the Breit-Wigner propagators
\beq\label{eq_BW}
{\mathrm BW}^{\mathrm GS}_{\rho(m_\rho)}(s) 
               \;=\; 
          \frac{m_\rho^2(\,1+d\cdot\Gamma_\rho/m_\rho)}
               {m_\rho^2\:-\:s\:+\:f(s)\:-\:i\sqrt{s}\,\Gamma_\rho(s)}~,
\eeq
where
\beq
f(s) \;=\; \Gamma_\rho \frac{m_\rho^2}{k^3(m_\rho^2)}\,
           \Bigg[ 
 \, k^2(s) \left( h(s)-h(m_\rho^2)\right) \:+\:
                 (\,m_\rho^2-s)\,k^2(m_\rho^2)\,
                   \frac{d h}{d s}\bigg|_{s=m_\rho^2}
         \,\Bigg] ~.
\eeq
The P-wave energy-dependent width is given by
\beq\label{width}
\Gamma_\rho(s) \;=\; \Gamma_\rho(m_\rho^2)\,
                    \frac{m_\rho}{\sqrt{s}}
                    \left(\frac{k(s)}{k(m_\rho^2)}\right)^{\!\!3}~,
\eeq
where $k(s)=\frac{1}{2}\,\sqrt{s}\,\beta^-(s)$ and $k(m_\rho^2)$ are 
pion momenta in the $\rho$ rest frame. 
The function $h(s)$ is def\/ined as
\beq
h(s) \;=\; \frac{2}{\pi}\,\frac{k(s)}{\sqrt{s}}\,
           {\mathrm ln}\frac{\sqrt{s}+2k(s)}{2m_\pi}~,
\eeq
with $dh/ds|_{m_\rho^2} = 
h(m_\rho^2)\left[(8k^2(m_\rho^2))^{-1}-(2m_\rho^2)^{-1}\right]
\,+\, (2\pi m_\rho^2)^{-1}$.
Interference with the isospin-violating electromagnetic 
$\omega\rightarrow\pi^+\pi^-$ decay occurs only in \ee\ annihilation.
Consequently, $\delta$ is f\/ixed to zero when f\/itting $\tau$ data.
The normalization BW$^{\mathrm GS}_{\rho(m_\rho)}(0)= 1$ f\/ixes the
parameter $d=f(0)/(\Gamma_\rho m_\rho)$, which is found to 
be~\cite{gounarissak}
\beq
d \;=\; \frac{3}{\pi}\frac{m_\pi^2}{k^2(m_\rho^2)}\,
        {\mathrm ln}\frac{m_\rho+2k(m_\rho^2)}{2m_\pi} \:+\:
        \frac{m_\rho}{2\pi\,k(m_\rho^2)} \:-\: 
        \frac{m_\pi^2 m_\rho}{\pi\,k^3(m_\rho^2)}~.
\eeq

\subsubsection{Fit to the ALEPH data}

\begin{table}[t]
\caption[.]{{Fit results of the ALEPH pion form factor in \tauto\pipiz\nut\
          decays using the Gounaris-Sakurai (GS) parametrization. The 
          parameters $m_{\rho^\pm(1700)}$ and $\Gamma_{\rho(1700)}$ are 
          kept fixed to values obtained from fits of \ee\ data extending 
          in mass-squared up to 3.6~GeV$^2$.}}
\begin{center}
\begin{tabular}{|c||c|} \hline 
  Parameter                   & ALEPH GS   \\ \hline \hline 
$m_{\rho^\pm(770)}$ (MeV)     & 775.5 $\pm$ 0.7    \\
$\Gamma_{\rho^\pm(770)}$ (MeV)& 149.0 $\pm$ 1.2    \\ \hline
$\beta$                       & 0.120 $\pm$ 0.008  \\
$\phi_\beta$ (degrees)        & 153 $\pm$ 7        \\
$m_{\rho^\pm(1450)}$ (MeV)    & 1328  $\pm$ 15     \\
$\Gamma_{\rho(1450)}$(MeV)    & 468 $\pm$ 41       \\ \hline
$\gamma$                      & 0.023 $\pm$ 0.008  \\ 
$m_{\rho^\pm(1700)}$ (MeV)    & [1713]             \\
$\Gamma_{\rho(1700)}$ (MeV)   & [235]              \\ \hline
\hline
$\chi^2$/DF                  & 119/110            \\ \hline
\end{tabular}
\end{center}
\label{tab_fit_tau}
\end{table} 

\begin{figure} 
   \centerline{\psfig{file=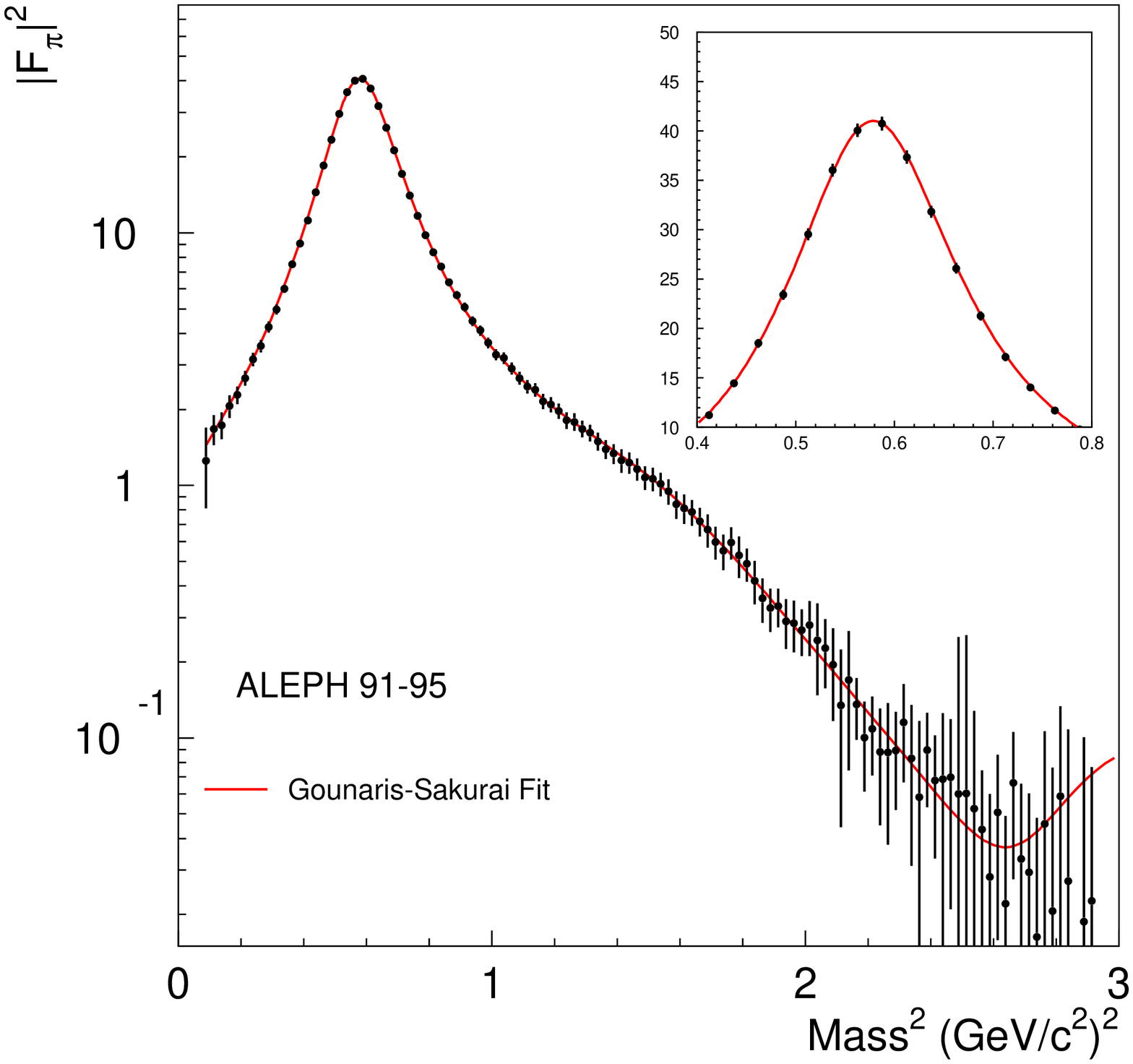,width=80mm}}
\caption{ Fit of the ALEPH \pipiz\ form factor squared using
             the Gounaris-Sakurai parametrization.}
\label{rho_fit_aleph}
\end{figure}

The result of the fit to the ALEPH data using the GS parametrization is given 
in Table~\ref{tab_fit_tau} and illustrated in Fig~\ref{rho_fit_aleph}.
The $\rho$ mass uncertainty is dominated by systematic effects, the largest being
the knowledge of the $\pi^0$ energy scale (calibration).

Concluding from Table~\ref{tab_fit_tau}, the f\/it establishes a need
for the $\rho(1450)$ contribution to the weak pion form factor. 
Weak evidence is found for a $\rho(1700)$ contribution which is expected
since the resonance lies close to the spectrum end-point. 
Most of the fitted parameters exhibit large correlations as seen in 
Table~\ref{fit_rho_corr}. 

\begin{table}
  \caption[.]{ The correlation coefficients between the fitted parameters
     in the fit of the ALEPH $\pi \pi^0$ \sf.}
  \begin{center}
\begin{tabular}{|l||ccccccc|} \hline
ALEPH
& $m_{\rho(770)}$ &$\Gamma_{\rho(770)}$ 
& $\beta$ & $\phi_\beta$ & $m_{\rho(1450)}$ & $\Gamma_{\rho(1450)}$
& $\gamma$\\ \hline\hline
$m_{\rho(770)}$      &  1  &0.38&0.29&$-0.28$&  0.42 &  0.46 &$-0.36$ \\
$\Gamma_{\rho(770)}$ & --  &  1 &0.52&$-0.02$&  0.16 &  0.71 &$-0.38$ \\
$\beta$              & --  & -- &  1 &  0.31 &$-0.08$&  0.63 &  0.03  \\
$\phi_\beta$         & --  & -- & -- &  1    &$-0.87$&$-0.40$&  0.77  \\
$m_{\rho(1450)}$     & --  & -- & -- & --    &  1    &  0.58 &$-0.94$ \\
$\Gamma_{\rho(1450)}$& --  & -- & -- & --    & --    &  1    &$-0.66$ \\
$\gamma$             & --  & -- & -- & --    & --    & --    &  1     \\
\hline
\end{tabular}
 \label{fit_rho_corr}
 \end{center}
\end{table}

\subsection{Combined fit to $\tau$ and $e^+e^-$ data}

It is interesting to perform a combined fit using  
$\tau$ (ALEPH and CLEO) and \ee\ data in order to better constrain 
the lesser known parameters in the phenomenological form factor. 
For this study, the $\tau$ \sf\ is duly corrected for the 
isospin-breaking effects identified in Section~\ref{vsf_cvc}. Also photon
vacuum polarization contributions are removed in the \ee\ \sf\ since they
are absent in the $\tau$ data. In this way, the mass and width 
of the dominant $\rho(770)$ resonance in the two isospin states can be 
determined. For the sub-leading amplitudes from the higher
vector mesons, isospin symmetry is assumed and therefore common masses 
and widths are used in the fit.

The result of the combined fit is given in Table~\ref{rho_common_fit}. 
The differences between the masses and widths of the charged 
and neutral $\rho(770)$'s can be extracted to yield
\begin{eqnarray}
   m_{\rho^-}-m_{\rho^0}~&=&~ (2.4 \pm 0.8)~{\rm MeV} \\
   \Gamma_{\rho^-}-\Gamma_{\rho^0}~&=&~ (0.2 \pm 1.0)~{\rm MeV}
\label{splitting}
\end{eqnarray}
The mass splitting is somewhat larger than the theoretical prediction 
($<0.7$ MeV)~\cite{bijnens}, but only at the $2\sigma$ level. The expected
width splitting, from known isospin breaking, but not taking into account any 
$\rho$ mass splitting, is $(0.7\pm0.3)$ MeV~\cite{ecker,dehz}. However,
if the mass difference is taken as an experimental fact, 
then a larger width difference would be expected. From
the chiral model of the $\rho$ resonance~\cite{pich-portoles,ecker},
one expects
\beq
\Gamma_{\rho^0}~=~\Gamma_{\rho^-}\left(\frac {m_{\rho^0}}{m_{\rho^-}}\right)^3 
              \left(\frac {\beta_0}{\beta_-}\right)^3~+~\Delta \Gamma_{EM}
\label{gamma_calc}
\eeq  
where $\Delta \Gamma_{EM}$ is the width difference from electromagnetic decays
(as discussed above),
leading to a total width difference of ($2.1 \pm 0.5$) MeV, marginally 
consistent with the observed value.

Mass splitting for the $\rho$ was in fact considered in our previous 
analysis~\cite{aleph13_h}. Using pre-CMD-2 \ee\ data, a combined 
fit was attempted which produced a mass splitting consitent with 0 within an
uncertainty of 1.1 MeV. However, the form factor from \ee\ data still 
contained the vacuum polarization contribution (producing a 1.1 MeV shift) 
and a normalization problem was discovered in the treatment 
of the $\tau$ data in the combined fit. 
With the advent of precise CMD-2 data~\cite{cmd2_old}, it became apparent that
differences were showing up between $\tau$ and \ee\ form factors. A large
part of the discrepancy was removed when CMD-2 re-analysed their 
data~\cite{cmd2_new}. Since the $\tau$ results from ALEPH, CLEO and OPAL have 
been shown to be consistent, and since the recent
results from the radiative return analysis of KLOE~\cite{kloe} are 
in fair agreement with the corrected CMD-2 results, the question of the
$\rho$ mass splitting is now more relevant. However a combined fit as shown
in Table~\ref{rho_common_constraint} using the $\tau$ data on one side and 
the \ee\ data on the other, and requiring for consistency the constraint 
from Eq.~(\ref{gamma_calc}) has a $\chi^2$ probability of only $0.6\%$. 
In fact, correcting for different masses extracted from the fit and using 
the corresponding constrained widths, improves the agreement between 
the $\tau$ and \ee\ line shapes, but at the expense of a significant 
discrepancy in normalization. It should be noted that an additional 
correction for this apparent $\rho$ mass splitting increases the present 
discrepancy for the muon anomalous magnetic moment between the estimates 
based on $\tau$ and \ee\ \sfs~\cite{davier_pisa_g-2}. Looking back at the
detailed comparison between spectral functions in Fig.~\ref{comp_eetau_2pi},
it is seen that the new KLOE data shows a trend below and above the $\rho$
resonance which is not in good agreement with CMD-2 (and even worse with
$\tau$ data). This behaviour prevents one at the moment from attributing the
\ee-$\tau$ difference to a simple isospin-breaking correction resulting
from different $\rho$ masses and widths.  
The origin of the discrepancy thus remains ununderstood.

\begin{table}
\caption{
         Combined fit to the pion form factor-squared to ALEPH, CLEO 
         $\tau$ and all \ee\ data, where vacuum polarization has been excluded in the 
         latter data. The parametrization of the $\rho(770)$, $\rho(1450)$, $\rho(1700)$ line 
         shapes follows the Gounaris-Sakurai formula. Separate masses and widths 
         are fitted for the $\rho(770)$, while common values are kept for the 
         higher vector mesons. All mass and width values are in units of MeV and 
         the phases are in degrees.}
\begin{center}
{\small
\begin{tabular}{|c|c|} \hline 
           &  $\tau$  and \ee\   \\
\hline \hline
$m_{\rho^-(770)}$      & 775.5 $\pm$ 0.6  \\
$m_{\rho^0(770)}$      & 773.1 $\pm$ 0.5  \\
$\Gamma_{\rho^-(770)}$ & 148.2 $\pm$ 0.8  \\
$\Gamma_{\rho^0(770)}$ & 148.0 $\pm$ 0.9  \\
\hline
$\alpha_{\rho\omega}$ & $ (2.03 \pm 0.10)~10^{-3}$  \\
$\phi_\alpha$         & $ (13.0 \pm 2.3)$  \\
\hline
$\beta$   & 0.166 $\pm$ 0.005 \\
$\phi_\beta$   & 177.8 $\pm$ 5.2   \\
$m_{\rho(1450)}$   & 1409 $\pm$ 12     \\
$\Gamma_{\rho(1450)}$  & 501 $\pm$ 37           \\   
\hline
$\gamma$   &   0.071 $\pm$ 0.006       \\
$\phi_\gamma$   & [0] \\
$m_{\rho(1700)}$      &  1740 $\pm$ 20         \\
$\Gamma_{\rho(1700)}$   &  [235]       \\
\hline\hline
$\chi^2$/DF             & 383/326      \\
\hline
\hline
\end{tabular}
}
\label{rho_common_fit}
\end{center}
\end{table}

\begin{table}
\caption{
         Combined fit to the pion form factor squared to ALEPH, CLEO $\tau$ 
         and \ee\ data, where vacuum polarization has been excluded for the latter data. 
         The parametrization of the $\rho(770)$, $\rho(1450)$, $\rho(1700)$ line shapes 
         follows the Gounaris-Sakurai formula. For the $\rho(770)$, only 
         $m_{\rho^-}$, $m_{\rho^0}$, and $\Gamma_{\rho^-}$ are fitted, while
         $\Gamma_{\rho^0}$ is computed from Eq.~(\ref{gamma_calc}).
         All mass and width values are in units of MeV and the phases
         are in degrees. The parameters related to $\rho(1450)$ and 
         $\rho(1700)$ amplitudes are fitted, assuming they are identical in
         both data sets.}
\begin{center}
{\small
\begin{tabular}{|c|c|} \hline 
           &  $\tau$  and \ee\   \\
\hline \hline
$m_{\rho^-(770)}$      & 775.4 $\pm$ 0.6  \\
$m_{\rho^0(770)}$      & 773.1 $\pm$ 0.5  \\
$\Gamma_{\rho^-(770)}$ & 148.8 $\pm$ 0.8  \\
$\Gamma_{\rho^0(770)}$ & (146.7)          \\
\hline
$\alpha_{\rho\omega}$ & $ (2.02 \pm 0.10)~10^{-3}$  \\
$\phi_\alpha$         & $ (15.3 \pm 2.0)$  \\
\hline
$\beta$   & 0.167 $\pm$ 0.006  \\
$\phi_\beta$   & 177.5 $\pm$ 6.0    \\
$m_{\rho(1450)}$   & 1410 $\pm$ 16    \\
$\Gamma_{\rho(1450)}$  & 505 $\pm$ 53          \\   
\hline
$\gamma$   &   0.071 $\pm$ 0.007       \\
$\phi_\gamma$   & [0] \\
$m_{\rho(1700)}$      &  1748 $\pm$ 21         \\
$\Gamma_{\rho(1700)}$   &  [235]       \\
\hline\hline
$\chi^2$/DF             & 390/327      \\
\hline
\hline
\end{tabular}
}
\label{rho_common_constraint}
\end{center}
\end{table}

\section{QCD analysis of \boldmath$\tau$ spectral functions}
\label{qcd}

The measurement of \asm\ presented in this section is an update with 
the full $\tau$ sample of the analysis published in Ref.~\cite{alephasf}.
The method is based on a simultaneous f\/it of QCD parametrizations including
perturbative and nonperturbative components to the measured ratio \Rt\
\beq
     R_\tau = \frac{\Gamma(\tau^-\rightarrow{\rm hadrons}^-\,\nu_\tau)}
                   {\Gamma(\tau^-\rightarrow e^-\,\bar{\nu}_e\nu_\tau)}~,
\eeq
and to the spectral moments def\/ined below (Section~\ref{sec_defmoments}). 
It was proposed by Le Diberder and Pich~\cite{pichledib} and has 
been employed in previous analyses by the ALEPH~\cite{alephsf1,alephasf}, 
CLEO~\cite{cleo_as} and OPAL~\cite{opal_2pi} Collaborations. 

\subsection{Theoretical prediction}

According to Eq.~(\ref{eq_imv}) the absorptive parts of the vector 
and axial-vector two-point correlation functions 
$\Pi^{(J)}_{\bar{u}d,V/A}(s)$, with the spin $J$ of the hadronic 
system, are proportional to the $\tau$ hadronic \sfs\ with 
corresponding quantum numbers. The non-strange ratio \Rt\
can be written as an integral of these \sfs\ over the 
invariant mass-squared $s$ of the f\/inal state hadrons~\cite{bnp}:
\beq
\label{eq_rtauth1}
   R_\tau \;=\;
12\pi S_{\rm EW}\intl_0^{m_\tau^2}\frac{ds}{m_\tau^2}\left(1-\frac{s}{m_\tau^2}
                                    \right)^{\!\!2}
     \left[\left(1+2\frac{s}{m_\tau^2}\right){\rm Im}\Pi^{(1)}(s+i\e)
      \,+\,{\rm Im}\Pi^{(0)}(s+i\e)\right]~,
\eeq
where $\Pi^{(J)}$ can be decomposed as
$\Pi^{(J)}=|V_{ud}|^2\left(\Pi_{ud,V}^{(J)}+\Pi_{ud,A}^{(J)}\right)$.
The correlation function $\Pi^{(J)}$ is analytic in the complex $s$ plane 
everywhere except on the positive real axis where singularities exist.
Hence by Cauchy's theorem, the imaginary part of $\Pi^{(J)}$ is 
proportional to the discontinuity across the positive real axis. 
\vs
The energy scale $s_0= m_\tau^2$ is large enough that
contributions from nonperturbative ef\/fects are small. It is therefore 
assumed that one can use the Operator Product Expansion (OPE) 
to organize perturbative and nonperturbative contributions to 
\Rts. The factor $(1-s/m_\tau^2)^2$ suppresses the contribution 
from the region near the positive real axis where $\Pi^{(J)}(s)$ has 
a branch cut and the OPE validity is restricted~\cite{braaten88}. 
\vs
The theoretical prediction of the vector and axial-vector
ratio \RtVA\ can thus be written as:
\beq
\label{eq_delta}
   R_{\tau,V/A} \;=\;
     \frac{3}{2}|V_{ud}|^2S_{\rm EW}\left(1 + \delta^{(0)} + 
     \delta^\prime_{\rm EW} + \delta^{(2-\rm mass)}_{ud,V/A} + 
     \hm\hm\sum_{D=4,6,\dots}\hm\hm\hm\hm\delta_{ud,V/A}^{(D)}\right)~,
\eeq
with the residual non-logarithmic electroweak
correction $\delta^\prime_{\rm EW}=0.0010$~\cite{braaten}, 
neglected in the following, and the dimension $D=2$ 
contribution $\delta^{(2-\rm mass)}_{ud,V/A}$ 
from quark masses which is lower than $0.1\%$ for $u,d$ quarks.
The term $\delta^{(0)}$ is the massless perturbative 
contribution, while the $\delta^{(D)}$ are the OPE
terms in powers of $m_\tau^{-D}$:
\beq
\label{eq_ope}
    \delta_{ud,V/A}^{(D)} \;=\;
       \hm\hm\hm\sum_{{\rm dim}{\cal O}=D}\hm\hm\hm C_{ud,V/A}(s,\mu)
            \frac{\langle{\cal O}_{ud}(\mu)\rangle_{V/A}}
                 {(-m_\tau)^{D}}~,
\eeq
where the parameter $\mu$ separates the 
long-distance nonperturbative ef\/fects, absorbed into
the vacuum expectation elements $\langle{\cal O}_{ud}(\mu)\rangle$,
from the short-distance ef\/fects which are included in the
Wilson coef\/f\/icients $C_{ud,V/A}(s,\mu)$~\cite{wilson}.

The perturbative prediction adopted in this analysis follows 
in detail Ref.~\cite{pert}. The perturbative contribution is 
given in the chiral limit. Ef\/fects from quark masses have been 
calculated in Ref.~\cite{pertmass} and are found to be well below
1\pc\ for the light quarks. Thus the contributions from vector 
and axial-vector currents coincide to any given order of 
perturbation theory and the results are f\/lavour independent.
\vs
The perturbative contribution in Eq.~(\ref{eq_delta}) is then
given by~\cite{pert}
\beq 
\label{eq_knan}
   1+\delta^{(0)} \;=\; 
       \sum_{n=0}^3 K_n A^{(n)}(\alpha_s)~,
\eeq
with $K_0=K_1=1$, $K_2=1.63982$ and $K_3=6.37101$ for three active 
f\/lavours in the \MSbar\ scheme~\cite{3loop}. The coef\/f\/icients $K_n$ 
are known up to three-loop order $\alpha_s^3$ and for $n\ge2$ they 
depend on the renormalization scheme employed. Although progress was recently made
in the evaluation of some contributions at the four-loop order~\cite{k4_cal}, 
the full calculation of $K_4$ is not yet available. The functions 
$A^{(n)}(\alpha_s)$ in Eq.~(\ref{eq_knan}) are the contour integrals
\beq
\label{eq_an}
   A^{(n)}(\alpha_s) \;=\;
      \frac{1}{2\pi i}\hm\ointl_{|s|=m_\tau^2}\hm\hm\hm
      \frac{ds}{s}
       \left[1-2\frac{s}{m_\tau^2} + 2\left(\frac{s}{m_\tau^2}\right)^{\!\!3}
             - \left(\frac{s}{m_\tau^2}\right)^{\!\!4}
       \right]\left(\frac{\alpha_s(-s)}{\pi}\right)^{\!\!n}~,
\eeq
where the contour runs counter clockwise around the
circle from $m_\tau^2+i\epsilon$ to $m_\tau^2-i\epsilon$.
The strong coupling constant in the vicinity of $m_\tau^2$
can be expanded in powers of \assz, with 
coef\/f\/icients that are polynomials in ln$(s/m_\tau^2)$~\cite{bnp}. 
The perturbative prediction becomes then a function of
the $K_n$ coef\/f\/icients and elementary integrals. Up
to fourth order the f\/ixed-order perturbation theory (FOPT) 
expansion reads
\beqn
\label{eq_delta0exp}
   1+\delta^{(0)} 
   &=&
      1+ \frac{\alpha_s(m_\tau^2)}{\pi} 
      + 5.2023\left(\frac{\alpha_s(m_\tau^2)}{\pi}\right)^{\!\!2}
      + 26.366\left(\frac{\alpha_s(m_\tau^2)}{\pi}\right)^{\!\!3} 
         \nonumber\\
   & &    +\,(K_4+78.00)\left(\frac{\alpha_s(m_\tau^2)}{\pi}\right)^{\!\!4}~,
\eeqn
with the unknown $K_4$ coef\/f\/icient.
\vs
Another approach to the solution of the contour integral~(\ref{eq_an}) 
is to perform a direct numerical evaluation using the solution of the 
renormalization group equation (RGE) to four loops~\cite{rit} 
as input for the running $\alpha_s(-s)$~\cite{pivo,pert}. 
It implicitly provides a resummation of all known higher order 
logarithmic integrals and improves the convergence 
of the perturbative series. While, for instance, the third order 
term in the expansion~(\ref{eq_delta0exp}) contributes with $17\%$ 
to the total (truncated) perturbative prediction, the corresponding 
term of the numerical solution amounts only to $6.6\%$ 
(assuming $\alpha_s(m_\tau^2)=0.35$). This numerical solution of 
Eq.~(\ref{eq_knan}) will be referred as {\it contour-improved} 
f\/ixed-order perturbation theory (\FOPTCI) in the following.
\vs
Despite a number of arguments expressed in Ref.~\cite{pert}, the 
intrinsic ambiguity between FOPT and \FOPTCI\ is unresolvable at 
present. This is due to the truncation of the perturbative 
approximation of $\delta^{(0)}$ at f\/inite order in \as.
A conservative measure of this ambiguity is obtained from the 
deviation in $R_\tau$ found when cutting all additional orders 
in \as\ (which is FOPT) and keeping them (\FOPTCI), respectively. 
Both methods are likewise considered in this analysis.

The leading $D=2$ corrections induced by the light-quark masses are
computed using the running quark masses evaluated at the two-loop level.
The following values are used for the renormalization-invariant quark
mass parameters $\hat{m}_i$ defined in Ref.~\cite{bnp}:
\beq
\hat{m}_u = (8.7\pm1.5)~{\rm MeV}~,~~ \hat{m}_d = (15.4\pm1.5)~{\rm MeV}~,
~~\hat{m}_s = (270\pm30)~{\rm MeV}~.
\eeq

The dimension $D=4$ operators have dynamical 
contributions from the gluon condensate \GG\ and quark condensates 
$m_u\langle0|{\bar u}u|0\rangle$, $m_d\langle0|{\bar d}d|0\rangle$
of the light $u,d$ quarks. The contribution 
of the gluon condensate to \RtVA\ vanishes in f\/irst order in 
$\alpha_s(m_\tau^2)$. However, there appear second order terms in the
Wilson coef\/f\/icients due to the logarithmic $s$ dependence of 
\ass\ which after performing the integral~(\ref{eq_rtauth1}) becomes 
$\alpha_s^2(m_\tau^2)$. The quark condensates are parametrized~\cite{bnp} by 
$\langle m_j{\bar \psi_i} \psi_i\rangle = -\hat{m}_j \hat{\mu}_i^3$ with
\beq
\hat{\mu}_u = \hat{\mu}_d =(189\pm7)~{\rm MeV}~,~~ \hat{\mu}_s = (160\pm10)~{\rm MeV}~.
\eeq

The contributions from dimension $D=6$ and 8 operators are rather 
complex. They are treated phenomenologically in this analysis.
Higher order contributions from $D\ge10$ operators 
are expected to be small as, equivalent to the gluon 
condensate, constant terms and terms in leading order 
\as\ vanish in Eq.~(\ref{eq_rtauth1}) after integration.
\vs
The formulae are taken entirely from Ref.~\cite{bnp}, in which
Eq.~(\ref{eq_rtauth1}) is evaluated after the power terms~(\ref{eq_ope})
are inserted into the integral.

\subsection{Spectral moments}
\label{sec_defmoments}

It was shown in Ref.~\cite{pichledib} that it is possible
to benef\/it from the information provided by the 
explicit shape of the \sfs\ in order to obtain 
additional constraints on \assz\ and --- more 
importantly --- on the nonperturbative condensates.
The spectral moments at $m_\tau^2$ are def\/ined as:
\beq
\label{eq_moments}
   R_{\tau,V/A}^{kl} \;\equiv\; 
       \intl_0^{m_\tau^2} ds\,\left(1-\frac{s}{m_\tau^2}\right)^{\!\!k}
                              \left(\frac{s}{m_\tau^2}\right)^{\!\!l}
       \frac{dR_{\tau,V/A}}{ds}~,
\eeq
with $R_{\tau,V/A}^{00}=R_{\tau,V/A}$. The factor 
$(1-s/m_\tau^2)^k$ suppresses the integrand at the crossing of 
the positive real axis where the validity of the OPE is 
less certain and the experimental accuracy is statistically limited. 
Its counterpart $(s/m_\tau^2)^l$ projects out higher energies. 
The spectral information is used to f\/it simultaneously \asm\ 
and the phenomenological effective operators 
$\langle(\alpha_s/\pi)GG_{D=4}\rangle$, $\langle{\cal O}_{D=6}\rangle$ 
and $\langle{\cal O}_{D=8}\rangle$. Due to the intrinsic strong 
correlations only f\/ive moments are used as input to the f\/it.
\vs
In analogy to \Rt, the contributions to the moments originating from 
perturbative and nonperturbative QCD are separated \via\ the OPE.
The prediction of the perturbative contribution takes the form
\beq 
   \delta^{(0,kl)} \;=\; 
       \sum_{n=1}^3 K_n A^{(n,kl)}(\alpha_s)~,
\eeq
with contour integrals $A^{(n,kl)}(\alpha_s)$~\cite{pichledib}
that are expanded up to $\alpha_s^3(s)$ (FOPT) or numerically resolved 
for the running $\alpha_s(-s)$ obtained from the RGE (\FOPTCI).

For practical purpose it is more convenient to def\/ine moments 
that are normalized to the corresponding \RtVA\ in order 
to decouple the normalization from the shape of the $\tau$ \sfs:
\beq
\label{eq_dkl}
   D_{\tau,V/A}^{kl} \equiv
     \frac{R_{\tau,V/A}^{kl}}{R_{\tau,V/A}} =
       \intl_0^{m_\tau^2} ds\,\left(1-\frac{s}{m_\tau^2}\right)^{\!\!k}
                              \left(\frac{s}{m_\tau^2}\right)^{\!\!l}
       \frac{1}{N_{V/A}}\frac{dN_{V/A}}{ds}~.
\eeq

The two sets of experimentally almost uncorrelated observables 
--- \RtVA\ on one hand and the spectral moments on the other hand--- yield 
independent constraints on \asm\ and thus provide an important test of 
consistency.

The results given in Section~\ref{vasep} are recalled:
\beqn
\label{e_rtauv}
    R_{\tau,V}      &=& 1.787 \,\pm\, 0.013~, \\
\label{e_rtaua}
    R_{\tau,A}      &=& 1.695 \,\pm\, 0.013~, \\
\label{e_rtauvpa}
    R_{\tau,V+A}    &=& 3.482 \,\pm\, 0.014~.
\eeqn

The normalization according to Eq.~(\ref{eq_dkl}) reduces considerably 
the correlations between \Rt\ and the moments. It is completely
negligible in the $(V+A)$ case where \RtVpA\ is 
calculated from the dif\/ference $R_\tau-R_{\tau,S}$, which
has no correlation with the hadronic invariant mass spectrum.

\begin{table}[t]
  \caption[.]{\label{tab_moments}
              Spectral Moments of vector ($V$), axial-vector ($A$) and
              vector plus axial-vector $(V+A)$ inclusive $\tau$ decays.
              The errors give the total experimental uncertainties
              including statistical and systematic effects.}
  \begin{center}
  \begin{tabular}{|l||cccc|} \hline 
 ALEPH          &   $l=0$   &   $l=1$   &   $l=2$   &   $l=3$   \\ \hline\hline
 $D_V^{1l}$     &   0.7160  &   0.1694  &   0.0533  &   0.0227  \\ 
 $\Delta^{\mathrm{exp}} D_V^{1l}$      
                &   0.0025  &   0.0004  &   0.0006  &   0.0004  \\ \hline
 $D_A^{1l}$     &   0.7107  &   0.1489  &   0.0657  &   0.0317  \\ 
 $\Delta^{\mathrm{exp}} D_A^{1l}$      
                &   0.0018  &   0.0006  &   0.0004  &   0.0003  \\ \hline
 $D_{V+A}^{1l}$ &   0.7134  &   0.1595  &   0.0594  &   0.0271  \\ 
 $\Delta^{\mathrm{exp}} D_{V+A}^{1l}$  
                &   0.0016  &   0.0004  &   0.0003  &   0.0003  \\ \hline
  \end{tabular}
  \end{center}
\vspace{1cm}
  \caption[.]{\label{tab_momcorr}
              Experimental correlations between the
              moments $D_{\tau,V/A/V+A}^{kl}$. There are no correlations 
              between \RtVpA\ and the corresponding moments.}
\centerline{
\begin{tabular}{|l||cccc|} \hline
ALEPH 
& $D_{\tau,V}^{10}$ & $D_{\tau,V}^{11}$ 
& $D_{\tau,V}^{12}$ & $D_{\tau,V}^{13}$ \\ \hline\hline
$R_{\tau,V}$      &$-0.34$ & $ 0.20$ & $ 0.33$ & $ 0.35$ \\
$D_{\tau,V}^{10}$ &  1     & $-0.82$ & $-0.98$ & $-0.99$ \\
$D_{\tau,V}^{11}$ & --     &  1      &  0.90   &    0.82 \\
$D_{\tau,V}^{12}$ & --     & --      &  1      &    0.99 \\
$D_{\tau,V}^{13}$ & --     & --      & --      &    1    \\ \hline
\end{tabular}
\begin{tabular}{|l||cccc|} \hline
ALEPH 
& $D_{\tau,A}^{10}$ & $D_{\tau,A}^{11}$ 
& $D_{\tau,A}^{12}$ & $D_{\tau,A}^{13}$ \\ \hline\hline
$R_{\tau,A}$      &$-0.23$ & $ 0.02$ & $ 0.16$ & $ 0.25$ \\
$D_{\tau,A}^{10}$ &  1     & $-0.78$ & $-0.97$ & $-0.98$ \\
$D_{\tau,A}^{11}$ & --     &  1      &    0.87 &  0.66   \\
$D_{\tau,A}^{12}$ & --     & --      &    1    &  0.94   \\
$D_{\tau,A}^{13}$ & --     & --      & --      &  1      \\ \hline
\end{tabular}
}
  \begin{center}
\begin{tabular}{|l||cccc|} \hline
ALEPH
& $D_{\tau,V+A}^{10}$ & $D_{\tau,V+A}^{11}$ 
& $D_{\tau,V+A}^{12}$ & $D_{\tau,V+A}^{13}$ \\ \hline\hline
$D_{\tau,V+A}^{10}$ &  1     & $-0.73$ & $-0.98$ & $-0.99$ \\
$D_{\tau,V+A}^{11}$ & --     &  1      &    0.80 &  0.67   \\
$D_{\tau,V+A}^{12}$ & --     & --      &    1    &  0.98   \\
$D_{\tau,V+A}^{13}$ & --     & --      & --      &  1      \\ \hline
\end{tabular}
  \end{center}
\end{table}

The measured values of the moments for the $V$, $A$ and $(V+A)$ 
\sfs\ are given in Table~\ref{tab_moments}, and their correlation 
matrices in Table~\ref{tab_momcorr}. The correlations between the 
moments are computed analytically from the contraction of the 
derivatives of two involved moments with the covariance matrices 
of the respective normalized invariant mass-squared spectra. In all cases, 
the negative sign for the correlations between the $k=1,l=0$ and the $k=1,l\ge1$
moments is understood to be due to the $\rho$ and the
$\pi$, $a_1$ peaks which determine the major part of the
$k=1,l=0$ moments. They are much less important for higher moments
and consequently the amount
of negative correlation increases with $l=1,2,3$. This also 
explains the large and increasing positive correlations between the 
$k=1,l\ge1$ moments, in which, with growing $l$, the high energy tail 
becomes more important than the low energy peaks. The total errors 
for the $(V+A)$ case are dominated by the uncertainties on the
hadronic fractions.

\subsection{Results of the fits: $\alpha_s(m^2_\tau)$}

The combined f\/its to the measured $V$, $A$ and $(V+A)$ ratios \Rt\ 
and moments adjust the parameters \asm, \GG, \Osix\ and \Oeight\ of
the OPE in the theoretical predictions~(\ref{eq_delta}) and 
(\ref{eq_moments}) of the above quantities. 
\vs
The uncertainties entering the theoretical predictions have been 
estimated in great detail in the previous analysis~\cite{alephasf}.
Minor contributions originate from the CKM matrix element $|V_{ud}|$,
the electroweak radiative correction factor $S_{EW}$, the light quark
masses $m_{u,d}$, and the quark condensates. The largest contributions
have their origin in the truncation of the perturbative expansion.
Although it might introduce double-counting, the procedure used
considers separate variations of the unknown higher order coefficient
$K_4$ and the choice of the renormalization scale. Thus the estimate is 
expected to be conservative. The renormalization scale is varied around
$m_\tau$ from 1.1 to 2.3~GeV with the variation over half of the range
taken as systematic uncertainty. Concerning the coefficient $K_4$, some
theoretical progress has been achieved for its direct calculation. Part of 
the contributions have been recently calculated~\cite{k4_cal}, lending 
support to the estimate ($K_4 \sim 27.5$) using the principle of minimal 
sensitivity~\cite{pms}, which allows the computation of a renormalization 
scheme with optimal convergence~\cite{k4_pms}. Another estimate of $K_4$
was obtained~\cite{k4_fld} using the a priori freedom of the choice of the
renormalization scale $\mu$ to increase the sensitivity of the perturbative
series on $K_4$, yielding $K_4=27\pm5$. Taking advantage of the new 
theoretical developments, the value $K_4 = 25 \pm 25$ is assumed in this 
analysis, instead of $50\pm50$ used in the previous analysis~\cite{alephasf}.

\begin{figure}[t]
   \centerline{\psfig{file=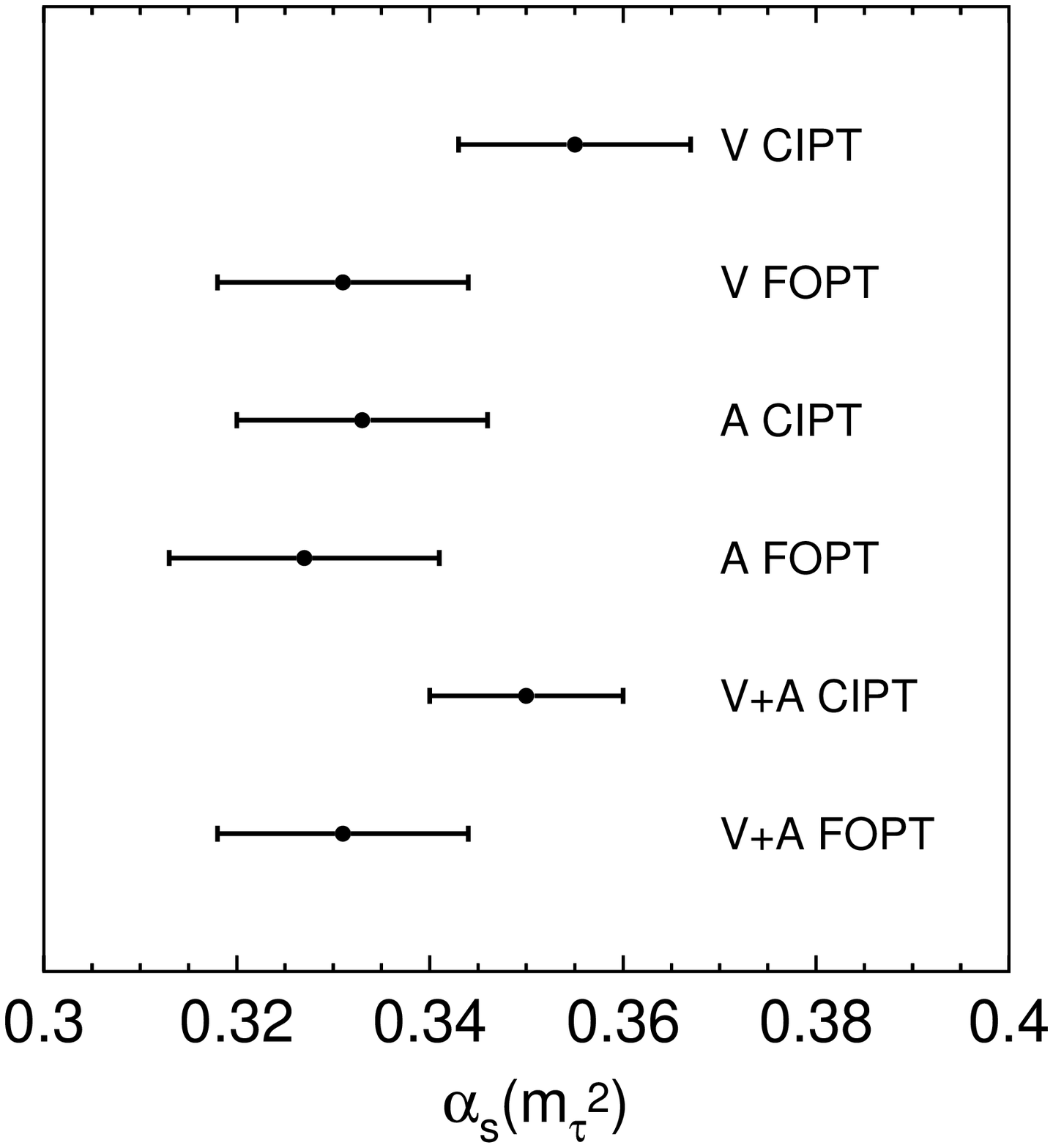,width=80mm}}
  \caption[.]{\label{alphas_res}
              Results for \asm\ from the fits of $R_{\tau,V,A,V+A}$ and 
              the moments $D^{kl}_{V,A,V+A}$ using the \FOPTCI and FOPT
              perturbative expansions. The measurements are correlated due 
              to the theoretical errors (see Table~\ref{tab_asresults}).}
\end{figure}

The f\/it minimizes the $\chi^2$ of the dif\/ferences between measured 
and f\/itted quantities contracted with the inverse of the sum of the 
experimental and theoretical covariance matrices. 
\vs
\begin{table}[p]
  \caption[.]{\label{tab_asresults}
              F\/it results of \asm\ and the OPE nonperturbative 
              contributions from vector, axial-vector and $(V+A)$ combined
              fits using the corresponding ratios \Rt\ and the spectral 
              moments as input parameters. Where two errors are given they 
              denote experimental (first number) and theoretical 
              uncertainties (second number). The $\delta^{(2)}$ term is the 
              pure theoretical prediction with quark masses varying within 
              their prescribed range (see text). The quark condensates in
              the $\delta^{(4)}$ term are fixed to their theoretical
              values~\cite{alephasf} within errors and only the gluon 
              condensate is varied as a free parameter. The total 
              nonperturbative contribution is the sum 
              $\delta_{\rm NP}=\delta^{(4)}+\delta^{(6)}+\delta^{(8)}$.
              Results are given separately for the \FOPTCI\ and the FOPT
              perturbative expansion (see text). }
  \begin{center}
{\small
  \begin{tabular}{|l||c|c|c|} \hline 
     & & &  \\
  \rs{fitted variable}     &\rs{Vector ($V$)} &\rs{Axial-Vector ($A$)}&  \rs{$V\,+\,A$}\\
\hline \hline
 \asm\ (\FOPTCI) &  $0.355\pm0.008\pm0.009$  
                 &  $0.333\pm0.009\pm0.009$   
                 &  $0.350\pm0.005\pm0.009$   
\\
 \asm\ (FOPT)    &  $0.331\pm0.006\pm0.012$  
                 &  $0.327\pm0.007\pm0.012$   
                 &  $0.331\pm0.004\pm0.012$ 
\\ \hline  
 \gluonc\ (GeV$^4$) (\FOPTCI)   &  $(0.4\pm0.3) \times 10^{-2}$  
                                &  $(-1.3\pm0.4) \times 10^{-2}$   
                                &  $(-0.5\pm0.3) \times 10^{-2}$   
\\ 
 \gluonc\ (GeV$^4$) (FOPT)    &  $(1.5\pm0.3) \times 10^{-2}$  
                              &  $(-0.2\pm0.4) \times 10^{-2}$   
                              &  $(0.6\pm0.2) \times 10^{-2}$  
\\ \hline
 $\delta^{(6)}$ (\FOPTCI) &  $(2.85\pm0.22) \times 10^{-2}$  
                          &  $(-3.23\pm0.26) \times 10^{-2}$   
                          &  $(-2.1\pm2.2) \times 10^{-3}$   
\\
 $\delta^{(6)}$ (FOPT)    &  $(2.70\pm0.25) \times 10^{-2}$  
                          &  $(-2.96\pm0.31) \times 10^{-2}$   
                          &  $(-1.6\pm2.5) \times 10^{-3}$    
\\ \hline
 $\delta^{(8)}$ (\FOPTCI) &  $(-9.0\pm0.5) \times 10^{-3}$  
                          &  $(8.9\pm0.6) \times 10^{-3}$   
                          &  $(-0.3\pm4.8) \times 10^{-4}$   
\\
 $\delta^{(8)}$ (FOPT)    &  $(-8.6\pm0.6) \times 10^{-3}$  
                          &  $(8.6\pm0.6) \times 10^{-3}$   
                          &  $(1.2\pm5.2) \times 10^{-4}$
\\ \hline \hline
 $\chi^2/$1 DF (\FOPTCI)       & 0.52            & 5.71            & 3.84    
 \\ 
 $\chi^2/$1 DF (FOPT)          & 0.01            & 0.63            & 0.11 
 \\ \hline \hline
 $\delta^{(2)}$ (\FOPTCI)    & $(-3.3\pm3.0) \times 10^{-4}$
                              & $(-5.1\pm3.0) \times 10^{-4}$
                              & $(-4.4\pm2.0) \times 10^{-4}$
 \\
 $\delta^{(2)}$ (FOPT)       & $(-3.0\pm3.0) \times 10^{-4}$
                              & $(-5.0\pm3.0) \times 10^{-4}$
                              & $(-4.0\pm2.0) \times 10^{-4}$
 \\ \hline
 $\delta^{(4)}$ (\FOPTCI)    & $(4.1\pm1.2) \times 10^{-4}$
                              & $(-5.7\pm0.1) \times 10^{-3}$
                              & $(-2.7\pm0.1) \times 10^{-3}$
 \\      
 $\delta^{(4)}$ (FOPT)       & $(6.8\pm1.0) \times 10^{-4}$
                              & $(-5.3\pm0.1) \times 10^{-3}$
                              & $(-2.4\pm0.1) \times 10^{-3}$
 \\ \hline 
 Total $\delta_{\rm NP}$ (\FOPTCI)    & $(1.99\pm0.27) \times 10^{-2}$  
                                       & $(-2.91\pm0.20) \times 10^{-2}$
                                       & $(-4.8\pm1.7) \times 10^{-3}$
 \\
 Total $\delta_{\rm NP}$ (FOPT)       & $(1.91\pm0.31) \times 10^{-2}$  
                                       & $(-2.63\pm0.25) \times 10^{-2}$
                                       & $(-3.9\pm2.0) \times 10^{-3}$
 \\ \hline \hline

  \end{tabular}
}
  \end{center}
\vspace{1.0cm}
  \caption[.]{\label{tab_rescorr}
              Correlation matrices according to the fits presented
              in Table~\ref{tab_asresults} for vector (left table),
              axial-vector (middle) and $(V+A)$ (right table) using
              \FOPTCI. As the gluon condensate contributes only 
              insignificantly to $\delta^{(4)}$, the correlations 
              to the total $\delta^{(4)}$ term are small.}
{\footnotesize
\begin{center}
  \begin{tabular}{|l||ccc|ccc|ccc|} \hline 
\mc{1}{|c||}{ALEPH} &  
$\langle GG\rangle_V$ &  $\delta^{(6)}_V$ & $\delta^{(8)}_V$ &
$\langle GG\rangle_A$ &  $\delta^{(6)}_A$ & $\delta^{(8)}_A$ &
$\langle GG\rangle_{V+A}$
                     &  $\delta^{(6)}_{V+A}$ & $\delta^{(8)}_{V+A}$ \\
\hline\hline
\asm\         & $-0.39$  & $-0.28$  & $-0.34$
              & $-0.57$  & $0.52$  & $-0.55$   
              & $-0.37$  & $0.38$  & $-0.45$ \\
$\langle GG\rangle_{V/A/V+A}$
              &  $   1$  & $0.44$  &  $0.46$
              &  $   1$  & $-0.81$  &  $0.80$   
              &  $   1$  & $-0.65$  &  $0.65$ \\
$\delta^{(6)}_{V/A/V+A}$ 
              &     --   &  $   1$  & $-0.98$
              &     --   &  $   1$  & $-0.99$   
              &     --   &  $   1$  & $-0.98$ \\
$\delta^{(8)}_{V/A/V+A}$ 
              &     --   &     --   &  $   1$
              &     --   &     --   &  $   1$
              &     --   &     --   &  $   1$ \\
\hline
  \end{tabular}
\end{center}
}
\end{table}
The results are listed in Table~\ref{tab_asresults}, separately for
the FOPT and \FOPTCI\ approaches. It is to be noted that even if the fits are
better with the FOPT expansion, the two sets of results are
considered on the same footing and averaged later on.
Table~\ref{tab_rescorr} gives the corresponding correlation matrices
between the f\/itted parameters. The limited number of observables
and the strong correlations between the spectral moments explain the
large correlations observed, especially between the f\/itted 
nonperturbative operators. The precision of \asm\ obtained with the 
two perturbative methods employed is comparable, however their central
values dif\/fer by  0.01-0.02.  The $\delta^{(2)}$ term 
is the pure theoretical contribution from the known masses (within errors)
of the light $u, d$ quarks. In the $\delta^{(4)}$ term the quark 
condensates and the quark masses are rather well known and are 
f\/ixed theoretically within their errors, while the gluon condensate 
is adjusted in the f\/it. 
\vs
As mentioned before, there exists no constraining 
prescription which allows a resolution of the ambiguity between 
\FOPTCI\ and FOPT. The f\/inal result on \asm\ is thus taken as the arithmetic 
average of the two values given in Table~\ref{tab_asresults}, with half of 
their dif\/ference added as additional theoretical error: 
\beq
\label{eq_asres}
   \alpha_s(m_\tau^2)    = 0.340 \pm 0.005_{\rm exp} 
                                      \pm 0.014_{\rm th}~.
\eeq
The f\/irst error accounts for the experimental uncertainty, the 
second number gives the uncertainty of the theoretical prediction of
$R_\tau$ and the spectral moments as well as the ambiguity of the 
theoretical approaches employed.
\vs
In Table~\ref{tab_asresults}, a remarkable agreement within statistical 
errors is found between the \asm\ values using vector and axial-vector data,
providing a first important consistency check of the analysis, since
the two corresponding spectral functions are independently measured and
manifest a quite different resonant behaviour. The main results are displayed
in Fig.~\ref{alphas_res}.

\subsection{Results of the fits: nonperturbative contributions}

The advantage of 
separating the vector and axial-vector channels and comparing to
the inclusive $(V+A)$ f\/it becomes obvious in the adjustment of the 
leading nonperturbative contributions of dimension $D=6$ and $D=8$, 
which approximately cancel in the inclusive sum. This cancellation of the 
nonperturbative terms increases the confidence in the \asm\ 
determination from the inclusive $(V+A)$ observables. The gluon 
condensate is determined by the f\/irst $l=0,1$ moments, which receive 
lowest order contributions. It is observed that the values obtained in 
the $V$ and $A$ fits are not very consistent, which could indicate some
problem in the validity of the OPE approach, since nonperturbative
contributions are significant in these cases. Taking the value obtained in 
the $(V+A)$ inclusive f\/it, where nonperturbative effects are negligible, 
and adding as systematic uncertainties half of the dif\/ference between 
the vector and axial-vector f\/its as well as between 
the \FOPTCI\ and FOPT results, the gluon condensate is found to be
\beq
\langle(\alpha_s/\pi) GG\rangle=(0.001\pm0.012)~{\rm GeV}^4.
\label{gluon_cond}
\eeq
This result does not provide evidence for a nonzero gluon condensate, but it 
is consistent with and has comparable accuracy to the independent value 
obtained using charmonium sum rules and \ee\ data in the charm 
region, $(0.011\pm0.009)~{\rm GeV}^4$ in a combined determination with 
the $c$ quark mass~\cite{ioffe}. 
\vs
The $D=6,8$ nonperturbative contributions are obtained after averaging the
FOPT and \FOPTCI\ values:
\beq
\begin{array}{rclrcl}
  \delta^{(6)}_V	&=& (2.8 \pm 0.3) \times 10^{-2}~, &
  \delta^{(8)}_V 	&=& (-8.8 \pm 0.6) \times 10^{-3}~,\\
  \delta^{(6)}_A 	&=& (-3.1 \pm 0.3) \times 10^{-2}~, & 
  \delta^{(8)}_A 	&=& (8.7 \pm 0.6) \times 10^{-3}~,  \\
  \delta^{(6)}_{V+A} 	&=& (-1.8 \pm 2.4) \times 10^{-3}~,  &
  \delta^{(8)}_{V+A} 	&=& (0.5 \pm 5.1) \times 10^{-4}~. 
\end{array}
\eeq
The remarkable feature is the approximate cancellation of these contributions
in the $V+A$ case, both for $D=6$ and $D=8$. This property was 
predicted~\cite{bnp} for $D=6$ using the simplifying assumption of vacuum
saturation for the matrix elements of four-quark operators, yieding
$\delta^{(6)}_V = -(7/11)~\delta^{(6)}_A$ and 
$\delta^{(6)}_V = (2.5 \pm 1.3) \times 10^{-2}$, in fair agreement with our results.
There is no reliable prediction for the $D=8$ contributions.

The total nonperturbative $V+A$ correction is 
$\delta_{{\rm NP},V+A}=(-4.3\pm1.9) \times 10^{-3}$,
much smaller than the corresponding values in the $V$ and $A$
components, $\delta_{{\rm NP},V}=(2.0\pm0.3) \times 10^{-2}$ and 
$\delta_{{\rm NP},V}=(-2.8\pm0.3) \times 10^{-2}$. After removing the
theoretically well-known contribution from the quark condensates ($D=4$),
the remaining non perturbative component in $V+A$ is 
$-(1.7\pm1.9)\times 10^{-3}$, one order of magnitude smaller compared 
to $V$ and $A$, and even consistent with zero.

\subsection{Evolution of $\alpha_s$ with the mass scale}

\subsubsection{Test of the running of $\alpha_s(s)$ at low energies}

Using the \sfs, one can simulate the physics of a hypothetical 
$\tau$ lepton with a mass $\sqrt{s_0}$ smaller than $m_\tau$
through Eq.~(\ref{eq_rtauth1}). Assuming quark-hadron duality, 
the evolution of $R_\tau(s_0)$ provides a direct test of the 
running of $\alpha_s(s_0)$, governed by the RGE $\beta$-function. 
On the other hand, it is a test of the validity of the OPE approach 
in $\tau$ decays. The studies performed in this section employ only 
\FOPTCI. Results obtained with FOPT are similar and dif\/fer only 
in the central \asm\ value.
\begin{figure}[t]
   \centerline{\psfig{file=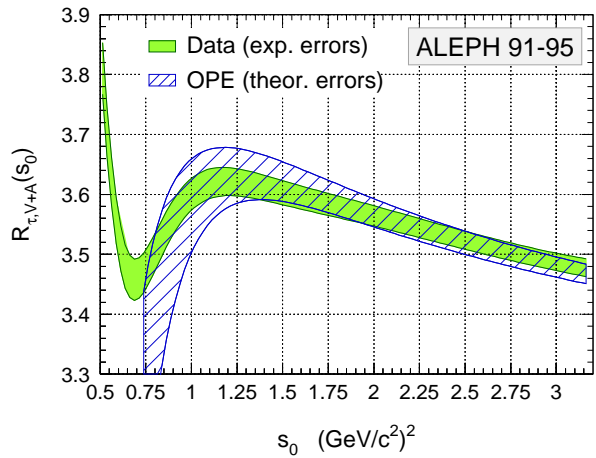,width=80mm}}
  \caption[.]{\label{vpa_runrtau}
              The ratio \RtVpA\ versus the square ``$\tau$ mass'' $s_0$. 
              The curves are plotted as error bands to emphasize their 
              strong point-to-point correlations in $s_0$. Also 
              shown is the theoretical prediction using \FOPTCI\ and
              the results for \RtVpA\ and the nonperturbative 
              terms from Table~\rm\ref{tab_asresults}.}
\end{figure}
\begin{figure}
   \centerline{\psfig{file=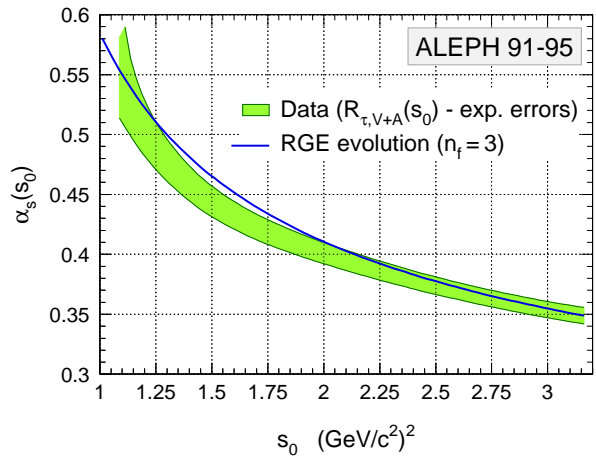,width=80mm}}
  \caption[.]{\label{vpa_runas}
              The running of $\alpha_s(s_0)$ obtained from the 
              fit of the theoretical prediction to \RtVpAs.
              The shaded band shows the data including experimental
              errors. The curve gives the four-loop RGE evolution 
              for three flavours.}
\end{figure}

The functional dependence of \RtVpAs\ is plotted in 
F\/ig.~\ref{vpa_runrtau} together with the theoretical 
prediction using the results of Table~\ref{tab_asresults}. The 
slight mismatch appearing at $s_0=m_\tau^2$ between the experimental 
and theoretical values of \RtVpA\ is due to the fact that the global
fits presented in Table~\ref{tab_asresults} involve not only the 
measured \RtVpA, but also the spectral moments.
The spread due to uncertainties are shown as bands. 
The correlations between two adjacent bins $s_1<s_2$ are large as 
the only new information is provided by the small mass dif\/ference 
between the two bins and the slightly dif\/ferent weight function. 
They are reinforced by the original experimental and theoretical 
correlations. Below $1~{\rm GeV}^2$ the error of the theoretical 
prediction of \RtVpAs\ starts to blow up due to the 
increasing uncertainty from the unknown $K_4$ perturbative term;
errors of the nonperturbative contributions are {\it not} contained 
in the theoretical error band. F\/igure~\ref{vpa_runas} shows the plot 
corresponding to F\/ig.~\ref{vpa_runrtau}, translated into the running 
of $\alpha_s(s_0)$, \ie, the experimental value for $\alpha_s(s_0)$ 
has been individually determined at every $s_0$ from the comparison 
of data and theory. Also plotted is the four-loop RGE evolution using 
three quark f\/lavours.

It is remarkable that the theoretical prediction using the parameters 
determined at the $\tau$ mass and \RtVpAs\ extracted from the measured
$V+A$ \sf\ agree down to $s_0 \sim 0.8$~GeV$^2$. The agreement is good to
about 2\% at 1~GeV$^2$. This result, even more directly illustrated by 
Fig.~\ref{vpa_runas}, demonstrates the validity of the perturbative approach
down to masses around 1~GeV, well below the $\tau$ mass scale. The
agreement with the expected scale evolution between 1 and 1.8~GeV is an
interesting result, considering the relatively low mass range, where
$\alpha_s$ is seen to decrease by a factor of 1.6 and reaches rather 
large values $\sim 0.55$ at the lowest masses. This behaviour provides
confidence that the \asm\ measurement is on solid phenomenological ground.

\subsubsection{Evolution to $M_Z$}

The evolution of the 
\asm\ measurement from the inclusive $(V+A)$ observables to the $Z$ mass scale
can be performed, based on the 
Runge-Kutta integration of the dif\/ferential equation to four loops and 
matching conditions at heavy quark ($c$, $b$) thresholds to three 
loops~\cite{rit,chet1,chet2,wetzel,pichsanta}. There is some ambiguity in
the procedure followed to satisfy the matching conditions, in particular
whether the matching should be performed at a scale given by the quark mass
or twice the quark mass, and also on the choice of quark mass, {\it i.e.}
the running mass or the pole mass. Different procedures have been tried with
the relevant parameters varied within their uncertainties. The results are
consistent within the quoted systematic uncertainty and for definiteness 
the final value is obtained with the matching realized at the quark-pair
thresholds: 
\beqn
\label{eq_asres_mz}
   \alpha_s(M_Z^2) &=& 0.1209 \pm 0.0006_{\rm exp} 
                                       \pm 0.0016_{\rm th} 
                                       \pm 0.0005_{\rm evol} \nonumber \\
                         &=& 0.1209 \pm 0.0018~.
\eeqn

The f\/irst two errors originate from the \asm\ determination given
in Eq.~(\ref{eq_asres}), while the last error stands for
possible ambiguities in the evolution due to uncertainties in the 
matching scales of the quark thresholds. The uncertainty on the evolution
has main components from the $c$-quark mass (0.00020), the $b$-quark mass
(0.00005), the renormalization scale (taken from 0.7~$m_q$ to 3.0~$m_q$, 
0.00023), the 3-loop truncation in the matching expansion (0.00026) and the
four-loop truncation in the RGE equation (0.00031).
\vs
The result~(\ref{eq_asres_mz}) is a 1.5\% determination of the strong 
coupling constant at the $M_Z$ scale, limited in accuracy by theoretical
uncertainties in the perturbative expansion. The significant improvement in
precision compared to our previous determination~\cite{alephasf},
$0.1202 \pm 0.008_{\rm exp} \pm 0.0024_{\rm th} \pm 0.0010_{\rm evol}$, 
results from the higher statistics and the more detailed experimental analysis,
but mostly from the increased knowledge from theory, essentially at the level 
of the perturbative expansion~\cite{k4_cal}, and also for the scale 
evolution procedure.

\begin{figure}[t]
   \centerline{\psfig{file=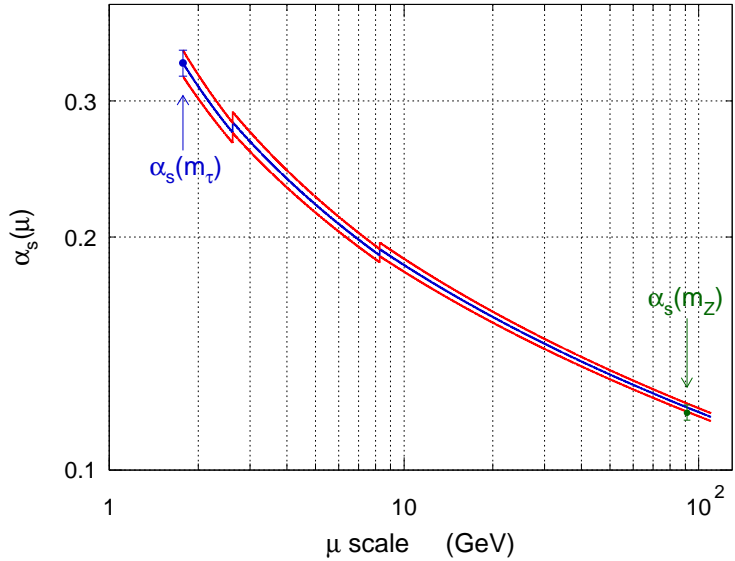,width=100mm}}
  \caption[.]{\label{evol_mz}
              The evolution of the measured \asm\ to higher scales $\mu$
              using the 4-loop RGE and the 3-loop matching conditions
              applied at the heavy quark-pair thresholds (hence the 
              discontinuities at 2~$m_c$ and 2~$m_b$). The evolution of
              the ALEPH \asm\ determination is in agreement with the direct
              value obtained from the Z width through the global electroweak
              fit~\cite{ewfit}.}
\end{figure}

The $\alpha_s(M_Z^2)$ result from the present analysis of $\tau$ 
decays is in agreement with the determination from the $Z$ width as 
obtained by the global fit to all electroweak data~\cite{ewfit}:
\beq
\label{eq_ewfit}
   \alpha_s(M_Z^2)_{\rm Z~width} = 0.1186 \pm 0.0027~,
\eeq
as demonstrated in Fig.~\ref{evol_mz}.
Both determinations have somewhat comparable accuracies, 
which are however very different in nature. 
On one hand, the $\tau$ value is dominated by 
theoretical uncertainties. On the other hand, the determination 
at the $Z$ resonance, benefiting from the much larger energy scale 
and the correspondingly small uncertainties from the truncated 
perturbative expansion, is limited by the experimental precision 
on the electroweak observables, essentially the ratio of leptonic 
to hadronic peak cross sections. The consistency between the two results,
Eqs.~(\ref{eq_asres_mz}) and (\ref{eq_ewfit}), provides a powerful test,
over a range of $s$ spanning more than three orders of magnitude, 
of the evolution of the strong interaction coupling as predicted by the
nonabelian nature of the QCD gauge theory. It is therefore a precise test 
of asymptotic freedom at the 2.7\% level.

\section{Conclusions}

A final analysis of $\tau$ decay branching fractions has been presented
using all LEP 1 data collected with the ALEPH detector. As in the 
publication based on the 1991-1993 data set, it uses a global analysis 
of all modes, classified according to charged particle 
identification, and charged particle and $\pi^0$ multiplicity 
up to 4 $\piz$s in the final state. Major improvements 
are introduced with respect to the published analysis and a better 
understanding is achieved, in particular in the separation between 
genuine and fake photons. In this process shortcomings and small biases
of the previous method were discovered and corrected, leading to more 
robust results. As modes with kaons ($K^\pm$, $K^0_S$, and $K^0_L$) 
have already been studied and published with the full statistics, 
the non-strange modes without kaons are emphasized in this paper. 
Taken together these results provide a complete description of $\tau$ decays 
up to 6 hadrons in the final state. All decay channels with branching 
fractions larger than 0.1\% are considered in this analysis.

The measured branching fractions are internally consistent and agree with
known constraints from other measurements in the framework of the
Standard Model. More specifically,
the results on the leptonic and pionic fractions lead to powerful tests 
of universality in the charged leptonic weak current, showing that 
the $e-\mu-\tau$ couplings are equal within 0.3\%. The branching ratio 
of $\tau \rightarrow \nu_\tau \pi \piz$, which is of particular interest
to the accurate determination of vacuum polarization effects, is determined
with a precision of 0.5\% to be $(25.47 \pm 0.13)~\%$. Also the ratio of 
$a_1$ branching fractions into $\pi 2\piz$ and $3\pi$ final states is 
measured to be $0.979 \pm 0.018$, in agreement with expectation from partial
wave analyses of these decays. Separating nonstrange hadronic channels into
vector (V) and axial-vector (A) components and normalizing to the electronic
width yields the ratios $R_{\tau,V} = 1.787 \pm 0.013$, 
$R_{\tau,A} = 1.695 \pm 0.013$, $R_{\tau,V+A} = 3.482 \pm 0.014$ and 
$R_{\tau,V-A} = 0.092 \pm 0.023$.

Spectral functions for the hadronic modes are determined through the
unfolding of the measured invariant mass distributions, taking into account
detector effects. As for the branching fractions, the spectral functions
are separated into vector and axial-vector contributions. The vector
spectral functions are compared to the corresponding ones obtained through
$e^+e^-$ annihilation, providing accurate tests of isospin invariance of
the electroweak current. Both spectral functions are fitted within QCD
to perturbative and nonperturbative components. The latter are determined
to be very small, while the dominant perturbative part yields a precise
value for the strong coupling at the $\tau$ mass scale,
$\alpha_s(m^2_\tau) = 0.340 \pm (0.005)_{exp} \pm (0.014)_{th}$. The evolution
to the Z mass scale yields $\alpha_s(M_Z^2) = 0.1209 \pm 0.0018$. It
agrees well with the direct determination from the Z width and provides the 
most accurate test to date of asymptotic freedom in the QCD gauge theory.

\section*{Acknowledgements}

We wish to thank our colleagues from the accelerator divisions for the
successful operation of LEP. We are 
indebted to the engineers and technicians in all our institutions for 
their 
contribution to the excellent performance of ALEPH. Those of us from 
non-member countries thank CERN for its hospitality. It is a pleasure to
thank W.~Marciano and A.~Pich for clarifying many theoretical aspects.

\end{document}

%% file: authb_plus.tex
\pagestyle{empty}
\newpage
\small
%
%
\newlength{\saveparskip}
\newlength{\savetextheight}
\newlength{\savetopmargin}
\newlength{\savetextwidth}
\newlength{\saveoddsidemargin}
\newlength{\savetopsep}
\setlength{\saveparskip}{\parskip}
\setlength{\savetextheight}{\textheight}
\setlength{\savetopmargin}{\topmargin}
\setlength{\savetextwidth}{\textwidth}
\setlength{\saveoddsidemargin}{\oddsidemargin}
\setlength{\savetopsep}{\topsep}
%
%
\setlength{\parskip}{0.0cm}
\setlength{\textheight}{25.0cm}
\setlength{\topmargin}{-1.5cm}
\setlength{\textwidth}{16 cm}
\setlength{\oddsidemargin}{-0.0cm}
\setlength{\topsep}{1mm}
\pretolerance=10000
\centerline{\large\bf The ALEPH Collaboration}
\footnotesize
\vspace{0.5cm}
{\raggedbottom
\begin{sloppypar}
\samepage\noindent
S.~Schael
\nopagebreak
\begin{center}
\parbox{15.5cm}{\sl\samepage
Physikalisches Institut das RWTH-Aachen, D-52056 Aachen, Germany}
\end{center}\end{sloppypar}
\vspace{2mm}
\begin{sloppypar}
\noindent
R.~Barate,
R.~Bruneli\`ere,
I.~De~Bonis,
D.~Decamp,
C.~Goy,
S.~J\'ez\'equel,
J.-P.~Lees,
F.~Martin,
E.~Merle,
\mbox{M.-N.~Minard},
B.~Pietrzyk,
B.~Trocm\'e
\nopagebreak
\begin{center}
\parbox{15.5cm}{\sl\samepage
Laboratoire de Physique des Particules (LAPP), IN$^{2}$P$^{3}$-CNRS,
F-74019 Annecy-le-Vieux Cedex, France}
\end{center}\end{sloppypar}
\vspace{2mm}
\begin{sloppypar}
\noindent
S.~Bravo,
M.P.~Casado,
M.~Chmeissani,
J.M.~Crespo,
E.~Fernandez,
M.~Fernandez-Bosman,
Ll.~Garrido,$^{15}$
M.~Martinez,
A.~Pacheco,
H.~Ruiz
\nopagebreak
\begin{center}
\parbox{15.5cm}{\sl\samepage
Institut de F\'{i}sica d'Altes Energies, Universitat Aut\`{o}noma
de Barcelona, E-08193 Bellaterra (Barcelona), Spain$^{7}$}
\end{center}\end{sloppypar}
\vspace{2mm}
\begin{sloppypar}
\noindent
A.~Colaleo,
D.~Creanza,
N.~De~Filippis,
M.~de~Palma,
G.~Iaselli,
G.~Maggi,
M.~Maggi,
S.~Nuzzo,
A.~Ranieri,
G.~Raso,$^{24}$
F.~Ruggieri,
G.~Selvaggi,
L.~Silvestris,
P.~Tempesta,
A.~Tricomi,$^{3}$
G.~Zito
\nopagebreak
\begin{center}
\parbox{15.5cm}{\sl\samepage
Dipartimento di Fisica, INFN Sezione di Bari, I-70126 Bari, Italy}
\end{center}\end{sloppypar}
\vspace{2mm}
\begin{sloppypar}
\noindent
X.~Huang,
J.~Lin,
Q. Ouyang,
T.~Wang,
Y.~Xie,
R.~Xu,
S.~Xue,
J.~Zhang,
L.~Zhang,
W.~Zhao
\nopagebreak
\begin{center}
\parbox{15.5cm}{\sl\samepage
Institute of High Energy Physics, Academia Sinica, Beijing, The People's
Republic of China$^{8}$}
\end{center}\end{sloppypar}
\vspace{2mm}
\begin{sloppypar}
\noindent
D.~Abbaneo,
T.~Barklow,$^{26}$
O.~Buchm\"uller,$^{26}$
M.~Cattaneo,
B.~Clerbaux,$^{23}$
H.~Drevermann,
R.W.~Forty,
M.~Frank,
F.~Gianotti,
J.B.~Hansen,
J.~Harvey,
D.E.~Hutchcroft,$^{30}$,
P.~Janot,
B.~Jost,
M.~Kado,$^{2}$
P.~Mato,
A.~Moutoussi,
F.~Ranjard,
L.~Rolandi,
D.~Schlatter,
F.~Teubert,
A.~Valassi,
I.~Videau
\nopagebreak
\begin{center}
\parbox{15.5cm}{\sl\samepage
European Laboratory for Particle Physics (CERN), CH-1211 Geneva 23,
Switzerland}
\end{center}\end{sloppypar}
\vspace{2mm}
\begin{sloppypar}
\noindent
F.~Badaud,
S.~Dessagne,
A.~Falvard,$^{20}$
D.~Fayolle,
P.~Gay,
J.~Jousset,
B.~Michel,
S.~Monteil,
D.~Pallin,
J.M.~Pascolo,
P.~Perret
\nopagebreak
\begin{center}
\parbox{15.5cm}{\sl\samepage
Laboratoire de Physique Corpusculaire, Universit\'e Blaise Pascal,
IN$^{2}$P$^{3}$-CNRS, Clermont-Ferrand, F-63177 Aubi\`{e}re, France}
\end{center}\end{sloppypar}
\vspace{2mm}
\begin{sloppypar}
\noindent
J.D.~Hansen,
J.R.~Hansen,
P.H.~Hansen,
A.C.~Kraan,
B.S.~Nilsson
\nopagebreak
\begin{center}
\parbox{15.5cm}{\sl\samepage
Niels Bohr Institute, 2100 Copenhagen, DK-Denmark$^{9}$}
\end{center}\end{sloppypar}
\vspace{2mm}
\begin{sloppypar}
\noindent
A.~Kyriakis,
C.~Markou,
E.~Simopoulou,
A.~Vayaki,
K.~Zachariadou
\nopagebreak
\begin{center}
\parbox{15.5cm}{\sl\samepage
Nuclear Research Center Demokritos (NRCD), GR-15310 Attiki, Greece}
\end{center}\end{sloppypar}
\vspace{2mm}
\begin{sloppypar}
\noindent
A.~Blondel,$^{12}$
\mbox{J.-C.~Brient},
F.~Machefert,
A.~Roug\'{e},
H.~Videau
\nopagebreak
\begin{center}
\parbox{15.5cm}{\sl\samepage
Laoratoire Leprince-Ringuet, Ecole
Polytechnique, IN$^{2}$P$^{3}$-CNRS, \mbox{F-91128} Palaiseau Cedex, France}
\end{center}\end{sloppypar}
\vspace{2mm}
\begin{sloppypar}
\noindent
V.~Ciulli,
E.~Focardi,
G.~Parrini
\nopagebreak
\begin{center}
\parbox{15.5cm}{\sl\samepage
Dipartimento di Fisica, Universit\`a di Firenze, INFN Sezione di Firenze,
I-50125 Firenze, Italy}
\end{center}\end{sloppypar}
\vspace{2mm}
\begin{sloppypar}
\noindent
A.~Antonelli,
M.~Antonelli,
G.~Bencivenni,
F.~Bossi,
G.~Capon,
F.~Cerutti,
V.~Chiarella,
P.~Laurelli,
G.~Mannocchi,$^{5}$
G.P.~Murtas,
L.~Passalacqua
\nopagebreak
\begin{center}
\parbox{15.5cm}{\sl\samepage
Laboratori Nazionali dell'INFN (LNF-INFN), I-00044 Frascati, Italy}
\end{center}\end{sloppypar}
\vspace{2mm}
\begin{sloppypar}
\noindent
J.~Kennedy,
J.G.~Lynch,
P.~Negus,
V.~O'Shea,
A.S.~Thompson
\nopagebreak
\begin{center}
\parbox{15.5cm}{\sl\samepage
Department of Physics and Astronomy, University of Glasgow, Glasgow G12
8QQ,United Kingdom$^{10}$}
\end{center}\end{sloppypar}
\vspace{2mm}
\begin{sloppypar}
\noindent
S.~Wasserbaech
\nopagebreak
\begin{center}
\parbox{15.5cm}{\sl\samepage
Utah Valley State College, Orem, UT 84058, U.S.A.}
\end{center}\end{sloppypar}
\vspace{2mm}
\begin{sloppypar}
\noindent
R.~Cavanaugh,$^{4}$
S.~Dhamotharan,$^{21}$
C.~Geweniger,
P.~Hanke,
V.~Hepp,
E.E.~Kluge,
A.~Putzer,
H.~Stenzel,
K.~Tittel,
M.~Wunsch$^{19}$
\nopagebreak
\begin{center}
\parbox{15.5cm}{\sl\samepage
Kirchhoff-Institut f\"ur Physik, Universit\"at Heidelberg, D-69120
Heidelberg, Germany$^{16}$}
\end{center}\end{sloppypar}
\vspace{2mm}
\begin{sloppypar}
\noindent
R.~Beuselinck,
W.~Cameron,
G.~Davies,
P.J.~Dornan,
M.~Girone,$^{1}$
N.~Marinelli,
J.~Nowell,
S.A.~Rutherford,
J.K.~Sedgbeer,
J.C.~Thompson,$^{14}$
R.~White
\nopagebreak
\begin{center}
\parbox{15.5cm}{\sl\samepage
Department of Physics, Imperial College, London SW7 2BZ,
United Kingdom$^{10}$}
\end{center}\end{sloppypar}
\vspace{2mm}
\begin{sloppypar}
\noindent
V.M.~Ghete,
P.~Girtler,
E.~Kneringer,
D.~Kuhn,
G.~Rudolph
\nopagebreak
\begin{center}
\parbox{15.5cm}{\sl\samepage
Institut f\"ur Experimentalphysik, Universit\"at Innsbruck, A-6020
Innsbruck, Austria$^{18}$}
\end{center}\end{sloppypar}
\vspace{2mm}
\begin{sloppypar}
\noindent
E.~Bouhova-Thacker,
C.K.~Bowdery,
D.P.~Clarke,
G.~Ellis,
A.J.~Finch,
F.~Foster,
G.~Hughes,
R.W.L.~Jones,
M.R.~Pearson,
N.A.~Robertson,
M.~Smizanska
\nopagebreak
\begin{center}
\parbox{15.5cm}{\sl\samepage
Department of Physics, University of Lancaster, Lancaster LA1 4YB,
United Kingdom$^{10}$}
\end{center}\end{sloppypar}
\vspace{2mm}
\begin{sloppypar}
\noindent
O.~van~der~Aa,
C.~Delaere,$^{28}$
G.Leibenguth,$^{31}$
V.~Lemaitre$^{29}$
\nopagebreak
\begin{center}
\parbox{15.5cm}{\sl\samepage
Institut de Physique Nucl\'eaire, D\'epartement de Physique, Universit\'e Catholique de Louvain, 1348 Louvain-la-Neuve, Belgium}
\end{center}\end{sloppypar}
\vspace{2mm}
\begin{sloppypar}
\noindent
U.~Blumenschein,
F.~H\"olldorfer,
K.~Jakobs,
F.~Kayser,
A.-S.~M\"uller,
B.~Renk,
H.-G.~Sander,
S.~Schmeling,
H.~Wachsmuth,
C.~Zeitnitz,
T.~Ziegler
\nopagebreak
\begin{center}
\parbox{15.5cm}{\sl\samepage
Institut f\"ur Physik, Universit\"at Mainz, D-55099 Mainz, Germany$^{16}$}
\end{center}\end{sloppypar}
\vspace{2mm}
\begin{sloppypar}
\noindent
A.~Bonissent,
P.~Coyle,
C.~Curtil,
A.~Ealet,
D.~Fouchez,
P.~Payre,
A.~Tilquin
\nopagebreak
\begin{center}
\parbox{15.5cm}{\sl\samepage
Centre de Physique des Particules de Marseille, Univ M\'editerran\'ee,
IN$^{2}$P$^{3}$-CNRS, F-13288 Marseille, France}
\end{center}\end{sloppypar}
\vspace{2mm}
\begin{sloppypar}
\noindent
F.~Ragusa
\nopagebreak
\begin{center}
\parbox{15.5cm}{\sl\samepage
Dipartimento di Fisica, Universit\`a di Milano e INFN Sezione di
Milano, I-20133 Milano, Italy.}
\end{center}\end{sloppypar}
\vspace{2mm}
\begin{sloppypar}
\noindent
A.~David,
H.~Dietl,$^{32}$
G.~Ganis,$^{27}$
K.~H\"uttmann,
G.~L\"utjens,
W.~M\"anner$^{32}$,
\mbox{H.-G.~Moser},
R.~Settles,
M.~Villegas,
G.~Wolf
\nopagebreak
\begin{center}
\parbox{15.5cm}{\sl\samepage
Max-Planck-Institut f\"ur Physik, Werner-Heisenberg-Institut,
D-80805 M\"unchen, Germany\footnotemark[16]}
\end{center}\end{sloppypar}
\vspace{2mm}
\begin{sloppypar}
\noindent
J.~Boucrot,
O.~Callot,
M.~Davier,
L.~Duflot,
\mbox{J.-F.~Grivaz},
Ph.~Heusse,
A.~H\"ocker,
A.~Jacholkowska,$^{6}$
L.~Serin,
\mbox{J.-J.~Veillet},
C.~Z.~Yuan,
Z.~Q.~Zhang
\nopagebreak
\begin{center}
\parbox{15.5cm}{\sl\samepage
Laboratoire de l'Acc\'el\'erateur Lin\'eaire, Universit\'e de Paris-Sud,
IN$^{2}$P$^{3}$-CNRS, F-91898 Orsay Cedex, France}
\end{center}\end{sloppypar}
\vspace{2mm}
\begin{sloppypar}
\noindent
P.~Azzurri, 
G.~Bagliesi,
T.~Boccali,
L.~Fo\`a,
A.~Giammanco,
A.~Giassi,
F.~Ligabue,
A.~Messineo,
F.~Palla,
G.~Sanguinetti,
A.~Sciab\`a,
G.~Sguazzoni,
P.~Spagnolo,
R.~Tenchini,
A.~Venturi,
P.G.~Verdini
\samepage
\begin{center}
\parbox{15.5cm}{\sl\samepage
Dipartimento di Fisica dell'Universit\`a, INFN Sezione di Pisa,
e Scuola Normale Superiore, I-56010 Pisa, Italy}
\end{center}\end{sloppypar}
\vspace{2mm}
\begin{sloppypar}
\noindent
O.~Awunor,
G.A.~Blair,
G.~Cowan,
A.~Garcia-Bellido,
M.G.~Green,
T.~Medcalf,
A.~Misiejuk,
J.A.~Strong,
P.~Teixeira-Dias
\nopagebreak
\begin{center}
\parbox{15.5cm}{\sl\samepage
Department of Physics, Royal Holloway \& Bedford New College,
University of London, Egham, Surrey TW20 OEX, United Kingdom$^{10}$}
\end{center}\end{sloppypar}
\vspace{2mm}
\begin{sloppypar}
\noindent
R.W.~Clifft,
T.R.~Edgecock,
P.R.~Norton,
I.R.~Tomalin,
J.J.~Ward
\nopagebreak
\begin{center}
\parbox{15.5cm}{\sl\samepage
Particle Physics Dept., Rutherford Appleton Laboratory,
Chilton, Didcot, Oxon OX11 OQX, United Kingdom$^{10}$}
\end{center}\end{sloppypar}
\vspace{2mm}
\begin{sloppypar}
\noindent
\mbox{B.~Bloch-Devaux},
D.~Boumediene,
P.~Colas,
B.~Fabbro,
E.~Lan\c{c}on,
\mbox{M.-C.~Lemaire},
E.~Locci,
P.~Perez,
J.~Rander,
B.~Tuchming,
B.~Vallage
\nopagebreak
\begin{center}
\parbox{15.5cm}{\sl\samepage
CEA, DAPNIA/Service de Physique des Particules,
CE-Saclay, F-91191 Gif-sur-Yvette Cedex, France$^{17}$}
\end{center}\end{sloppypar}
\vspace{2mm}
\begin{sloppypar}
\noindent
A.M.~Litke,
G.~Taylor
\nopagebreak
\begin{center}
\parbox{15.5cm}{\sl\samepage
Institute for Particle Physics, University of California at
Santa Cruz, Santa Cruz, CA 95064, USA$^{22}$}
\end{center}\end{sloppypar}
\vspace{2mm}
\begin{sloppypar}
\noindent
C.N.~Booth,
S.~Cartwright,
F.~Combley,$^{25}$
P.N.~Hodgson,
M.~Lehto,
L.F.~Thompson
\nopagebreak
\begin{center}
\parbox{15.5cm}{\sl\samepage
Department of Physics, University of Sheffield, Sheffield S3 7RH,
United Kingdom$^{10}$}
\end{center}\end{sloppypar}
\vspace{2mm}
\begin{sloppypar}
\noindent
A.~B\"ohrer,
S.~Brandt,
C.~Grupen,
J.~Hess,
A.~Ngac,
G.~Prange
\nopagebreak
\begin{center}
\parbox{15.5cm}{\sl\samepage
Fachbereich Physik, Universit\"at Siegen, D-57068 Siegen, Germany$^{16}$}
\end{center}\end{sloppypar}
\vspace{2mm}
\begin{sloppypar}
\noindent
C.~Borean,
G.~Giannini
\nopagebreak
\begin{center}
\parbox{15.5cm}{\sl\samepage
Dipartimento di Fisica, Universit\`a di Trieste e INFN Sezione di Trieste,
I-34127 Trieste, Italy}
\end{center}\end{sloppypar}
\vspace{2mm}
\begin{sloppypar}
\noindent
H.~He,
J.~Putz,
J.~Rothberg
\nopagebreak
\begin{center}
\parbox{15.5cm}{\sl\samepage
Experimental Elementary Particle Physics, University of Washington, Seattle,
WA 98195 U.S.A.}
\end{center}\end{sloppypar}
\vspace{2mm}
\begin{sloppypar}
\noindent
S.R.~Armstrong,
K.~Berkelman,
K.~Cranmer,
D.P.S.~Ferguson,
Y.~Gao,$^{13}$
S.~Gonz\'{a}lez,
O.J.~Hayes,
H.~Hu,
S.~Jin,
J.~Kile,
P.A.~McNamara III,
J.~Nielsen,
Y.B.~Pan,
\mbox{J.H.~von~Wimmersperg-Toeller}, 
W.~Wiedenmann,
J.~Wu,
Sau~Lan~Wu,
X.~Wu,
G.~Zobernig
\nopagebreak
\begin{center}
\parbox{15.5cm}{\sl\samepage
Department of Physics, University of Wisconsin, Madison, WI 53706,
USA$^{11}$}
\end{center}\end{sloppypar}
\vspace{2mm}
\begin{sloppypar}
\noindent
G.~Dissertori
\nopagebreak
\begin{center}
\parbox{15.5cm}{\sl\samepage
Institute for Particle Physics, ETH H\"onggerberg, 8093 Z\"urich,
Switzerland.}
\end{center}\end{sloppypar}
}
\footnotetext[1]{Also at CERN, 1211 Geneva 23, Switzerland.}
\footnotetext[2]{Now at Fermilab, PO Box 500, MS 352, Batavia, IL 60510, USA}
\footnotetext[3]{Also at Dipartimento di Fisica di Catania and INFN Sezione di
 Catania, 95129 Catania, Italy.}
\footnotetext[4]{Now at University of Florida, Department of Physics, Gainesville, Florida 32611-8440, USA}
\footnotetext[5]{Also IFSI sezione di Torino, INAF, Italy.}
\footnotetext[6]{Also at Groupe d'Astroparticules de Montpellier, Universit\'{e} de Montpellier II, 34095, Montpellier, France.}
\footnotetext[7]{Supported by CICYT, Spain.}
\footnotetext[8]{Supported by the National Science Foundation of China.}
\footnotetext[9]{Supported by the Danish Natural Science Research Council.}
\footnotetext[10]{Supported by the UK Particle Physics and Astronomy Research
Council.}
\footnotetext[11]{Supported by the US Department of Energy, grant
DE-FG0295-ER40896.}
\footnotetext[12]{Now at Departement de Physique Corpusculaire, Universit\'e de
Gen\`eve, 1211 Gen\`eve 4, Switzerland.}
\footnotetext[13]{Also at Department of Physics, Tsinghua University, Beijing, The People's Republic of China.}
\footnotetext[14]{Supported by the Leverhulme Trust.}
\footnotetext[15]{Permanent address: Universitat de Barcelona, 08208 Barcelona,
Spain.}
\footnotetext[16]{Supported by Bundesministerium f\"ur Bildung
und Forschung, Germany.}
\footnotetext[17]{Supported by the Direction des Sciences de la
Mati\`ere, C.E.A.}
\footnotetext[18]{Supported by the Austrian Ministry for Science and Transport.}
\footnotetext[19]{Now at SAP AG, 69185 Walldorf, Germany}
\footnotetext[20]{Now at Groupe d' Astroparticules de Montpellier, Universit\'e de Montpellier II, 34095 Montpellier, France.}
\footnotetext[21]{Now at BNP Paribas, 60325 Frankfurt am Mainz, Germany}
\footnotetext[22]{Supported by the US Department of Energy,
grant DE-FG03-92ER40689.}
\footnotetext[23]{Now at Institut Inter-universitaire des hautes Energies (IIHE), CP 230, Universit\'{e} Libre de Bruxelles, 1050 Bruxelles, Belgique}
\footnotetext[24]{Now at Dipartimento di Fisica e Tecnologie Relative, Universit\`a di Palermo, Palermo, Italy.}
\footnotetext[25]{Deceased.}
\footnotetext[26]{Now at SLAC, Stanford, CA 94309, U.S.A}
\footnotetext[27]{Now at CERN, 1211 Geneva 23, Switzerland}
\footnotetext[28]{Research Fellow of the Belgium FNRS}
\footnotetext[29]{Research Associate of the Belgium FNRS} 
\footnotetext[30]{Now at Liverpool University, Liverpool L69 7ZE, United Kingdom} 
\footnotetext[31]{Supported by the Federal Office for Scientific, Technical and Cultural Affairs through
the Interuniversity Attraction Pole P5/27} 
\footnotetext[32]{Now at Henryk Niewodnicznski Institute of Nuclear Physics, Polish Academy of Sciences, Cracow, Poland}   
\setlength{\parskip}{\saveparskip}
\setlength{\textheight}{\savetextheight}
\setlength{\topmargin}{\savetopmargin}
\setlength{\textwidth}{\savetextwidth}
\setlength{\oddsidemargin}{\saveoddsidemargin}
\setlength{\topsep}{\savetopsep}
\normalsize
\newpage
\pagestyle{plain}
\setcounter{page}{1}

%% file: brpap.bbl
\begin{thebibliography}{***}

\bibitem{aleph01}    ALEPH Coll., \ZP\ {\bf C54} (1992) 211.
\bibitem{aleph13_l}  ALEPH Coll., \ZP\ {\bf C70} (1996) 561.
\bibitem{aleph13_h}  ALEPH Coll., \ZP\ {\bf C70} (1996) 579.
\bibitem{alephk1a}   ALEPH Coll., \PL\ {\bf B332} (1994) 209.
\bibitem{alephk1b}   ALEPH Coll., \PL\ {\bf B332} (1994) 219.
\bibitem{alephk3}    ALEPH Coll., \EPJ\ {\bf C1} (1998) 65.
\bibitem{alephks}    ALEPH Coll., \EPJ\ {\bf C4} (1998) 29.
\bibitem{alephkl}    ALEPH Coll., \EPJ\ {\bf C10} (1999) 1.
\bibitem{alephksum}  ALEPH Coll., \EPJ\ {\bf C11} (1999) 599.
\bibitem{alephsf1}   ALEPH Coll., \PL\ {\bf B307} (1993) 209.
\bibitem{alephvsf}   ALEPH Coll., \ZP\ {\bf C76} (1997) 15.
\bibitem{alephasf}   ALEPH Coll., \EPJ\ {\bf C4} (1998) 409.
\bibitem{alepheta}   ALEPH Coll., \ZP\ {\bf C74} (1997) 263.
\bibitem{alephdet}   ALEPH Coll., \NIM\ {\bf A294} (1990) 127.
\bibitem{alephperf}  ALEPH Coll., \NIM\ {\bf A360} (1995) 481.
\bibitem{alephpol}   ALEPH Coll., \EPJ\ {\bf C20} (2001) 401.
\bibitem{was}        S. Jadach, B.F.L. Ward, and Z. Was, 
                     \CPC\ {\bf 79} (1994) 503.
\bibitem{tauola}     S. Jadach {\it et al.},
                     \CPC\ {\bf 76} (1993) 361.
\bibitem{aleph94}    ALEPH Coll., \ZP\ {\bf C62} (1994) 539.
\bibitem{unibab}     H.~Anlauf \ea, \CPC\ {\bf 79} (1994) 466.
\bibitem{lund}       T.~Sj\"ostrand, \CPC\ {\bf 39} (1986) 347.
\bibitem{photos}     E.~Barberio, B.~van~Eijk and Z.~Was, 
                     \CPC\ {\bf 66} (1991) 115; E.~Barberio and Z.~Was,
                     \CPC\ {\bf 79} (1994) 291.
\bibitem{aleph_undetect}   S.~Snow, Proceedings of the $2^{\rm nd}$
                        International Workshop on $\tau$ Lepton Physics, 
                        Columbus 1992, K.~K.~Gan ed., World Scientific (1993).
\bibitem{ecker}      V.~Cirigliano, G.~Ecker and H.~Neufeld,
                      \PL\ {\bf B513} (2001) 361;
	              {\em JHEP} {\bf 0208} (2002) 002.
\bibitem{pdg2004}    Review of Particle Physics, S.~Eidelman {\it et al.}, 
                      \PL\ {\bf B592} (2004) 1.
\bibitem{cleoomega}  D.~Bortoletto \ea, CLEO Coll., \PRL\ {\bf 71} (1993) 1791.
\bibitem{cleoeta3pi} T.~Bergfeld \ea, CLEO Coll., \PRL\ {\bf 79} (1997) 2406;
                     A.~Weinstein, Proceedings of the $6^{\rm th}$
                     International Workshop on $\tau$ Lepton Physics, 
                     Victoria 2000, R.~J.~Sobie and J.~M.~Roney eds., 
                     North Holland (2001).
\bibitem{zielinski}  M.~Zielinski \ea, \PRL\ {\bf 52} (1984) 1195.
\bibitem{cleo6pi}    A.~Anastassov \ea, CLEO Coll., \PRL\ {\bf 86} (2001) 4467.
\bibitem{argus_be}   H.~Albrecht \ea, ARGUS Coll., \PL\ {\bf B316} (1993) 608.
\bibitem{cleo_be}    A.~Anastassov \ea, CLEO Coll., \PR\ {\bf D55} (1997) 2559.
\bibitem{delphi_be}  P.~Abreu \ea, DELPHI Coll., \EPJ\ {\bf C10} (1999) 201.
\bibitem{opal_be}    G.~Abbiendi \ea, OPAL Coll., \PL\ {\bf B447} (1999) 134.
\bibitem{l3_be}      M.~Acciarri \ea, L3 Coll., \PL\ {\bf B507} (2001) 47.
\bibitem{opal_bmu}   G.~Abbiendi \ea, OPAL Coll., \PL\ {\bf B551} (2003) 35.
\bibitem{opal_bh}    K.~Ackerstaff \ea, OPAL Coll., \EPJ\ {\bf C4} (1998) 193.
\bibitem{cleo_bhpi0} M.~Artuso \ea, CLEO Coll., \PRL\ {\bf 72} (1994) 3762.
\bibitem{cleo_bh2pi0} M.~Procario \ea, CLEO Coll., \PRL\ {\bf 70} (1993) 1207.
\bibitem{cleo_b3h}   R.~Balest \ea, CLEO Coll., \PRL\ {\bf 75} (1995) 3809.
\bibitem{opal_b3h}   R.~Akers \ea, OPAL Coll., \ZP\ {\bf C68} (1995) 555.
\bibitem{cleo_b3h2-3pi0} D.~Bortoletto \ea, CLEO Coll., 
                     \PRL\ {\bf 71} (1993) 3762;
                     A.~Anastassov \ea, CLEO Coll., \PRL\ {\bf 86} (2001) 4467.
\bibitem{hrs_b5h}    B.~Bylsma \ea, HRS Coll., \PR\ {\bf D35} (1987) 2269.
\bibitem{argus_b5h}  H.~Albrecht \ea, ARGUS Coll., \PL\ {\bf B202} (1988) 149.
\bibitem{cleo_b5h}   O.~Gibaut \ea, CLEO Coll., \PRL\ {\bf 73} (1984) 934.
\bibitem{opal_b5h}   K.~Ackerstaff \ea, OPAL Coll., \EPJ\ {\bf C8} (1999) 183.
\bibitem{babar_b5h}  B.~Aubert \ea, BaBar Coll., hep-ex/0505004
\bibitem{cleo_b5hpi0}A.~Anastassov \ea, CLEO Coll., \PRL\ {\bf 86} (2001) 4467.
\bibitem{delphitopol}  P.~Abreu \ea, DELPHI Coll., \EPJ\ {\bf C20} (2001) 617.
\bibitem{cleo7pr}    K.~Edwards \ea, CLEO Coll., \PR\ {\bf D56} (1997) 5297.
\bibitem{marciano-sirlin}  W.~Marciano and A.~Sirlin, \PRL\
                           {\bf 61} (1988) 1815.
\bibitem{triumf}     D.I.~Britton \ea, \PRL\ {\bf 68} (1992) 3000.
\bibitem{psi}        C.~Czapek \ea, \PRL\ {\bf 70} (1993) 17.
\bibitem{decker-fink}  R.~Decker and M.~Finkemeier, \PR\ {\bf D48} (1993) 4203.
\bibitem{adh}        R.~Alemany, M.~Davier and A.~H\"ocker,
                     \EPJ\ {\bf C2} (1998) 123.
\bibitem{dh97a}      M.~Davier and A.~H\"ocker, \PL\ {\bf B419} (1998) 419.
\bibitem{dh97b}      M.~Davier and A.~H\"ocker, \PL\ {\bf B435} (1998) 427.
\bibitem{dehz}       M.~Davier, S.~Eidelman, A.~H\"ocker and Z.~Zhang,
                     \EPJ\ {\bf C27} (2003) 497.
\bibitem{cleo_2pi}   S.~Anderson \ea, CLEO Coll., \PR\ {\bf D61} (2000) 112002.
\bibitem{cmd2_old}   R.R. Akhmetshin \ea, CMD-2 Coll., 
                     \PL\ {\bf B527} (2002) 161.
\bibitem{cmd2_new}   R.R. Akhmetshin \ea, CMD-2 Coll., 
                     \PL\ {\bf B578} (2004) 285.
\bibitem{dehz03}     M.~Davier, S.~Eidelman, A.~H\"ocker and Z.~Zhang,
                     \EPJ\ {\bf C31} (2003) 503.
\bibitem{bnlg-2}     G.~W.~Bennett, {\it et al.}, Muon $(g-2)$ Coll., 
                     \PRL\ {\bf 92} (2004) 161802.
\bibitem{cleoa1}     D.~Asner \ea, CLEO Coll., \PR\ {\bf D61} (2000) 012002.
\bibitem{cleoksteta} M.~Bishai \ea, CLEO Coll., \PRL\ {\bf 82} (1999) 281.
\bibitem{cleo_kkpi}  T.~E.~Coan \ea, CLEO Coll., \PRL\ {\bf 92} (2004) 232001.
\bibitem{besmtau}    J.~Z.~Bai \ea, \PRL\ {\bf 88} (2002) 101802.
\bibitem{bnp}        E.~Braaten, S.~Narison and A.~Pich, 
                     \NP\ {\bf B373} (1992) 581.
\bibitem{pichtau94}  A.~Pich, \NP\ {\bf B39} (Proc. Suppl.) (1995) 326.
\bibitem{unfold}     A.~H\"ocker and V.~Kartvelishvili, 
                     \NIM\ {\bf A372} (1996) 469.
\bibitem{opal_2pi}   K. Ackerstaff \ea, OPAL Coll., \EPJ\ {\bf C7} (1999) 571.
\bibitem{3loop}      L.R.~Surguladze and M.A.~Samuel, 
                     \PRL\ {\bf 66} (1991) 560; \\
                     S.G.~Gorishny, A.L.~Kataev and S.A.~Larin,
                     \PL\ {\bf B259} (1991) 144.
\bibitem{svz}        M.A.~Shifman, A.L.~Vainshtein and V.I.~Zakharov,
                     \NP\ {\bf B147} (1979) 385; 
                     {\it ibid} {\bf B147} (1979) 448; 
                     {\it ibid} {\bf B147} (1979) 519.
\bibitem{braaten}    E.~Braaten and C.S.~Li, \PR\ {\bf D42} (1990) 3888.
\bibitem{czyz}       H.~Czy\.{z} and J.~H.~K\"uhn, \EPJ\ {\bf C18} (2001) 497.
\bibitem{bijnens}    J.~Bijnens and P.~Gosdzinsky, \PL\ {\bf B388} (1996) 203.
\bibitem{tof_2pi}    I.B.~Vasserman \ea, TOF Coll.,
                     {\it Sov. J. Nucl. Phys.} {\bf 33} (1981) 709.
\bibitem{olya_2pi}   I.B.~Vasserman \ea, OLYA Coll.,
                     {\it Sov. J. Nucl. Phys.} {\bf 30} (1979) 519.
\bibitem{cmd_2pi}    L.M.~Barkov \ea, OLYA and CMD Coll.,
                     \NP\ {\bf B256} (1985) 365.
\bibitem{dm1_2pi}    A.~Quenzer \ea, DM1 Coll., \PL\ {\bf B76} (1987) 512.
\bibitem{dm2_2pi}    D.~Bisello \ea, DM2 Coll., \PL\ {\bf B220} (1989) 321.
\bibitem{na7}      S.~R.~Amendolia \ea, NA7 Coll., \PL\ {\bf B138} (1984) 454.
\bibitem{kloe}       A.~Aloisio \ea, KLOE Coll., \PL\ {\bf B606} (2005) 12.
\bibitem{cleo_kpi0}  M.~Battle \ea, CLEO Coll., \PRL\ {\bf 73} (1994) 1079.
\bibitem{eidelman_tau04} S.~Eidelman, {\it Recent results from Novosibirsk},
                     International Workshop on $\tau$ Lepton Physics, 
                     Nara (Sept. 2004).
\bibitem{gounarissak}G.J.~Gounaris and J.J.~Sakurai,
                     \PRL\ {\bf 21} (1968) 244. 
\bibitem{pich-portoles} F.~Guerrero and A.~Pich, \PL\ {\bf B412} (1997) 382;
                     D.~G\'omez Gumm, A.~Pich and J.~Portol\'es,
                     \PR\ {\bf 62} (2000) 054014.
\bibitem{davier_pisa_g-2} M.~Davier, \NP\ (Proc. Suppl.) 
                     {\bf B 131} (2004) 192.
\bibitem{pichledib}  F.~Le Diberder and A.~Pich, \PL\ {\bf B289} (1992) 165.
\bibitem{cleo_as}    T.~Coan \ea, CLEO \Cl, \PL\ {\bf B356} (1995) 580.
\bibitem{braaten88}  E. Braaten, \PRL\ {\bf 60} (1988) 1606.

\bibitem{wilson}     K.G.~Wilson, \PR\ {\bf 179} (1969) 1499.
\bibitem{pert}        F.~Le Diberder and A.~Pich, \PL\ {\bf B286} (1992) 147.
\bibitem{pertmass}   K.G.~Chetyrkin and A.~Kwiatkowski, 
                     \ZP\ {\bf C59} (1993) 525.
\bibitem{k4_cal}     P.A.~Baikov, K.G.~Chetyrkin and J.H.~K\"uhn, 
                     \PR\ {\bf D67} (2003) 074026.
\bibitem{pms}        P.~M.~Stevenson, \PR\ {\bf D23} (1981) 2916.
\bibitem{k4_pms}     A.~L.~Kataev and V.~V.~Starshenko,
                     {\it Mod. Phys. Lett.} {\bf A10} (1995) 235.
\bibitem{k4_fld}    F.~Le Diberder, \NP\ {\bf B39} (Proc. Suppl.) (1995) 318.
\bibitem{rit}        S.A.~Larin, T.~van Ritbergen and J.A.M.~Vermaseren,
                     \PL\ {\bf B400} (1997) 379.
\bibitem{pivo}       A.A.~Pivovarov, 
                     {\it Sov. J. Nucl. Phys.} {\bf 54} (1991) 676,
                     \ZP\ {\bf C53} (1992) 461.
\bibitem{chet1}      K.G.~Chetyrkin, B.A.~Kniehl and M.~Steinhauser,
                     \PRL\ {\bf 79} (1997) 2184.
\bibitem{chet2}      K.G.~Chetyrkin, B.A.~Kniehl and M.~Steinhauser,
                     \NP\ {\bf B510} (1998) 61.
\bibitem{ioffe}      B.~L.~Ioffe and K.~N.~Zyablyuk,
                     \EPJ\ {\bf C27} (2003) 229.
\bibitem{wetzel}     W.~Bernreuther, W.~Wetzel, \NP\ {\bf B197} (1982) 228; \\
                     W.~Wetzel, \NP\ {\bf B196} (1982) 259; \\
                     W.~Bernreuther, {\it Ann. Phys.} {\bf 151} (1983) 127.
\bibitem{pichsanta}  G.~Rodrigo, A.~Pich, A.~Santamaria, 
                     \PL\ {\bf B424} (1998) 367.
\bibitem{ewfit}       LEPEWWG/2004-01, hep-ex/0412015.

\end{thebibliography}
